%% file: MLreview-v3.tex
\documentclass[aps,rmp,reprint,amsmath,graphix,amssymb,longbibliography]{revtex4-1}

\pdfoutput=1

\usepackage{color,comment}
\usepackage{bbold} 

\usepackage[pdftex]{graphicx}
\graphicspath{{figures/},{./}}
\usepackage{subfigure}
\usepackage{hyperref}
\hypersetup{ 
	colorlinks   = true
}

\usepackage{bm}

\usepackage{listings}
\usepackage{textcomp}
\usepackage[space=true]{accsupp}

\newcommand{\pdfactualhex}[3]{\newcommand{#1}{%
		\BeginAccSupp{method=hex,ActualText=#2}#3\EndAccSupp{}}}

\pdfactualhex{\pdfactualdspace}{2020}{\textperiodcentered\textperiodcentered}
\pdfactualhex{\pdfactualsquote}{27}{'}
\pdfactualhex{\pdfactualbtick}{60}{`}

\definecolor{deepblue}{rgb}{0,0,0.8}
\definecolor{deepred}{rgb}{1.0,0,0}
\definecolor{deepgreen}{rgb}{0,0.7,0}
\definecolor{blueviolet}{RGB}{138,43,226}
\definecolor{darkyellow}{RGB}{204,204,0}
\definecolor{codegray}{rgb}{0.6,0.6,0.6}
\definecolor{weborange}{RGB}{255,165,0}
\definecolor{gold}{RGB}{255,205,0}
\definecolor{codegreen}{rgb}{0,0.6,0}
\definecolor{codepurple}{rgb}{0.58,0,0.82}

\definecolor{backcolour}{rgb}{0.95,0.95,0.92}

\lstdefinestyle{sublime}{
	backgroundcolor=\color{backcolour},   
	commentstyle=\color{deepgreen},
	keywordstyle=\color{deepred},
	numberstyle=\tiny,
	stringstyle=\color{weborange},
	basicstyle=\small\ttfamily, 
	breakatwhitespace=false,         
	breaklines=true,                 
	captionpos=t,                    
	keepspaces=true,                 
	numbers=left,                    
	numbersep=5pt,                  
	showspaces=false,                
	showstringspaces=false,
	showtabs=false,                  
	tabsize=4,
	columns=flexible,
	emptylines=10000,
	literate={'}{\pdfactualsquote}1{`}{\pdfactualbtick}1{\ \ }{\pdfactualdspace}2,
	inputpath=./code_snippets,
	keywords={lambda,xrange,abs,for,return},
}
\usepackage{marvosym,etoolbox}
\usepackage[T1]{fontenc}
\usepackage{graphicx}
\newcommand\0{\scalebox{-1}[1]{0}}
\let\svttfamily\ttfamily

\catcode`0=\active
\def0{\0}
\renewcommand\ttfamily{\svttfamily\catcode`0=\active }
\renewcommand\texttt{\bgroup\ttfamily\texttthelp}
\def\texttthelp#1{#1\egroup}
\catcode`0=12 %

\newcommand{\bdxi}{\boldsymbol{x}^{(i)}}
\newcommand{\mbbE}{\mathbb{E}}
\DeclareMathOperator*{\argmin}{arg\,min}
\DeclareMathOperator*{\argmax}{arg\,max}

\definecolor{shiningblue}{rgb}{0.3,0.68,0.89}
\definecolor{pastel_red}{rgb}{1,0.41,0.38}
\definecolor{pastel_green}{rgb}{0.18,0.65,0.34}
\definecolor{pastel_blue}{rgb}{0.47,0.7,0.9}
\definecolor{soft_blue}{rgb}{0,0,0}
\definecolor{deeplilac}{rgb}{0.6, 0.33, 0.73}

\newcommand \bea {\begin{eqnarray}} 
\newcommand \eea {\end{eqnarray}} 
\newcommand \be {\begin{equation}} 
\newcommand \ee {\end{equation}}

\newcommand{\bs} {\boldsymbol}
\newcommand{\bd} {\boldsymbol}
\newcommand{\mbf}{\mathbf}
\begin{document}

\title{A high-bias, low-variance introduction to Machine Learning for physicists}

\author{Pankaj Mehta}
\affiliation{Department of Physics, Boston University, Boston, MA 02215, USA}
\email{pankajm@bu.edu}
\author{Marin Bukov}
\affiliation{Department of Physics, University of California, Berkeley, CA 94720, USA}
\email{mgbukov@berkeley.edu}
\author{Ching-Hao Wang}
\affiliation{Department of Physics, Boston University, Boston, MA 02215, USA}
\author{Alexandre G.~R.~Day}
\affiliation{Department of Physics, Boston University, Boston, MA 02215, USA}
\author{Clint Richardson}
\affiliation{Department of Physics, Boston University, Boston, MA 02215, USA}
\author{Charles K.~Fisher}
\affiliation{Unlearn.AI, San Francisco, CA 94108}
\author{David J.~Schwab}
\affiliation{Initiative for the Theoretical Sciences, The Graduate Center,
City University of New York, 365 Fifth Ave., New York, NY 10016}

\date{\today{}}

\begin{abstract}
Machine Learning (ML) is one of the most exciting and dynamic areas of modern research and application. The purpose of this review is to provide an introduction to the core concepts and tools of machine learning in a manner easily understood and intuitive to physicists. The review begins by covering fundamental concepts in ML and modern statistics such as the bias-variance tradeoff, overfitting, regularization,  generalization, and gradient descent before moving on to more advanced topics in both supervised and unsupervised learning. Topics covered in the review include ensemble models, deep learning and neural networks, clustering and data visualization, energy-based models (including MaxEnt models and Restricted Boltzmann Machines), and variational methods. Throughout, we emphasize the many natural connections between ML and statistical physics. A notable aspect of the review is the use of \href{https://physics.bu.edu/~pankajm/MLnotebooks.html}{Python Jupyter notebooks} to introduce modern ML/statistical packages to readers using physics-inspired datasets (the Ising Model and Monte-Carlo simulations of supersymmetric decays of proton-proton collisions). We conclude with an extended outlook discussing possible uses of machine learning for furthering our understanding of the physical world as well as open problems in ML where physicists may be able to contribute.
\end{abstract}


\maketitle

\tableofcontents{}

\section{Introduction}
\label{sec:intro}
\input{sections/introduction.tex}

\section{Why is Machine Learning difficult?}
\label{sec:why_is_ML_difficult}
\input{sections/MLisdifficult.tex}

\section{Basics of Statistical Learning Theory}
\label{sec:stat_learn_theory}
\input{sections/StatisticalLearningTheory}

\section{Gradient Descent and its Generalizations}\label{sec:gradient_descent}
\input{sections/gd.tex}

\section{Overview of Bayesian Inference}
\label{sec:bayesian_inference}
\input{sections/BayesianInference}

\section{Linear Regression}
\label{sec:lin_reg}
\input{sections/lin_reg.tex}

\section{Logistic Regression}
\label{sec:log_reg}
\input{sections/log_reg.tex}

\section{Combining Models}
\label{sec:combining}
\input{sections/ensemble.tex}

\input{sections/DNNs.tex}

\section{Dimensional Reduction and Data Visualization}
\label{sec:dim-red}
\input{sections/dim_reduce.tex}

\section{Clustering}
\label{sec:clustering}
\input{sections/Clustering.tex}

\section{Variational Methods and Mean-Field Theory (MFT)}
\label{sec:varl_MFT}
\input{sections/Variational.tex}

\input{sections/energyPM.tex}

\section{Variational AutoEncoders (VAEs) and Generative Adversarial Networks (GANs)}
\label{sec:vae}
\input{sections/VAE_PM.tex}

\section{Outlook}
\label{sec:outlook}
\input{sections/outlook.tex}

\section{Acknowledgments}

PM and DJS would like to thank Anirvan Sengupta, Justin Kinney, and Ilya Nemenman for useful conversations during the ACP working group. The authors are also grateful to all readers who provided valuable feedback on this manuscript while it was under peer review. We encourage readers to help keep the Notebooks which accompany the review up-to-date, by contributing to them on Github at \url{https://github.com/drckf/mlreview_notebooks}.
PM, CHW, and AD were supported by Simon's Foundation in the form of a Simons Investigator in the MMLS and NIH MIRA program grant: 1R35GM119461.
MB acknowledges support from the Emergent Phenomena in Quantum Systems initiative of the Gordon and Betty Moore Foundation, the ERC synergy grant UQUAM, and the U.S. Department of Energy, Office of Science, Office of Advanced Scientific Computing Research, Quantum Algorithm Teams Program. 
DJS was supported as a Simons Investigator in the MMLS and by NIH K25 grant GM098875-02.
PM and DJS would like to thank the NSF grant: PHYD1066293 for supporting the Aspen Center for Physics (ACP) for facilitating discussions leading to this work. 
This research was supported in part by the National Science Foundation under Grant No. NSF PHY-1748958.
The authors are pleased to acknowledge that the computational work reported on in this paper was performed on the Shared Computing Cluster which is administered by \href{https://www.bu.edu/tech/support/research/}{Boston University's Research Computing Services}.

\appendix
\section{Overview of the Datasets used in the Review}
\input{sections/dataset.tex}
\label{single}

\bibliography{MLreview.bib}

\end{document}

%% file: sections/introduction.tex
Machine Learning (ML), data science, and statistics are fields that describe how to learn from, and make predictions about, data. The availability of big datasets is a hallmark of modern science, including physics, where data analysis has become an important component of diverse areas, such as experimental particle physics, observational astronomy and cosmology, condensed matter physics, biophysics, and quantum computing. Moreover, ML and data science are playing increasingly important roles in many aspects of modern technology, ranging from biotechnology to the engineering of self-driving cars and smart devices. Therefore, having a thorough grasp of the concepts and tools used in ML is an important skill that is increasingly relevant in the physical sciences.

The purpose of this review is to serve as an introduction to foundational and state-of-the-art techniques in ML and data science for physicists. The review seeks to find a middle ground between a short overview and a full-length textbook. While there exist many wonderful ML textbooks~\cite{friedman2001elements,murphy2012machine,bishop2006pattern,abu2012learning}, they are lengthy and use specialized language that is often unfamiliar to physicists.  This review builds upon the considerable knowledge most physicists already possess in statistical physics in order to introduce many of the major ideas and techniques used in modern ML. We take a physics-inspired pedagogical approach, emphasizing simple examples (e.g., regression and clustering), before delving into more advanced topics. The intention of this review and the accompanying \href{https://physics.bu.edu/~pankajm/MLnotebooks.html}{Jupyter notebooks} (available at \url{https://physics.bu.edu/~pankajm/MLnotebooks.html}) is to give the reader the requisite background knowledge to follow and apply these techniques to their own areas of interest.

While this review is written with a physics background in mind, we aim for it to be useful to anyone with some background in statistical physics, and it is suitable for both graduate students and researchers as well as advanced undergraduates. The review is based on an advanced topics graduate course taught at Boston University in Fall of 2016. As such, it assumes a level of familiarity with several topics found in graduate physics curricula (partition functions, statistical mechanics) and a fluency in mathematical techniques such as linear algebra, multi-variate calculus, variational methods, probability theory, and Monte-Carlo methods. It also assumes a familiarity with basic computer programming and algorithmic design.

\subsection{What is Machine Learning?}

Most physicists learn the basics of classical statistics early on in undergraduate laboratory courses.  Classical statistics is primarily concerned with how to use data to estimate the value of an unknown quantity. For instance, estimating the speed of light using measurements obtained with an interferometer is one such example that relies heavily on techniques from statistics. 

Machine Learning is a subfield of artificial intelligence with the goal of developing algorithms capable of learning from data automatically. In particular, an artificially intelligent agent needs to be able to recognize objects in its surroundings and predict the behavior of its environment in order to make informed choices. Therefore, techniques in ML tend to be more focused on prediction rather than estimation. For example, how do we use data from the interferometry experiment to predict what interference pattern would be observed under a different experimental setup? In addition, methods from ML tend to be applied to more complex high-dimensional problems than those typically encountered in a classical statistics course. 

Despite these differences, estimation and prediction problems can be cast into a common conceptual framework. In both cases, we choose some observable quantity $\boldsymbol{x}$ of the system we are studying (e.g., an interference pattern) that is related to some parameters $\boldsymbol{\theta}$ (e.g., the speed of light) of a model $p( \boldsymbol{x}|\boldsymbol{\theta})$ that describes the probability of observing $\boldsymbol{x}$ given $\boldsymbol{\theta}$. Now, we perform an experiment to obtain a dataset $\boldsymbol{X}$ and use these data to fit the model. Typically, ``fitting'' the model involves finding $\hat{\boldsymbol{\theta}}$ that provides the best explanation for the data. In the case when ``fitting'' refers to the method of least squares, the estimated parameters maximize the probability of observing the data (i.e., $\hat{\boldsymbol{\theta}} = \text{argmax}_{\boldsymbol{\theta}} \left\{  p(\boldsymbol{X} | \boldsymbol{\theta}) \right\}$). \textit{Estimation problems} are concerned with the accuracy of $\hat{\boldsymbol{\theta}}$, whereas \textit{prediction problems} are concerned with the ability of the model to predict new observations (i.e., the accuracy of $ p(\boldsymbol{x} | \hat{\boldsymbol{\theta}})$). Although the goals of estimation and prediction are related, they often lead to different approaches. As this review is aimed as an introduction to the concepts of ML, we will focus on prediction problems and refer the reader to one of many excellent textbooks on classical statistics for more information on estimation~\cite{lehmann2006theory,lehmann2006testing,witte2013statistics,wasserman2013all}. 

\subsection{Why study Machine Learning?}

The last three decades have seen an unprecedented increase in our ability to generate and analyze large data sets. This ``big data'' revolution has been spurred by an exponential increase in computing power and memory commonly known as Moore's law. Computations that were unthinkable a few decades ago can now be routinely performed on laptops. Specialized computing machines (such as GPU-based machines) are continuing this trend towards cheap, large-scale computation, suggesting that the ``big data''  revolution is here to stay. 

This increase in our computational ability has been accompanied by new techniques for analyzing and learning from large datasets. These techniques draw heavily from ideas in statistics, computational neuroscience, computer science, and physics.  Similar to physics, modern ML places a premium on empirical results and intuition over the more formal treatments common in statistics, computer science, and mathematics. This is not to say that proofs are not important or undesirable. Rather, many of the advances of the last two decades -- especially in fields like deep learning -- do not have formal justifications (much like there still exists no mathematically well-defined concept of the Feynman path-integral in $d>1$).

Physicists are uniquely situated to benefit from and contribute to ML. Many of the core concepts and techniques used in ML  -- such as Monte-Carlo methods, simulated annealing, variational methods -- have their origins in physics. Moreover, ``energy-based models'' inspired by statistical physics are the backbone of many deep learning methods. For these reasons, there is much in modern ML that will be familiar to physicists.

Physicists and astronomers have also been at the forefront of using ``big data''. For example, experiments such as CMS and ATLAS at the LHC generate petabytes of data per year. In astronomy, projects such as the Sloan Digital Sky Survey (SDSS) routinely analyze and release hundreds of terabytes of data measuring the properties of nearly a billion stars and galaxies. Researchers in these fields are increasingly incorporating recent advances in ML and data science, and this trend is likely to accelerate in the future.

Besides applications to physics, part of the goal of this review is to serve as an introductory resource for those looking to transition to more industry-oriented projects. Physicists have already made many important contributions to modern big data applications in an industrial setting \cite{wiredphysicists}. Data scientists and ML engineers in industry use concepts and tools developed for ML to gain insight from large datasets. A familiarity with ML is a prerequisite for many of the most exciting employment opportunities in the field, and we hope this review will serve as a useful introduction to ML for physicists beyond an academic setting.

\subsection{Scope and structure of the review}

Any review on ML must simultaneously accomplish two related but distinct goals. First, it must convey the rich theoretical foundations underlying modern ML. This task is made especially difficult because ML is very broad and interdisciplinary, drawing on ideas and intuitions from many fields including statistics, computational neuroscience, and physics. Unfortunately, this means making choices about what theoretical ideas to include in the review. This review emphasizes connections with statistical physics, physics-inspired Bayesian inference, and computational neuroscience models. Thus, certain ideas (e.g., gradient descent, expectation maximization, variational methods, and deep learning and neural networks) are covered extensively, while other important ideas are given less attention or even omitted entirely (e.g., statistical learning, support vector machines, kernel methods, Gaussian processes). Second, any ML review must give the reader the practical know-how to start using the tools and concepts of ML for practical problems. To accomplish this, we have written a series of \href{https://physics.bu.edu/~pankajm/MLnotebooks.html}{Jupyter notebooks} to accompany this review. These python notebooks introduce the nuts-and-bolts of how to use, code, and implement the methods introduced in the main text. Luckily, there are numerous great ML software packages available in Python (scikit-learn, tensorflow, Pytorch, Keras) and we have made extensive use of them. We have also made use of a new package, Paysage, for energy-based generative models which has been co-developed by one of the authors (CKF) and maintained by \href{https://unlearn.ai/}{Unlearn.AI} (a company affiliated with two of the authors: CKF and PM). The purpose of the notebooks is to both familiarize physicists with these resources and to serve as a starting point for experimenting and playing with ideas.

ML can be divided into three broad categories: supervised learning, unsupervised learning, and reinforcement learning. Supervised learning concerns learning from labeled data (for example, a collection of pictures labeled as \emph{containing a cat} or \emph{not containing a cat}). Common supervised learning tasks include classification and regression. Unsupervised learning is concerned with finding patterns and structure in unlabeled data. Examples of unsupervised learning include clustering, dimensionality reduction, and generative modeling. Finally, in reinforcement learning an agent learns by interacting with an environment and changing its behavior to maximize its reward. For example, a robot can be trained to navigate in a complex environment by assigning a high reward to actions that help the robot reach a desired destination. We refer the interested reader to the classic book by Sutton and Barto \emph{Reinforcement Learning: an Introduction} \cite{sutton1998reinforcement}. While useful, the distinction between the three types of ML is sometimes fuzzy and fluid, and many applications often combine them in novel and interesting ways. For example, the recent success of Google DeepMind in developing ML algorithms that excel at tasks such as playing Go and video games employ deep reinforcement learning, combining reinforcement learning with supervised learning methods based on deep neural networks.

Here, we limit our focus to supervised and unsupervised learning. The literature on reinforcement learning is extensive and uses ideas and concepts that, to a large degree, are distinct from supervised and unsupervised learning tasks. For this reason, to ensure cohesiveness and limit the length of this review, we have chosen not to discuss reinforcement learning. However, this omission should not be mistaken for a value judgement on the utility of reinforcement learning for solving physical problems. For example, some of the authors have used  inspiration from reinforcement learning to tackle difficult problems in quantum control~\cite{bukov_17RL,bukov2018reinforcement}.
 
In writing this review, we have tried to adopt a style that reflects what we consider to be the best of the physics tradition. Physicists understand the importance of well-chosen examples for furthering our understanding. It is hard to imagine a graduate course in statistical physics without the Ising model. Each new concept that is introduced in statistical physics (mean-field theory, transfer matrix techniques, high- and low-temperature expansions, the renormalization group, etc.) is applied to the Ising model. This allows for the progressive building of intuition and ultimately a coherent picture of statistical physics. We have tried to replicate this pedagogical approach in this review by focusing on a few well-chosen techniques --  linear and logistic regression in the case of supervised learning and clustering in the case of unsupervised learning -- to introduce the major theoretical concepts.

In this same spirit, we have chosen three interesting datasets with which to illustrate the various algorithms discussed here. (i) The SUSY data set consists of $5,000,000$ Monte-Carlo samples of proton-proton collisions decaying to either signal or background processes, which are both parametrized with $18$ features. The signal process is the production of electrically-charged supersymmetric particles, which decay to $W$ bosons and an electrically-neutral supersymmetric particle, invisible to the detector, while the background processes are various decays involving only Standard Model particles~\cite{baldi2014searching}. (ii) The Ising data set consists of $10^4$ states of the 2D Ising model on a $40\times 40$ square lattice, obtained using Monte-Carlo (MC) sampling at a few fixed temperatures $T$. (iii) The MNIST dataset comprises $70000$ handwritten digits, each of which comes in a square image, divided into a $28\times 28$ pixel grid. The first two datasets were chosen to reflect the various sub-disciplines of physics (high-energy experiment, condensed matter) where we foresee techniques from ML becoming an increasingly important tool for research. The MNIST dataset, on the other hand, introduces the flavor of present-day ML problems. By re-analyzing the same datasets with multiple techniques, we hope readers will be able to get a sense of the various, inevitable trade-offs involved in choosing how to analyze data. Certain techniques work better when data is limited while others may be better suited to large data sets with many features. A short description of these datasets are given in the Appendix.

This review draws generously on many wonderful textbooks on ML and we encourage the reader to consult them for further information. They include Abu Mostafa's masterful \emph{Learning from Data}, which introduces the basic concepts of statistical learning theory \cite{abu2012learning}, the more advanced but equally good \emph{The Elements of Statistical Learning} by Hastie, Tibshirani, and Friedman \cite{friedman2001elements}, Michael Nielsen's indispensable \emph{Neural Networks and Deep Learning} which serves as a wonderful introduction to the neural networks and deep learning \cite{nielsen2015neural} and David MacKay's outstanding \emph{Information Theory, Inference, and Learning Algorithms} which introduced Bayesian inference and information theory to a whole generation of physicists \cite{mackay2003information}. More comprehensive (and much longer) books on modern ML techniques include Christopher Bishop's classic \emph{Pattern Recognition and Machine Learning} \cite{bishop2006pattern} and the more recently published \emph{Machine Learning: A Probabilistic Perspective} by Kevin Murphy \cite{murphy2012machine}. Finally, one of the great successes of modern ML is deep learning, and some of the pioneers of this field have written a textbook for students and researchers entitled \emph{Deep Learning} \cite{Goodfellow-et-al-2016}. In addition to these textbooks, we have consulted numerous research papers, reviews, and web resources. Whenever possible, we have tried to point the reader to key papers and other references that we have found useful in preparing this review. However, we are neither capable of nor have we made any effort to make a comprehensive review of the literature.

The review is organized as follows. We begin by introducing polynomial regression as a simple example that highlights many of the core ideas of ML. The next few chapters introduce the language and major concepts needed to make these ideas more precise including tools from statistical learning theory such as overfitting, the bias-variance tradeoff, regularization, and the basics of Bayesian inference. The next chapter builds on these examples to discuss stochastic gradient descent and its generalizations. We then apply these concepts to linear and logistic regression, followed by a detour to discuss how we can combine multiple statistical techniques to improve supervised learning, introducing bagging, boosting, random forests, and XG Boost. These ideas, though fairly technical, lie at the root of many of the advances in ML over the last decade. The review continues with a thorough discussion of supervised deep learning and neural networks, as well as convolutional nets. We then turn our focus to unsupervised learning. We start with data visualization and dimensionality reduction before proceeding to a detailed treatment of clustering. Our discussion of clustering naturally leads to an examination of variational methods and their close relationship with mean-field theory. 
The review continues with a discussion of deep unsupervised learning, focusing on energy-based models, such as Restricted Boltzmann Machines (RBMs) and Deep Boltzmann Machines (DBMs). Then we discuss two new and extremely popular modeling frameworks for unsupervised learning, generative adversarial networks (GANs) and variational autoencoders (VAEs). We conclude the review with an outlook and discussion of promising research directions at the intersection physics and ML.

%% file: sections/MLisdifficult.tex
\begin{figure*}[t!] 
 \begin{minipage}[b]{0.5\linewidth}
    \centering
    \includegraphics[width=0.99\linewidth]{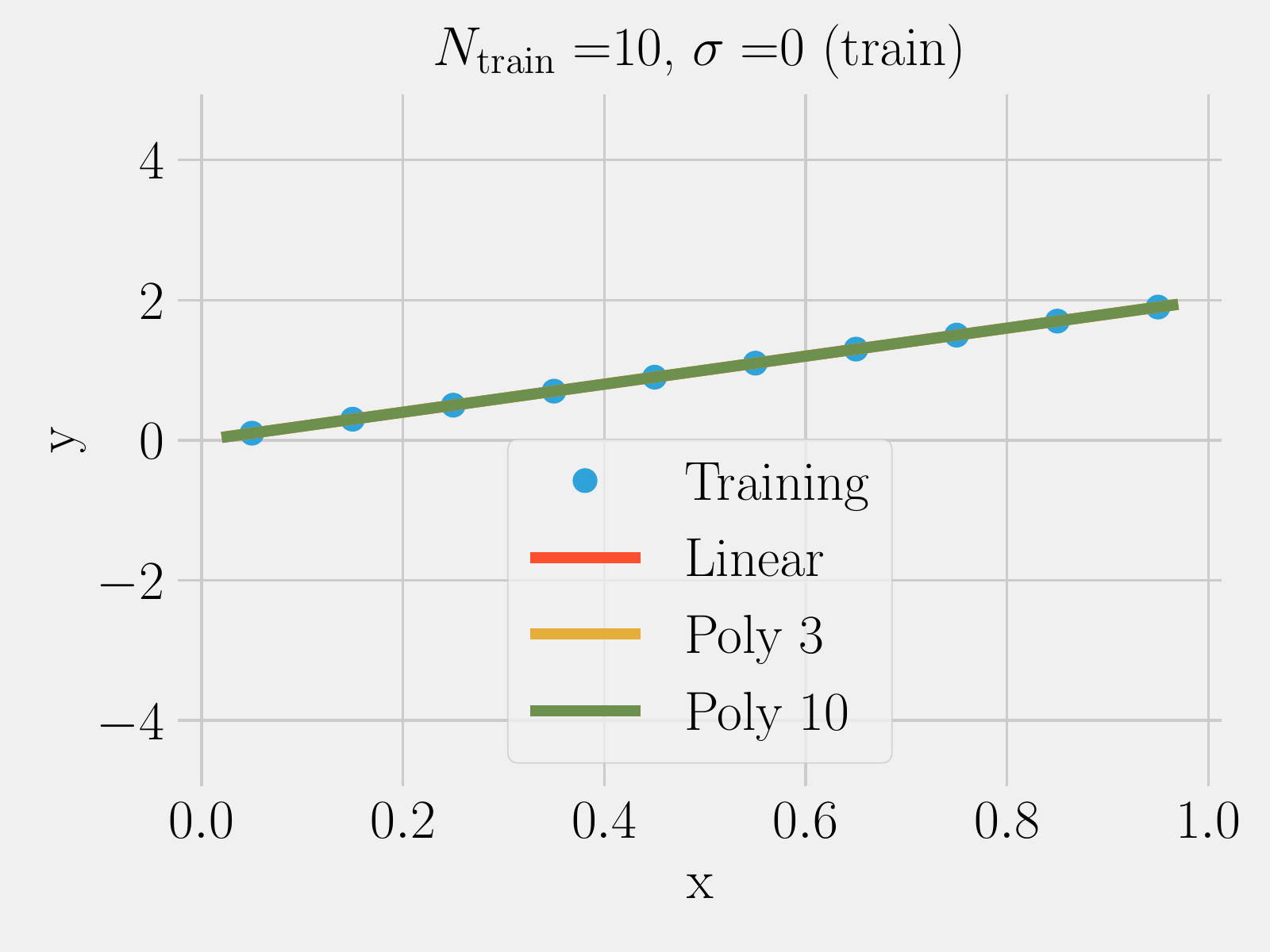} 
    \vspace{4ex}
  \end{minipage}%
  \begin{minipage}[b]{0.5\linewidth}
    \centering
    \includegraphics[width=0.99\linewidth]{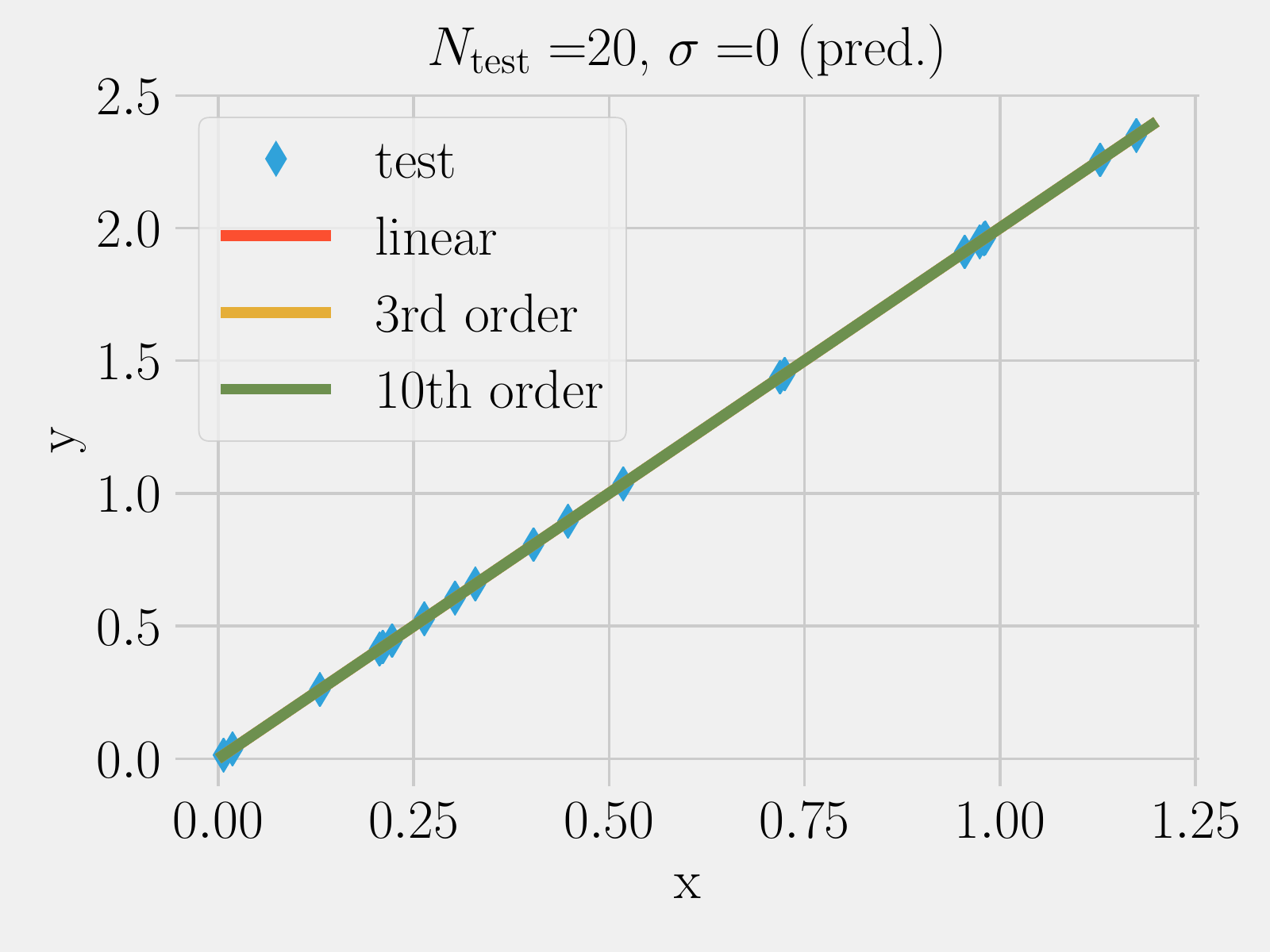} 
    \vspace{4ex}
 \end{minipage} %
 \begin{minipage}[b]{0.5\linewidth}
    \centering
    \includegraphics[width=0.99\linewidth]{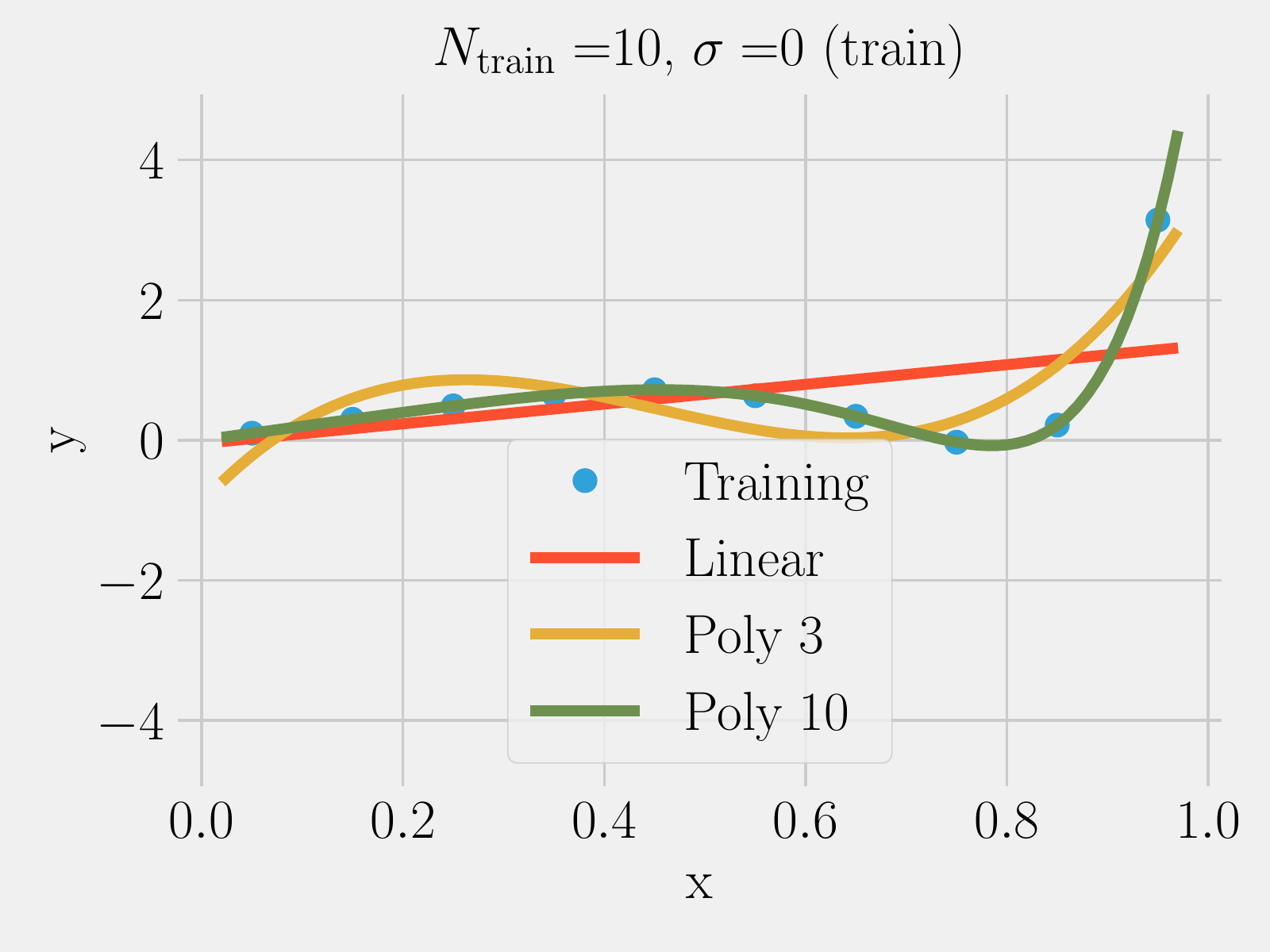} 
 \end{minipage}%
 \begin{minipage}[b]{0.5\linewidth}
    \centering
    \includegraphics[width=0.99\linewidth]{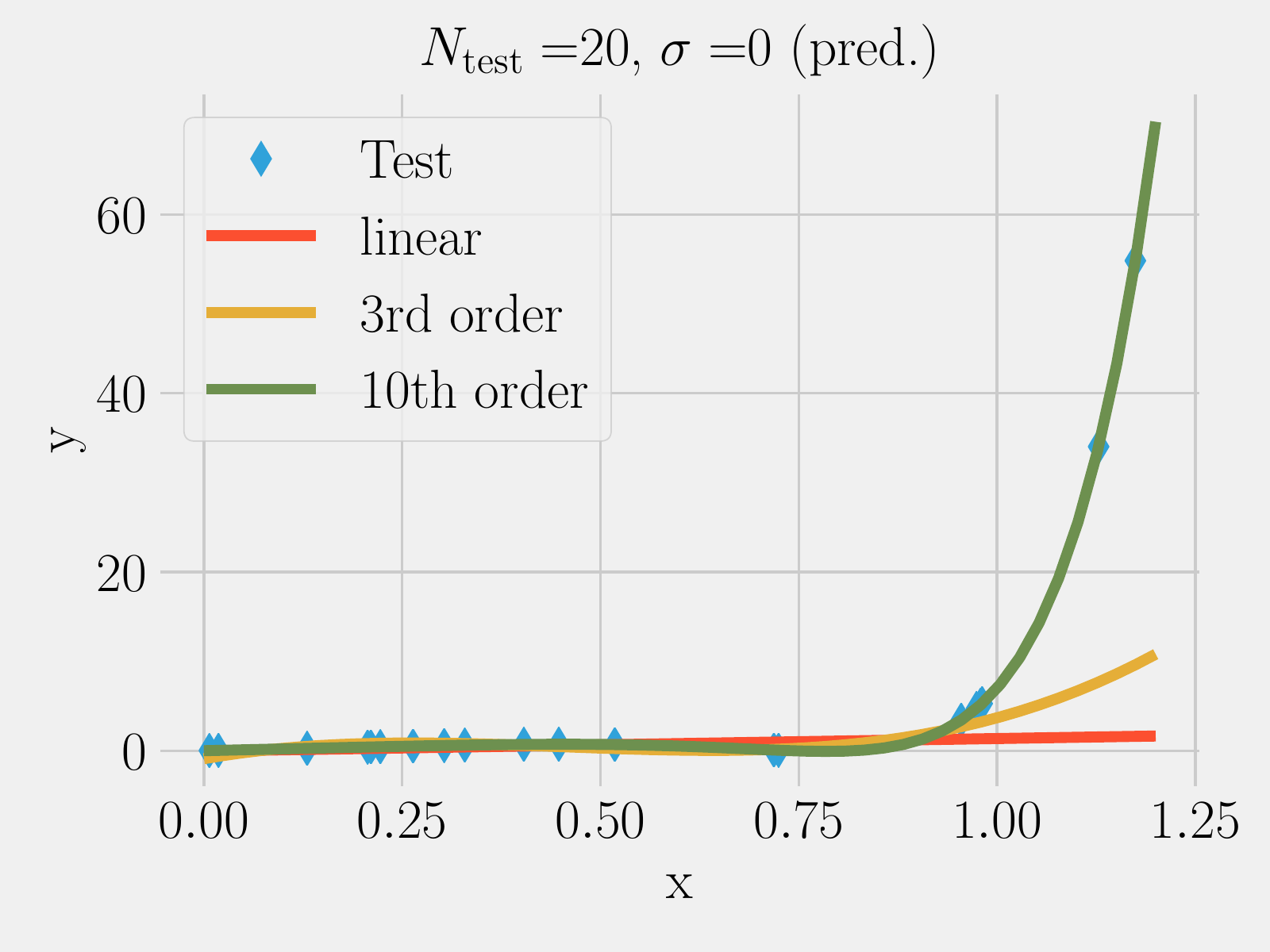} 
  \end{minipage} %
  \caption{ {\bf Fitting versus predicting for noiseless data}. $N_{\mathrm{train}}=10$ points in the range $x \in [0,1]$ were generated from a linear model (top) or tenth-order polynomial (bottom). This data was fit using three model classes: linear models (red), all polynomials of order 3 (yellow), all polynomials of order 10 (green)   and used to make prediction on $N_{\mathrm{test}}=20$ new data points with $x_{\mathrm{test}} \in [0,1.2]$ (shown on right). Notice that in the absence of noise ($\sigma=0$), given enough data points that fitting and predicting are identical.}
  \label{fig:II.1} 
\end{figure*}

\subsection{Setting up a problem in ML and data science}

Many problems in ML and data science starts with the same ingredients. The first ingredient is the dataset $\mathcal{D} = (\boldsymbol{X}, \boldsymbol{y})$ where $\boldsymbol{X}$ is a matrix of independent variables and $\boldsymbol{y}$ is a vector of dependent variables. The second is the model $f(\boldsymbol{x}; \boldsymbol{\theta})$, which is a function $f: \boldsymbol{x} \rightarrow y$ of the parameters $\boldsymbol{\theta}$. That is, $f$ is a function used to predict an output from a vector of input variables. The final ingredient is the cost function $\mathcal{C}( \boldsymbol{y}, f(\boldsymbol{X}; \boldsymbol{\theta}))$ that allows us to judge how well the model performs on the observations $\boldsymbol{y}$. The model is fit by finding the value of $\boldsymbol{\theta}$ that minimizes the cost function. For example, one commonly used cost function is the squared error. Minimizing the squared error cost function is known as the method of least squares, and is typically appropriate for experiments with Gaussian measurement errors. 

ML researchers and data scientists follow a standard recipe to obtain models that are useful for prediction problems. We will see why this is necessary in the following sections, but it is useful to present the recipe up front to provide context. The \emph{first step} in the analysis is to {\it randomly} divide the dataset $\mathcal{D}$ into two mutually exclusive groups $\mathcal{D}_\mathrm{train}$ and $\mathcal{D}_\mathrm{test}$ called the training and test sets. The fact that this must be the first step should be heavily emphasized -- performing some analysis (such as using the data to select important variables) before partitioning the data is a common pitfall that can lead to incorrect conclusions. Typically, the majority of the data are partitioned into the training set (e.g., 90\%) with the remainder going into the test set. The model is fit by minimizing the cost function using only the data in the training set $\hat{\boldsymbol{\theta}} = \argmin_{\boldsymbol{\theta}} \left \{ \mathcal{C}( \boldsymbol{y}_\mathrm{train}, f(\boldsymbol{X}_\mathrm{train}; \boldsymbol{\theta})) \right\}$. Finally, the performance of the model is evaluated by computing the cost function using the test set $\mathcal{C}( \boldsymbol{y}_\mathrm{test}, f(\boldsymbol{X}_\mathrm{test}; \hat{\boldsymbol{\theta}}))$. The value of the cost function for the best fit model on the training set is called the in-sample error $E_\mathrm{in} = \mathcal{C}( \boldsymbol{y}_\mathrm{train}, f(\boldsymbol{X}_\mathrm{train}; \boldsymbol{\theta}))$ and the value of the cost function on the test set is called the out-of-sample error $E_\mathrm{out} = \mathcal{C}( \boldsymbol{y}_\mathrm{test}, f(\boldsymbol{X}_\mathrm{test}; \boldsymbol{\theta}))$.

One of the most important observations we can make is that the out-of-sample error is almost always greater than the in-sample error, i.e. $E_\mathrm{out} \geq E_\mathrm{in}$. We explore this point further in Sec.~\ref{sec:lin_reg} and its \href{https://physics.bu.edu/~pankajm/MLnotebooks.html}{accompanying notebook}. Splitting the data into mutually exclusive training and test sets provides an unbiased estimate for the predictive performance of the model -- this is known as cross-validation in the ML and statistics literature. In many applications of classical statistics, we start with a mathematical model that we assume to be true (e.g., we may assume that Hooke's law is true if we are observing a mass-spring system) and our goal is to estimate the value of some unknown model parameters (e.g., we do not know the value of the spring stiffness). Problems in ML, by contrast, typically involve inference about complex systems where we do not know the exact form of the mathematical model that describes the system. Therefore, it is not uncommon for ML researchers to have multiple candidate models that need to be compared. This comparison is usually done using $E_\mathrm{out}$; the model that minimizes this out-of-sample error is chosen as the best model (i.e. model selection). Note that once we select the best model on the basis of its performance on $E_\mathrm{out}$, the real-world performance of the winning model should be expected to be slightly worse because the test data was now used in the fitting procedure.

\subsection{Polynomial Regression}

In the previous section, we mentioned that multiple candidate models are typically compared using the out-of-sample error $E_\mathrm{out}$. It may be at first surprising that the model that has the lowest out-of-sample error $E_\mathrm{out}$ usually \emph{does not} have the lowest in-sample error $E_\mathrm{in}$. Therefore, if our goal is to obtain a model that is useful for prediction we may not want to choose the model that provides the best explanation for the current observations. At first glance, the observation that the model providing the best explanation for the current dataset probably will not provide the best explanation for future datasets is very counter-intuitive.  

Moreover, the discrepancy between $E_\mathrm{in}$ and $E_\mathrm{out}$ becomes more and more important, as the complexity of our data, and the models we use to make predictions, grows. As the number of parameters in the model increases, we are forced to work in high-dimensional spaces. The ``curse of dimensionality'' ensures that many phenomena that are absent or rare in low-dimensional spaces become generic. For example, the nature of distance changes in high dimensions, as evidenced in the derivation of the Maxwell distribution in statistical physics where the fact that all the volume of a $d$-dimensional sphere of radius $r$ is contained in a small spherical shell around $r$ is exploited. Almost all critical points of a function (i.e.,~the points where all derivatives vanish) are saddles rather than maxima or minima (an observation first made in physics in the context of the $p$-spin spherical spin glass). For all these reasons, it turns out that for complicated models studied in ML, predicting and fitting are very different things \cite{bickel2006regularization}.

\begin{figure*}[t!] 
	\begin{minipage}[b]{0.5\linewidth}
		\centering
		\includegraphics[width=0.99\linewidth]{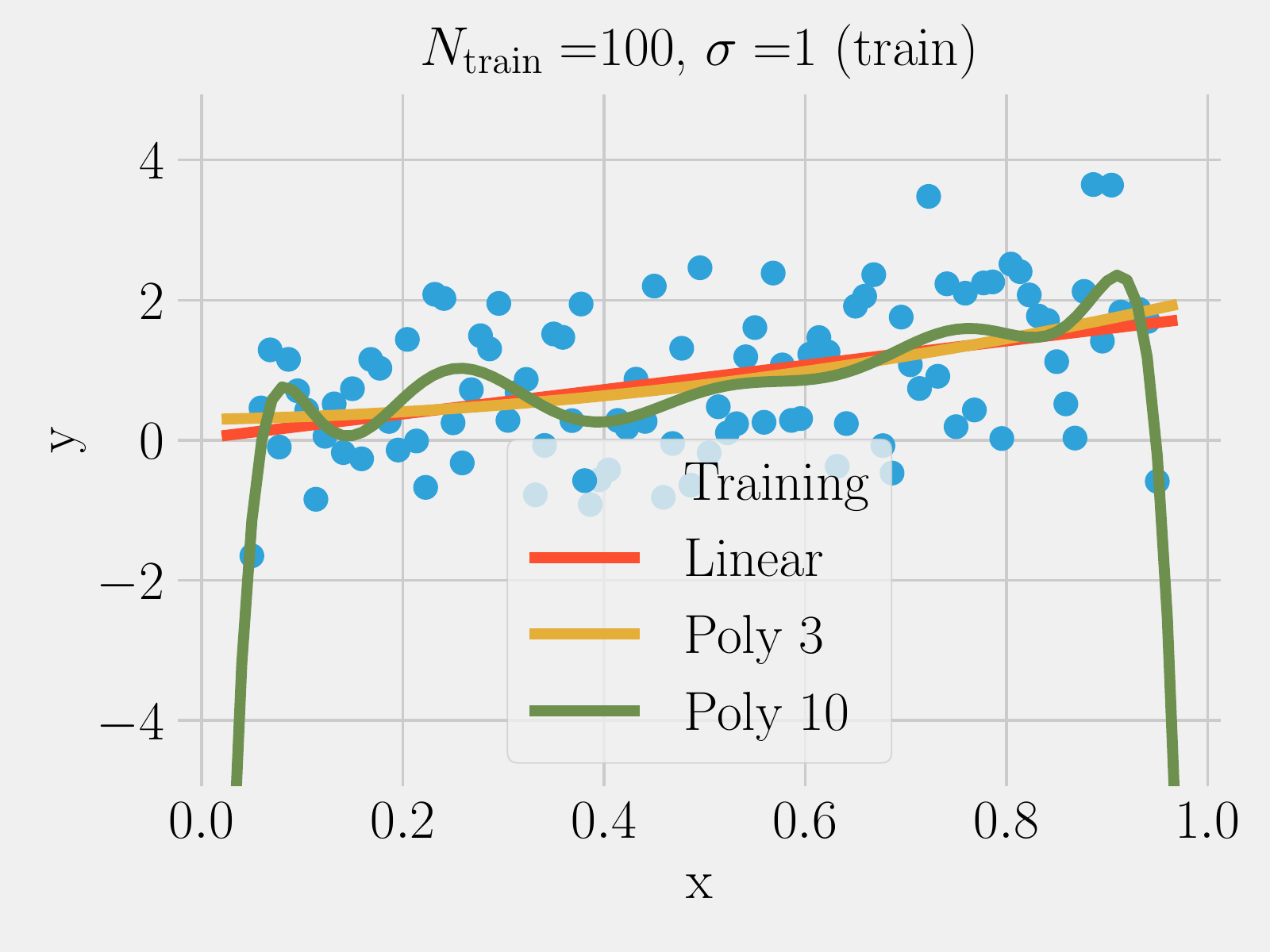} 
		\vspace{4ex}
	\end{minipage}%
	\begin{minipage}[b]{0.5\linewidth}
		\centering
		\includegraphics[width=0.99\linewidth]{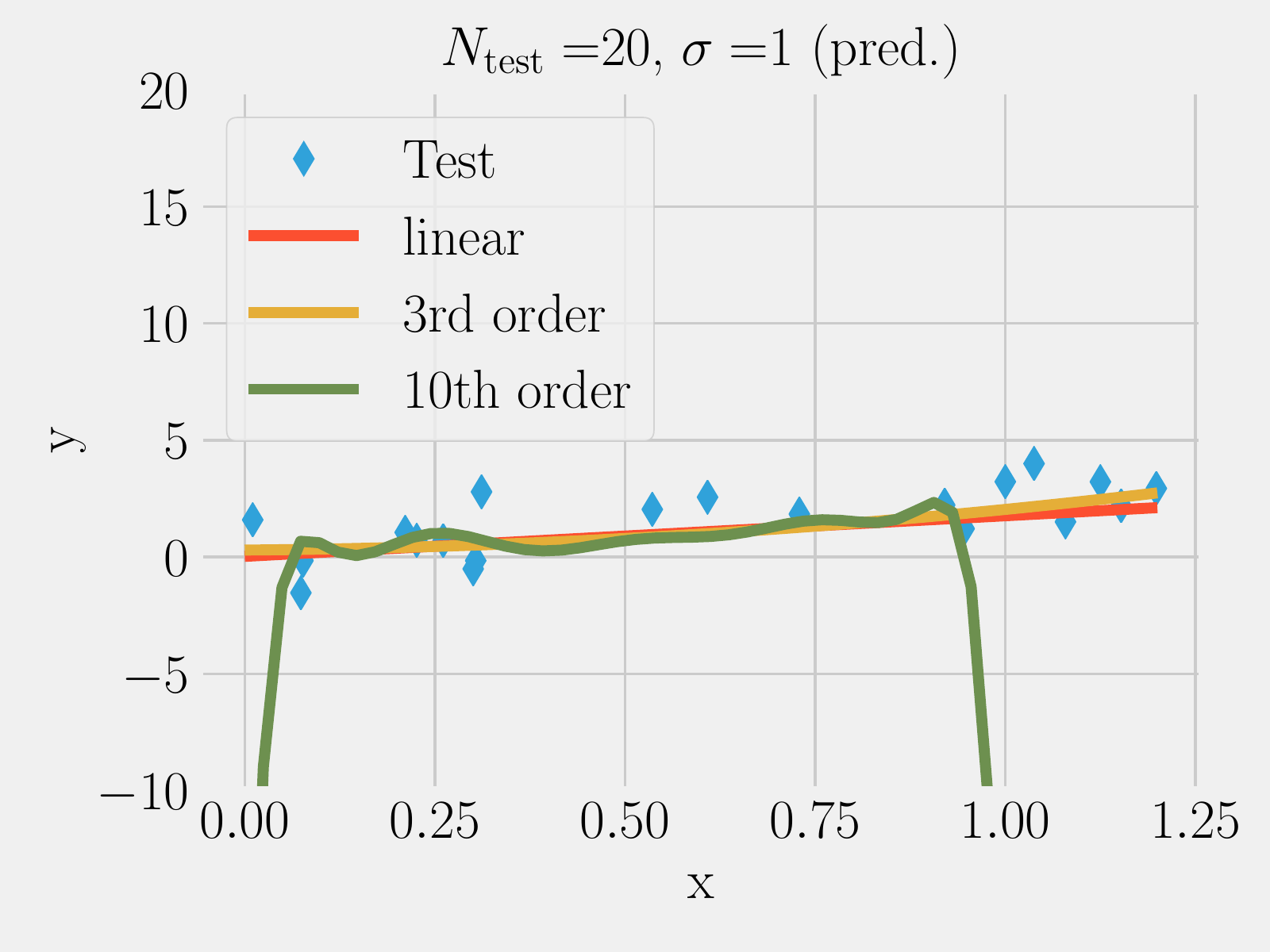} 
		\vspace{4ex}
	\end{minipage} %
	\begin{minipage}[b]{0.5\linewidth}
		\centering
		\includegraphics[width=0.99\linewidth]{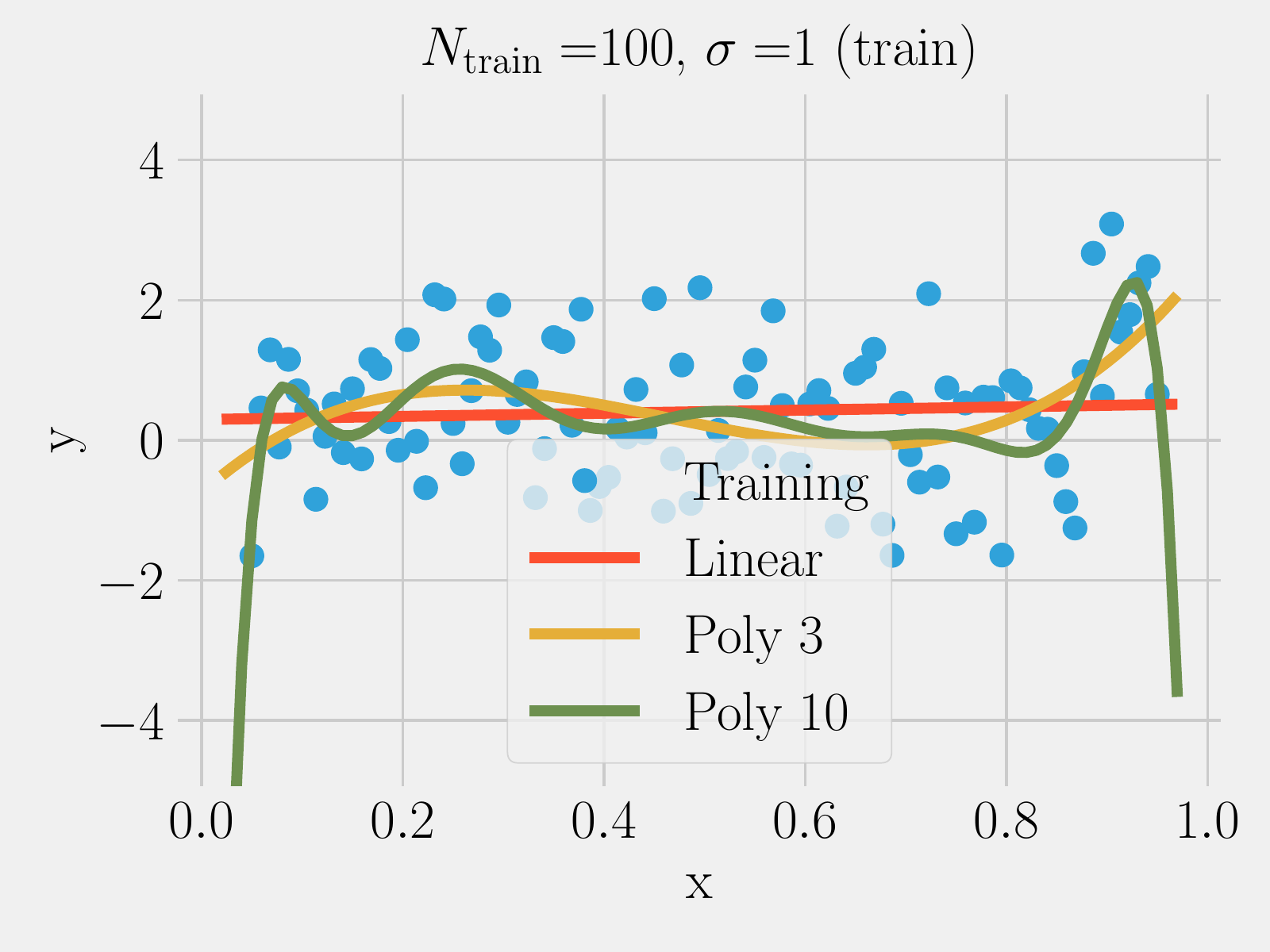} 
	\end{minipage}%
	\begin{minipage}[b]{0.5\linewidth}
		\centering
		\includegraphics[width=0.99\linewidth]{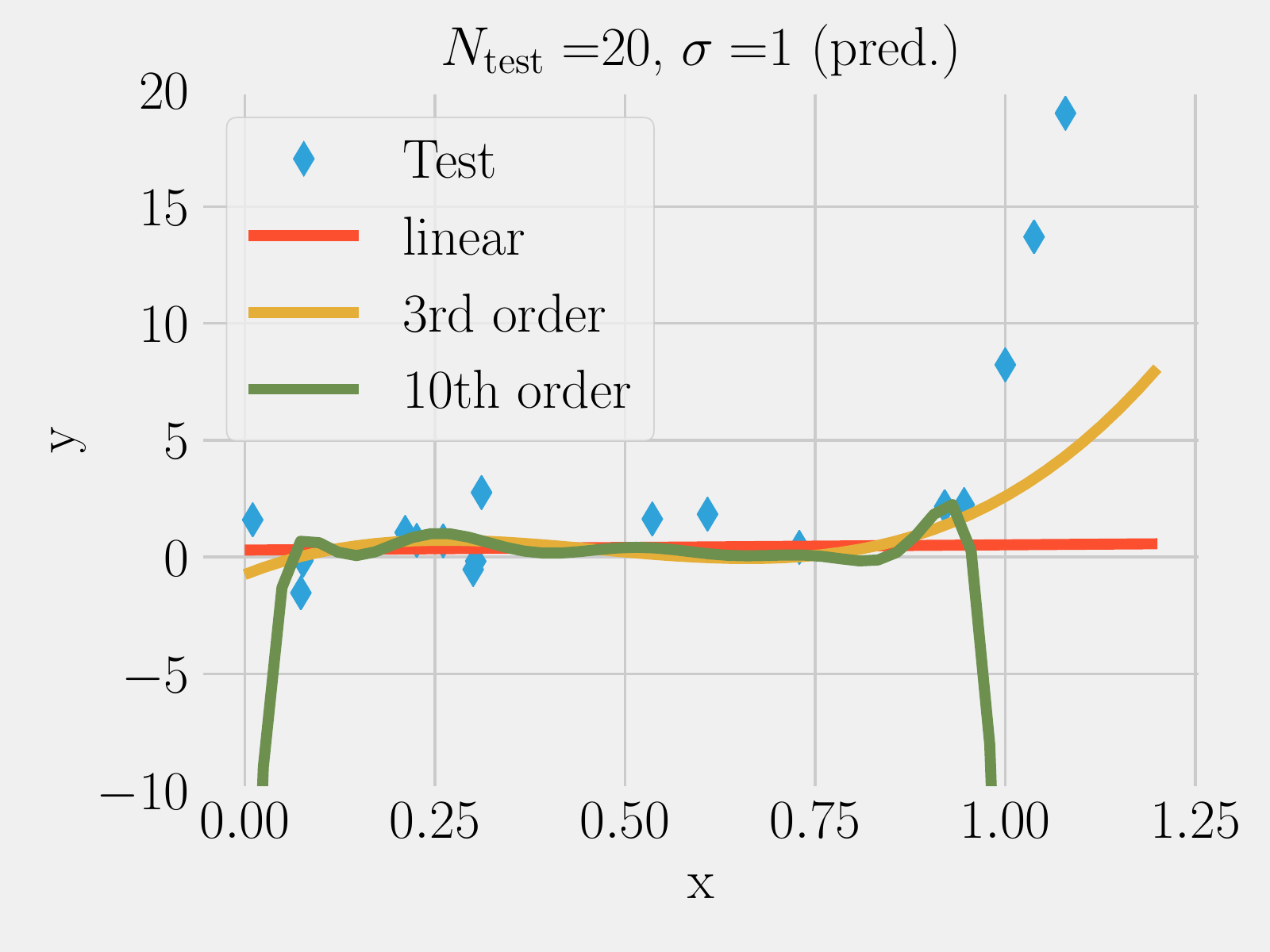} 
	\end{minipage} %
	\caption{ {\bf Fitting versus predicting for noisy data}. $N_{\mathrm{train}}=100$ noisy data points ($\sigma=1$) in the range $x \in [0,1]$ were generated from a linear model (top) or tenth-order polynomial (bottom). This data was fit using three model classes: linear models (red), all polynomials of order 3 (yellow), all polynomials of order 10 (green)   and used to make prediction on $N_{\mathrm{test}}=20$ new data points with  $x_{\mathrm{test}} \in [0,1.2]$(shown on right). Notice that even when the data was generated using a tenth order polynomial, the linear and third order polynomials give better out-of-sample predictions, especially beyond the $x$ range over which the model was trained.}
	\label{fig:II.2} 
\end{figure*}

To develop some intuition about why we need to pay close attention to out-of-sample performance, we will consider a simple one-dimensional problem -- polynomial regression. Our task is a simple one, fitting data with polynomials of different order. We will explore how our ability to predict depends on the number of data points we have, the ``noise'' in the data generation process, and our prior knowledge about the system. The goal is to build intuition about why prediction is difficult in preparation for introducing general strategies that overcome these difficulties. 

\emph{Before reading the rest of the section, we strongly encourage the reader to read \href{https://physics.bu.edu/~pankajm/MLnotebooks.html}{Notebook 1} and complete the accompanying exercises.}
 
Consider a probabilistic process that assigns a label $y_i$ to an observation $x_i$. The data are generated by drawing samples from the equation
\bea
    y_i= f(x_i) + \eta_i,
\label{II.1}
\eea
where $f(x_i)$ is some fixed (but possibly unknown) function, and $\eta_i$ is a Gaussian, uncorrelated noise variable, such that
\bea
\langle \eta_i \rangle=0, \nonumber  \\
\langle \eta_i \eta_j \rangle = \delta_{ij} \sigma^2. \nonumber
\eea
We will refer to the $f(x_i)$ as the function used to generate the data, and $\sigma$ as the noise strength. The larger $\sigma$ is the noisier the data; $\sigma=0$ corresponds to the noiseless case.

To make predictions, we will consider a family of functions $f_\alpha(x;\boldsymbol{\theta}_\alpha)$ that depend on some parameters $\boldsymbol{\theta}_\alpha$. These functions represent the \emph{model class} that we are using to model the data and make predictions. Note that we choose the model class without knowing the function $f(x)$. The $f_\alpha(x;\boldsymbol{\theta}_\alpha)$ encode the \emph{features} we choose to represent the data. In the case of polynomial regression we will consider three different model classes: (i) all polynomials of order 1 which we denote by $f_1(x;\boldsymbol{\theta}_1)$, (ii) all polynomials up to order 3 which we denote by $f_3(x;\boldsymbol{\theta}_3)$, and (iii) all polynomials of order 10, $f_{10}(x;\boldsymbol{\theta}_{10})$. Notice that these three model classes contain different number of  parameters. Whereas $f_1(x;\boldsymbol{\theta}_1)$ has only two parameters (the coefficients of the zeroth and first order terms in the polynomial), $f_3(x;\boldsymbol{\theta}_3)$ and $f_{10}(x; \boldsymbol{\theta}_{10})$ have four and eleven parameters, respectively. This reflects the fact that these three models have different \emph{model complexities}. If we think of each term in the polynomial as a ``feature'' in our model, then increasing the order of the polynomial we fit increases the number of features. Using a more complex model class may give us better predictive power, but only if we have a large enough sample size to accurately learn the model parameters associated with these extra features from the training dataset.

To learn the parameters $\boldsymbol{\theta}_\alpha$, we will train our models on a \emph{training dataset} and then test the effectiveness of the model on a \underline{different} dataset, the \emph{test dataset}. Since we are interested only in gaining intuition, we will simply plot the fitted polynomials and compare the predictions of our fits for the test data with the true values. As we will see below, the models that give the best fit to existing data do not necessarily make the best predictions even for a simple task like polynomial regression. 

To illustrate these ideas, we encourage the reader to experiment with the accompanying notebook to generate data using a linear function $f(x)=2x$ and a tenth order polynomial $f(x)=2x-10x^5+15x^{10}$ and ask how the size of the training dataset $N_{\mathrm{train}}$ and the noise strength $\sigma$ affect the ability to make predictions. Obviously, more data and less noise leads to better predictions. To train the models (linear, third-order, tenth-order), we uniformly sampled the interval $x \in [0,1]$ and constructed  $N_{\mathrm{train}}$ training examples using (\ref{II.1}). We then fit the models on these training samples using standard least-squares regression. To visualize the performance of the three models, we plot the predictions using the best fit parameters for a test set where $x$ are drawn uniformly from the interval $x \in [0,1.2]$. Notice that the test interval is slightly larger than the training interval.

Figure \ref{II.1} shows the results of this procedure for the noiseless case, $\sigma=0$. Even using a small training set with $N_{\mathrm{train}}=10$ examples, we find that the model class that generated the data also provides the best fit and the most accurate out-of-sample predictions. That is, the linear model performs the best for data generated from a linear polynomial (the third and tenth order polynomials perform similarly), and the tenth order model performs the best for data generated from a tenth order polynomial. While this may be expected, the results are quite different for larger noise strengths.

\begin{figure*}[t!] 
	\begin{minipage}[b]{0.5\linewidth}
		\centering
		\includegraphics[width=0.99\linewidth]{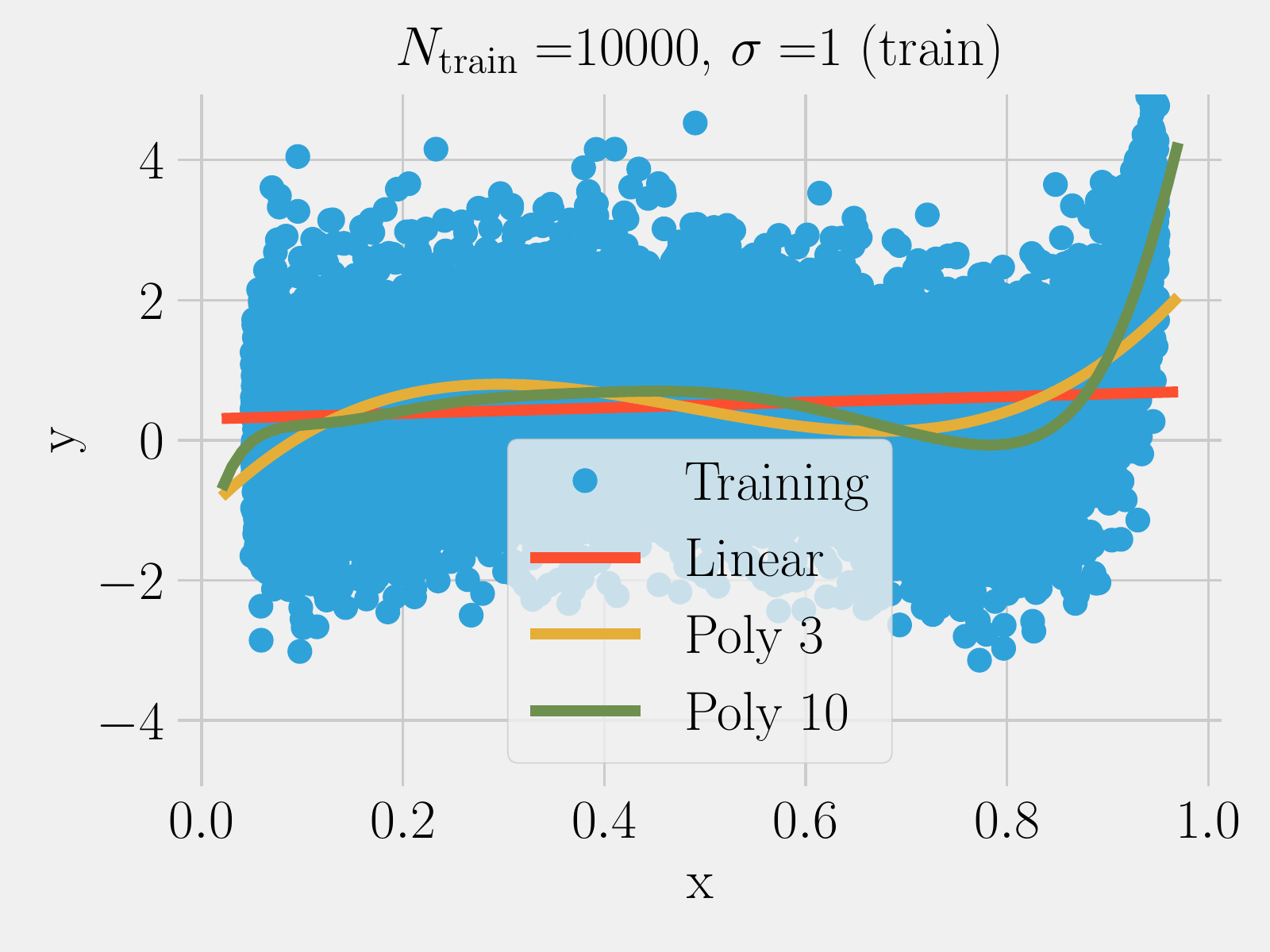} 
		\vspace{4ex}
	\end{minipage}%
	\begin{minipage}[b]{0.5\linewidth}
		\centering
		\includegraphics[width=0.99\linewidth]{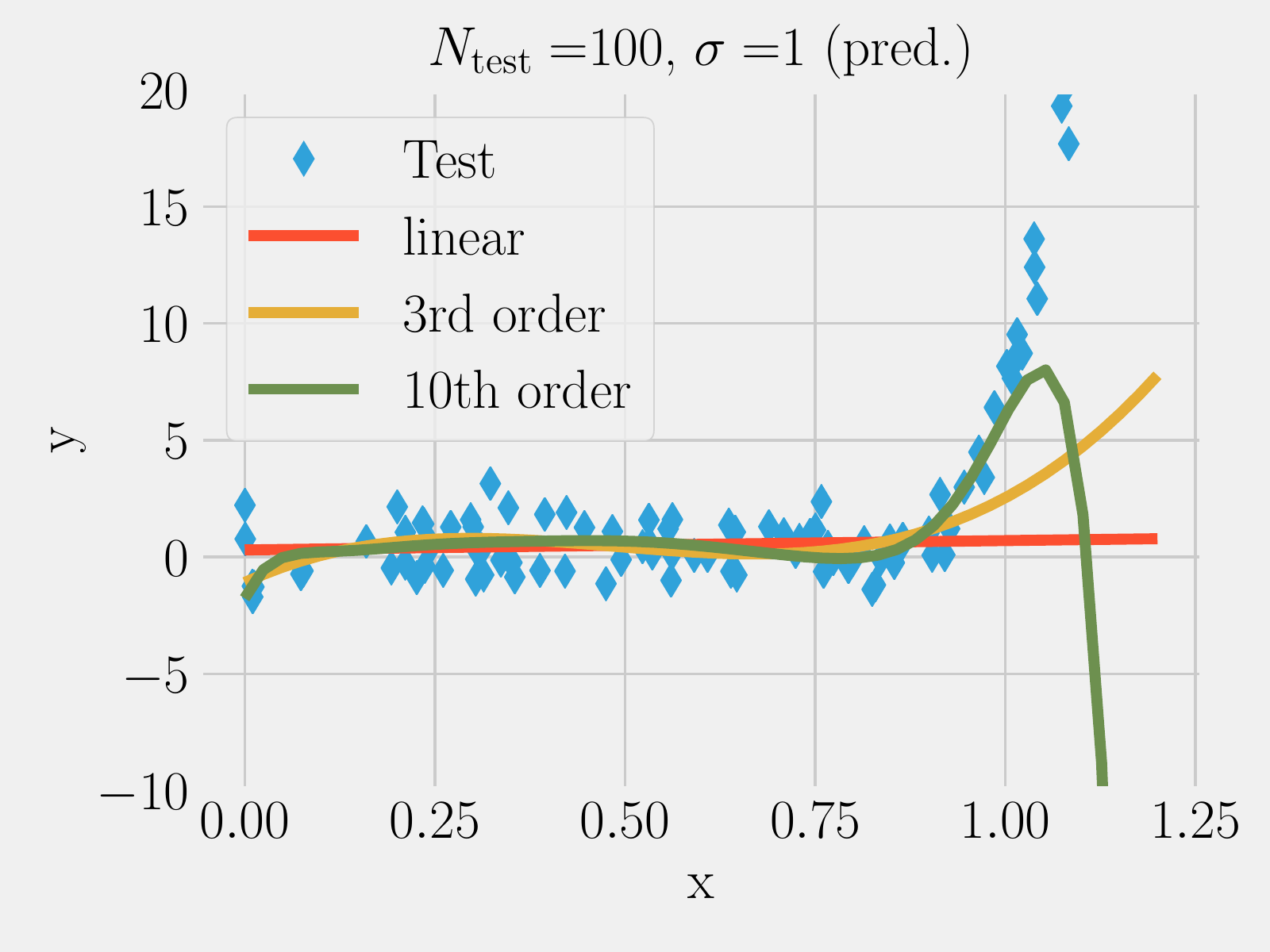} 
		\vspace{4ex}
	\end{minipage} %
\caption{ {\bf Fitting versus predicting for noisy data}. $N_{\mathrm{train}}=10^4$ noisy data points ($\sigma=1$) in the range $x \in [0,1]$ were generated from a tenth-order polynomial. This data was fit using three model classes: linear models (red), all polynomials of order 3 (yellow), all polynomials of order 10 (green)   and used to make prediction on $N_{\mathrm{test}}=100$ new data points with  $x_{\mathrm{test}} \in [0,1.2]$(shown on right). The tenth order polynomial gives good predictions but the model's predictive power quickly degrades beyond the training data range.}
\label{fig:II.3} 
\end{figure*}

Figure \ref{fig:II.2} shows the results of the same procedure for noisy data, $\sigma=1$, and a larger training set, $N_{\mathrm{train}}=100$. As in the noiseless case, the tenth order model provides the best fit to the data (i.e., the lowest $E_\mathrm{in}$). In contrast, the tenth order model now makes the worst out-of-sample predictions (i.e., the highest $E_\mathrm{out}$). Remarkably, this is true even if the data were generated using a tenth order polynomial. 

At small sample sizes, noise can create fluctuations in the data that look like genuine patterns. Simple models (like a linear function) cannot represent complicated patterns in the data, so they are forced to ignore the fluctuations and to focus on the larger trends. Complex models with many parameters, such as the tenth order polynomial in our example, can capture both the global trends and noise-generated patterns at the same time. In this case, the model can be tricked into thinking that the noise encodes real information. This problem is called ``overfitting'' and leads to a steep drop-off in predictive performance. 

We can guard against overfitting in two ways: we can use less expressive models with fewer parameters, or we can collect more data so that the likelihood that the noise appears patterned decreases. Indeed, when we increase the size of the training data set by two orders of magnitude to $N_{\mathrm{train}}=10^4$ (see Figure \ref{fig:II.3}) the tenth order polynomial clearly gives both the best fits and the most predictive power over the entire training range $x \in [0,1]$, and even slightly beyond to approximately $x \approx 1.05$. This is our first experience with what is known as the \emph{bias-variance} tradeoff, c.f.~Sec.~\ref{subsec:bias_variance_tradeoff}. When the amount of training data is limited as it is when $N_{\mathrm{train}}=100$, one can often get better predictive performance by using a less expressive model (e.g., a lower order polynomial) rather than the more complex model (e.g., the tenth-order polynomial). The simpler model has more ``bias'' but is less dependent on the particular realization of the training dataset, i.e. less ``variance''. Finally we note that even with ten thousand data points, the model's performance quickly degrades beyond the original training data range. This demonstrates the difficulty of predicting beyond the training data we mentioned earlier.

This simple example highlights why ML is so difficult and holds some universal lessons that we will encounter repeatedly in this review:
\begin{itemize}
\item {Fitting is not predicting.} Fitting existing data well is fundamentally different from making predictions about new data.
\item{Using a complex model can result in overfitting.} Increasing a model's complexity (i.e number of fitting parameters) will usually yield better results on the training data. However when the training data size is small and the data are noisy, this results in \emph{overfitting} and can substantially degrade the predictive performance of the model. 
\item{For complex datasets and small training sets, simple models can be better at prediction than complex models due to the bias-variance tradeoff.}  It takes less data to train a simple model than a complex one. Therefore, even though the correct model is guaranteed to have better predictive performance for an infinite amount of training data (less bias), the training errors stemming from finite-size sampling (variance) can cause simpler models to outperform the more complex model when sampling is limited.
\item{It is difficult to generalize beyond the situations encountered in the training data set.}
 \end{itemize}

%% file: sections/StatisticalLearningTheory.tex
 In this section, we briefly summarize and discuss the sense in which learning is possible, with a focus on supervised learning. We begin with an unknown function $y=f(x)$ and fix a \emph{hypothesis set} $\mathcal{H}$ consisting of all functions we are willing to consider, defined also on the domain of $f$. This set may be uncountably infinite (e.g.~if there are real-valued parameters to fit). The choice of which functions to include in $\mathcal{H}$ usually depends on our intuition about the problem of interest. The function $f(x)$ produces a set of pairs $(x_i,y_i)$, $i=1\dots N$, which serve as the observable data. Our goal is to select a function from the hypothesis set $h\in\mathcal{H}$ that approximates $f(x)$ as best as possible, namely, we would like to find $h\in\mathcal{H}$ such that $h\approx f$ in some strict mathematical sense which we specify below. If this is possible, we say that we \emph{learned} $f(x)$. But if the function $f(x)$ can, in principle, take any value on \emph{unobserved} inputs, how is it possible to learn in any meaningful sense?

The answer is that learning is possible in the restricted sense that the fitted model will probably perform approximately as well on new data as it did on the training data. Once an appropriate error function $E$ is chosen for the problem under consideration (e.g.~sum of squared errors in linear regression), we can define two distinct performance measures of interest. The in-sample error, $E_\mathrm{in}$, and the out-of-sample or generalization error, $E_\mathrm{out}$. Recall from Sec~\ref{sec:why_is_ML_difficult} that both metrics are required due to the distinction between fitting and predicting.

This raises a natural question: \emph{Can we say something general about the relationship between $E_{\mathrm{in}}$ and $E_{\mathrm{out}}$?} Surprisingly, the answer is `Yes'. We can in fact say quite a bit. This is the domain of statistical learning theory, and we give a brief overview of the main results in this section. Our goal is to briefly introduce some of the major ideas from statistical learning theory because of the important role they have played in shaping how we think about machine learning. However, this is a highly technical and theoretical field, so we will just skim over some introductory topics. A more thorough introduction to statistical learning theory can be found in the introductory textbook by Abu Mostafa~\cite{abu2012learning}.

\subsection{Three simple schematics that summarize the basic intuitions from Statistical Learning Theory}

\begin{figure}[t!] 
 \includegraphics[width=1.0\columnwidth]{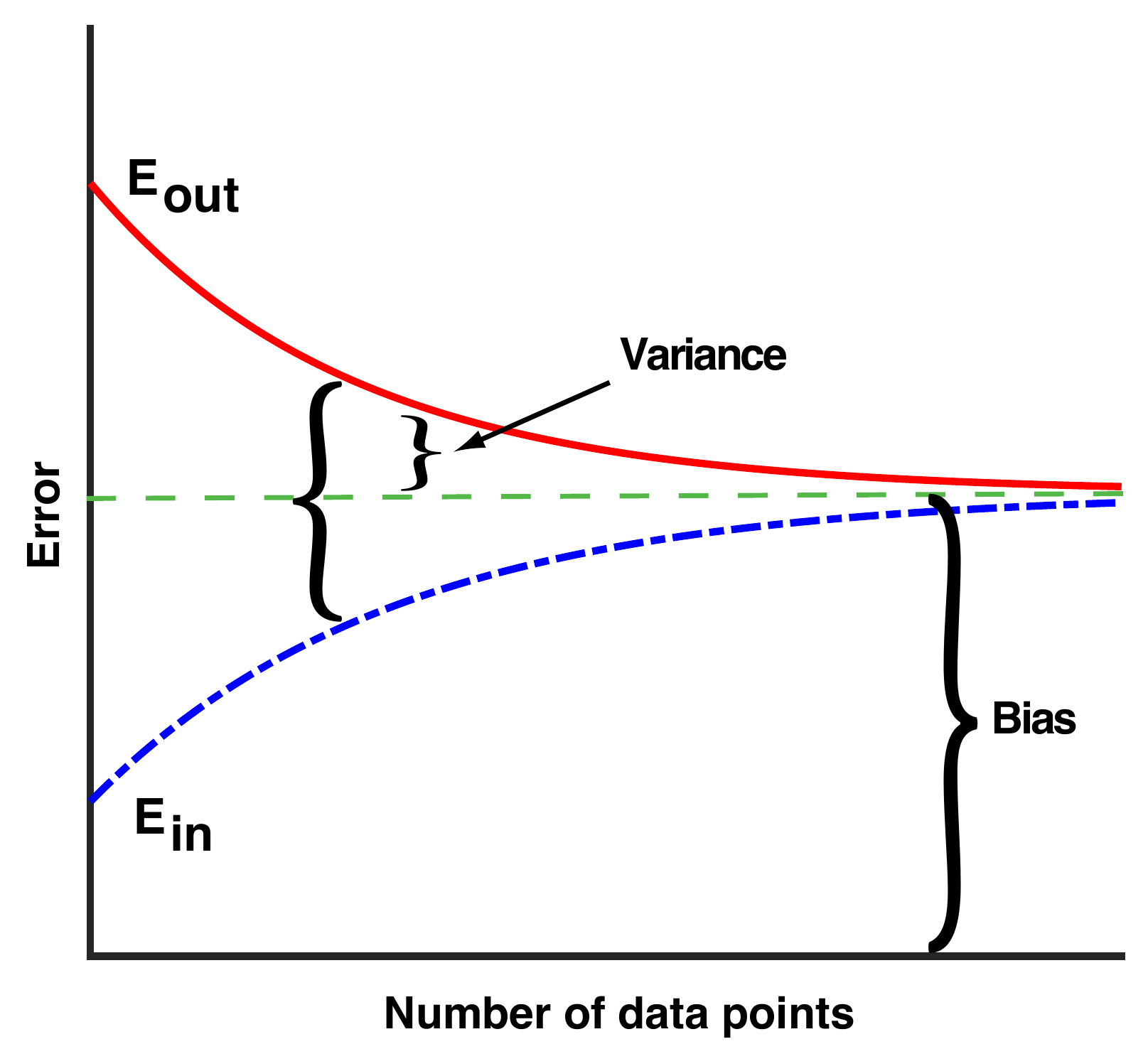} 
  \caption{ {\bf Schematic of typical in-sample and out-of-sample error as a function of training set size}. The typical in-sample or training error, $E_\mathrm{in}$, out-of-sample or generalization error, $E_\mathrm{out}$, bias, variance, and difference of errors as a
  function of the number of training data points. The schematic assumes that the number of data points is large (in particular, the schematic does not show the initial drop in $E_\mathrm{in}$ for small amounts of data), and that our model cannot exactly fit the true function $f(x)$. 
}
 \label{fig:III.1} 
\end{figure}

The basic intuitions of statistical learning can be summarized in three simple schematics. The first schematic, shown in Figure \ref{fig:III.1}, shows the typical out-of-sample error, $E_\mathrm{out}$, and in-sample error, $E_\mathrm{in}$, as a function of the amount of training data. In making this graph, we have assumed that the true data is drawn from a sufficiently complicated distribution, so that we cannot exactly learn the function $f(x)$. Hence, after a quick initial drop (not shown in figure), the in-sample error will increase with the number of data points, because our models are not powerful enough to learn the true function we are seeking to approximate. In contrast, the out-of-sample error will decrease with the number of data points. As the number of data points gets large, the sampling noise decreases and the training data set becomes more representative of the true distribution from which the data is drawn. For this reason, in the infinite data limit, the in-sample and out-of-sample errors must approach the same value, which is called the ``bias'' of our model. 

The bias represents the best our model could do if we had an infinite amount of training data to beat down sampling noise. The bias is a property of the kind of functions, or model class, we are using to approximate $f(x)$. In general, the more complex the model class we use, the smaller the bias. However, we do not generally have an infinite amount of data. For this reason, to get best predictive power it is better to minimize the out-of-sample error, $E_\mathrm{out}$, rather than the bias. As shown in Figure \ref{fig:III.1}, $E_\mathrm{out}$ can be naturally decomposed into a bias, which measures how well we can hypothetically do in the infinite data limit, and a variance, which measures the typical errors introduced in training our model due to sampling noise from having a finite training set.
 
The final quantity shown in Figure \ref{fig:III.1} is the difference between the generalization and training error. It measures how well our in-sample error reflects the out-of-sample error, and measures how much worse we would do on a new data set compared to our training data. For this reason, the difference between these errors is precisely the quantity that measures the difference between fitting and predicting. Models with a large difference between the in-sample and out-of-sample errors are said to ``overfit'' the data. One of the lessons of statistical learning theory is that it is not enough to simply minimize the training error, because the out-of-sample error can still be large. As we will see in our discussion of regression in Sec.~\ref{sec:lin_reg}, this insight naturally leads to the idea of ``regularization''. 

\begin{figure}[t!] 
 \includegraphics[width=1.0\columnwidth]{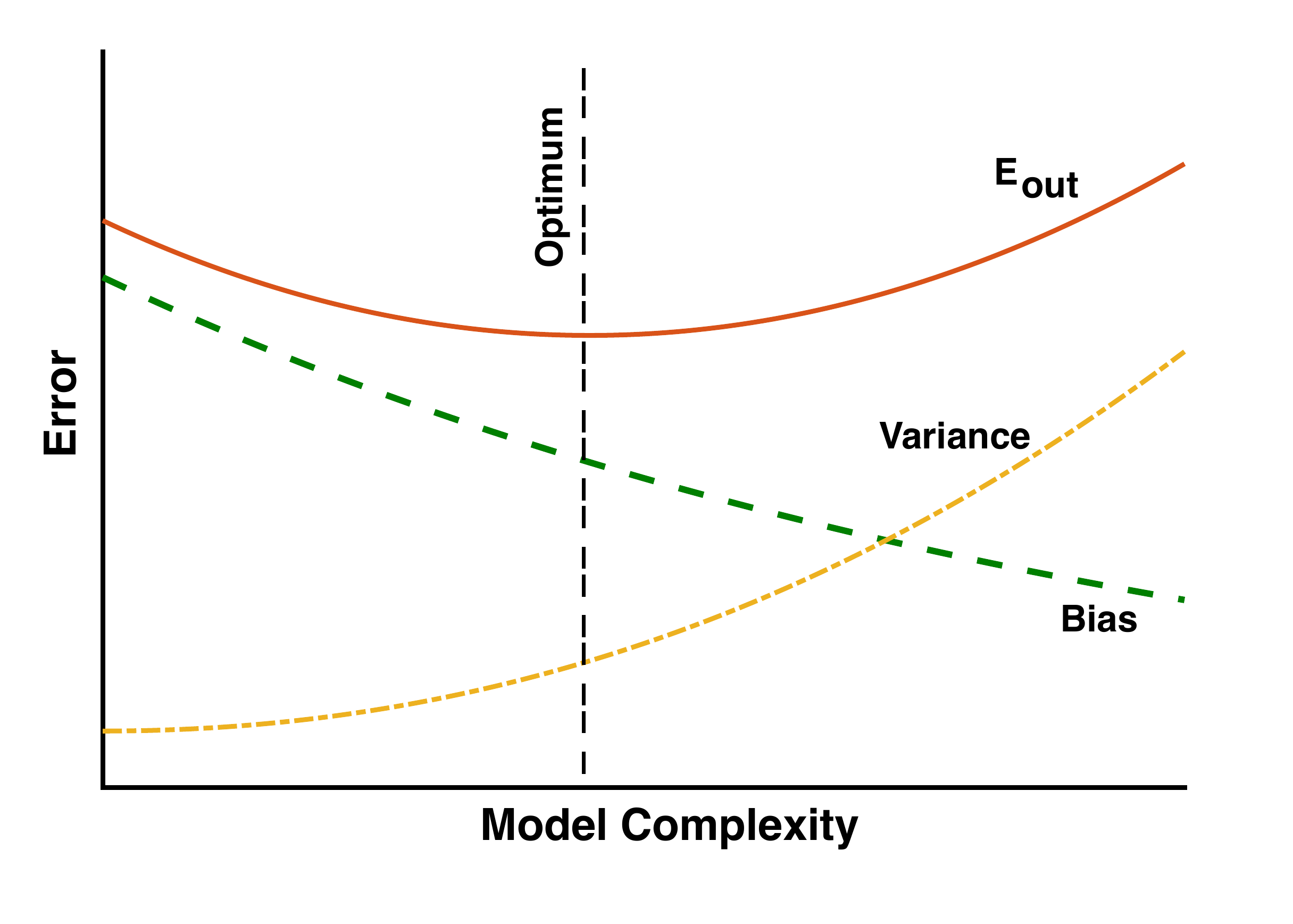}
  \caption{ {\bf Bias-Variance tradeoff and model complexity.} This schematic shows the typical out-of-sample error $E_{out}$ as function of the model complexity for a training dataset of fixed size. Notice how the bias always decreases with
  model complexity, but the variance, i.e. fluctuation in performance due to finite size sampling effects, increases with model complexity. Thus, optimal performance is achieved at intermediate levels of model complexity.}
 \label{fig:III.2} 
\end{figure}

The second schematic, shown in Figure \ref{fig:III.2}, shows the out-of-sample, or test, error $E_\mathrm{out}$ as a function of ``model complexity''. Model complexity is a very subtle idea and defining it precisely is one of the great achievements of statistical learning theory. In many cases, model complexity is related to the number of parameters we are using to approximate the true function $f(x)$\footnote{There are, of course, exceptions. One neat example in the context of one-dimensional regression in given in ~\cite{friedman2001elements}, Figure 7.5.}. In the example of polynomial regression discussed above, higher-order polynomials are more complex than the linear model. If we consider a training dataset of a fixed size, $E_\mathrm{out}$ will be a non-monotonic function of the model complexity, and is generally minimized for models with \emph{intermediate} complexity. The underlying reason for this is that, even though using a more complicated model always reduces the bias, at some point the model becomes too complex for the amount of training data and the generalization error becomes large due to high variance. Thus, to minimize $E_\mathrm{out}$ and maximize our predictive power, it may be more suitable to use a more biased model with small variance than a less-biased model with large variance. This important concept is commonly called the bias-variance tradeoff and gets at the heart of why machine learning is difficult.

\begin{figure}[t!] 
 \includegraphics[width=1.0\columnwidth]{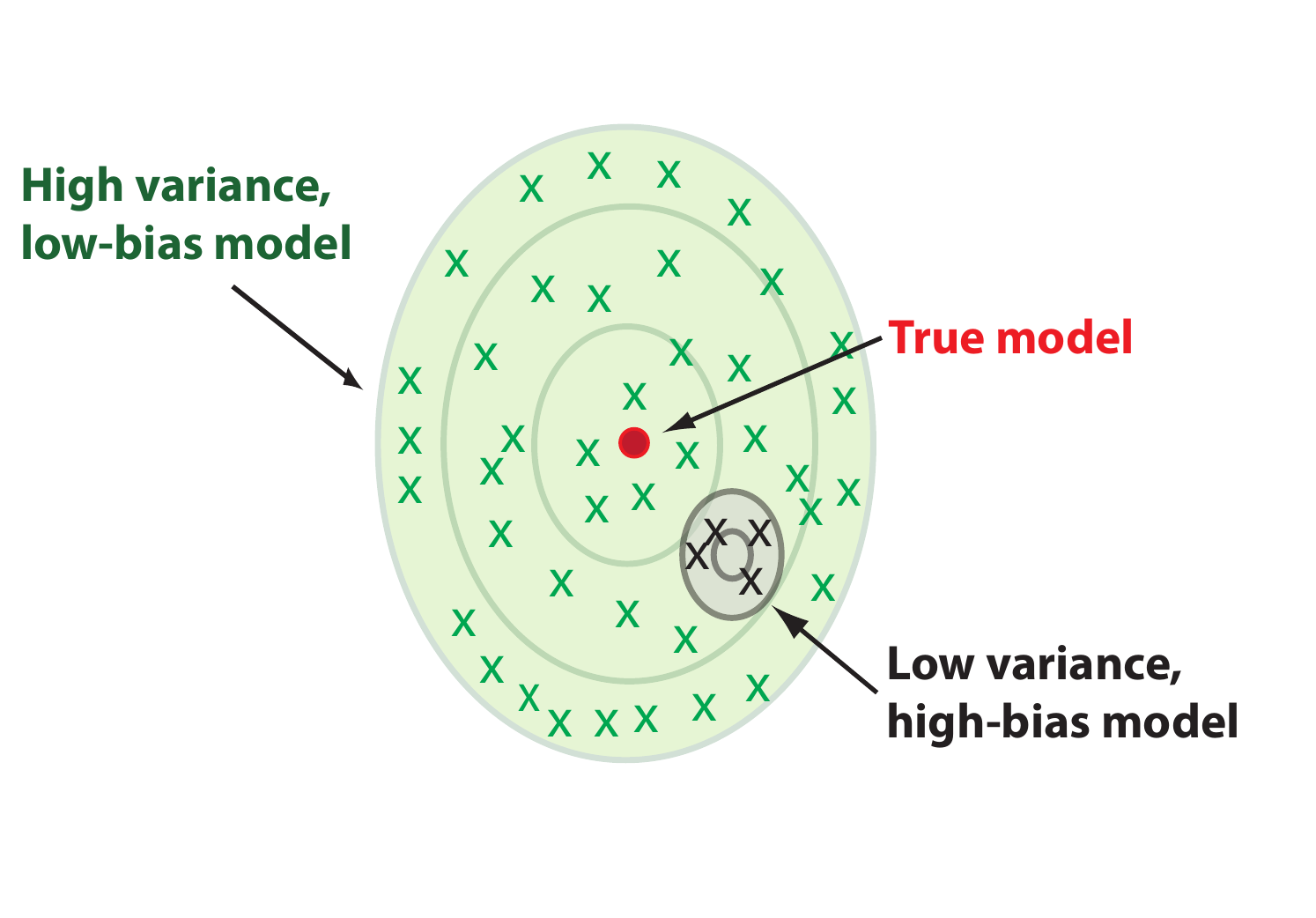} 
  \caption{ {\bf Bias-Variance tradeoff.} Another useful depiction of the bias-variance tradeoff is to think about how $E_\mathrm{out}$ varies as we consider different training data sets of a fixed size. A more complex model (green) will exhibit larger fluctuations (variance) due to finite size sampling effects than the simpler model (black). However, the average over all the trained models (bias) is closer to the true model for the more complex model.}
 \label{fig:III.3} 
\end{figure}

Another way to visualize the bias-variance tradeoff is shown in Figure \ref{fig:III.3}. In this figure, we imagine training a complex model (shown in green) and a simpler model (shown in black) many times on different training sets of a fixed size $N$. Due to the sampling noise from having finite size data sets, the learned models will differ for each choice of training sets. In general, more complex models need a larger amount of training data. For this reason, the fluctuations in the learned models (variance) will be much larger for the more complex model than the simpler model. However, if we consider the asymptotic performance as we increase the size of the training set (the bias), it is clear that the complex model will eventually perform better than the simpler model. Thus, depending on the amount of training data, it may be more favorable to use a less complex, high-bias model to make predictions.

\subsection{\label{subsec:bias_variance_tradeoff}Bias-Variance Decomposition}
 
In this section, we dig further into the central principle that underlies much of machine learning: the bias-variance tradeoff. We will discuss the bias-variance tradeoff in the context of continuous predictions such as regression. However, many of the intuitions and ideas discussed here also carry over to classification tasks. Consider a dataset $\mathcal{D} = (\boldsymbol{X}, \boldsymbol{y})$ consisting of the $N$ pairs of independent and dependent variables. Let us assume that the true data is generated from a noisy model
\be
y=f(\bd{x}) + \epsilon
\ee
where $\epsilon$ is normally distributed with mean zero and standard deviation $\sigma_\epsilon$.

Assume that we have a statistical procedure (e.g. least-squares regression) for forming a predictor $f(\bd{x}; \hat{\boldsymbol{\theta}})$ that gives the prediction of our model for a new data point $\bd{x}$.  This estimator is chosen by minimizing a cost function which we take to be the squared error
\be
 \mathcal{C}( \boldsymbol{y}, f(\boldsymbol{X}; \boldsymbol{\theta})) =  \sum_i (y_i - f(\bd{x}_i; \boldsymbol{\theta}))^2. 
\ee
Therefore, the estimates for the parameters,  
\be
\hat{\boldsymbol{\theta}}_{\mathcal{D}} = \argmin_{\theta}  \mathcal{C}( \boldsymbol{y}, f(\boldsymbol{X}; \boldsymbol{\theta})).
\ee
are a function of the dataset, $\mathcal{D}$. We would obtain a different error $\mathcal{C}( \boldsymbol{y}_j, f(\boldsymbol{X}_j; \hat{\boldsymbol{\theta}}_{\mathcal{D}_j}))$ for each dataset $\mathcal{D}_j = (\boldsymbol{y}_j, \boldsymbol{X}_j)$ in a universe of possible datasets obtained by drawing $N$ samples from the true data distribution. We denote an expectation value over all of these datasets as $\mathbb{E}_{\mathcal{D}}$.

We would also like to average over different instances of the ``noise'' $\epsilon$ and we denote the expectation value over the noise by $\mathbb{E}_\epsilon$. Thus, we can decompose the expected generalization error as
\begin{widetext}
\bea
\mathbb{E}_\mathcal{D, \epsilon}[\mathcal{C}( \boldsymbol{y}, f(\boldsymbol{X}; \hat{\boldsymbol{\theta}}_{\mathcal{D}})) ]
&=& \mathbb{E}_\mathcal{D,\epsilon}\left[ \sum_i ({y}_i - f(\bd{x}_i; \hat{\boldsymbol{\theta}}_{\mathcal{D}}))^2 \right] \nonumber \\
 &=& \mathbb{E}_\mathcal{D, \epsilon}\left[ \sum_{i}({y}_i -f(\bd{x}_i) +f(\bd{x}_i)- f(\bd{x}_i; \hat{\boldsymbol{\theta}}_{\mathcal{D}}))^2\right] \nonumber \\
 &=& \sum_i \mathbb{E}_\epsilon[ ({y}_i -f(\bd{x}_i))^2 ] + \mathbb{E}_\mathcal{D, \epsilon}[(f(\bd{x}_i) - f(\bd{x}_i; \hat{\boldsymbol{\theta}}_{\mathcal{D}}))^2] + 2\mathbb{E}_\epsilon[{y}_i - f(\bd{x}_i)] \mathbb{E}_\mathcal{D}[f(\bd{x}_i)- f(\bd{x}_i; \hat{\boldsymbol{\theta}}_{\mathcal{D}})] \nonumber \\
 &=&\sum_i \sigma_\epsilon^2 + \mathbb{E}_\mathcal{D}[(f(\bd{x}_i) - f(\bd{x}_i; \hat{\boldsymbol{\theta}}_{\mathcal{D}}))^2],
\eea
where in the last line we used the fact that our noise has zero mean and variance $\sigma_\epsilon^2$ and the sum over $i$ applies to all terms. It is also helpful to further decompose the second term as follows:
\bea
\mathbb{E}_\mathcal{D}[(f(\bd{x}_i)- f(\bd{x}_i; \hat{\boldsymbol{\theta}}_{\mathcal{D}}))^2] 
&=& \mathbb{E}_\mathcal{D}[\{f(\bd{x}_i) - \mathbb{E}_\mathcal{D}[f(\bd{x}_i; \hat{\boldsymbol{\theta}}_{\mathcal{D}})] + \mathbb{E}_\mathcal{D}[f(\bd{x}_i; \hat{\boldsymbol{\theta}}_{\mathcal{D}})]- f(\bd{x}_i; \hat{\boldsymbol{\theta}}_{\mathcal{D}})\}^2] \nonumber \\
&=&\mathbb{E}_\mathcal{D}[\{f(\bd{x}_i) - \mathbb{E}_\mathcal{D}[f(\bd{x}_i; \hat{\boldsymbol{\theta}}_{\mathcal{D}})]\}^2] + \mathbb{E}_\mathcal{D}[\{f(\bd{x}_i; \hat{\boldsymbol{\theta}}_{\mathcal{D}}) - \mathbb{E}_\mathcal{D}[f(\bd{x}_i; \hat{\boldsymbol{\theta}}_{\mathcal{D}})]\}^2] \nonumber \\
&&+2\mathbb{E}_\mathcal{D}[\{f(\bd{x}_i) - \mathbb{E}_\mathcal{D}[f(\bd{x}_i; \hat{\boldsymbol{\theta}}_{\mathcal{D}})]\}\{f(\bd{x}_i; \hat{\boldsymbol{\theta}}_{\mathcal{D}}) - \mathbb{E}_\mathcal{D}[f(\bd{x}_i; \hat{\boldsymbol{\theta}}_{\mathcal{D}})]\}] \nonumber \\
&=&(f(\bd{x}_i) - \mathbb{E}_\mathcal{D}[f(\bd{x}_i; \hat{\boldsymbol{\theta}}_{\mathcal{D}})])^2 + 
\mathbb{E}_\mathcal{D}[\{f(\bd{x}_i; \hat{\boldsymbol{\theta}}_{\mathcal{D}}) - \mathbb{E}_\mathcal{D}[f(\bd{x}_i; \hat{\boldsymbol{\theta}}_{\mathcal{D}})]\}^2].
\eea
\end{widetext}
The first term is called the bias
\be
Bias^2= \sum_i (f(\bd{x}_i) - \mathbb{E}_\mathcal{D}[f(\bd{x}_i; \hat{\boldsymbol{\theta}}_{\mathcal{D}})])^2
\ee
and measures the deviation of the expectation value of our estimator (i.e.~the asymptotic value of our estimator in the infinite data limit) from the true value. The second term is called the variance
\be
Var=\sum_i \mathbb{E}_\mathcal{D}[( f(\bd{x}_i; \hat{\boldsymbol{\theta}}_{\mathcal{D}}) - \mathbb{E}_\mathcal{D}[f(\bd{x}_i; \hat{\boldsymbol{\theta}}_{\mathcal{D}})])^2],
\ee
and measures how much our estimator fluctuates due to finite-sample effects. Combining these expressions, we see that the expected out-of-sample error, $E_\mathrm{out} := \mathbb{E}_\mathcal{D, \epsilon}[\mathcal{C}( \boldsymbol{y}, f(\boldsymbol{X}; \hat{\boldsymbol{\theta}}_{\mathcal{D}})) ] $, can be decomposed as
\be
E_\mathrm{out} = Bias^2 + Var + Noise,
\ee
with $Noise = \sum_i \sigma_\epsilon^2$.

The bias-variance tradeoff  summarizes the fundamental tension in machine learning, particularly supervised learning, between the complexity of a model and the amount of training 
data needed to train it.  Since data is often limited, in practice it is often useful to use a less-complex model with  higher bias -- a model whose asymptotic performance is worse than another model --  because it is easier to train and less sensitive to sampling noise arising from having a finite-sized training dataset (smaller variance). This is the basic intuition behind the schematics in Figs.~\ref{fig:III.1},~\ref{fig:III.2}, and~\ref{fig:III.3}.

%% file: sections/gd.tex
                                   
Almost every problem in ML and data science starts with the same ingredients: a dataset $\boldsymbol{X}$, a model $g(\boldsymbol{\theta})$, which is a function of the parameters $\boldsymbol{\theta}$, and a cost function $\mathcal{C}( \boldsymbol{X}, g(\boldsymbol{\theta}))$ that allows us to judge how well the model $g(\boldsymbol{\theta})$ explains the observations $\boldsymbol{X}$. The model is fit by finding the values of $\boldsymbol{\theta}$ that minimize the cost function. 

In this section, we discuss one of the most powerful and widely used classes of methods for performing this minimization -- gradient descent and its generalizations. The basic idea behind these methods is straightforward: iteratively adjust the parameters$\boldsymbol{\theta}$ in the direction where the gradient of the cost function is large and negative. In this way, the training procedure ensures the parameters flow towards a \emph{local} minimum of the cost function. However, in practice gradient descent is full of surprises and a series of ingenious tricks have been developed by the optimization and machine learning communities to improve the performance of these algorithms. 

The underlying reason why training a machine learning algorithm is difficult is that the cost functions we wish to optimize are usually complicated, rugged, non-convex functions in a high-dimensional space with many local minima. To make things even more difficult, we almost never have access to the true function we wish to minimize: instead, we must estimate this function directly from data. In modern applications, both the size of the dataset and the number of parameters we wish to fit is often enormous (millions of parameters and examples). The goal of this chapter is to explain how gradient descent methods can be used to train machine learning algorithms even in these difficult settings.

This chapter seeks to both introduce commonly used methods and give intuition for why they work. We also include some practical tips for improving the performance of stochastic gradient descent \cite{lecun1998efficient, bottou2012stochastic}. To help the reader gain more intuition about gradient descent and its variants, we have developed a \href{https://physics.bu.edu/~pankajm/MLnotebooks.html}{Jupyter notebook} that allows the reader to visualize how these algorithms perform on two dimensional surfaces. The reader is encouraged to experiment with the accompanying notebook whenever a new method is introduced (especially to explore how changing hyper-parameters can affect performance). The reader may also wish to consult useful reviews that cover these topics \cite{ruder2016overview} and this blog \url{http://ruder.io/optimizing-gradient-descent/}.

\subsection{Gradient Descent and Newton's method}

We begin by introducing a simple first-order gradient descent method and comparing and contrasting it with another algorithm, Newton's method. Newton's method is intimately related to many algorithms (conjugate gradient, quasi-Newton methods) commonly used in physics for optimization problems. Denote the function we wish to minimize by  $E(\boldsymbol{\theta})$. 

In the context of machine learning, $E(\boldsymbol{\theta})$ is just the cost function $E(\boldsymbol{\theta})=\mathcal{C}( \boldsymbol{X}, g(\boldsymbol{\theta}))$. As we shall see for linear and logistic regression in Secs.~\ref{sec:lin_reg},~\ref{sec:log_reg}, this energy function can almost always be written as a sum over $n$ data points,
\be
E(\boldsymbol{\theta}) = \sum_{i=1}^n e_i(\mathbf{x}_i, \boldsymbol{\theta}).
\ee
For example, for linear regression $e_i$ is just the mean square-error for data point $i$; for logistic regression, it is the cross-entropy. To make analogy with physical systems, we will often refer to this function as the ``energy''.

\begin{figure}
\includegraphics[width=1.0\linewidth]{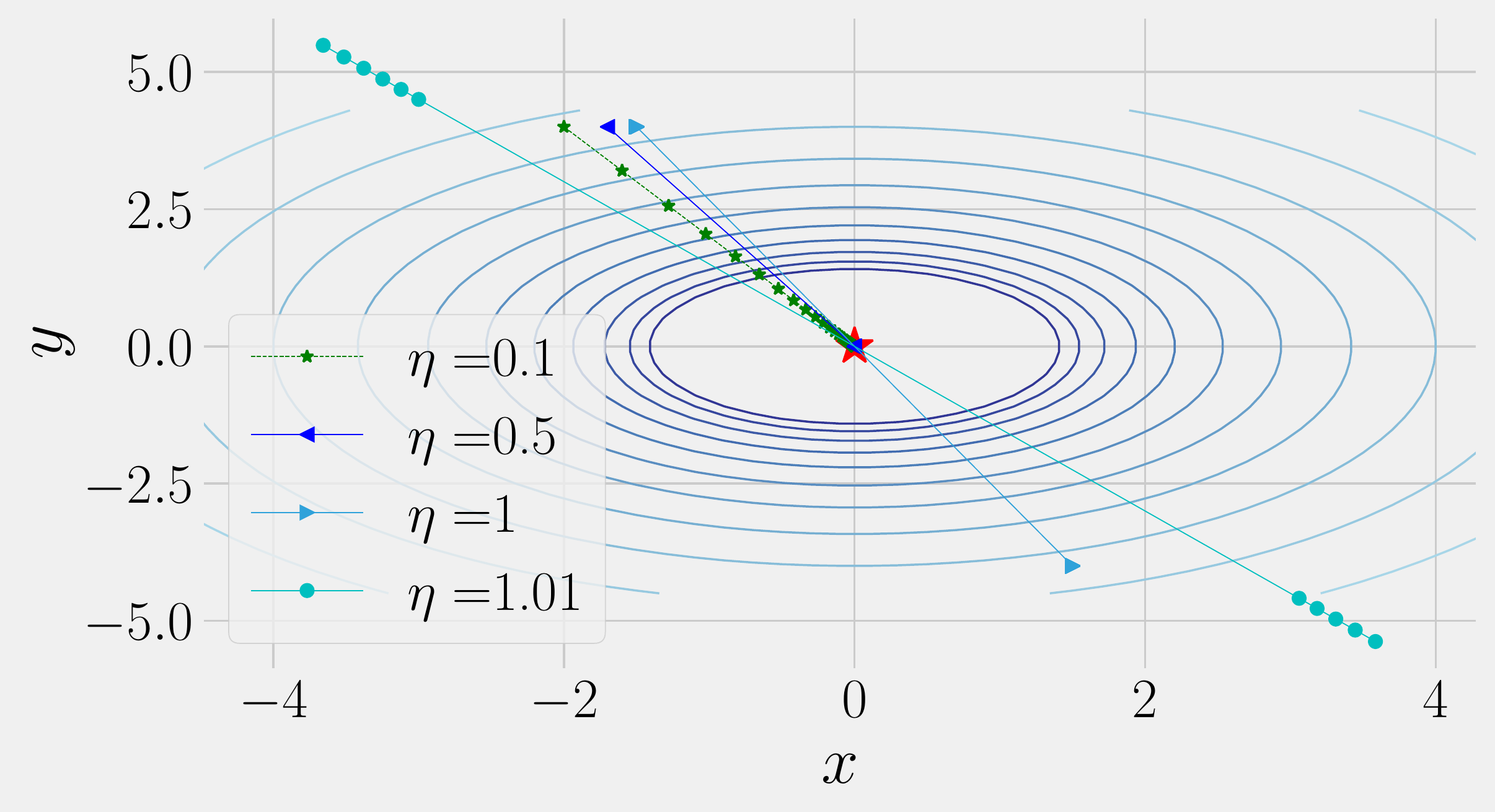} 
\caption{{\bf Gradient descent exhibits three qualitatively different regimes as a function of the learning rate.} Result of gradient descent on surface $z=x^2+y^2-1$ for learning rate of  $\eta=0.1, 0.5, 1.01$.
Notice that the trajectory converges to the global minima in multiple steps for small learning rates ($\eta=0.1$). Increasing the learning rate further ($\eta=0.5$) causes the trajectory to oscillate around the global minima before converging. For even larger learning rates ($\eta=1.01$) the trajectory diverges from the minima.  See corresponding notebook for details.}\label{GD:3regimes}
\end{figure}

In the simplest gradient descent (GD) algorithm, we update the parameters as follows. Initialize the parameters to some value $\boldsymbol{\theta}_0$ and iteratively update the parameters according to the equation
\bea
 \mathbf{v}_t&=& \eta_t \nabla_\theta E(\boldsymbol{\theta}_t), \nonumber \\
 \boldsymbol{\theta}_{t+1}&=& \boldsymbol{\theta}_t -\mathbf{v}_t
\label{GD:gd_def}
\eea
where $\nabla_\theta  E(\boldsymbol{\theta})$ is the gradient of $E(\boldsymbol{\theta})$ w.r.t.~$\boldsymbol{\theta}$ and we have introduced
a \emph{learning rate}, $\eta_t$, that controls how big a step we should take in the direction of the gradient at time step $t$.
It is clear that for sufficiently small choice of the learning rate $\eta_t$  this methods will converge to a \emph{local minimum} (in all directions) of the cost function. However, choosing a small $\eta_t$ comes at a huge computational cost. The smaller $\eta_t$, the more steps we have to take to reach the local minimum. In contrast, if $\eta_t$ is too large, we can overshoot the minimum and the algorithm becomes unstable (it either oscillates or even moves away from the minimum). This is shown in Figure \ref{GD:3regimes}. In practice, one usually specifies a ``schedule'' that decreases $\eta_t$  at long times. Common schedules include power law and exponential decay in time.

To better understand this behavior and highlight some of the shortcomings of GD, it is useful to contrast GD with Newton's method which is the inspiration for many widely employed optimization methods. In Newton's method, we choose the step $\mathbf{v}$ for the parameters in such a way as to minimize a second-order Taylor expansion to the energy function
\be
E(\boldsymbol{\theta}+\mathbf{v}) \approx E(\boldsymbol{\theta} )+ \nabla_\theta E(\boldsymbol{\theta})\mathbf{v} + {1 \over 2} \mathbf{v}^T H(\mathbf{\theta})\mathbf{v},
\nonumber
\ee
where $H(\boldsymbol{\theta})$ is the Hessian matrix of second derivatives.  Differentiating this equation respect to $\mathbf{v}$ and noting that for the optimal value $\mathbf{v}_\mathrm{opt}$ we expect $\nabla_\theta E(\boldsymbol{\theta} +\mathbf{v}_{\mathrm{opt}})=0$, yields  the following equation
\be
0=\nabla_\theta E(\boldsymbol{\theta})+ H(\mathbf{\theta})\mathbf{v}_\mathrm{opt}.
\ee
Rearranging this expression results in the desired update rules for Newton's method
\bea\label{eq:Newton-def1}
\mathbf{v}_t = H^{-1}(\mathbf{\theta}_t)\nabla_\theta E(\boldsymbol{\theta}_t)\\\label{eq:Newton-def2}
 \boldsymbol{\theta}_{t+1}= \boldsymbol{\theta}_t -\mathbf{v}_t.
\eea
Since we have no guarantee that the Hessian is well conditioned, in almost all applications of Netwon's method, one replaces the inverse of the Hessian $ H^{-1}(\mathbf{\theta}_t)$ by some suitably regularized pseudo-inverse such as $[H(\mathbf{\theta}_t)+\epsilon I]^{-1}$ with $\epsilon$ a small parameter \cite{battiti1992first}.

For the purposes of machine learning, Newton's method is not practical for two interrelated reasons. First, calculating a Hessian is an extremely expensive numerical computation. Second, even if we employ  first-order approximation methods to approximate the Hessian (commonly called quasi-Newton methods), we must store and invert a matrix with $n^2$ entries, where $n$ is the number of parameters. For models with millions of parameters such as those commonly employed in the neural network literature, this is close to impossible with present-day computational power.
Despite these practical shortcomings, Newton's method gives many important intuitions about how to modify GD algorithms to improve their performance. Notice that, unlike in GD where the learning rate is the same for all parameters, Newton's method automatically ``adapts'' the learning rate of different parameters depending on the Hessian matrix.
Since the Hessian encodes the curvature of the surface we are trying to find the minimum of -- more specifically, the singular values of the Hessian are inversely proportional to the squares of the local curvatures of the surface -- Newton's method automatically adjusts the step size so that one takes larger steps in flat directions with small curvature and smaller steps in steep directions with large curvature. 

\begin{figure}
\includegraphics[width=1.0\linewidth]{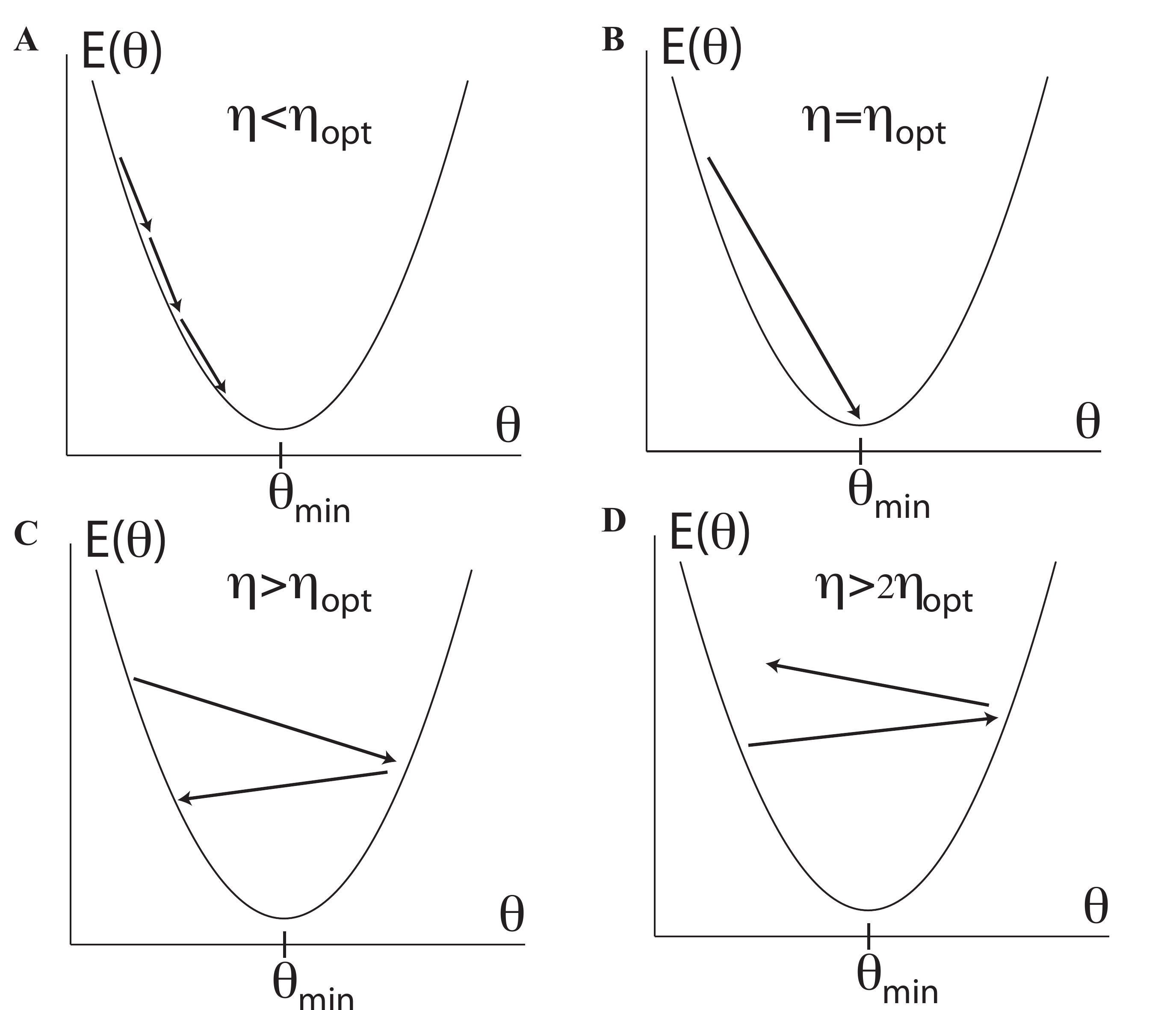} 

\caption{{\bf Effect of learning rate on convergence}. For a one dimensional quadratic potential, one can show that there exists four different qualitative behaviors for gradient descent (GD) as a function of the learning rate $\eta$ depending on the relationship between $\eta$ and $\eta_{\mathrm{opt}}= [\partial_\theta^2 E(\theta)]^{-1}$. (a) For $\eta <\eta_{\mathrm{opt}}$, GD converges to the minimum. (b) For $\eta =\eta_{\mathrm{opt}}$, GD converges in a single step. (c) For $\eta_{\mathrm{opt}} <\eta < 2\eta_{\mathrm{opt}}$, GD oscillates around the minima and eventually converges. (d) For $\eta  > 2\eta_{\mathrm{opt}}$, GD moves away from the minima.  This figure is adapted from \cite{lecun1998efficient}. }
\label{GD:lecunfigure}
\end{figure}

Our derivation of Newton's method also allows us to develop intuition about the role of the learning rate in GD. Let us first consider the special case of using GD to find the minimum of a quadratic energy function of a single parameter $\theta$ \cite{lecun1998efficient}. Given the current value of our parameter $\theta$, we can ask what is the optimal choice of the learning rate $\eta_{\mathrm{opt}}$, where $\eta_{\mathrm{opt}}$ is defined as the value of $\eta$ that allows us to reach the minimum of the quadratic energy function in a single step (see Figure \ref{GD:lecunfigure}). To find $\eta_{\mathrm{opt}}$, we expand the energy function to second order around the current value
\be
E(\theta +v) =E(\theta_c )+ \partial_\theta E(\theta)v+ {1 \over 2}{\partial_\theta^2 E(\theta)}v^2.
\ee
Differentiating with respect to $v$ and setting $\theta_{\mathrm{min}}=\theta-v$  yields
\be
\theta_{\mathrm{min}}=\theta- [\partial_\theta^2 E(\theta)]^{-1} \partial_\theta E(\theta).
\ee
Comparing with (\ref{GD:gd_def}) gives,
\be
\eta_{\mathrm{opt}}= [\partial_\theta^2 E(\theta)]^{-1}.
\label{GD:eta_opt}
\ee

One can show that there are four qualitatively different regimes possible (see Fig. \ref{GD:lecunfigure}) \cite{lecun1998efficient}. If $\eta <\eta_{\mathrm{opt}}$, then GD will take multiple small steps to reach the bottom of the potential. For $\eta=\eta_{\mathrm{opt}}$, GD reaches the bottom of the potential in a single step. If $ \eta_{\mathrm{opt}} < \eta < 2 \eta_{\mathrm{opt}}$, then the GD algorithm will  oscillate across both sides of the potential before eventually converging to the minimum. However, when $\eta  > 2 \eta_{\mathrm{opt}}$, the algorithm actually diverges!

It is straightforward to generalize this to the multidimensional case. The natural multidimensional generalization of the second derivative is the Hessian $H(\theta)$. We can always perform a singular value decomposition (i.e. a rotation by an orthogonal matrix for quadratic minima where the Hessian is symmetric, see Sec.~\ref{subsec:ridge_reg} for a brief introduction to SVD) and consider the singular values $\{\lambda\}$ of the Hessian. If we use a single learning rate for all parameters, in analogy with (\ref{GD:eta_opt}), convergence requires that
\be
\eta < {2 \over \lambda_{\mathrm{max}}},
\ee
where $\lambda_{\mathrm{max}}$ is the largest singular value of the Hessian. If the minimum eigenvalue $\lambda_{\mathrm{min}}$ differs significantly from the largest value $\lambda_{\mathrm{max}}$, then convergence in the $\lambda_{\mathrm{min}}$-direction will be extremely slow! One can actually show that the convergence time scales with the condition number $\kappa= \lambda_{\mathrm{max}} /\lambda_{\mathrm{min}}$ \cite{lecun1998efficient}.

\subsection{Limitations of the simplest gradient descent algorithm}

The last section hints at some of the major shortcomings of the simple GD algorithm described in (\ref{GD:gd_def}). Before proceeding, we briefly summarize these limitations and discuss general strategies for modifying GD to overcome these deficiencies.
\begin{itemize}
	\item  \emph{GD finds local minima of the cost function.} Since the GD algorithm is deterministic, if it converges, it will converge to a local minimum of our energy function. Because in ML we are often dealing with extremely rugged landscapes with many local minima, this can lead to poor performance. A similar problem is encountered in physics. To overcome this, physicists often use methods like simulated annealing that introduce a  fictitious ``temperature'' which is eventually taken to zero. The ``temperature'' term introduces stochasticity in the form of thermal fluctuations that allow the algorithm to thermally tunnel over energy barriers. This suggests that, in the context of ML, we should modify GD to include stochasticity. 
	
	\item \emph{Gradients are computationally expensive to calculate for large datasets}. In many cases in statistics and ML, the energy function is a sum of terms, with one term for each data point. For example, in linear regression, $E \propto \sum_{i=1}^n (y_i - \mathbf{w}^T\cdot\mathbf{x}_i)^2$; for logistic regression, the square error is replaced by the cross entropy, see Secs.~\ref{sec:lin_reg},~\ref{sec:log_reg}. Thus, to calculate the gradient we have 
    to sum over \emph{all} $n$ data points. Doing this at every GD step becomes extremely computationally expensive. An ingenious solution to this, discussed below, is to calculate the gradients using small subsets of the data called ``mini batches''. This has the added benefit of introducing stochasticity into our algorithm.

	\item \emph{GD is very sensitive to choices of the learning rates}. As discussed above, GD is extremely sensitive to the choice of learning rates. If the learning rate is very small, the training process takes an extremely long time. For larger learning rates, GD can diverge and give poor results. Furthermore, depending on what the local landscape looks like, we have to modify the learning rates to ensure convergence. Ideally, we would ``adaptively'' choose the learning rates to match the landscape.

	\item \emph{GD treats all directions in parameter space uniformly.} Another major drawback of GD is that unlike Newton's method, the learning rate for GD is the same in all directions in parameter space. For this reason, the maximum learning rate is set by the behavior of the steepest direction and this can significantly slow down training. Ideally, we would like to take large steps in flat directions and small steps in steep directions. Since we are exploring rugged landscapes where curvatures change, this requires us to keep track of not only the gradient but second derivatives of the energy function (note as discussed above, the ideal scenario would be to calculate the Hessian but this proves to be too computationally expensive). 
	
	\item \emph{ GD is sensitive to initial conditions.} One consequence of the local nature of GD is that initial conditions matter. Depending on where one starts, one will end up at a different local minimum. Therefore, it is very important to think about how one initializes the training process. This is true for GD as well as more complicated variants of GD introduced below.
	
	\item \emph{GD can take exponential time to escape saddle points, even with random initialization.} As we mentioned, GD is extremely sensitive to the initial condition since it determines the particular local minimum GD would eventually reach. However, even with a good initialization scheme, through randomness (to be introduced later), GD can still take exponential time to escape saddle points, which are prevalent in high-dimensional spaces, even for non-pathological objective functions \cite{du2017gradient}. Indeed, there are modified GD methods developed recently to accelerate the escape. The details of these boosted method are beyond the scope of this review, and we refer avid readers to \cite{jin2017accelerated} for details.

\end{itemize}

In the next few subsections, we will introduce variants of GD that address many of these shortcomings. These generalized gradient descent methods form the backbone of much of modern deep learning and neural networks, see Sec~\ref{sec:DNNs}. For this reason, the reader is encouraged to really experiment with different methods in landscapes of varying complexity using the accompanying notebook.

\subsection{Stochastic Gradient Descent (SGD) with mini-batches}

One of the most widely-applied variants of the gradient descent algorithm is stochastic gradient descent (SGD)\cite{bottou2012stochastic, williams1986learning}. As the name suggests, unlike ordinary GD, the algorithm is stochastic. Stochasticity is incorporated by approximating the gradient on a subset of the data called a \emph{minibatch}\footnote{Traditionally, SGD was reserved for the case where you train on a single example -- in other words minibatches of size $1$. However, we will use SGD to mean any approximation to the gradient on a subset of the data.}. The size of the minibatches is almost always much smaller than the total number of data points $n$, with typical minibatch sizes ranging from ten to a few hundred data points. If there are $n$ points in total, and the mini-batch size is $M$, there will be $n/M$ minibatches. Let us denote these minibatches by $B_k$ where $k=1,\ldots,n/M$. Thus, in SGD, at each gradient descent step we approximate the gradient using a single minibatch $B_k$,
\be
\nabla_\theta E(\boldsymbol{\theta})= \sum_{i=1}^n \nabla_\theta e_i(\mathbf{x}_i, \boldsymbol{\theta})\longrightarrow \sum_{i \in B_k} \nabla_\theta e_i(\mathbf{x}_i, \boldsymbol{\theta}).
\ee
We then cycle over all $k=1,\dots,n/M$ minibatches one at a time, and use the mini-batch approximation to the gradient to update the parameters $\boldsymbol{\theta}$ at every step $k$. A full iteration over all $n$ data points  -- in other words using all $n/M$ minibatches -- is called an \emph{epoch}. For notational convenience, we will denote the mini-batch approximation to the gradient by 
\be
\nabla_\theta E^{MB}(\boldsymbol{\theta}) =\sum_{i \in B_k} \nabla_\theta e_i(\mathbf{x}_i, \boldsymbol{\theta}).
\ee
With this notation, we can rewrite the SGD algorithm as
\bea
 \mathbf{v}_t= \eta_t \nabla_\theta E^{MB}(\boldsymbol{\theta}), \nonumber \\
 \boldsymbol{\theta}_{t+1}= \boldsymbol{\theta}_t -\mathbf{v}_t.
\label{GD:sgd_def}
\eea

Thus, in SGD, we replace the actual gradient over the full data at each gradient descent step by an approximation to the gradient computed using a minibatch. This has two important benefits. First, it introduces stochasticity and decreases the chance that our fitting algorithm gets stuck in isolated local minima. Second, it significantly speeds up the calculation as one does not have to use all $n$ data points to approximate the gradient. Empirical and theoretical work suggests that SGD has additional benefits. Chief among these is that introducing
stochasticity is thought to  act as a natural regularizer that prevents overfitting in deep, isolated minima \cite{bishop1995training, keskar2016large}.

\subsection{Adding Momentum}

In practice, SGD is almost always used with a ``momentum'' or inertia term that serves as a memory of the direction we are moving in parameter space.  This is typically 
implemented as follows
\bea
\mathbf{v}_{t}&=&\gamma \mathbf{v}_{t-1}+\eta_{t}\nabla_\theta E(\boldsymbol{\theta}_t) \nonumber \\
\boldsymbol{\theta}_{t+1}&=& \boldsymbol{\theta}_t -\mathbf{v}_{t},
\label{GD:sgd_mom_def}
\eea
where we have introduced a momentum parameter $\gamma$, with $0 \le \gamma \le 1$, and for brevity we dropped the explicit notation to indicate the gradient is to be taken over a different mini-batch at each step. We call this algorithm gradient descent with momentum (GDM). From these equations, it is clear that $\mathbf{v}_t$ is a running average of recently encountered gradients and $(1-\gamma)^{-1}$ sets the characteristic time scale for the memory used in the averaging procedure. Consistent with this, when $\gamma=0$, this just reduces down to ordinary SGD as described in Eq.~\eqref{GD:sgd_def}. An equivalent way of writing the updates is
\be
\Delta \boldsymbol{\theta}_{t+1} = \gamma \Delta \boldsymbol{\theta}_t -\ \eta_{t}\nabla_\theta E(\boldsymbol{\theta}_t),
\label{GD:sgd_mom_def2}
\ee
where we have defined $\Delta \boldsymbol{\theta}_{t}= \boldsymbol{\theta}_t-\boldsymbol{\theta}_{t-1}$. In what should be a familiar scenario to many physicists, momentum based methods were first introduced in old, largely forgotten (until recently) Soviet papers \cite{polyak1964some, nesterov1983method}.

Before proceeding further, let us try to get  more intuition from these equations. It is helpful to consider a simple physical analogy with a particle of mass $m$ moving in a viscous medium with viscous damping coefficient $\mu$ and potential 
$E(\mathbf{w})$ \cite{qian1999momentum}. If we denote the particle's position by $\mathbf{w}$, then its motion is described by 
\be
m {d^2 \mathbf{w} \over dt^2} + \mu {d \mathbf{w} \over dt }= -\nabla_w E(\mathbf{w}).
\ee
We can discretize this equation in the usual way to get
\be
m { \mathbf{w}_{t+\Delta t}-2 \mathbf{w}_{t} +\mathbf{w}_{t-\Delta t} \over (\Delta t)^2}+\mu {\mathbf{w}_{t+\Delta t}- \mathbf{w}_{t} \over \Delta t} = -\nabla_w E(\mathbf{w}).
\ee
Rearranging this equation, we can rewrite this as
\be
\Delta \mathbf{w}_{t +\Delta t}= - { (\Delta t)^2 \over m +\mu \Delta t} \nabla_w E(\mathbf{w})+ {m \over m +\mu \Delta t} \Delta \mathbf{w}_t.
\ee
Notice that this equation is identical to Eq.~\eqref{GD:sgd_mom_def2} if we identify the position of the particle, $\mathbf{w}$, with the parameters $\boldsymbol{\theta}$. This allows
us to identify the momentum parameter and learning rate with the mass of the particle and the viscous damping as:
\begin{equation}
\gamma= {m \over m +\mu \Delta t }, \qquad \eta = {(\Delta t)^2 \over m +\mu \Delta t}.
\end{equation}
Thus, as the name suggests, the momentum parameter is proportional to the mass of the particle and effectively provides inertia.  Furthermore, in the large viscosity/small learning rate limit, our memory time scales as $(1-\gamma)^{-1} \approx  m/(\mu \Delta t)$.

Why is momentum useful? SGD momentum helps the gradient descent algorithm gain speed in directions with persistent but small gradients even in the presence of stochasticity, while suppressing oscillations in high-curvature directions. This becomes especially important in situations where the landscape is shallow and flat in some directions and narrow and steep in others. It has been argued that first-order methods (with appropriate initial conditions) can perform comparable to more expensive second order methods, especially in the context of complex deep learning models \cite{sutskever2013importance}. Empirical 
studies suggest that the benefits of including momentum are especially pronounced in complex models in the initial ``transient phase'' of training, rather than during a subsequent fine-tuning of a coarse minimum. The reason for this is that, in this transient phase, correlations in the gradient persist across many gradient descent steps, accentuating the role of inertia and memory.

These beneficial properties of momentum can sometimes become even more pronounced by using a slight modification of the classical momentum algorithm called Nesterov Accelerated Gradient (NAG)
\cite{sutskever2013importance, nesterov1983method}. In the NAG algorithm, rather than calculating the gradient at the current parameters, $\nabla_\theta E(\boldsymbol{\theta}_t)$, one calculates the gradient at the expected value of the parameters given our current momentum,   $\nabla_\theta E(\boldsymbol{\theta}_t +\gamma \mathbf{v}_{t-1})$. This yields the NAG update rule
\bea
\mathbf{v}_{t}&=&\gamma \mathbf{v}_{t-1}+\eta_{t}\nabla_\theta E(\boldsymbol{\theta}_t +\gamma \mathbf{v}_{t-1}) \nonumber \\
\boldsymbol{\theta}_{t+1}&=& \boldsymbol{\theta}_t -\mathbf{v}_{t}.
\label{GD:NAG_def}
\eea
One of the major advantages of NAG is that it allows for the use of a larger learning rate than GDM for the same choice of $\gamma$.

\subsection{Methods that use the second moment of the gradient}

In stochastic gradient descent, with and without momentum, we still have to specify a  ``schedule'' for tuning the learning rate $\eta_t$ as a function of time.  As discussed in the context of Newton's method, this presents a number of dilemmas. The learning rate is limited by the steepest direction which can change depending on the current position in the landscape. To circumvent this problem, ideally our algorithm would keep track of curvature and take large steps in shallow, flat directions and small steps in steep, narrow directions.  Second-order methods accomplish this by calculating or approximating the Hessian and normalizing the learning rate by the curvature. However, this is very computationally expensive for models with extremely large number of parameters. Ideally, we would like to be able to adaptively change the step size to match the landscape without paying the steep computational price of calculating or approximating Hessians.

Recently, a number of methods have been introduced that accomplish this by tracking not only the gradient, but also the second moment of the gradient. These methods include AdaGrad \cite{duchi2011adaptive}, AdaDelta \cite{zeiler2012adadelta}, RMSprop \cite{tieleman2012lecture}, and ADAM \cite{kingma2014adam}. Here, we discuss the last two as representatives of this class of algorithms. 

In RMSprop, in addition to keeping a running average of the first moment of the gradient, we also keep track of the second moment denoted by $\mathbf{s}_t=\mathbb{E}[\mathbf{g}_t^2]$. The update rule for RMSprop is given by
\bea
\mathbf{g}_t &=& \nabla_\theta E(\boldsymbol{\theta}) \\
\mathbf{s}_t &=&\beta \mathbf{s}_{t-1} +(1-\beta)\mathbf{g}_t^2 \nonumber \\
\boldsymbol{\theta}_{t+1}&=&\boldsymbol{\theta}_t - \eta_t { \mathbf{g}_t \over \sqrt{\mathbf{s}_t +\epsilon}}, \nonumber
\label{GD:RMSprop}
\eea
where $\beta$ controls the averaging time of the second moment and is typically taken to be about $\beta=0.9$, $\eta_t$ is a learning rate typically chosen to be $10^{-3}$, and $\epsilon\sim 10^{-8} $ is a small regularization constant to prevent divergences. Multiplication and division by vectors is understood as an element-wise operation. It is clear from this formula that the learning rate is reduced in directions where the gradient is consistently large. This greatly speeds up the convergence by allowing us to use a larger learning rate for flat directions.

A related algorithm is the  ADAM optimizer. In ADAM, we keep a running average of both the first and second moment of the gradient and use this information to adaptively change the learning rate for different parameters. In addition to keeping a running average of the first and second moments of the gradient (i.e. $\mathbf{m}_t=\mathbb{E}[\mathbf{g}_t]$ and $\mathbf{s}_t=\mathbb{E}[\mathbf{g}^2_t]$, respectively), ADAM performs an additional bias correction to account for the fact that we are estimating the first two moments of the gradient using a running average (denoted by the hats in the update rule below). The update rule for ADAM is given by (where multiplication and division are once again understood to be element-wise operations)
\bea
\mathbf{g}_t &=& \nabla_\theta E(\boldsymbol{\theta}) \\
\mathbf{m}_t &=& \beta_1 \mathbf{m}_{t-1} + (1-\beta_1) \mathbf{g}_t \nonumber \\
\mathbf{s}_t &=&\beta_2 \mathbf{s}_{t-1} +(1-\beta_2)\mathbf{g}_t^2 \nonumber \\
\hat{\mathbf{m}}_t&=&{\mathbf{m}_t \over 1-(\beta_1)^t} \nonumber \\
\hat{\mathbf{s}}_t &=&{\mathbf{s}_t \over1-(\beta_2)^t} \nonumber \\
\boldsymbol{\theta}_{t+1}&=&\boldsymbol{\theta}_t - \eta_t { \hat{\mathbf{m}}_t \over \sqrt{\hat{\mathbf{s}}_t} +\epsilon}, \nonumber \\
\label{GD:ADAM}
\eea
where $\beta_1$ and $\beta_2$ set the memory lifetime of the first and second moment and are typically taken to be $0.9$ and $0.99$ respectively, and $(\beta_j)^t$ denotes $\beta_j$ to the power $t$. The parameters $\eta$ and $\epsilon$ have the same role as in RMSprop. 

Like in RMSprop, the effective step size of a parameter depends on the magnitude of its gradient squared.  To understand this better, let us rewrite this expression in terms of the variance $\boldsymbol{\sigma}_t^2 = \hat{\mathbf{s}}_t - (\hat{\mathbf{m}}_t)^2$. Consider a single parameter $\theta_t$. The update rule for this parameter is given by
\be
\Delta \theta_{t+1}= -\eta_t { \hat{m}_t \over \sqrt{\sigma_t^2 +  \hat m_t^2 }+\epsilon}.
\ee
We now examine different limiting cases of this expression. Assume that our gradient estimates are consistent so that the variance is small. In this case our update rule tends to $\Delta \theta_{t+1} \rightarrow -\eta_t$ (here we have assumed that $\hat{m}_t \gg \epsilon$). This is equivalent to cutting off large persistent gradients at $1$ and limiting the maximum step size in steep directions. On the other hand, imagine that the gradient is widely fluctuating between gradient descent steps. In this case $\sigma^2 \gg \hat{m}_t^2$ so that our update becomes $ \Delta \theta_{t+1} \rightarrow -\eta_t \hat{m}_t/\sigma_t$. In other words, we adapt our learning rate so that it is proportional to the signal-to-noise ratio  (i.e.~the mean in units of the standard deviation). From a physics standpoint, this is extremely desirable: the standard deviation serves as a natural adaptive scale for deciding whether a gradient is large or small. Thus, ADAM has the beneficial effects of (i) adapting our step size so that we cut off large gradient directions (and hence prevent oscillations and divergences), and (ii) measuring gradients in terms of a natural length scale, the standard deviation $\sigma_t$. The discussion above also explains empirical observations showing that  the performance of both ADAM and RMSprop is drastically reduced if the square root is omitted in the update rule. It is also worth noting that recent studies have shown adaptive methods like RMSProp, ADAM, and AdaGrad to generalize worse than SGD in classification tasks, though they achieve smaller training error. Such discussion is beyond the scope of this review so we refer readers to~\cite{wilson2017marginal} for more details.

\subsection{Comparison of various methods}

\begin{figure}
\includegraphics[width=1.0\linewidth]{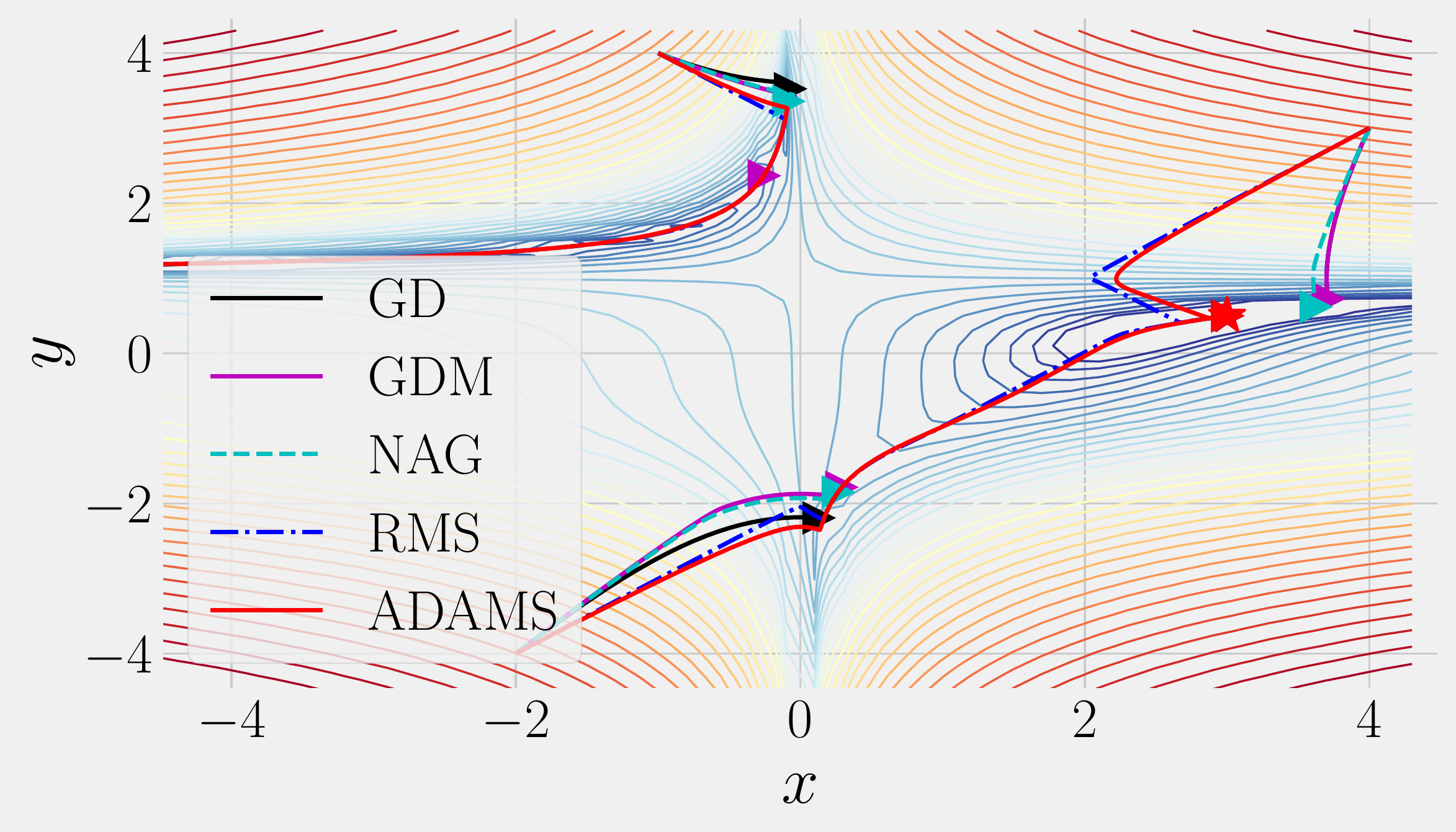} 
\caption{{\bf Comparison of GD and its generalization for Beale's function}. Trajectories from gradient descent (GD; black line), gradient descent with momentum (GDM; magenta line), NAG (cyan-dashed line),  RMSprop (blue dash-dot line), and ADAM (red line) for $N_\mathrm{steps}=10^4$. The learning rate for GD, GDM, NAG is $\eta=10^{-6}$ and $\eta=10^{-3}$ for ADAM and RMSprop.
$\beta=0.9$ for RMSprop, $\beta_1=0.9$ and $\beta_2=0.99$ for ADAM, and $\epsilon=10^{-8}$ for both methods. Please see corresponding notebook for details.}
\label{GD:Beales}
\end{figure}

To better understand these methods, it is helpful to visualize the performance of the five methods discussed above -- gradient descent (GD), gradient descent with momentum (GDM),  NAG, ADAM, and RMSprop. To do so, we will use Beale's function:
\begin{eqnarray}
f(x,y)&=&(1.5-x+xy)^2\\
&& + (2.25-x+xy^2)^2+(2.625-x+xy^3)^2.\nonumber
\end{eqnarray}
This function has a global minimum at $(x,y)=(3,0.5)$ and an interesting structure that can be seen in Fig. \ref{GD:Beales}. The figure shows the results of using all five methods for $N_\mathrm{steps}=10^4$ steps for three different initial conditions. In the figure, the learning rate for GD, GDM, and NAG are set to $\eta=10^{-6}$ whereas RMSprop and ADAM have a learning rate of $\eta=10^{-3}$. The learning rates for RMSprop and ADAM can be set significantly higher than the other methods due to their adaptive step sizes. For this reason, ADAM and RMSprop tend to be much quicker at navigating the landscape than simple momentum based methods (see Fig. \ref{GD:Beales}). Notice that in some cases (e.g.~initial condition of $(-1,4)$), the trajectories do not find the global minimum but instead follow the deep, narrow ravine that occurs along $y=1$.  This kind of landscape structure is generic in high-dimensional spaces where saddle points proliferate. Once again, the adaptive step size and momentum of ADAM and RMSprop allows these methods to traverse the landscape faster than the simpler first-order methods. The reader is encouraged to consult the corresponding \href{https://physics.bu.edu/~pankajm/MLnotebooks.html}{Jupyter notebook} and experiment with changing initial conditions, the cost function surface being minimized, and hyper-parameters to gain more intuition about all these methods.

\subsection{Gradient descent in practice: practical  tips}

We conclude this chapter by compiling some practical tips from experts for getting the best performance from gradient descent based algorithms, especially in the context of deep neural networks discussed later in the review, see Secs.~\ref{sec:DNNs}, and~\ref{sec:RBMs}. This section draws heavily on best practices laid out in \cite{lecun1998efficient, bottou2012stochastic,tieleman2012lecture}.

\begin{itemize}
	\item \emph{Randomize the data when making mini-batches}. It is always important to randomly shuffle the data when forming mini-batches. Otherwise, the gradient descent method can fit spurious correlations resulting from the order in which data is presented.

	\item \emph{Transform your inputs}. As we discussed above, learning becomes difficult when our landscape has a mixture of steep and flat directions. 
	One simple trick for minimizing these situations is to standardize the data by subtracting the mean and normalizing the variance of input variables. Whenever possible, also decorrelate the inputs. To understand why this is helpful, consider the case of linear regression. It is easy to show that for the squared error cost function, the Hessian of the energy matrix is just the correlation matrix between the inputs. Thus, by standardizing the inputs, we are ensuring that the landscape looks homogeneous in all directions in parameter space. Since most deep networks can be viewed as linear transformations followed by a non-linearity at each layer, we expect this intuition to hold beyond the linear case.

	\item \emph{Monitor the out-of-sample performance.} Always monitor the performance of your model on a validation set (a small portion of the training data that is held out of the training process to serve as a proxy for the test set -- see Sec. \ref{sec:DNN_concepts} for more on validation sets). If the validation error starts increasing, then the model is beginning to overfit. Terminate the learning process. This \emph{early stopping} significantly improves performance in many settings.
	
	\item \emph{Adaptive optimization methods do not always have good generalization.} As we mentioned, recent studies have shown that adaptive methods such as ADAM, RMSprop, and AdaGrad tend to have poor generalization compared to SGD or SGD with momentum, particularly in the high-dimensional limit (i.e. the number of parameters exceeds the number of data points) \cite{wilson2017marginal}. Although it is not clear at this stage why sophisticated methods, such as ADAM, RMSprop, and AdaGrad, perform so well in training deep neural networks such as generative adversarial networks (GANs)~\cite{goodfellow2014generative} [see Sec.~\ref{sec:vae}], simpler procedures like properly-tuned plain SGD may work equally well or better in some applications.

\end{itemize}

%% file: sections/BayesianInference.tex
Statistical modeling usually revolves around estimation or prediction~\cite{jaynes1996probability}. Bayesian methods are based on the fairly simple premise that probability can be used as a mathematical framework for describing uncertainty. This is not that different in spirit from the main idea of statistical mechanics in physics, where we use probability to describe the behavior of large systems where we cannot know the positions and momenta of all the particles even if the system itself is fully deterministic (at least classically). In practice, Bayesian inference provides a set of principles and procedures for learning from data and for describing uncertainty. In this section, we give a gentle introduction to Bayesian inference, with special emphasis on its logic (i.e.~Bayesian reasoning) and provide a connection to ML discussed in Sec.~\ref{sec:why_is_ML_difficult} and~\ref{sec:stat_learn_theory}. For a technical account of Bayesian inference in general, we refer readers to ~\cite{gelman2014bayesian,barber2012bayesian}.
\subsection{Bayes Rule}

To solve a problem using Bayesian methods, we have to specify two functions: the \textit{likelihood function} $p(\boldsymbol{X} | \boldsymbol{\theta})$, which describes the probability of observing a dataset $\boldsymbol{X}$ for a given value of the unknown parameters $\boldsymbol{\theta}$, and the \textit{prior distribution} $p(\boldsymbol{\theta})$, which describes any knowledge we have about the parameters before we collect the data. Note that the likelihood should be considered as a function of the parameters $\boldsymbol{\theta}$ with the data $\boldsymbol{X}$ held fixed. The prior distribution and the likelihood function are used to compute the \textit{posterior distribution} $p(\boldsymbol{\theta} | \boldsymbol{X})$ via Bayes' rule:
\begin{equation}
p(\boldsymbol{\theta} | \boldsymbol{X}) = \frac{p(\boldsymbol{X} | \boldsymbol{\theta} ) p(\boldsymbol{\theta}) }{ \int \mathrm{d} \boldsymbol{\theta}' \,p(\boldsymbol{X} | \boldsymbol{\theta}' ) p(\boldsymbol{\theta}')}.
\end{equation}
The posterior distribution describes our knowledge about the unknown parameter $\boldsymbol{\theta}$ after observing the data $\boldsymbol{X}$. In many cases, it will not be possible to analytically compute the normalizing constant in the denominator of the posterior distribution, i.e.~$p(\boldsymbol{X}) = \int d \boldsymbol{\theta}\, p(\boldsymbol{X} | \boldsymbol{\theta} ) p(\boldsymbol{\theta})$, and Markov Chain Monte Carlo (MCMC) methods are needed to draw random samples from $p(\boldsymbol{\theta} | \boldsymbol{X})$. 

The likelihood function $p(\boldsymbol{X} | \boldsymbol{\theta})$ is a common feature of both classical statistics and Bayesian inference, and is determined by the model and the measurement noise. Many common statistical 
procedures such as least-square fitting can be cast as \emph{Maximum Likelihood Estimation} (MLE). In MLE, one chooses the parameters $\hat{\boldsymbol{\theta}}$ that maximize the likelihood (or equivalently the log-likelihood since log is a monotonic function) of the observed data:
\be 
\hat{\boldsymbol{\theta}} = \argmax_{\boldsymbol{\theta}} \log{p(\boldsymbol{X} | \boldsymbol{\theta})}.
\ee
In other words, in MLE we choose the parameters that maximize the probability of seeing the observed data given our generative model. MLE is an important concept in both frequentist and Bayesian statistics.

The prior distribution, by contrast, is uniquely Bayesian. There are two general classes of priors: if we do not have any specialized knowledge about $\boldsymbol{\theta}$ before we look at the data then we would like to select an \textit{uninformative} prior that reflects our ignorance, otherwise we should select an \textit{informative} prior that accurately reflects the knowledge we have about $\boldsymbol{\theta}$. This review will focus on \textit{informative} priors that are commonly used for ML applications. However, there is a large literature on \textit{uninformative} priors, including reparameterization invariant priors, that would be of interest to physicists and we refer the interested reader to~\cite{jeffreys1946invariant,jaynes1996probability,berger1992development,gelman2014bayesian,Mattinglyignorance2018}.

Using an informative prior tends to decrease the variance of the posterior distribution while, potentially, increasing its bias. This is beneficial if the decrease in variance is larger than the increase in bias. In high-dimensional problems, it is reasonable to assume that many of the parameters will not be strongly relevant. Therefore, many of the parameters of the model will be zero or close to zero. We can express this belief using two commonly used priors: the Gaussian prior $p(\boldsymbol{\theta} | \lambda) = \prod_{j} \sqrt{\frac{\lambda}{2 \pi}} e^{-\lambda \theta_j^2}$ is used to express the assumption that many of the parameters will be small, and the Laplace prior $p(\boldsymbol{\theta} | \lambda) = \prod_{j} \frac{\lambda}{2} e^{-\lambda | \theta_j|}$ is used to express the assumption that many of the parameters will be zero. We'll come back to this point later in Sec.~\ref{subsec:bayes_vs_freq_reg}.

\subsection{Bayesian Decisions}

The above section presents the tools for computing the posterior distribution $p(\boldsymbol{\theta} | \boldsymbol{X})$, which uses probability as a framework for expressing our knowledge about the parameters $\boldsymbol{\theta}$. In most cases, however, we need to summarize our knowledge and pick a single ``best'' value for the parameters. In principle, the specific value of the parameters should be chosen to maximize a utility function. In practice, however, we usually use one of two choices: the posterior mean $\langle \boldsymbol{\theta} \rangle = \int \mathrm{d} \boldsymbol{\theta}\, \boldsymbol{\theta} p(\boldsymbol{\theta} | \boldsymbol{X})$, or the posterior mode $\hat{\boldsymbol{\theta}}_\text{MAP} = \argmax_{\boldsymbol{\theta}} p(\boldsymbol{\theta} | \boldsymbol{X})$.  Often, $\langle \boldsymbol{\theta} \rangle$ is called the Bayes estimate and $\hat{\boldsymbol{\theta}}_\text{MAP}$ is called the \textit{maximum-a-posteriori} or MAP estimate. While the Bayes estimate minimizes the mean-squared error, the MAP estimate is often used instead because it is easier to compute. 

\subsection{Hyperparameters}
\label{subsec:BayesianHyperpara}
The Gaussian and Laplace prior distributions, used to express the assumption that many of the model parameters will be small or zero, both have an extra parameter $\lambda$. This \textit{hyperparameter} or \textit{nuisance variable} has to be chosen somehow. One standard Bayesian approach is to define another prior distribution for $\lambda$ -- usually using an uninformative prior -- and to average the posterior distribution over all choices of $\lambda$. This is called a hierarchical prior. Computing averages, however, often requires long Markov Chain Monte Carlo simulations that are computationally intensive. Therefore, it is simpler if we can find a good value of $\lambda$ using an optimization procedure instead. We will discuss how this is done in practice when discussing linear regression in Sec.~\ref{sec:lin_reg}.

%% file: sections/lin_reg.tex
In Section~\ref{sec:why_is_ML_difficult}, we performed our first numerical ML experiments by fitting datasets generated by polynomials in the presence of different levels of additive noise. We used the fitted parameters to make predictions on `unseen' observations,  allowing us to gauge the performance of our model on new data. These experiments highlighted the fundamental tension common to all ML models between how well we fit the training dataset and  predictions on new data. The optimal choice of predictor depended on, among many other things, the functions used to fit the data and the underlying noise level. In Section~\ref{sec:stat_learn_theory}, we formalized this by introducing the notion of model complexity and the bias-variance decomposition, and discussed the statistical meaning of learning. In this section, we take a closer look at these ideas in the simple setting of linear regression. 

As in Section~\ref{sec:why_is_ML_difficult}, fitting a given set of samples $(y_i, {\bf x}_i)$ means relating the independent variables ${\bf x}_i$ to their responses $y_i$. For example, suppose we want to see how the voltage across two sides of a metal slab $V$ changes in response to the applied electric current $I$. Normally we would first make a bunch of measurements labeled by $i$ and plot them on a two-dimensional scatterplot, $(V_i,I_i)$. The next step is to assume, either from an oracle or from theoretical reasoning, some \emph{models} that might explain the measurements and measuring their performance. Mathematically, this amounts to finding some function $f$ such that $V_i=f(I_i; w)$, where $w$ is some parameter  (e.g.~the electrical resistance $R$ of the metal slab in the case of Ohm's law). We then try to minimize the errors made in explaining the given set of measurements based on our model $f$ by tuning the parameter $w$. To do so, we need to first define the error function (formally called the \emph{loss function}) that characterizes the deviation of our prediction from the actual response. 

Before formulating the problem, let us set up the notation. Suppose we are given a dataset with $n$ samples $\mathcal{D}=\{(y_i,\bd{x}^{(i)})\}_{i=1}^n$, where $\bd{x}^{(i)}$ is the $i$-th observation vector while $y_i$ is its corresponding (scalar) response. We assume that every sample has  $p$ features, namely, $\bd{x}^{(i)}\in\mathbb{R}^p$. Let $f$ be the true function/model that generated these samples via $y_i=f(\bd{x}^{(i)};\bd{w}_{\text{true}})+\epsilon_i$, where $\bd{w}_{\text{true}}\in\mathbb{R}^p$ is a parameter vector and $\epsilon_i$ is some i.i.d.~white noise with zero mean and finite variance. Conventionally, we cast all samples into an $n\times p$ matrix, $\bd{X}\in\mathbb{R}^{n\times p}$, called the \emph{design matrix}, with the rows $\bd{X}_{i,:} = \bd{x}^{(i)}\in\mathbb{R}^p,, i=1,\cdots, n$ being observations and the columns $\bd{X}_{:,j}\in\mathbb{R}^n, j = 1,\cdots p$ being measured features. Bear in mind that this function $f$ is never known to us explicitly, though in practice we usually presume its functional form. For example, in \emph{linear regression}, we assume $y_i=f(\bd{x}^{(i)};\bd{w}_\text{true}) +\epsilon_i = \bd{w}_\text{true}^\text{T} \bd{x}^{(i)}+\epsilon_i$ for some unknown but fixed $\bd{w}_\text{true}\in\mathbb{R}^p$.

We want to find a function $g$ with parameters $\bd{w}$ fit to the data $\mathcal{D}$ that can best approximate $f$. When this is done, meaning we have found a $\hat{\bd{w}}$ such that $g(\bd{x};\hat{\bd{w}})$ yields our best estimate of $f$, we can use this $g$ to make predictions about the response $y_0$ for a new data point $\bd{x}_0$, as we did in Section~\ref{sec:why_is_ML_difficult}. 

It will be helpful for our discussion of linear regression to define one last piece of notation. For any real number $p\ge 1$, we define the L$^p$ norm of a vector $\bd{x}=(x_1,\cdots, x_d)\in\mathbb{R}^d$ to be
\be
||\bd{x}||_p=(|x_1|^p+\cdots+ |x_d|^p)^{\frac{1}{p}}
\ee

\subsection{Least-square regression}
\label{subsec:least_sq_reg}

\begin{figure}[h!]
	\includegraphics[width=0.5\columnwidth]{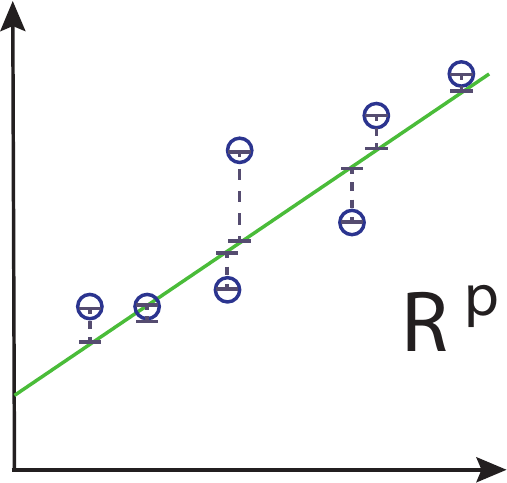}
	\caption{\label{fig:LSGeo} Geometric interpretation of least squares regression. The regression function $g$ defines a hyperplane in $\mathbb{R}^p$ (green solid line, here we have $p=2$) while the residual of data point $\bd{x}^{(i)}$ (hollow circles) is its projection onto this hyperplane (bar-ended dashed line).  }
\end{figure}

\begin{figure}[h!]
\includegraphics[width=0.6\columnwidth]{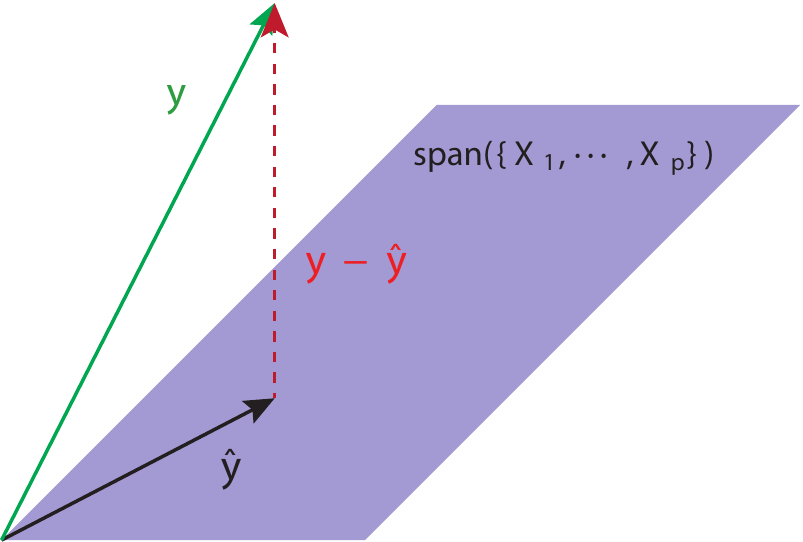}
\caption{\label{fig:LSprojection} The projection matrix $P_{\bd{X}}$ projects the response vector $\bd{y}$ onto the column space spanned by the columns of $\bd{X}$, span$(\{\bd{X}_{:,1},\cdots, \bd{X}_{:,p}\})$ (purple area), thus forming a fitted vector $\hat{\bd{y}}$. The residuals in Eq.~\eqref{eq:LSminimize} are illustrated by the red vector $\bd{y}-\hat{\bd{y}}$.}
\end{figure}

\emph{Ordinary least squares linear regression} (OLS) is defined as the minimization of the L$_2$ norm of the difference between the response $y_i$ and the predictor $g(\bd{x}^{(i)};\bd{w}) = \bd{w}^\text{T}\bd{x}^{(i)}$:
\be\label{eq:LSminimize}
\min _{\bd{w}\in\mathbb{R}^p} ||\bd{X}\bd{w}-\bd{y}||_2^2=\min _{\bd{w}\in\mathbb{R}^p}\sum_{i=1}^n ( \bd{w}^\text{T}\bd{x}^{(i)}-y_i)^2.
\ee
In other words, we are looking to find the $\bd{w}$ which minimizes the $L_2$ error. Geometrically speaking, the predictor function $g(\bd{x}^{(i)};\bd{w}) =  \bd{w}^\text{T}\bd{x}^{(i)}$ defines a hyperplane in $\mathbb{R}^p$. Minimizing the least squares error is therefore equivalent to minimizing the sum of all projections (i.e.~residuals) for all points $\bd{x}^{(i)}$ to this hyperplane (see Fig.~\ref{fig:LSGeo}). Formally, we denote the solution to this problem as $\hat{\bd{w}}_\text{LS}$:

\be\label{eq:LSproblem}
\hat{\bd{w}}_\text{LS}=\argmin _{\bd{w}\in\mathbb{R}^p} ||\bd{Xw}-\bd{y}||_2^2,
\ee
which, {after straightforward} differentiation, {leads to}
\be\label{eq:LSsoln}
\hat{\bd{w}}_\text{LS} =(\bd{X}^{\text{T}}\bd{X})^{-1}\bd{X}^\text{T}\bd{y}.
\ee
Note that we have assumed that $\bd{X}^{\text{T}}\bd{X}$ is invertible, which is often the case when $n\gg p$. Formally speaking, if rank$(\bd{X})=p$, namely, the predictors $\bd{X}_{:,1},\dots, \bd{X}_{:,p}$ (i.e. columns of $\bd{X}$) are linearly independent, then $\hat{\bd{w}}_\text{LS}$ is unique. {In the case of} rank$(\bd{X})<p$, which happens when $p>n$, $\bd{X}^{\text{T}}\bd{X}$ is singular, implying there are infinitely many solutions to the least squares problem, Eq.~\eqref{eq:LSproblem}. In this case, one can easily show that if ${\bd{w}_0}$ is a solution, ${\bd{w}_0}+\bd{\eta}$ is also a solution for any {$\bd{\eta}$ which satisfies $\bd{X\eta}=\bd{0}$} (i.e. $\bd{\eta}\in$ null$(\bd{X})$). {Having determined} the least squares solution, we can calculate {$\hat{\bd{y}}$}, the {best fit} of {our data} $\bd{X}$, {as} $\hat{\bd{y}}=\bd{X}\hat{\bd{w}}_\text{LS}=P_{\bd{X}} \bd{y}$, where $P_{\bd{X}}=\bd{X}(\bd{X}^{\text{T}}\bd{X})^{-1}\bd{X}^{\text{T}}$, c.f.~Eq.~\eqref{eq:LSminimize}. Geometrically, $P_{\bd{X}}$ is the projection matrix which acts on $\bd{y}$ and projects it onto the column space of $\bd{X}$, which is spanned by the predictors $\bd{X}_{:,1},\cdots, \bd{X}_{:,p}$ (see FIG.~\ref{fig:LSprojection}). Notice that we found the optimal solution $\hat{\boldsymbol{w}}_\text{LS}$ in one shot, without doing any sort of iterative optimization like that discussed in Section~\ref{sec:gradient_descent}.

In Section~\ref{sec:stat_learn_theory} we explained that the difference between learning and fitting lies in the prediction on ``unseen" data. It is therefore necessary to examine the out-of-sample error. 
For a more refined argument on {the role of} out-of-sample errors {in linear regression}, we encourage the reader to do the exercises in the corresponding \href{https://physics.bu.edu/~pankajm/MLnotebooks.html}{Jupyter notebooks}. The upshot is, following our definition of $\bar{E}_\text{in}$ and $\bar{E}_\text{out}$ in Section~\ref{sec:stat_learn_theory}, the average in-sample and out-of-sample error can be shown to be
\bea
\bar{E}_\text{in}&=&\sigma^2\left(1-\frac{p}{n}\right)\\
\bar{E}_\text{out}&=&\sigma^2\left(1+\frac{p}{n}\right),
\eea
provided  we obtain the least squares solution $\hat{\bd{w}}_\text{LS}$ from i.i.d.~samples $\bd{X}$ and $\bd{y}$ generated through $\bd{y}= \bd{X}\bd{w}_\text{true} +\bd{\epsilon}$~\footnote{This requires that $\bd{\epsilon}$ is a noise vector whose elements are i.i.d.~of zero mean and variance $\sigma^2$, and is independent of the samples $\bd{X}$.}. Therefore, we can calculate the average generalization error explicitly: 
\be
|\bar{E}_\text{in}-\bar{E}_\text{out}|=2\sigma^2 \frac{p}{n}.
\ee
This imparts an important message: if we have $p\gg n$ (i.e.~high-dimensional data), the generalization error is extremely large, meaning the model is not learning. Even when we have $p\approx n$, we might still not learn well due to the intrinsic noise $\sigma^2$. One way to ameliorate this is, as we shall see in the following few sections, to use regularization. We will mainly focus on two forms of regularization: {the first one employs an} $L_2$ penalty {and is} called \emph{Ridge regression}, while the second uses an $L_1$ penalty {and is} called \emph{LASSO}.

\subsection{Ridge-Regression}
\label{subsec:ridge_reg}
In this section, we study the effect of adding to the least squares loss function a \emph{regularizer} defined as the $L_2$ norm of the parameter vector we wish to optimize over. In other words, we {want to} solve the following {\it penalized }regression problem called \emph{Ridge regression}:
\be\label{eq:RidgeProbPenalized}
\hat{\bd{w}}_\text{Ridge}(\lambda)=\argmin _{\bd{w}\in\mathbb{R}^p} \left( ||\bd{Xw}-\bd{y}||_2^2 + \lambda ||\bd{w}||_2^2 \right).
\ee
This problem is equivalent to the following {\it constrained} optimization problem
\be\label{eq:RidgeProbConstrained}
\hat{\bd{w}}_\text{Ridge}(t)=\argmin _{\bd{w}\in\mathbb{R}^p:\ ||\bd{w}||_2^2 \le t} ||\bd{Xw}-\bd{y}||_2^2.
\ee
This means that for any $t\ge 0$ and solution $\hat{\bd{w}}_\text{Ridge}$ in Eq.~\eqref{eq:RidgeProbConstrained}, there exists a value $\lambda\ge 0$ such that $\hat{\bd{w}}_\text{Ridge}$ solves Eq.~\eqref{eq:RidgeProbPenalized}, and vice versa\footnote{Note that the equivalence between the penalized and the constrained (regularized) form of least square optimization does not always hold. It holds for Ridge and LASSO (introduced later), but not for best subset selection which is defined by choosing a $L^0$ norm: $\lambda||\bd{w}||_0$. In this case, for every $\lambda>0$ and any $\hat{\bd{w}}_{\text{BS}}$ that solves the penalized form of best subset selection, there is a value $t\ge 0$ such that $\hat{\bd{w}}_{\text{BS}}$ also solves that constrained form of best subset selection, but the converse is not true.}. With this equivalence, it is obvious that by adding a {regularization term}, $\lambda ||\bd{w}||_2^2$, to our least squares loss function, we are effectively constraining the magnitude of the parameter vector learned from the data.

To see this, let us solve Eq.~\eqref{eq:RidgeProbPenalized} explicitly. {Differentiating w.r.t.~$\bd{w}$, we obtain},
\be\label{eq:RidgeSoln}
\hat{\bd{w}}_\mathrm{Ridge}(\lambda) = (\bd{X}^{\text{T}}\bd{X} + \lambda I_{p\times p})^{-1}\bd{X}^{\text{T}}\bs{y}.
\ee
In fact, when $\bd{X}$ is orthogonal, one can simplify this expression further:
\be\label{eq:RidgesolnXorthog}
\hat{\bd{w}}_\mathrm{Ridge}(\lambda)=\frac{\hat{\bd{w}}_\mathrm{LS}}{1+\lambda},\quad\text{ for orthogonal } \bd{X},
\ee
where $\hat{\bd{w}}_\mathrm{LS}$ is the least squares solution given by Eq.~\eqref{eq:LSsoln}. This implies that the ridge estimate is merely the least squares estimate scaled by a factor $(1+\lambda)^{-1}$. 

Can we derive a similar relation between the fitted vector $\hat{\bd{y}}=\bd{X}\hat{\bd{w}}_\text{Ridge}$ and the prediction made by least squares linear regression? To answer this, let us do a singular value decomposition (SVD) on $X$. Recall that the SVD of an $n\times p$ matrix $\bd{X}$ has the form
\be\label{eq:SVDdef}
\bd{X}=\bd{UDV}^{\text{T}},
\ee
where $\bd{U}\in \mathbb{R}^{n\times p}$ and $\bd{V}\in\mathbb{R}^{p\times p}$ are orthogonal matrices such that the columns of $\bd{U}$ span the column space of $\bd{X}$ while the columns of $\bd{V}$ span the row space of $\bd{X}$. $\bd{D}\in \mathbb{R}^{p\times p}=$diag$(d_1,d_2,\cdots, d_p)$ is a diagonal matrix with entries $d_1\ge d_2 \ge\cdots d_p\ge 0$ called the singular values of $\bd{X}$. Note that $\bd{X}$ is singular if there is {at least} one  $d_j=0$. By writing $\bd{X}$ in terms of its SVD, one can recast the Ridge estimator Eq.~\eqref{eq:RidgeSoln} as 
\be
\hat{\bd{w}}_\text{Ridge}=\bd{V}(\bd{D}^2+\lambda \bd{I})^{-1}\bd{DU}^{\text{T}}\bd{y},
\ee 
which implies that the Ridge predictor satisfies
\bea
\hat{\bd{y}}_\text{Ridge}&=&\bd{X}\hat{\bd{w}}_\text{Ridge}\nonumber\\ 
&=&\bd{UD}(\bd{D}^2+\lambda \bd{I})^{-1}\bd{DU}^{\text{T}} \bd{y}\nonumber \\
&=& \sum_{j=1}^p \bd{U}_{:,j} \frac{d_j^2}{d_j^2+\lambda}\bd{U}_{:j}^{\text{T}}\bd{y}\label{eq:RidgePredictSVD}\\
&\le& \bd{UU}^{\text{T}}\bd{y}\\
&=&\bd{X}\hat{\bd{y}}\equiv\hat{\bd{y}}_\text{LS}\label{eq:LSPredictSVD},
\eea
where $\bd{U}_{:,j}$ are the columns of $\bd{U}$. Note that in the inequality step we assumed $\lambda\ge 0$ and used SVD to simplify Eq.~\eqref{eq:LSsoln}. By comparing Eq.~\eqref{eq:RidgePredictSVD} with Eq.~\eqref{eq:LSPredictSVD}, it is clear that in order to compute the fitted vector $\hat{\bs{y}}$, both Ridge and least squares linear regression have to project $\bd{y}$ to the column space of $\bd{X}$. The only difference is that Ridge regression further shrinks each basis component $j$ by a factor $d_j^2/(d_j^2+\lambda)$. We encourage the reader to do the exercises in \href{https://physics.bu.edu/~pankajm/MLnotebooks.html}{Notebook 3} to develop further intuition about {how Ridge regression works}.

\subsection{LASSO and Sparse Regression}
\label{subsec:lasso_sparse_reg}
\begin{figure}[h!]
		\includegraphics[width=1.0\columnwidth]{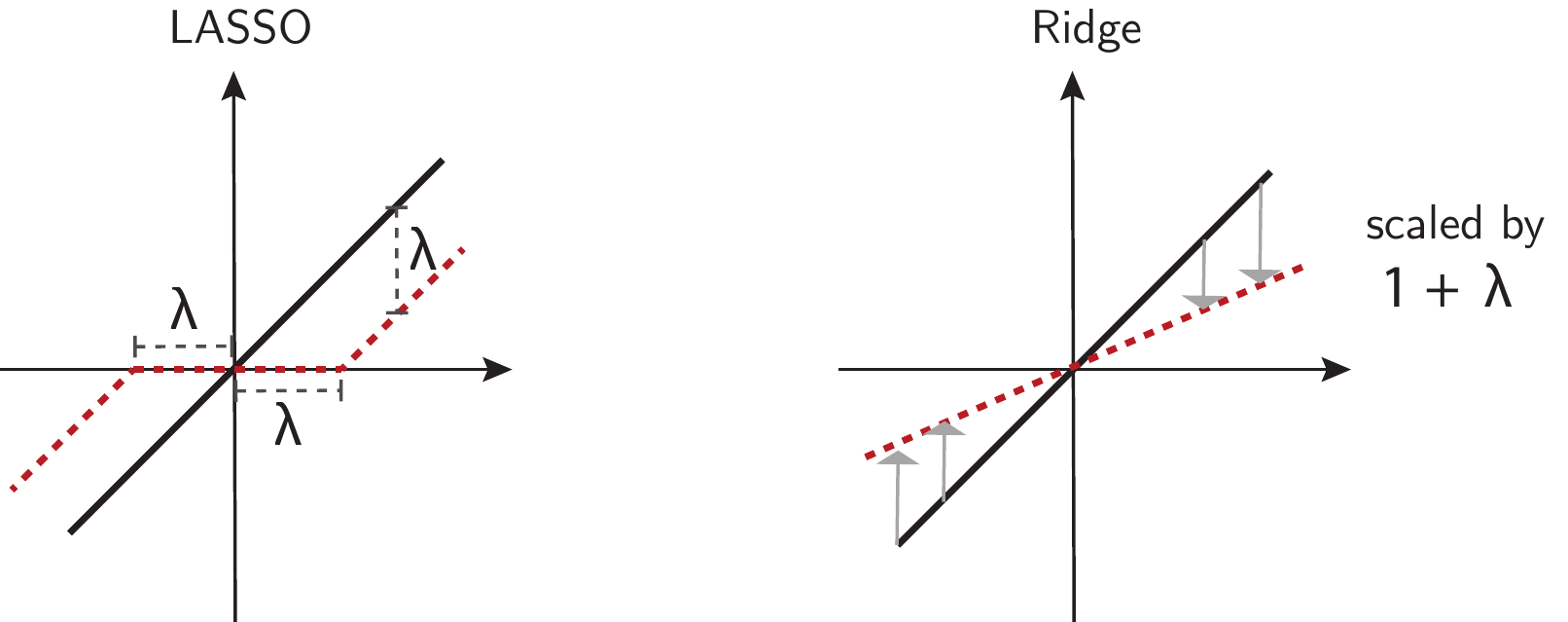}
		\caption{\label{fig:comparesol} [Adapted from \cite{friedman2001elements}] Comparing LASSO and Ridge regression. The black 45 degree line is the unconstrained estimate for reference. The estimators are shown by red dashed lines. For LASSO, this corresponds to the soft-thresholding function Eq.~\eqref{eq:LASSOsolnXorthog} while for Ridge regression the solution is given by Eq.~\eqref{eq:RidgesolnXorthog}}
\end{figure}

\begin{figure}[h!]
		\includegraphics[width=1.0\columnwidth]{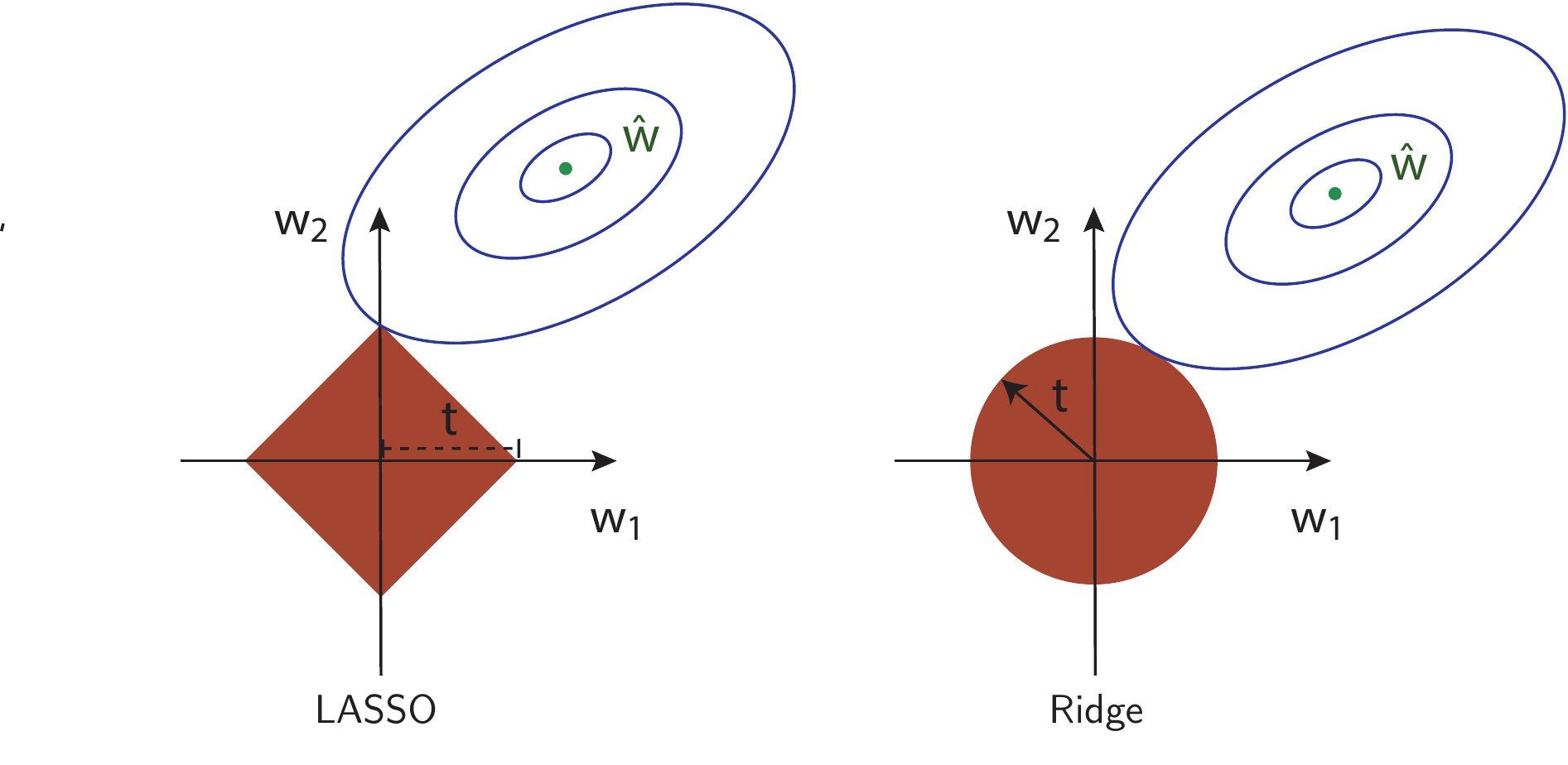}
		\caption{\label{fig:L2vsL1} [Adapted from \cite{friedman2001elements}] Illustration of LASSO (left) and Ridge regression (right). The blue concentric ovals are the contours of the regression function while the red shaded regions represent the constraint functions: (left) $|w_1|+|w_2|\le t$ and (right) $w_1^2+w_2^2\le t$. Intuitively, since the constraint function of LASSO has more protrusions, the ovals tend to intersect the constraint at the vertex, as shown on the left. Since the vertices correspond to parameter vectors $\bd{w}$ with only one non-vanishing component, LASSO tends to give sparse solution.}
\end{figure}

In this section, we study the effects of adding an $L_1$ regularization penalty, conventionally called \emph{LASSO}, which stands for ``least absolute shrinkage and selection operator''. Concretely, LASSO {in the penalized form}  is defined by the following regularized regression problem:
\be\label{eq:RLASSOProbPenalized}
\hat{\bd{w}}_\text{LASSO}(\lambda)=\argmin _{\bd{w}\in\mathbb{R}^p} ||\bd{Xw}-\bd{y}||_2^2 + \lambda ||\bd{w}||_1.
\ee
As in Ridge regression, there is another formulation for LASSO based on constrained optimization, namely,  
\be\label{eq:LASSOProbConstrained}
\hat{\bd{w}}_\text{LASSO}(t)=\argmin _{\bd{w}\in\mathbb{R}^p:\ ||\bd{w}||_1 \le t} ||\bd{Xw}-\bd{y}||_2^2.
\ee
The equivalence interpretation is the same as in Ridge regression, namely, for any $t\ge 0$ and solution $\hat{\bd{w}}_\text{LASSO}$ in Eq.~\eqref{eq:LASSOProbConstrained}, there is a value $\lambda\ge 0$ such that $\hat{\bd{w}}_\text{LASSO}$ solves Eq.~\eqref{eq:RLASSOProbPenalized}, and vice versa. {However,} to get the analytic solution of LASSO, we cannot simply take the gradient of Eq.~\eqref{eq:RLASSOProbPenalized} with respect to $\bd{w}$, since the $L_1$-regularizer is not everywhere differentiable, in particular at any point where $w_j=0$ (see Fig.~\ref{fig:L2vsL1}). Nonetheless, LASSO is a convex problem. Therefore, we can invoke the so-called ``subgradient optimality condition" \cite{boyd2004convex, rockafellar2015convex} in optimization theory to obtain the solution. {To keep the notation simple}, we only show the solution assuming $X$ is orthogonal:
\be\label{eq:LASSOsolnXorthog}
\hat{{w}}_j^\mathrm{LASSO}(\lambda)=\text{sign}(\hat{w}_j^\text{LS})(|\hat{w}_j^\text{LS}|-\lambda)_+,\,\text{ for orthogonal } \bd{X},
\ee
where $(x)_+$ denotes the positive part of $x$ and $\hat{{w}}_j^\mathrm{LS}$ is the $j$-th component of least squares solution. In Fig.~\ref{fig:comparesol}, we compare the Ridge solution Eq.~\eqref{eq:RidgesolnXorthog} with LASSO solution Eq.~\eqref{eq:LASSOsolnXorthog}. As we mentioned above, the Ridge solution is the least squares solution scaled by a factor of $(1+\lambda)$. Here LASSO does something conventionally called ``soft-thresholding" (see Fig.~\ref{fig:comparesol}). We encourage interested readers to work out the exercises in Notebook 3 to explore what this function does. 

How different are the solutions found using LASSO and Ridge regression? In general, LASSO tends to give sparse solutions, meaning many components of $\hat{\bd{w}}_\text{LASSO}$ are zero. An intuitive justification for this result is provided in Fig.~\ref{fig:L2vsL1}. In short, to solve a constrained optimization problem with a fixed regularization strength $t\ge 0$, for example, Eq.~\eqref{eq:RidgeProbConstrained} and Eq.~\eqref{eq:LASSOProbConstrained}, one first carves out the ``feasible region" specified by the regularizer in the $\{w_1,\cdots, w_d\}$ space. This means that a solution $\hat{\bd{w}}_0$ is legitimate only if it falls in this region. Then one proceeds by plotting the contours of the least squares regressors in an increasing manner until the contour touches the feasible region. The point where this occurs is the solution to our optimization problem (see Fig.~\ref{fig:L2vsL1} for illustration). Loosely speaking, since the $L_1$ regularizer of LASSO has sharp protrusions (i.e. vertices) along the axes, and {because} the regressor contours are in the shape of ovals (it is quadratic in $\bd{w}$), their intersection tends to occur at the vertex of the feasibility region, implying the solution vector will be sparse.

\begin{figure}[h!]
	\includegraphics[width=1.0\columnwidth]{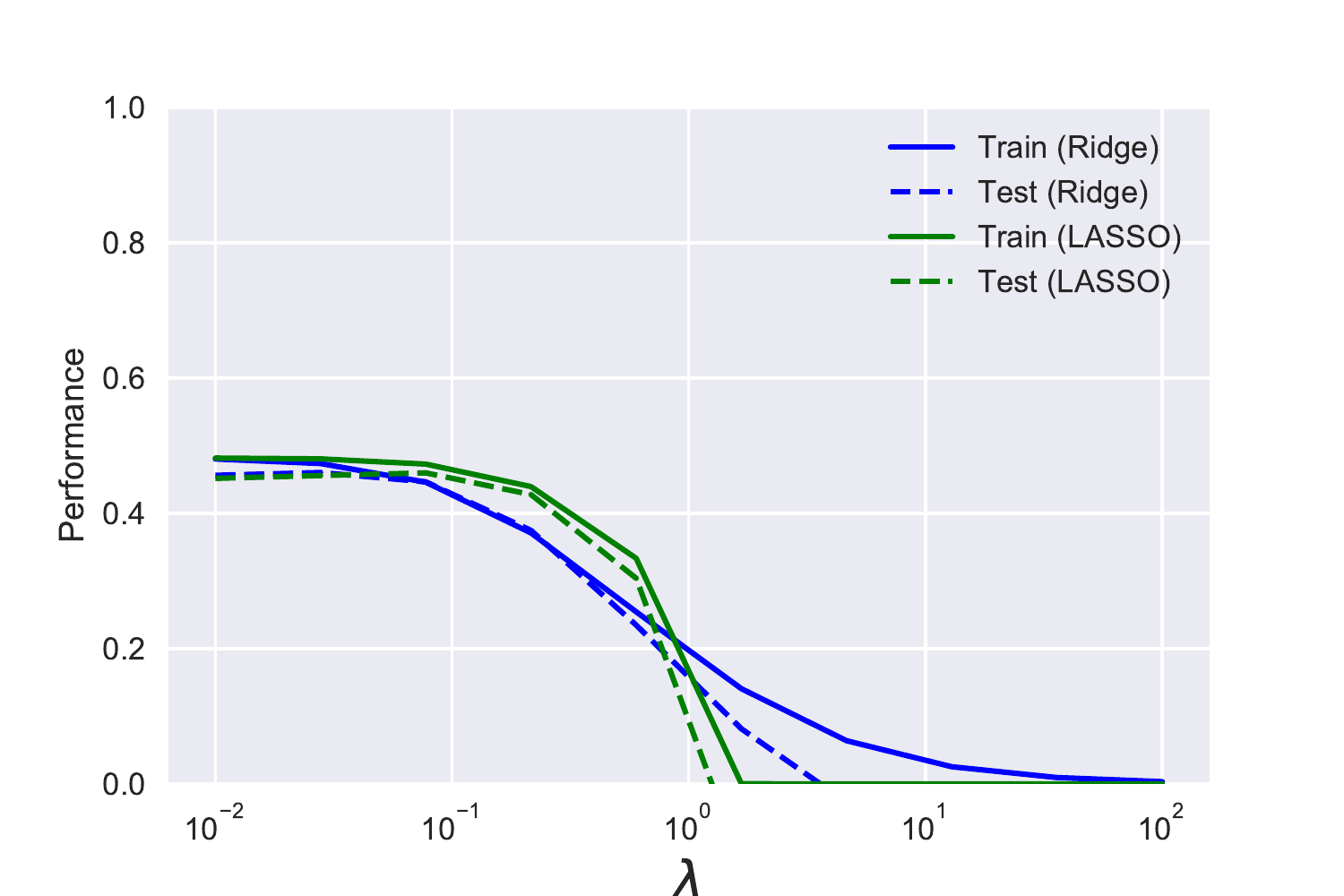}
	\caption{\label{fig:lin_reg_diabetes_performance} Performance of LASSO and ridge regression on the diabetes dataset measured by the $R^2$ coefficient of determination.  The best possible performance is $R^2=1$. See \href{https://physics.bu.edu/~pankajm/MLnotebooks.html}{Notebook 3}.}
\end{figure}

\begin{figure}
	\includegraphics[width=1.0\columnwidth]{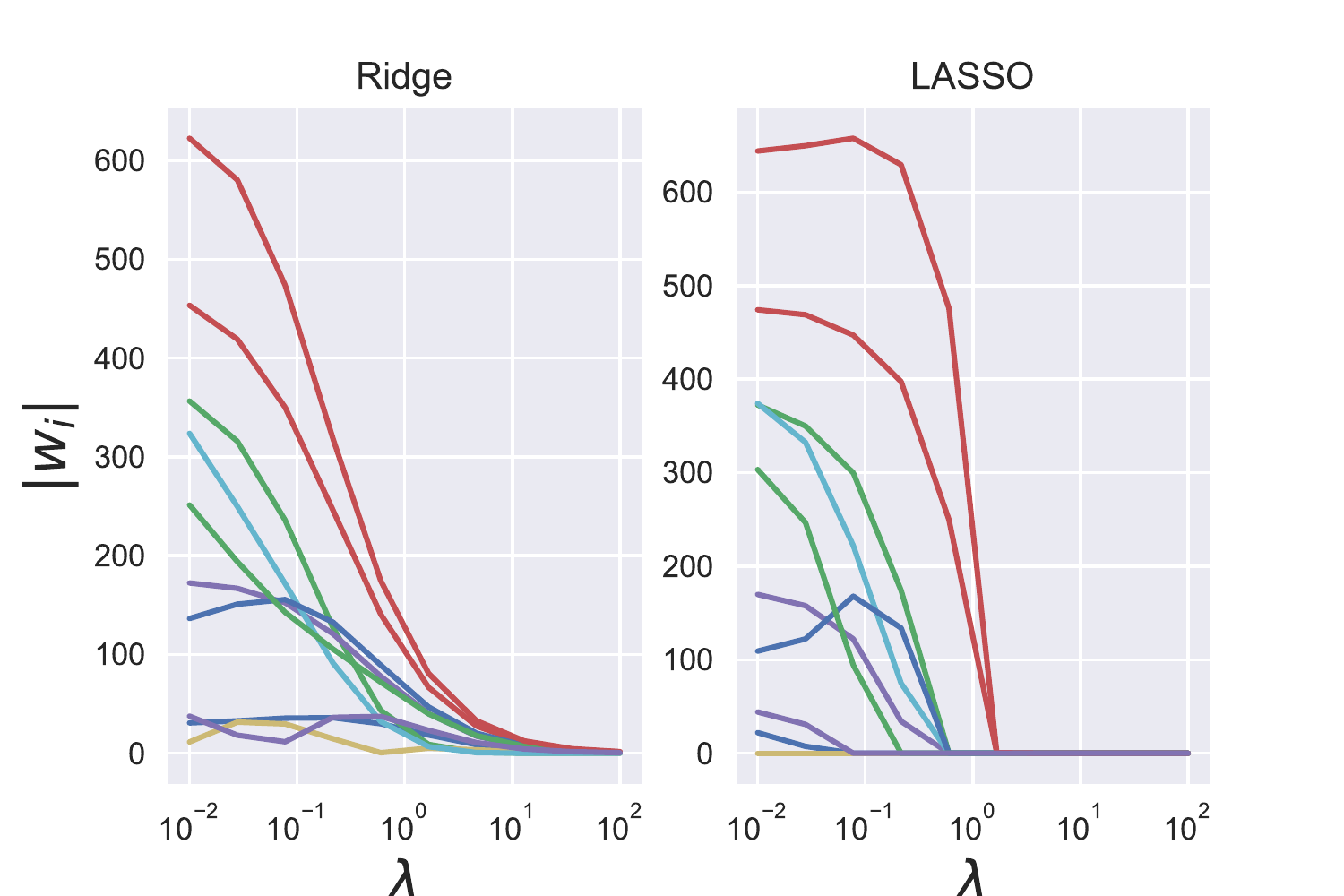}
	\caption{\label{fig:lin_reg_diabetes_sparsity} Regularization parameter $\lambda$ affects the weights (features) we learned in both Ridge regression (left) and LASSO regression (right) on the Diabetes dataset. Curves with different colors correspond to different $w_i$'s (features). Notice LASSO, unlike Ridge, sets feature weights to zero leading to sparsity. See \href{https://physics.bu.edu/~pankajm/MLnotebooks.html}{Notebook 3}.}
\end{figure}

In Notebook 3, we analyze a Diabetes dataset using both LASSO and Ridge regression to predict the diabetes outcome one year forward \cite{efron2004least}. In Figs.~\ref{fig:lin_reg_diabetes_performance}, \ref{fig:lin_reg_diabetes_sparsity}, we show the performance of both methods and the solutions $\hat{\bd{w}}_\text{LASSO}(\lambda)$, $\hat{\bd{w}}_\text{Ridge}(\lambda)$ explicitly. More details of this dataset and our regression implementation can be found in \href{https://physics.bu.edu/~pankajm/MLnotebooks.html}{Notebook 3}.

\subsection{\label{subsec:Ising_linreg}Using Linear Regression to Learn the Ising Hamiltonian}

To gain deeper intuition about what kind of physics problems linear regression allows us to tackle, consider the following problem of learning the Hamiltonian for the Ising model. Imagine you are given an ensemble of random spin configurations, and assigned to each state its energy, generated from the 1D Ising model:
\begin{equation}
H=-J\sum_{j=1}^L S_{j}S_{j+1}
\end{equation} 
where $J$ is the nearest-neighbor spin interaction, and $S_j\in\{\pm 1\}$ is a spin variable. Let's assume the data was generated with $J=1$. You are handed the data set $\mathcal{D}=(\{S_j\}_{j=1}^L, E_j)$ without knowledge of what the numbers $E_j$ mean, and the configuration $\{S_j\}_{j=1}^L$ can be interpreted in many ways: the outcome of coin tosses, black-and-white pixels of an image, the binary representation of integers, etc. Your goal is to learn a model that predicts $E_j$ from the spin configurations. 

Without any prior knowledge about the origin of the data set, physics intuition may suggest to look for a spin model with pairwise interactions between every pair of variables. That is, we choose the following model class:
\begin{equation}
H_\mathrm{model}[S^i] = - \sum_{j=1}^L \sum_{k=1}^L J_{j,k}S_{j}^iS_{k}^i,
\label{eq:Ising_linreg_ansatz} 
\end{equation}

The goal is to determine the interaction matrix $J_{j,k}$ by applying linear regression on the data set $\mathcal{D}$. This is a well-defined problem, since the unknown $J_{j,k}$ enters linearly into the definition of the Hamiltonian. To this end, we cast the above ansatz into the more familiar linear-regression form:
\begin{equation}
H_\mathrm{model}[S^i] =\mathbf{X}^i \cdot \mathbf{J}.
\end{equation}
The vectors $\mathbf{X}^i$ represent all two-body interactions $\{S_{j}^iS_{k}^i \}_{j,k=1}^L$, and the index $i$ runs over the samples in the dataset. To make the analogy complete, we can also represent the dot product by a single index $p = \{j,k\}$, i.e.~$\mathbf{X}^i \cdot \mathbf{J}=X^i_pJ_p$. Note that the regression model does not include the minus sign. In the following, we apply ordinary least squares, Ridge, and LASSO regression to the problem, and compare their performance.

\begin{figure}[h!]
	\includegraphics[width=1.0\columnwidth]{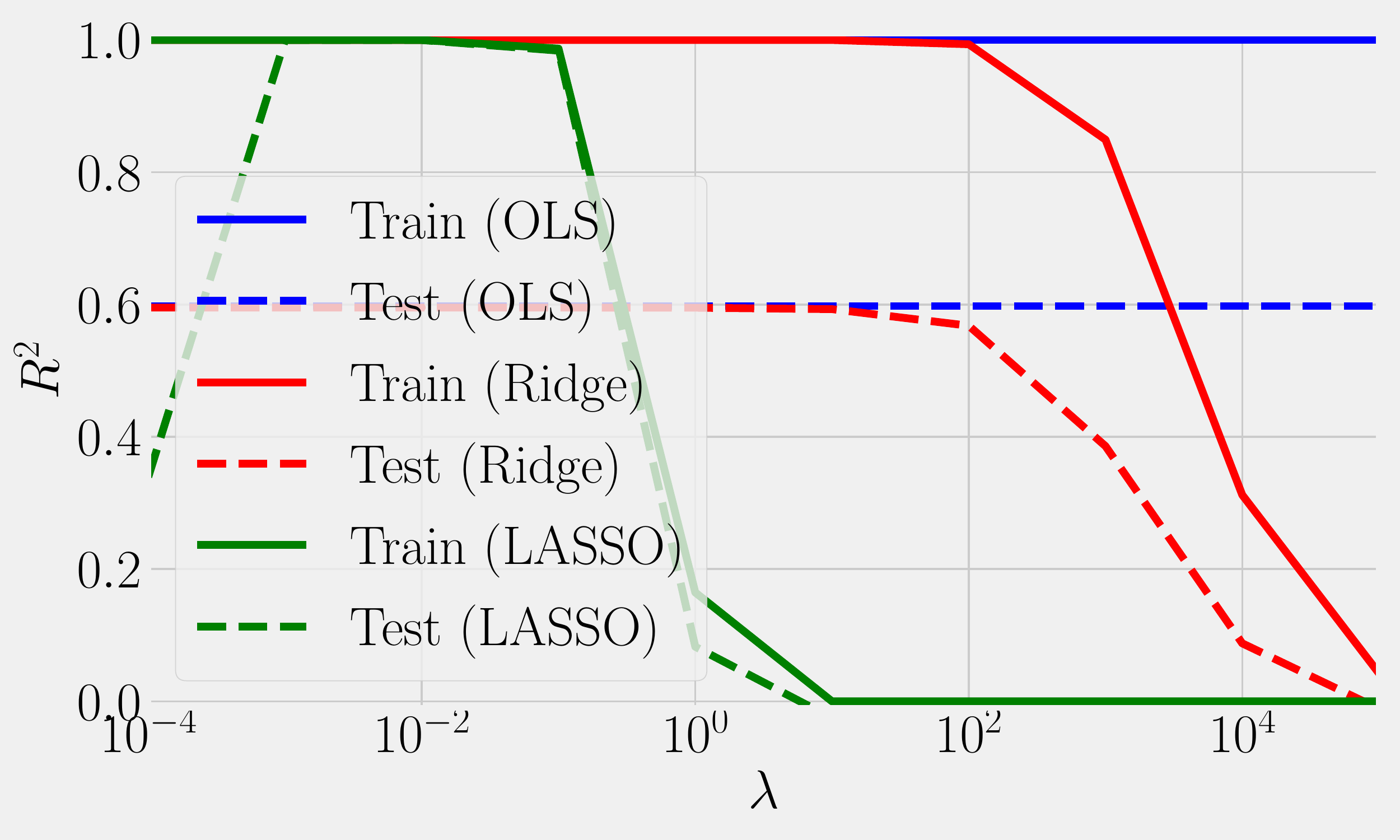}
	\caption{\label{fig:Ising_performance} Performance of OLS, Ridge and LASSO regression on the Ising model as measured by the $R^2$ coefficient of determination. Optimal performance is $R^2=1$.See \href{https://physics.bu.edu/~pankajm/MLnotebooks.html}{Notebook 4}.}
\end{figure} 

Figure.~\ref{fig:Ising_performance} shows the $R^2$ of the three regression models.
\begin{equation}
\label{eq:R^2_def}
R^2 = 1 - \frac{\sum_{i=1}^n \left| y^\mathrm{true}_i - y^\mathrm{pred}_i \right| ^2}{\sum_{i=1}^n \left| y^\mathrm{true}_i - \frac{1}{n}\sum_{i=1}^n y^\mathrm{pred}_i\right| ^2}.
\end{equation}
Let us make a few remarks: (i) the regularization parameter $\lambda$ affects the Ridge and LASSO regressions at scales separated by a few orders of magnitude. Notice that this is different for the data considered in the diabetes dataset, cf.~Fig.~\ref{fig:lin_reg_diabetes_performance}. Therefore, it is considered good practice to always check the performance for the given model and data as a function of $\lambda$. (ii) While the OLS and Ridge regression test curves are monotonic, the LASSO test curve is not -- suggesting an optimal LASSO regularization parameter is $\lambda\approx 10^{-2}$. At this sweet spot, the Ising interaction weights ${\bf J}$ contains only nearest-neighbor terms (as did the model the data was generated from).

\begin{figure*}
	\includegraphics[width=1.0\textwidth]{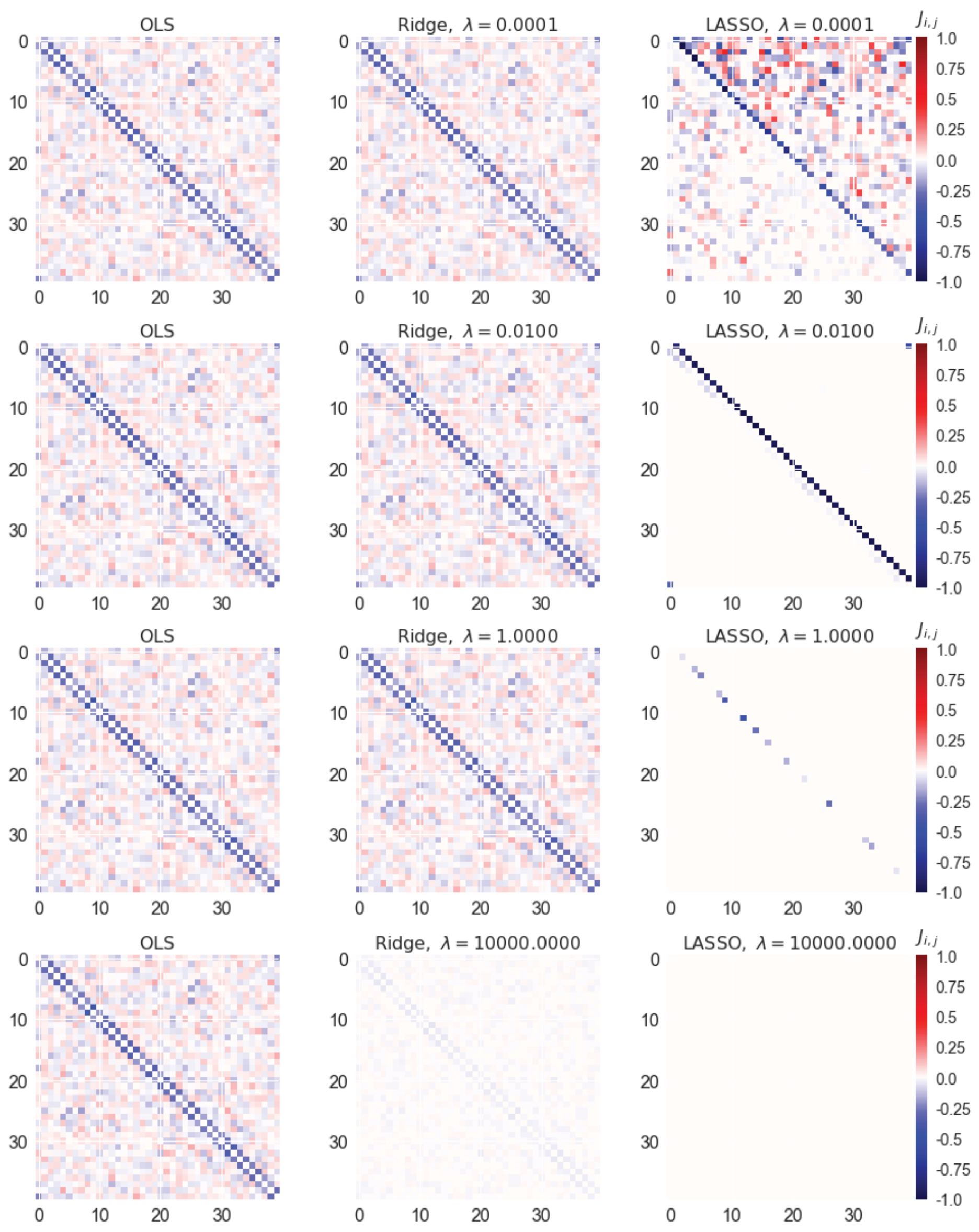}
	\caption{\label{fig:Ising_lin_reg} Learned interaction matrix $J_{ij}$ for the Ising model ansatz in Eq.~\eqref{eq:Ising_linreg_ansatz} for ordinary least squares (OLS) regression (left), Ridge regression (middle) and LASSO (right) at different regularization strengths $\lambda$. OLS is $\lambda$-independent but is shown for comparison throughout.See \href{https://physics.bu.edu/~pankajm/MLnotebooks.html}{Notebook 4}.}
\end{figure*}

Choosing whether to use Ridge or LASSO regression in this case turns out to be similar to fixing gauge degrees of freedom. Recall that the uniform nearest-neighbor interactions strength $J_{j,k}=J$ which we used to generate the data, was set to unity, $J=1$. Moreover, $J_{j,k}$ was NOT defined to be symmetric (we only used the $J_{j,j+1}$ but never the $J_{j,j-1}$ elements). Figure.~\ref{fig:Ising_lin_reg} shows the matrix representation of the learned weights $J_{j,k}$. Interestingly, OLS and Ridge regression learn nearly symmetric weights $J\approx-0.5$. This is not surprising, since it amounts to taking into account both the $J_{j,j+1}$ and the $J_{j,j-1}$ terms, and the weights are distributed symmetrically between them. LASSO, on the other hand, tends to break this symmetry (see matrix elements plots for $\lambda=0.01$)~\footnote{Look closer, and you will see that LASSO actually splits the weights rather equally for the periodic boundary condition element at the edges of the anti-diagonal.}. Thus, we see how different regularization schemes can lead to learning equivalent models but in different `gauges'. Any information we have about the symmetry of the unknown model that generated the data should be reflected in the definition of the model and the choice of regularization. In addition to the diabetes dataset in Notebook 3, we encourage the reader to work out \href{https://physics.bu.edu/~pankajm/MLnotebooks.html}{Notebook 4} in which linear regression is applied to the one-dimensional Ising model.

\subsection{Convexity of regularizer} 

In the previous section, we mentioned that the analytical solution of LASSO can be found by invoking its convexity. In this section, we provide a gentle introduction to convexity theory and highlight a few properties which can help us understand the differences between LASSO and Ridge regression. First, recall that a set $C\subseteq\mathbb{R}^n$ is called \emph{convex} if for any $x,y\in C$ and $t\in [0,1]$, 
\be
tx+(1-t)y \in C.
\ee
In other words, every line segment joining $x,y$ lies entirely in $C$. A function $f:\mathbb{R}^n\rightarrow \mathbb{R}$ is called convex if its domain, dom$(f)$, is a convex set, and for any $x,y\in$dom$(f)$ and $t\in [0,1]$ we have 
\be
f(tx+(1-t)y)\le tf(x)+(1-t)f(y),
\ee
That is, the function lies on or below the line segment joining its evaluation at $x$ and $y$. This function $f$ is called \emph{strictly convex} if this inequality holds strictly for $x\neq y$ and $t\in(0,1)$. Now, it turns out that \emph{for convex functions, any local minimizer is a global minimizer}. Algorithmically, this means that in the optimization procedure, as long as we are ``going down the hill'' and agree to stop when we reach a minimum, then we have hit the global minimum. In addition to this, there is an abundance of rich theory regarding convex duality and optimality, which allow us to understand the solutions even before solving the problem itself. We refer interested readers to~\cite{boyd2004convex, rockafellar2015convex}.

Now let us examine the two regularizers we introduced earlier. A close inspection reveals that LASSO and Ridge regressions are both convex problems but only Ridge regression is a strictly convex problem (assuming $\lambda >0$). From convexity theory, this means that we always have a unique solution for Ridge but not necessary for LASSO. In fact, it was recently shown that under mild conditions, such as demanding general position for columns of $X$, the LASSO solution is indeed unique \cite{tibshirani2013lasso}. Apart from this theoretical characterization, \cite{zou2005ElasticNet} introduced the notion of Elastic Net to retain the desirable properties of both LASSO and Ridge regression, which is now one of the standard tools for regression analysis and machine learning. We refer to reader to explore this in Notebook 2.

\subsection{Bayesian formulation of linear regression}
\label{subsec:bayes_vs_freq_reg}
In Section~\ref{sec:bayesian_inference}, we gave an overview of Bayesian inference and phrased it in the context of learning and uncertainty quantification. In this section we formulate least squares regression from a Bayesian point of view. We shall see that regularization in learning will emerge naturally as part of the Bayesian inference procedure. 
 
From the setup of linear regression, the data $\mathcal{D}$ used to fit the regression model is generated through $y=\bd{x}^T\bd{w}+\epsilon$. We often assume that $\epsilon$ is a Gaussian noise with mean zero and variance $\sigma^2$. To connect linear regression to the Bayesian framework, we often write the model as 
\be
\label{eq:bayes_Gaussian}
p(y|\bd{x},\bd{\theta})=\mathcal{N}(y|\mu(\bd{x}),\sigma^2(\bd{x})).
\ee
In other words, our regression model is defined by a conditional probability that depends not only on data $\bd{x}$ but on some model  parameters $\bd{\theta}$. For example, if the mean is a linear function of $\bd{x}$ given by $\mu=\bd{x}^T\bd{w}$, and the variance is fixed $\sigma^2(\bd{x})=\sigma^2$, then $\bd{\theta}=(\bd{w},\sigma^2)$. 

In statistics, many problems rely on estimation of some parameters of interest. For example, suppose we are given the height data of $20$ junior students from a regional high school, but what we are interested in is the average height of all high school juniors in the whole county. It is conceivable that the data we are given are not representative of the student population as a whole. It is therefore necessary to devise a systematic way to preform reliable estimation. Here we present the \emph{maximum likelihood estimation (MLE)}, and show that MLE for $\bd{\theta}$ is the one that minimizes the mean squared error (MSE) used in OLS, see Sec.\ref{subsec:least_sq_reg}. 

MLE is defined by maximizing the log-likelihood w.r.t.~the parameters $\bd{\theta}$:
\be
\hat{\bd{\theta}}\equiv \arg\max_{\bd{\theta}}\log p(\mathcal{D}|\bd{\theta}).
\ee
Using the assumption that samples are i.i.d., we can write the \emph{log-likelihood} as 
\be
l(\bd{\theta})\equiv \log p(\mathcal{D}|\bd{\theta}) =\sum_{i=1}^n \log p(y_i|\bd{x}^{(i)},\bd{\theta}).
\ee
Note that the conditional dependence of the response variable $y_i$ on the independent variable $\bd{x}^{(i)}$ in the likelihood function is made explicit since in regression the observed value of data, $y_i$, is predicted based on $\bd{x}^{(i)}$ using a model that is assumed to be a probability distribution that depends on unknown parameter $\bd{\theta}$. This distribution, when endowed with $\bd{\theta}$, can, as we hope, potentially explain our prediction on $y_i$. By definition, such distribution is the likelihood function we discussed in Sec.~\ref{sec:bayesian_inference}.  Note that this is consistent with the formal statistical treatment of regression where the goal is to estimate the \emph{conditional expectation} of the dependent variable given the value of the independent variable (sometimes called the covariate)~\cite{wasserman2013all}. We stress that this notation does \emph{not} imply $\bd{x}^{(i)}$ is unknown-- it is still part of the observed data!

Using Eq.~\eqref{eq:bayes_Gaussian}, we get
\bea
l(\bd{\theta}) &=&-\frac{1}{2\sigma^2}\sum_{i=1}^n \left(y_i-\bd{w}^{\text{T}}\bdxi\right)^2-\frac{n}{2}\log\left(2\pi\sigma^2\right) \nonumber\\
&=&-\frac{1}{2\sigma^2}||\bd{Xw}-\bd{y}||_2^2 +\text{ const.}
\label{MLEfinal}
\eea
By comparing Eq.~\eqref{eq:LSproblem} and Eq.~\eqref{MLEfinal}, it is clear that performing least squares is the same as maximizing the log-likelihood of this model.  

What about adding regularization? In Section~\ref{sec:bayesian_inference}, we introduced the \emph {maximum a posteriori probability (MAP) estimate}. Here we show that it actually corresponds to regularized linear regression, where the choice of prior determines the type of regularization. Recall Bayes' rule
\be\label{eq:bayes}
p(\bd{\theta}|D)\propto p(\mathcal{D}|\bd{\theta}) p(\bd{\theta}).
\ee
Now instead of maximizing the log-likelihood, $l(\bd{\theta})=\log  p(\mathcal{D}|\bd{\theta})$, let us maximize the log posterior, $\log p(\bd{\theta}|\mathcal{D})$. Invoking Eq.~\eqref{eq:bayes}, the MAP estimator becomes
\be
\hat{\bd{\theta}}_\text{MAP}\equiv \arg\max_{\bd{\theta}}\log p(\mathcal{D}|\bd{\theta})+\log p(\bd{\theta}).
\ee 
In Sec.~\ref{subsec:BayesianHyperpara}, we discussed that a common choice for the prior is a Gaussian distribution. Consider the Gaussian prior\footnote{Indeed, a Gaussian prior is the {\it conjugate prior} that gives a Gaussian posterior. For a given likelihood, conjugacy guarantees the preservation of prior distribution at the posterior level. For example, for a Gaussian (Geometric) likelihood with a Gaussian (Beta) prior, the posterior distribution is still Gaussian (Beta) distribution. } with zero mean and variance $\tau^2$, namely,  $p(\bd{w})=\prod_{j}\mathcal{N}(w_j|0,\tau^2)$. Then, we can recast the MAP estimator into
\bea\label{eq:MAPfinal}
\hat{\bd{\theta}}_\text{MAP}&\equiv& \arg\max_{\bd{\theta}}\left[-\frac{1}{2\sigma^2}\sum_{i=1}^n (y_i-\bd{w}^\text{T}\bdxi)^2-\frac{1}{2\tau^2}\sum_{j=1}^n w_j^2\right]\nonumber\\
&=&\arg\max_{\bd{\theta}}\left[-\frac{1}{2\sigma^2}||\bd{Xw}-\bd{y}||_2^2-\frac{1}{2\tau^2}||\bd{w}||_2^2\right].
\eea
Note that we dropped constant terms that do not depend on the maximization parameters $\bd{\theta}$. The equivalence between MAP estimation with a Gaussian prior and Ridge regression is established by comparing Eq.~\eqref{eq:MAPfinal} and Eq.~\eqref{eq:RidgeProbConstrained} with $\lambda\equiv \sigma^2/\tau^2$. We relegate the analogous derivation for LASSO to an exercise in Notebook 3.

\subsection{Recap and a general perspective on regularizers}
In this section, we explored least squares linear regression with and without regularization. We motivated the need for regularization due to poor generalization, in particular in the ``high-dimensional limit" ($p\gg n$). Instead of showing the average in-sample and out-of-sample errors for the regularized problem explicitly, we conducted numerical experiments in Notebook 3 on the diabetes dataset and showed that regularization typically leads to better generalization. Due to the equivalence between the constrained and penalized form of regularized regression (in LASSO and Ridge, but not generally true in cases such as L$^0$ penalization), we can regard the regularized regression problem as an un-regularized problem but on a constrained set of parameters. Since the size of the allowed parameter space (e.g. $\bd{w}\in\mathbb{R}^p$ when un-regularized vs. $\bd{w}\in C\subset\mathbb{R}^p$ when regularized) is roughly a proxy for model complexity, solving the regularized problem is in effect solving the un-regularized problem with a smaller model complexity class. This implies that we're less likely to overfit.

We also showed the connection between using a regularization function and the use of priors in Bayesian inference.
This connection can be used to develop more intuition about why regularization implies we are less likely to overfit the data: Let's say you are a young Physics student taking a laboratory class where the goal of the experiment is to measure the behavior of several different pendula and use that to predict the formula (i.e. model) that determines the period of oscillation.
In your investigation you would probably record many things (hopefully including the length and mass!) in an effort to give yourself the best possible chance of determining the unknown relationship, perhaps writing down the temperature of the room, any air currents, if the table were vibrating, etc.
What you have done is create a high-dimensional dataset for yourself.
However you actually possess an even higher-dimensional dataset than you probably would admit to yourself.
For example you are probably aware of the time of day, that it is a Wednesday, your friend Alice being in attendance, your friend Bob being absent with a cold, the country in which you are doing the experiment, and the planet you are on, but you almost assuredly haven't written these down in your notebook.
Why not? The reason is because you entered the classroom with strongly held prior beliefs that none of those things affect the physics which takes place in that room.
Even of the things you did write down in an effort to be a careful scientist you probably hold some doubt as to their importance to your result and what is serving you here is the intuition that probably only a few things matter in the physics of pendula.
Hence again you are approaching the experiment with prior beliefs about how many features you will need to pay attention to in order to predict what will happen when you swing an unknown pendulum.
This example might seem a bit contrived, but the point is that we live in a high-dimensional world of information and while we have good intuition about what to write down in our notebook for well-known problems, often in the field of ML we cannot say with any confidence a priori \textit{what} the small list of things to write down will be, but we can at least use regularization to help us enforce that the list not be too long so that we don't end up predicting that the period of a pendulum depends on Bob having a cold on Wednesdays.

Of course, in both LASSO and Ridge regression there is a parameter $\lambda$ involved. In principle, this \textbf{hyper-parameter} is usually predetermined, which means that it is not part of the regression process. As we saw in Fig.~\ref{fig:lin_reg_diabetes_sparsity}, our learning performance and solution depends strongly on $\lambda$, thus it is vital to choose it properly. As we discussed in Sec.~\ref{subsec:BayesianHyperpara}, one approach is to assume an {\it uninformative prior} on the hyper-parameters, $p(\lambda)$, and average the posterior over all choices of $\lambda$ following this distribution. However, this comes with a large computational cost. Therefore, it is simpler to choose the regularization parameter through some optimization procedure.

We'd like to emphasize that linear regression can be applied to model non-linear relationship between input and response. This can be done by replacing the input $\bd{x}$ with some nonlinear function $\bd{\phi}(\bd{x})$. Note that doing so preserves the linearity as a function of the parameters $\bd{w}$, since model is defined by the their inner product $\bd{\phi}^T(\bd{x})\bd{w}$. This method is known as \emph{basis function expansion} \cite{murphy2012machine,bishop2006pattern}. 

Recent years have also seen a surge of interest in understanding generalized linear regression models from a statistical physics perspective. Much of this research has focused on understanding high-dimensional linear regression and compressed sensing \cite{donoho2006compressed} (see \cite{advani2013statistical, zdeborova2016statistical} for accessible reviews for physicists). On a technical level, this research  imports and extends the machinery of spin glass physics (replica method, cavity method, and message passing) to analyze high-dimensional linear models \cite{krzakala2012statistical, krzakala2012probabilistic, krzakala2014variational, ramezanali2015critical,fisher2015bayesian, fisher2015bayesian2, zdeborova2016statistical, advani2016statistical}. This is a rich area of activity at the intersection of physics, computer science, information theory, and machine learning and interested readers are encouraged to consult the literature for further information (see also \cite{mezard2009information}).

%% file: sections/log_reg.tex
So far we have focused on learning from datasets for which there is a ``continuous'' output. For example, in linear regression we were concerned with learning the coefficients of a polynomial to predict the response of a continuous variable $y_i$ on unseen data based on its independent variables ${\bf x}_i$. However, a wide variety of problems, such as classification, are concerned with outcomes taking the form of discrete variables (i.e. categories). For example, we may want to detect if there is a cat or a dog in an image. Or given a spin configuration of, say, the $2D$ Ising model, we would like to identify its phase (e.g.~ordered/disordered). In this section, we introduce logistic regression which deals with binary, dichotomous outcomes (e.g.~True or False, Success or Failure, etc.). We encourage the reader to use the opportunity to build their intuition about the inner workings of logistic regression, as this will prove valuable later on in the study of modern supervised Deep Learning models (see Sec.~\ref{sec:DNNs}).

\begin{figure}[t!]
	\includegraphics[width=1.0\columnwidth]{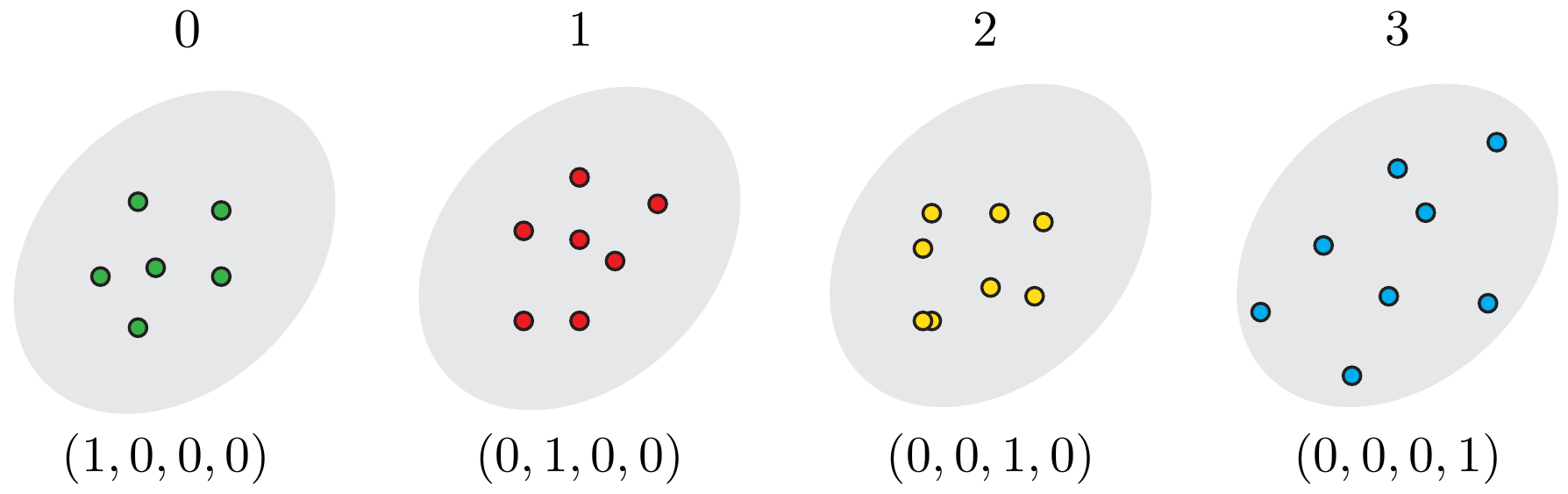}
	\caption{\label{fig:log_reg_intro} Pictorial representation of four data categories labeled by the integers $0$ through $3$ (above), or by one-hot vectors with binary inputs (below).}
\end{figure}

This section is structured as follows: first, we define logistic regression and derive its corresponding cost function (the cross entropy) using a Bayesian approach, and discuss its minimization. Then, we generalize logistic regression to the case of multiple categories which is called {\it SoftMax regression}. We demonstrate how to apply logistic regression using three different problems: (i) classifying phases of the $2D$ Ising model, (ii) learning features in the SUSY dataset, and (iii) MNIST handwritten digit classification.

Throughout this section, we consider the case where the dependent variables $y_i\in\mathbb{Z}$ are discrete and only take values from $m=0,\dots,M-1$ (which enumerate the $M$ classes), see Fig.~\ref{fig:log_reg_intro}. The goal is to predict the output classes from the design matrix $X\in\mathbb{R}^{n\times p}$ made of $n$ samples, each of which bears $p$ features. The primary goal is to identify the classes to which new unseen samples belong. 
 
Before delving into the details of logistic regression, it is helpful to consider a slightly simpler classifier: a linear classifier that categorizes examples using a weighted linear-combination of the features and an additive offset:
\begin{equation}
s_i = \bd{x}_i^T\bd{w} + b_0 \equiv  \mbf{x}_i^T\mbf{w},
\end{equation}
where we use the short-hand notation $\mbf{x}_i = (1,\bd{x}_i)$ and $\mbf{w} = (b_0,\bd{w})$. This function takes values on the entire real axis. In the case of logistic regression, however, the labels $y_i$ are discrete variables. One simple way to get a discrete output is to have sign functions that map the output of a linear regressor to $\{0,1\}$,  $\sigma(s_i)=$ sign$(s_i) = 1$ if $s_i\ge 0$ and $0$ otherwise. Indeed, this is commonly known as the ``perceptron'' in the machine learning literature. 

\begin{figure}[t!]
	\includegraphics[width=1.0\columnwidth]{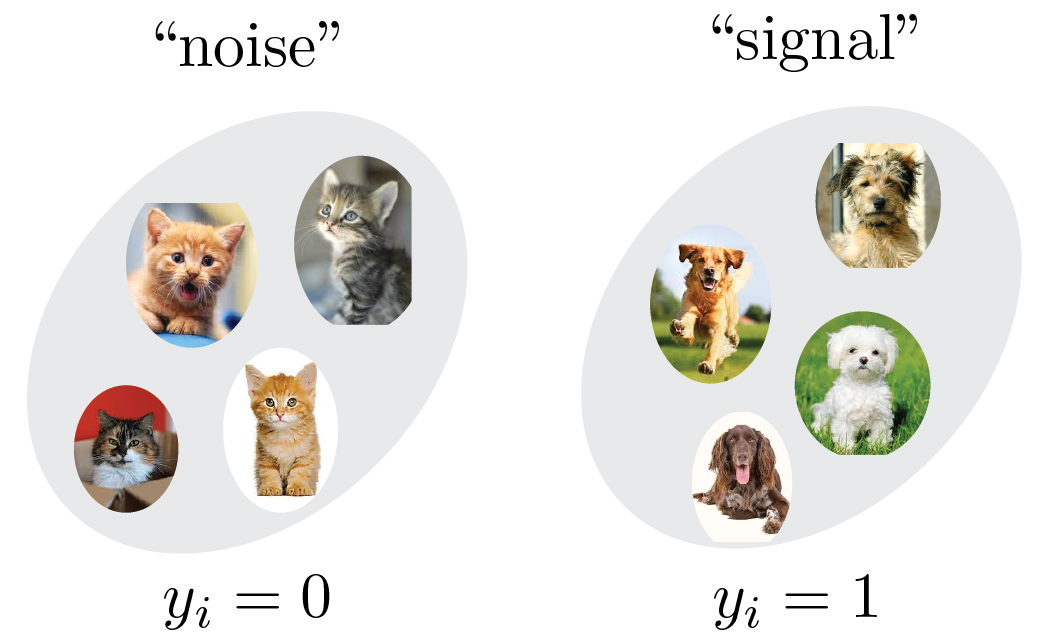}
	\caption{\label{fig:log_reg_ex} Classifying data in the simplest case of only two categories, labeled ``noise'' and ``signal'' (or ``cats'' and ``dogs''), is the subject of Logistic Regression.}
\end{figure} 

\subsection{The cross-entropy as a cost function for logistic regression}
\label{subsec:cross_entropy}
The perceptron is an example of a ``hard classification'': each datapoint is assigned to a category (i.e.~$y_i=0$ or $y_i=1$). Even though the perceptron is an extremely simple model, it is favorable in many cases (e.g.~when dealing with noisy data) to have a ``soft'' classifier that outputs the probability of a given category. For example, given $\mbf{x}_i$, the classifier returns the probability of being in category $m$. One such function is the logistic (or sigmoid) function:
\begin{equation}
\sigma(s) = \frac{1}{1+\mathrm e^{-s}}.
\label{eq:log_fun}
\end{equation} 
Note that $1-\sigma(s)= \sigma(-s)$, which will be useful shortly. In many cases, it is favorable to work with a ``soft'' classifier.

Logistic regression is the canonical example of a soft classifier. In logistic regression, the probability that a data point $\bd{x}_i$ belongs to a category $y_i=\{0,1\}$ is given by
\begin{eqnarray}
P(y_i=1|\bd{x}_i,\bd\theta) &=& \frac{1}{1+\mathrm{e}^{-\mbf{x}^T_i\bd{\theta}}},\nonumber\\
P(y_i=0|\bd{x}_i,\bd\theta) &=& 1 - P(y_i=1|\bd{x}_i,\bd\theta),
\label{eq:perceptron_P}	
\end{eqnarray}
where $\bd{\theta}=\mbf{w}$ are the weights we wish to learn from the data. To gain some intuition for these equations, consider a collection of non-interacting two-state systems coupled to a thermal bath (e.g.~a collection of atoms that can be in two states). Furthermore, denote the state of system $i$ by a binary variable: $y_i  \in \{0,1\}$. From elementary statistical mechanics, we know that if the two states have energies $\epsilon_0$ and $\epsilon_1$ the probability for finding the system in a state $y_i$ is:
\begin{eqnarray}
P(y_i=1) &=& \frac{\mathrm e^{-\beta \epsilon_0}}{\mathrm e^{-\beta \epsilon_0} + \mathrm e^{-\beta \epsilon_1}} = \frac{1}{1 + \mathrm e^{-\beta\Delta\epsilon}}, \nonumber\\
P(y_i=1) &=& 1 - P(y_i=0).
\label{eq:atom_P}
\end{eqnarray}
Notice that in these expressions, as is often the case in physics, only energy differences are observable. If the difference in energies between two states is given by  $\Delta
\epsilon= \mbf{x}_i^T\mbf{w}$, we recover the expressions for logistic regression. We shall use this mapping between partition functions and classification
to generalize the logistic regressor to SoftMax regression in Sec.~\ref{subsec:softmax}. Notice that in terms of the logistic function, we can write
\be
P(y_i=1) =\sigma(\mbf{x}_i^T\mbf{w})=1-P(y_i=0).
\ee

We now define the cost function for logistic regression using Maximum Likelihood Estimation (MLE). Recall, that in MLE we choose parameters to maximize the probability of seeing
the observed data. Consider a dataset $\mathcal{D}=\{(y_i,\bd{x}_i)\}$ with binary labels $y_i\in\{0,1\}$ from which the data points are drawn independently. The likelihood of observing the data under our model is just:
\bea
P(\mathcal{D}|\mbf{w})& = &\prod_{i=1}^n \left[\sigma(\mbf{x}_i^T\mbf{w})\right]^{y_i}\left[1-\sigma(\mbf{x}_i^T\mbf{w})\right]^{1-y_i}\nonumber \\
\label{eq:logreg_likelihood}
\eea
from which we can readily compute the log-likelihood:
\begin{equation}
l(\mbf{w}) = \sum_{i=1}^n  y_i\log \sigma(\mbf{x}_i^T\mbf{w}) + (1-y_i)\log\left[1-\sigma(\mbf{x}_i^T\mbf{w})\right].
\end{equation}
The maximum likelihood estimator is defined as the set of parameters that maximize the log-likelihood:
\begin{equation}
\hat{\mbf{w}} = \argmax_{\bd\theta} \sum_{i=1}^n y_i\log \sigma(\mbf{x}_i^T\mbf{w}) + (1-y_i)\log\left[1-\sigma(\mbf{x}_i^T\mbf{w})\right]. 
\end{equation}
Since the cost (error) function is just the negative log-likelihood, for logistic regression we find
\begin{eqnarray}
\label{eq:cross_entropy}
\mathcal{C}(\mbf w) &=& - l(\mbf{w}) \\
&=& \sum_{i=1}^n  -y_i\log \sigma(\mbf{x}_i^T\mbf{w}) - (1-y_i)\log\left[1-\sigma(\mbf{x}_i^T\mbf{w})\right].\nonumber
\end{eqnarray}
The right-hand side in Eq.~\eqref{eq:cross_entropy} is known in statistics as the \emph{cross entropy}. 

Having specified the cost function for logistic regression, we note that, just as in linear regression, 
in practice we usually supplement the cross-entropy with additional regularization terms, usually $L_1$ and $L_2$ regularization (see Sec.~\ref{sec:lin_reg} for discussion of these regularizers).

\subsection{Minimizing the cross entropy}
\label{subsec:GD_logreg}

The cross entropy is a convex function of the weights $\mbf w$ and, therefore, any local minimizer is  a global minimizer. Minimizing this cost function leads to the following equation
\begin{equation}
\bd 0=\bd{\nabla} \mathcal{C}(\mbf{w}) = \sum_{i=1}^n\left[\sigma(\mbf{x}_i^T\mbf{w})-y_i\right]\mbf{x}_i,
\label{eq:log_reg_wmin}
\end{equation} 
where we made use of the logistic function identity $\partial_z \sigma(s) = \sigma(s)[1-\sigma(s)]$. Equation~\eqref{eq:log_reg_wmin} defines a transcendental equation for $\mbf{w}$, the solution of which, unlike linear regression, cannot be written in a closed form. For this reason, one must use numerical methods such as those introduced in Sec.~\ref{sec:gradient_descent} to solve this optimization problem.

\subsection{Examples of binary classification}
\label{subsec:logreg_examples}

Let us now show how to use logistic regression in practice. In this section, we showcase two pedagogical examples to train a logistic regressor to classify binary data. Each example comes with a corresponding Jupyter notebook, see~\href{https://physics.bu.edu/~pankajm/MLnotebooks.html}{https://physics.bu.edu/~pankajm/MLnotebooks.html}.

\subsubsection{\label{subsubsec:ising_phases_logreg} Identifying the phases of the 2D Ising model} 

The goal of this example is to show how one can employ logistic regression to classify the states of the $2D$ Ising model according to their phase of matter. 

\begin{figure*}[t!]
	\includegraphics[width=1.0\textwidth]{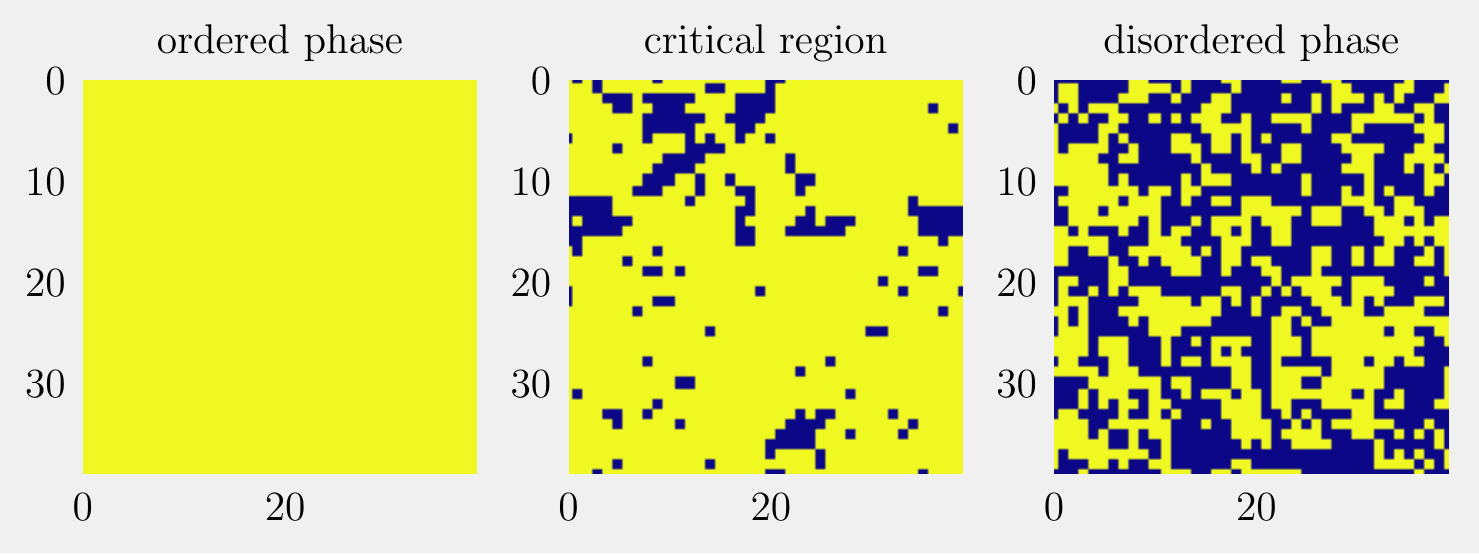}
	\caption{\label{fig:2D_Ising_states} Examples of typical states of the $2D$ Ising model for three different temperatures in the ordered phase ($T/J=0.75$, left), the critical region ($T/J=2.25$, middle) and the disordered phase ($T/J=4.0$, right). The linear system dimension is $L=40$ sites.}
\end{figure*}

The Hamiltonian for the classical Ising model is given by
\begin{equation}
H = -J\sum_{\langle ij\rangle}S_{i}S_j,\qquad \qquad S_j\in\{\pm 1\},
\end{equation}
where the lattice site indices $i,j$ run over all nearest neighbors of a $2D$ square lattice, and $J$ is an interaction energy scale. We adopt periodic boundary conditions. Onsager proved that this model undergoes a phase transition in the thermodynamic limit from an ordered ferromagnet with all spins aligned to a disordered phase at the critical temperature $T_c/J=2/\log(1+\sqrt{2})\approx 2.26$. For any finite system size, this critical point is smeared out to a critical region around $T_c$.

An interesting question to ask is whether one can train a statistical classifier to distinguish between the two phases of the Ising model. If successful, this can be used to locate the position of the critical point in more complicated models where an exact analytical solution has so far remained elusive~\cite{morningstar2017deep,zhang2017machine}. In other words, given an Ising state, we would like to classify whether it belongs to the ordered or the disordered phase, without any additional information other than the spin configuration itself. This categorical machine learning problem is well suited for logistic regression, and will thus consist of recognizing whether a given state is ordered by looking at its bit configurations. Notice that, for the purposes of applying logistic regression, the 2D spin state of the Ising model will be flattened out to a 1D array, so it will not be possible to learn information about the structure of the contiguous ordered 2D domains [see Fig.~\ref{fig:2D_Ising_states}]. Such information can be incorporated using deep convolutional neural networks, see Section~\ref{sec:DNNs}.

To this end, we consider the $2D$ Ising model on a $40\times 40$ square lattice, and use Monte-Carlo (MC) sampling to prepare $10^4$ states at every fixed temperature $T$ out of a pre-defined set. We furthermore assign a label to each state according to its phase: $0$ if the state is disordered, and $1$ if it is ordered. 

It is well-known that near the critical temperature $T_c$, the ferromagnetic correlation length diverges, which leads to, among other things, critical slowing down of the MC algorithm. Perhaps identifying the phases is also harder in the critical region. With this in mind, consider the following three types of states: ordered ($T/J<2.0$), near-critical ($2.0\leq T/J\leq 2.5)$ and disordered ($T/J>2.5$). We use both ordered and disordered states to train the logistic regressor and, once the supervised training procedure is complete, we will evaluate the performance of our classification model on unseen ordered, disordered, and near-critical states. 

Here, we deploy the {\it liblinear} routine (the default for Scikit's logistic regression) and stochastic gradient descent (SGD, see Sec.~\ref{sec:gradient_descent} for details) to optimize the logistic regression cost function with $L_2$ regularization. We define the accuracy of the classifier as the percentage of correctly classified data points. Comparing the accuracy on the training and test data, we can study the degree of overfitting. 
\begin{figure}[t!]
	\includegraphics[width=1.0\columnwidth]{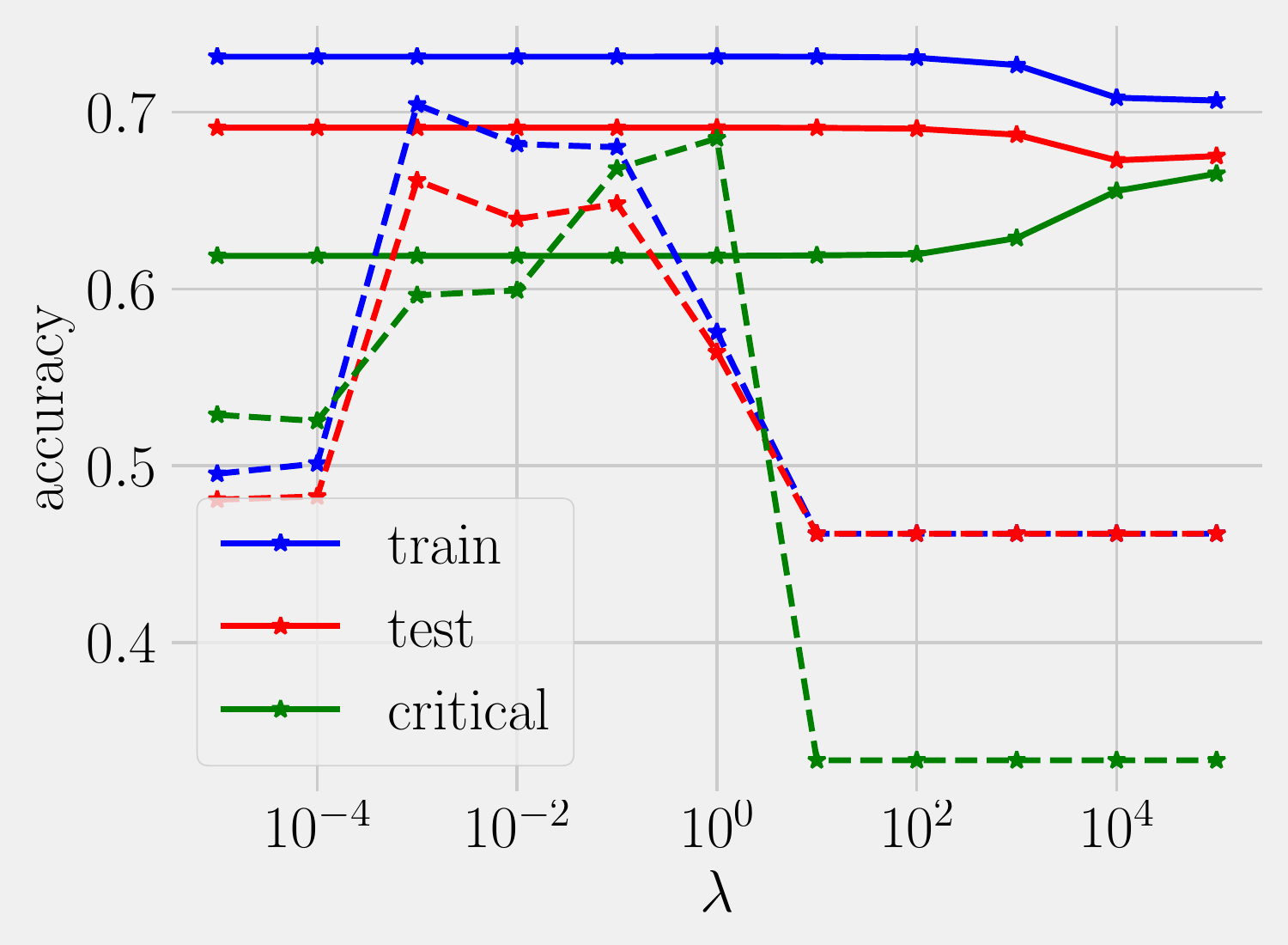}
	\caption{\label{fig:2D_Ising_accuracy} Accuracy as a function of the regularization parameter $\lambda$ in classifying the phases of the 2D Ising model on the training (blue), test (red), and critical (green) data. The solid and dashed lines compare the `liblinear' and `SGD' solvers, respectively.}
\end{figure}
The first thing to notice in Fig.~\ref{fig:2D_Ising_accuracy} is the small degree of overfitting, as suggested by the training (blue) and test (red) accuracy curves being very close to each other. Interestingly, the {\it liblinear} minimizer outperforms $SGD$ on the training and test data, but not on the near-critical data for certain values of the regularization strength $\lambda$. Moreover, similar to the linear regression examples, we find that there exists a sweet spot for the SGD regularization strength $\lambda$ that results in optimal performance of the logistic regressor, at about $\lambda\sim 10^{-1}$.
We might expect that the difficulty of the phase recognition problem depends on the temperature of the queried sample. Looking at the states in the near-critical region, c.f.~Fig.~\ref{fig:2D_Ising_states}, it is no longer easy for a trained human eye to distinguish between the ferromagnetic and the disordered phases close to $T_c$. Therefore, it is interesting to also compare the training and test accuracies to the accuracy of the near-critical state predictions. (Recall that the model is not trained on near-critical states.) Indeed, the {\it liblinear} accuracy is about $7\%$ smaller for the critical states (green curves) compared to the test data (red line).

Finally, it is important to note that all of Scikit's logistic regression solvers have in-built regularizers. We did not emphasize the role of the regularizers in this section, but they are crucial in order to prevent overfitting. We encourage the interested reader to play with the different regularization types and numerical solvers in \href{https://physics.bu.edu/~pankajm/MLnotebooks.html}{Notebook 6} and compare model performances.

\subsubsection{\label{subsubsec:susy_logreg} SUSY}
\begin{figure*}[t!]
\centering
\includegraphics[width=1.0\columnwidth]{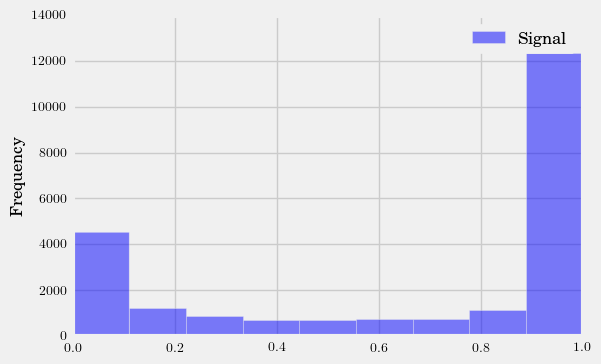} \hspace{2mm}
\includegraphics[width=1.0\columnwidth]{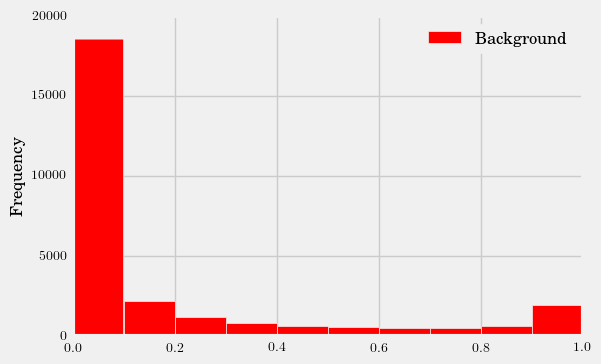}
\caption{The probability of an event being a classified as a signal event for true signal events (left, blue) and background events (right, red).}
\label{fig:LogProbSUSY}
\end{figure*}
In high energy physics experiments, such as the ATLAS and CMS detectors at the CERN LHC, one major hope is the discovery of new particles. 
To accomplish this task, physicists attempt to sift through events and classify them as either a signal of some new physical process or particle, or as a background event from already understood Standard Model processes. 
Unfortunately, we don't know for sure what underlying physical process occurred (the only information we have access to are the final state particles). 
However, we can attempt to define parts of phase space that will have a high percentage of signal events. 
Typically this is done by using a series of simple requirements on the kinematic quantities of the final state particles, for example having one or more leptons with large amounts of momentum that are transverse to the beam line ($p_{T}$). 
Instead, here we will use logistic regression in an attempt to find the relative probability that an event is from a signal or a background event. Rather than using the kinematic quantities of final state particles directly, we will use the output of our logistic regression to define a part of phase space that is enriched in signal events (see \href{https://physics.bu.edu/~pankajm/MLnotebooks.html}{Jupyter notebook}{Notebook 5}). 

The dataset we are using comes from the UC Irvine ML repository and has been produced using Monte Carlo simulations to contain events with two leptons (electrons or muons)~\cite{baldi2014searching}.
Each event has the value of 18 kinematic variables (``features''). 
The first 8 features are direct measurements of final state particles, in this case the $p_{T}$, pseudo-rapidity $\eta$, and azimuthal angle $\phi$ of two leptons in the event and the amount of missing transverse momentum (MET) together with its azimuthal angle. 
The last ten features are higher order functions of the first 8 features; these features are derived by physicists to help discriminate between the two classes. 
These high-level features can be thought of as the physicists' attempt to use non-linear functions to classify signal and background events, having been developed with formidable theoretical effort. Here, we will use only logistic regression to attempt to classify events as either signal (that is, coming from a SUSY process) or background (events from some already observed Standard Model process). Later on in the review, in Sec.~\ref{sec:DNNs}, we shall revisit the same problem with the tools of Deep Learning.

As stated before, we never know the true underlying process, and hence the goal in these types of analyses is to find regions enriched in signal events. If we find an excess of events above what is expected, we can have confidence that they are coming from the type of signal we are searching for.
Therefore, the two metrics of import are the efficiency of signal selection, and the background rejection achieved (also called detection/rejection rates and similar to recall/precision).
Oftentimes, rather than thinking about just a single working point, performance is characterized by Receiver Operator Charecteristic curves (ROC curves).
These ROC curves plot signal efficiency versus background rejection at various thresholds of some discriminating variable.
Here that variable will be the output signal probability of our logistic regression.
Figure~\ref{fig:LogProbSUSY} shows examples of these outputs for true signal events (left) and background events (right) using $L^2$  regularization with a regularization parameter of $10^{-5}$.

\begin{figure}[h!]
\includegraphics[width=1.0\columnwidth]{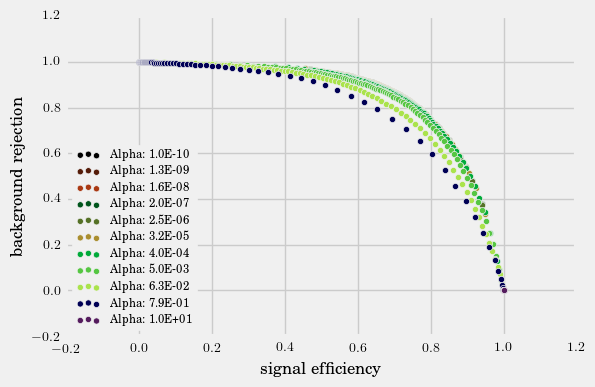}
\includegraphics[width=1.0\columnwidth]{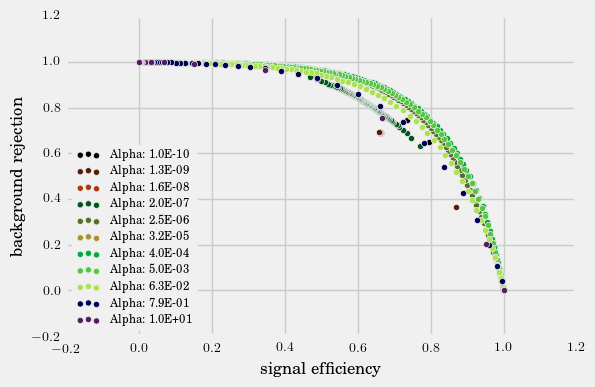}
\caption{ROC curves for a variety of regularization parameters with L2 regularization using TensorFlow (top) or Sci-Kit Learn (bottom).}
\label{SUSYROC}
\end{figure}

Notice that while the majority of signal events receive high probabilities of passing our discriminator and the majority of background events receive low probabilities, some signal events look background-like, and some background events look signal-like to our discriminator.
This is further reason to characterize performance of our selection in terms of ROC curves.
Figure~\ref{SUSYROC} shows examples of these curves using $L^2$ regularization for many different regularization parameters using two different ML python packages, either TensorFlow (top) or Sci-Kit Learn (bottom), when using the full set of $18$ input variables.
Notice there is minimal overfitting, in part because we trained on such a large dataset (4.5 million events).
More importantly, however, is the underlying data we are working with: each input variable is an important feature.

\begin{figure}[h!]
\includegraphics[width=1.0\columnwidth]{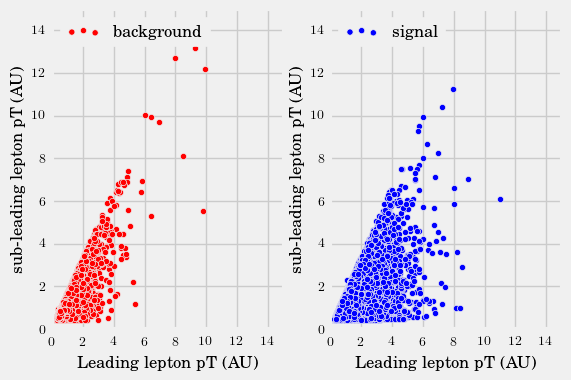}
\caption{Comparison of leading vs. sub-leading lepton $p_{T}$ for signal (blue) and background events (red). Recall that these variables have been scaled to have a mean of one.}
\label{fig:LeptonPTComparison}
\end{figure}

While figure~\ref{SUSYROC} shows nice discrimination power between signal and background events, the adoption of ML techniques adds complication to any analysis.
Given that we've already come up with a set of discriminating variables, including higher order ones derived from theories about SUSY particles, it's worth reflecting on whether there is utility to the increased sophistication of ML.
To show why we would want to use such a technique, recall that, even to the learning algorithm, some signal events and background events look similar.
We can illustrate this directly by looking at a plot comparing the $p_{T}$ spectrum of the two highest $p_{T}$ leptons (often referred to as the leading and sub-leading leptons) for both signal and background events.
Figure~\ref{fig:LeptonPTComparison} shows these two distributions, and one can see that while \emph{some} signal events are easily distinguished, many live in the same part of phase space as the background.
This effect can also be seen by looking at figure~\ref{fig:LogProbSUSY} where you can see that some signal events look like background events and vice-versa.

\begin{figure}[h!]
\includegraphics[width=1.0\columnwidth]{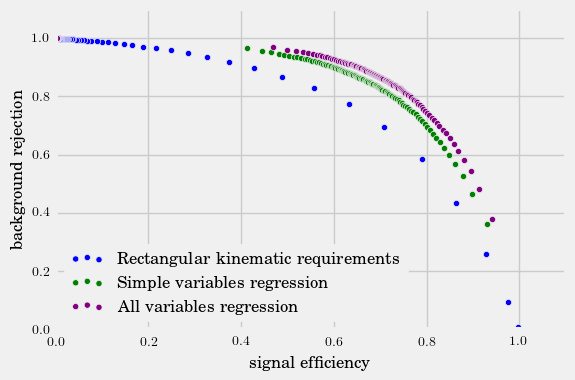}
\caption{A comparison of discrimination power from using logistic regression with only simple kinematic variables (green), logistic regression using both simple and higher-order kinematic variables (purple), and a cut-based approach that varies the requirements on the leading lepton $p_{T}$.}
\label{fig:SUSY-Comparison}
\end{figure}

One could then ask how much discrimination power is obtained by simply putting different requirements on the input variables rather than using ML techniques.
In order to compare this strategy (often referred to as cut-based in the field of HEP) to our regression results, different ROC curves have been made for each of the following cases: logistic regression with just the simple kinematic variables, logistic regression with the full set of variables, and simply putting a requirement on the leading lepton $p_{T}$.
Figure~\ref{fig:SUSY-Comparison} shows that there is a clear performance benefit from using logistic regression.
Note also that in the cut-based approach we have only used one variable where we could have put requirements on all of them.
While putting more requirements would indeed increase background rejection, it would also decrease signal efficiency. 
Hence, the cut-based approach will never yield as strong discrimination as the logistic regression we have performed.
One other interesting point about these results is that the higher-order variables noticeably help the ML techniques.
In later sections, we will return to this point to see if more sophisticated techniques can provide further improvement.

\subsection{Softmax Regression}
\label{subsec:softmax}

So far we have focused only on binary classification, in which the labels are dichotomous variables. Here we generalize logistic regression to multi-class classification. One approach is to treat the label as a vector $\bd{y}_i\in\mathbb{Z}_2^M$, namely a binary string of length $M$ with only one component of $y_i$ being $1$ and the rest zero. For example, $\bd{y}_i=(1,0,\cdots,0)$ means data the sample $\bd{x}_i$ belongs to class $1$\footnote{For an alternative mathematical description of the categories, which labels the classes by integers, see \url{http://ufldl.stanford.edu/wiki/index.php/Softmax\_Regression}.}, cf.~Fig.~\ref{fig:log_reg_intro}. Following the notation in Sec.~\ref{subsec:cross_entropy}, the probability of $\bd{x}_i$ being in class $m'$ is given by
\begin{equation}
P(y_{im^\prime}=1|\bd x_i,\{\mbf{w}_k\}_{k=0}^{M-1}) = \frac{\mathrm e^{-\mbf x_i^T\mbf w_{m'}}}{\sum_{m=0}^{M-1} \mathrm e^{-\mbf x_i^T\mbf w_m} },
\end{equation}
where ${y}_{im'}\equiv [\bd{y}_{i}]_{m'}$ refers to the $m'$-th component of vector $\bd{y}_i$. This is known as the {\bf SoftMax} function. Therefore, the likelihood of this $M$-class classifier is simply (cf.~Sec.~\ref{subsec:cross_entropy}):
\bea
P(\mathcal{D}|\{\mbf{w}_k\}_{k=0}^{M-1}) &=& \prod_{i=1}^n \prod_{m=0}^{M-1} \left[P(y_{im}=1|\bd x_i,\mbf{w}_m)\right]^{y_{im}}\nonumber \\
&\times&\left[1-P(y_{im}=1|\bd x_i,\mbf{w}_m)\right]^{1-y_{im}}
\label{eq:softmax_likelihood}
\eea
from which we can define the cost function in a similar fashion:
\bea
\label{eq:cross_entropy_softmax}
\mathcal{C}(\mbf w) &=& -\sum_{i=1}^n\sum_{m=0}^{M-1} y_{im}\log P(y_{im}=1|\bd x_i,\mbf{w}_m)\nonumber \\ 
&+& (1-y_{im})\log \left(1- P(y_{im}=1|\bd x_i,\mbf{w}_m)\right).
\eea
As expected, for $M=1$, we recover the cross entropy for logistic regression, cf.~Eq.~\eqref{eq:cross_entropy}.

\begin{figure}[t!]
	\includegraphics[width=1.0\columnwidth]{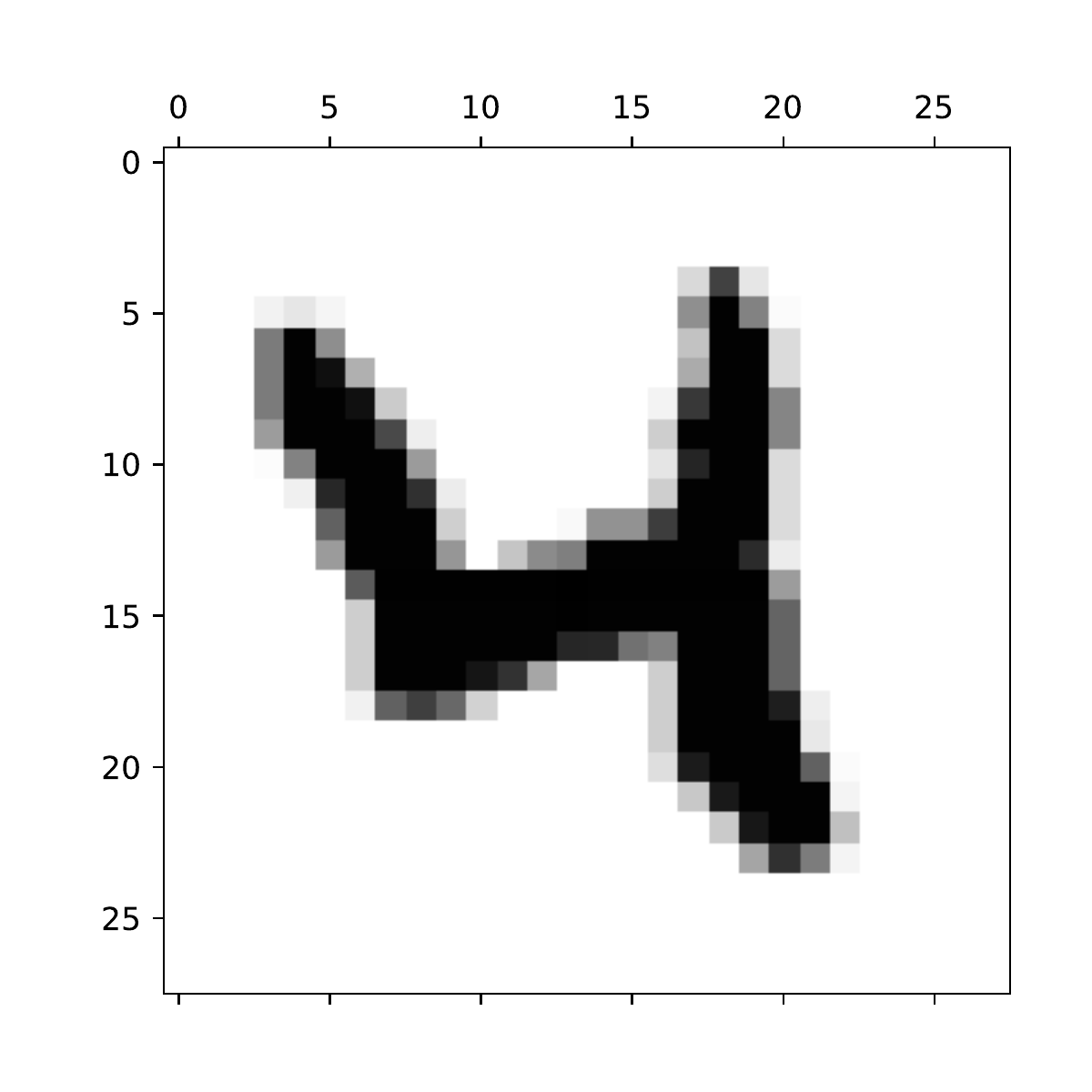}
	\caption{\label{fig:MNIST_data_vis} An example of an input datapoint from the MNIST data set. Each datapoint is a $28\times 28$-pixel image of a handwritten digit, with its corresponding label belonging to one of the $10$ digits. Each pixel contains a greyscale value represented by an integer between $0$ and $255$.}
\end{figure} 

\subsection{An Example of SoftMax Classification: MNIST Digit Classification}
\label{subsec:softmax_examples}

A paradigmatic example of SoftMax regression is to classify handwritten digits from the MNIST dataset. Yann LeCun and collaborators first collected and processed $70000$ handwritten digits, each of which is laid out on a $28\times 28$-pixel grid. Every pixel assumes one of $256$ grayscale values, interpolating between white and black. A representative input sample is show in Fig.~\ref{fig:MNIST_data_vis}.

\begin{figure*}[t!]
	\includegraphics[width=1.0\textwidth]{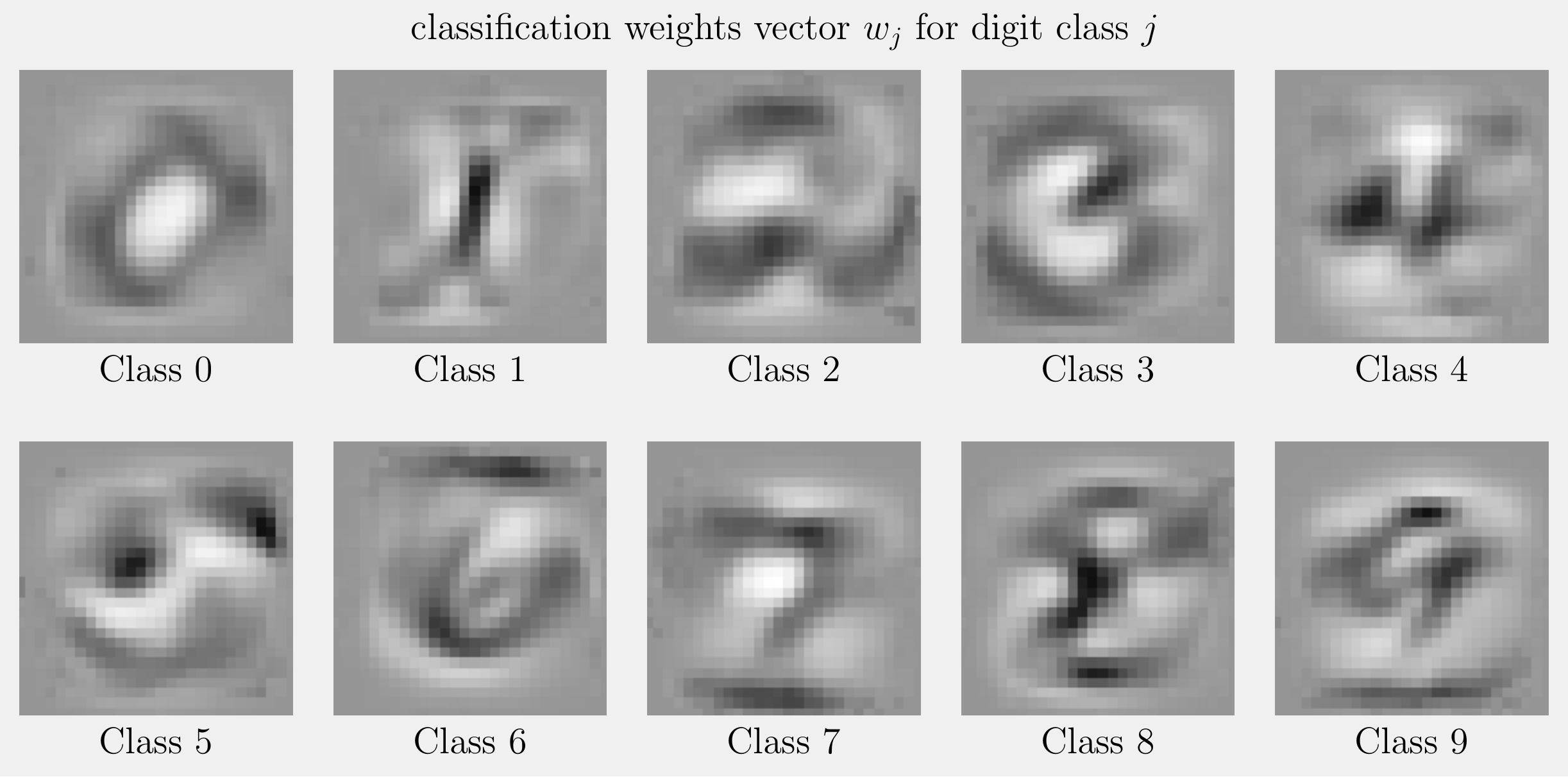}
	\caption{\label{fig:mnist_weights} Visualization of the weights $\mbf{w}_j$ after training a SoftMax Regression model on the MNIST dataset (see~\href{https://physics.bu.edu/~pankajm/MLnotebooks.html}{Notebook 7}). We emphasize that SoftMax Regression does not have explicit $2D$ spatial knowledge; the model learns from data points flattened out in a one-dimensional array.}
\end{figure*}

Since there are $10$ categories for the digits $0$ through $9$, this corresponds to SoftMax regression with $M=10$. We encourage readers to experiment with \href{https://physics.bu.edu/~pankajm/MLnotebooks.html}{Notebook 7} to explore SoftMax regression applied to MNIST. We include in Fig.~\ref{fig:mnist_weights} the learned weights $\bd{w}_k$, where $k$ corresponds to class labels (i.e.~digits). We shall come back to SoftMax regression in Sec.~\ref{sec:DNNs}.

%% file: sections/ensemble.tex
One of the most powerful and widely-applied ideas in modern machine learning is the use of ensemble methods that combine predictions from 
multiple, often weak, statistical models to improve predictive performance \cite{dietterich2000ensemble}. Ensemble methods, such as random forests 
\cite{breiman2001random, ho1998random, geurts2006extremely}, and boosted gradient trees, such as XGBoost \cite{friedman2001greedy, chen2016xgboost}, undergird many of the winning entries in 
data science competitions such as Kaggle, especially on structured datasets~\footnote{Neural networks generally perform better than ensemble methods on unstructured data, images, and audio.}.  Even in the context of neural networks, see Sec.~\ref{sec:DNNs}, it is common to combine predictions from multiple neural networks to 
increase performance on tough image classification tasks \cite{ioffe2015batch,he2015delving}.

In this section, we give an overview of ensemble methods and provide rules of thumb for when and why they work. On one hand, the idea of training multiple models and then using a weighted sum of the predictions of the all these models is very natural. After all, the idea of the ``wisdom of the crowds'' can be traced back, at least, to the writings of Aristotle in {\it{Politics}}. On the other hand, one can also imagine that the ensemble predictions can be much worse than the predictions from each of the individual models that constitute the ensemble, especially when pooling reinforces weak but correlated deficiencies in each of the individual predictors. Thus, it is important to understand when we expect ensemble methods to work.
 
In order to do this, we will revisit the bias-variance trade-off, discussed in Sec.~\ref{sec:stat_learn_theory}, and generalize it to consider an ensemble of classifiers. We will show that the key to determining when ensemble methods work is the degree of correlation between the models in the ensemble  \cite{louppe2014understanding}. Armed with this intuition, we will introduce some of the most widely-used and powerful ensemble methods including bagging  \cite{breiman1996bagging}, boosting \cite{freund1995desicion, freund1999short, schapire2012boosting}, random forests \cite{breiman2001random}, and gradient boosted trees such as XGBoost \cite{chen2016xgboost}.

\subsection{Revisiting the Bias-Variance Tradeoff for Ensembles}

The bias-variance tradeoff summarizes the fundamental tension in machine learning between the complexity of a model and the amount of training data needed to fit it (see Sec.~\ref{sec:stat_learn_theory}).  Since data is often limited, in practice it is frequently useful to use a less complex model with higher bias -- a model whose asymptotic performance is worse than another model --  because it is easier to train and less sensitive to sampling noise arising from having a finite-sized training dataset (i.e. smaller variance). Here, we will revisit the bias-variance tradeoff in the context of ensembles, drawing upon the beautiful discussion in Ref.~\cite{louppe2014understanding}.

A key property that will emerge from this analysis is the correlation between models that constitute the ensemble. The degree of correlation between models\footnote{For example, the correlation coefficient between the predictions made by two randomized models based on the same training set but with different random seeds, see Sec.~\ref{subsec:bias-variance-ensemble} for precise definition.} is important for two distinct reasons. First, holding the ensemble size fixed, averaging the predictions of correlated models reduces the variance less than averaging uncorrelated models. Second, in some cases, correlations between models within an ensemble can result in an \textit{increase} in bias, offsetting any potential reduction in variance gained from ensemble averaging. We will discuss this in the context of bagging below. One of the most dramatic examples of increased bias from correlations is the catastrophic predictive failure of almost all derivative models used by Wall Street during the 2008 financial crisis.

\subsubsection{Bias-Variance Decomposition for Ensembles}
\label{subsec:bias-variance-ensemble}
We will discuss the bias-variance tradeoff in the context of continuous predictions such as regression. However, many of the intuitions and ideas discussed here also carry over to classification tasks. Before discussing ensembles, let us briefly review the bias-variance tradeoff in the context of a single model. Consider a data set consisting of data $\mathbf{X}_\mathcal{L}=\{(y_j, \mathbf{x}_j), j=1\ldots N\}$. Let us assume that the true data is generated from a noisy model
\be
y=f(\mathbf{x}) + \epsilon,
\ee
where $\epsilon$ is a normally distributed with mean zero and standard deviation $\sigma_\epsilon$.

Assume that we have a statistical procedure (e.g.~least-squares regression) for forming a predictor $\hat{g}_{\mathcal{L}}(\mathbf{x})$ that gives the prediction of our model for a new data point $\mathbf{x}$ given that we trained the model using a dataset $\mathcal{L}$. This estimator is chosen by minimizing a cost function which, for the sake of concreteness, we take to be the squared error
\be
 \mathcal{C}( \boldsymbol{X}, g(\boldsymbol{x})) =  \sum_i (\mathbf{y}_i - \hat{g}_\mathcal{L}(\mathbf{x}_i))^2. 
\ee
The dataset $\mathcal{L}$ is drawn from some underlying distribution that describes the data. If we imagine drawing many datasets $\{\mathcal{L}_j\}$ of the same size as $\mathcal{L}$ from this distribution, we know that the corresponding estimators $\hat{g}_{\mathcal{L}_j}(\mathbf{x})$ will differ from each other due to stochastic effects arising from sampling noise. For this reason, we can view our estimator $\hat{g}_{\mathcal{L}}(\mathbf{x})$ as a random variable and define an expectation value $\mbbE_{\mathcal{L}}$ in the usual way. Note that the subscript denotes that the expectation is taken over $\mathcal{L}$.  In practice, $\mbbE_{\mathcal{L}}$ is computed by by drawing infinitely many different datasets $\{\mathcal{L}_j\}$ of the same size, fitting the corresponding estimator, and then averaging the results. We will also average over different instances of the ``noise'' $\epsilon$. The expectation value over the noise will be denoted by $E_\epsilon$. 

As discussed in Sec.~\ref{sec:stat_learn_theory}, we can decompose the expected generalization error as
\be
\mbbE_\mathcal{L, \epsilon}[\mathcal{C}( \boldsymbol{X}, g(\boldsymbol{x})) ] = Bias^2 + Var + Noise.
\ee
where the bias,
\be
Bias^2=\sum_i (f(\mathbf{x}_i)-\mbbE_\mathcal{L}[\hat{g}_\mathcal{L}(\mathbf{x}_i)])^2,
\ee
measures the deviation of the expectation value of our estimator (i.e.~the asymptotic value of our estimator in the limit of infinite data) from the true value. The variance
\be
Var=\sum_i \mbbE_\mathcal{L}[( \hat{g}_\mathcal{L}(\mathbf{x}_i)-\mbbE_\mathcal{L}[\hat{g}_\mathcal{L}(\mathbf{x}_i)])^2],
\ee
 measures how much our estimator fluctuates due to finite-sample effects. The noise term
\be
Noise= \sum_i \sigma_{\epsilon_i}^2
\ee
is the part of the error due to intrinsic noise in the data generation process that no statistical estimator can overcome.
 
Let us now generalize this to ensembles of estimators. Given a dataset $\mathbf{X}_\mathcal{L}$ and hyper-parameters $\theta$ that parameterize members of our ensemble, we will consider a procedure that deterministically generates a model $\hat{g}_\mathcal{L}(\mathbf{x}_i, \theta)$ given $\mathbf{X}_\mathcal{L}$ and $\theta$. We assume that the $\theta$ includes some random parameters that introduce stochasticity into our ensemble (e.g.~an initial condition for stochastic gradient descent or a random subset of features or data points used for training.) Concretely, with a giving dataset $\mathcal{L}$, one has a learning algorithm $\mathcal{A}$ that generates a model $\mathcal{A}(\theta, \mathcal{L})$ based on a deterministic procedure which introduced stochasticity through $\theta$ in its execution on dataset $\mathcal{L}$. We will be concerned with the expected prediction error of the \emph{aggregate ensemble predictor}
\be
\hat{g}_\mathcal{L}^A(\mathbf{x}_i, \{\theta\})= {1 \over M}\sum_{m=1}^M \hat{g}_\mathcal{L}(\mathbf{x}_i, \theta_m).
\label{Ens:def_aggregate}
\ee
For future reference, let us define the mean, variance, and covariance (i.e. the connected correlation function in the language of physics), and the normalized correlation coefficient of a single randomized model $ \hat{g}_\mathcal{L}(\mathbf{x}, \theta_m)$ as:
\bea
\mbbE_{\mathcal{L},\theta_m}[ \hat{g}_\mathcal{L}(\mathbf{x}, \theta_m)] &=& \mu_{\mathcal{L}, \theta_m}(\mathbf{x}) \nonumber \\
\mbbE_{\mathcal{L},\theta_m}[ \hat{g}_\mathcal{L}(\mathbf{x}, \theta_m)^2]-\mbbE_{\mathcal{L},\theta_m}[ \hat{g}_\mathcal{L}(\mathbf{x}, \theta)]^2 &=& \sigma_{\mathcal{L}, \theta_m}^2(\mathbf{x}) \nonumber \\
\mbbE_{\mathcal{L},\theta_m}[ \hat{g}_\mathcal{L}(\mathbf{x}, \theta_m) \hat{g}_\mathcal{L}(\mathbf{x},  \theta_{m^\prime})]-\mbbE_\theta[ \hat{g}_\mathcal{L}(\mathbf{x}, \theta_m)]^2 &=& C_{\mathcal{L}, \theta_m, \theta_{m^\prime}}(\mathbf{x}) \nonumber \\
\rho(\mathbf{x})= {C_{\mathcal{L}, \theta_m, \theta_{m^\prime}}(\mathbf{x}) \over \sigma_{\mathcal{L}, \theta}^2}.&&
\label{Ens:cor_aggregate}
\eea
Note that the expectation $\mbbE_{\mathcal{L},\theta_m}[\cdot]$ is computed over the joint distribution of $\mathcal{L}$ and $\theta_m$. Also, by definition, we assume $m \neq m^\prime$ in $C_{\mathcal{L}, \theta_m, \theta_{m\prime}}$.
 
We can now ask about the expected generalization (out-of-sample) error for the ensemble
\be
\mbbE_{\mathcal{L}, \epsilon, \theta}\left[\mathcal{C}( \boldsymbol{X}, \hat{g}_\mathcal{L}^A(\boldsymbol{x})) \right]=\mbbE_{\mathcal{L}, \epsilon, \theta} \left[\sum_i (\mathbf{y}_i - \hat{g}_\mathcal{L}^A(\mathbf{x}_i, \{\theta\}))^2\right].
\ee
As in the single estimator case, we decompose the error into a noise term, a bias-term, and a variance term. To see this, note that
\begin{widetext}
 \bea
\mbbE_{\mathcal{L}, \epsilon, \theta}[\mathcal{C}( \boldsymbol{X}, \hat{g}_\mathcal{L}^A(\mathbf{x})) ] 
&=&\mbbE_{\mathcal{L}, \epsilon, \theta} \left[\sum_i (\mathbf{y}_i - f(\mathbf{x}_i)+ f(\mathbf{x}_i)- \hat{g}_\mathcal{L}^A(\mathbf{x}_i, \{\theta\}))^2\right]  \nonumber \\
&=&\sum_i \mbbE_{\mathcal{L}, \epsilon, \theta}[(\mathbf{y}_i - f(\mathbf{x}_i))^2 + (f(\mathbf{x}_i)- \hat{g}_\mathcal{L}^A(\mathbf{x}_i, \{\theta\}))^2 + 2(\mathbf{y}_i - f(\mathbf{x}_i))(f(\mathbf{x}_i)- \hat{g}_\mathcal{L}^A(\mathbf{x}_i, \{\theta\}))] \nonumber \\
&=& \sum_i \sigma_{\epsilon_i}^2 +\sum_i \mbbE_{\mathcal{L}, \theta}[(f(\mathbf{x}_i)- \hat{g}_\mathcal{L}^A(\mathbf{x}_i, \{\theta\}))^2],
\eea
where in the last line we have used the fact that $\mbbE_\epsilon[y_i]=f(\mathbf{x}_i)$ to eliminate the last term.
We can further decompose the second term as
\bea
\mbbE_{\mathcal{L}, \theta}[(f(\mathbf{x}_i)- \hat{g}_\mathcal{L}^A(\mathbf{x}_i, \{\theta\}))^2] &=&  \mbbE_{\mathcal{L}, \theta}[(f(\mathbf{x}_i)- \mbbE_{\mathcal{L}, \theta}[  \hat{g}_\mathcal{L}^A(\mathbf{x}_i, \{\theta\})] + \mbbE_{\mathcal{L}, \theta}[  \hat{g}_\mathcal{L}^A(\mathbf{x}_i, \{\theta\})] -\hat{g}_\mathcal{L}^A(\mathbf{x}_i, \{\theta\}))^2] \nonumber \\
&=&\mbbE_{\mathcal{L}, \theta}[(f(\mathbf{x}_i)- \mbbE_{\mathcal{L}, \theta}[\hat{g}_\mathcal{L}^A(\mathbf{x}_i, \{\theta\})])^2]+  \mbbE_{\mathcal{L}, \theta}[(\mbbE_{\mathcal{L}, \theta}[  \hat{g}_\mathcal{L}^A(\mathbf{x}_i, \{\theta\})] -\hat{g}_\mathcal{L}^A(\mathbf{x}_i, \{\theta\}))^2] \nonumber \\
&+& 2\mbbE_{\mathcal{L}, \theta}[(E_{\mathcal{L}, \theta}[  \hat{g}_\mathcal{L}^A(\mathbf{x}_i, \{\theta\})] -\hat{g}_\mathcal{L}^A(\mathbf{x}_i, \{\theta\}))(f(\mathbf{x}_i)- \mbbE_{\mathcal{L}, \theta}[  \hat{g}_\mathcal{L}^A(\mathbf{x}_i, \{\theta\})])]\nonumber 
\\ &=&(f(\mathbf{x}_i)- \mbbE_{\mathcal{L}, \theta}[  \hat{g}_\mathcal{L}^A(\mathbf{x}_i, \{\theta\})])^2 +\mbbE_{\mathcal{L}, \theta}[  (\hat{g}_\mathcal{L}^A(\mathbf{x}_i, \{\theta\})-\mbbE_{\mathcal{L}, \theta}[  \hat{g}_\mathcal{L}^A(\mathbf{x}_i, \{\theta\})])^2]  \nonumber \\
&\equiv& Bias^2(\mathbf{x}_i) + Var(\mathbf{x}_i),
\eea
\end{widetext}
where we have defined the bias of an aggregate predictor as 
\be
Bias^2(\mathbf{x}) \equiv  ( f(\mathbf{x})- \mbbE_{\mathcal{L}, \theta}[  \hat{g}_\mathcal{L}^A(\mathbf{x}, \{\theta\}) ] )^2
\ee
and the variance as
\be
Var(\mathbf{x}) \equiv \mbbE_{\mathcal{L}, \theta}[ ( \hat{g}_\mathcal{L}^A(\mathbf{x}, \{\theta\})-\mbbE_{\mathcal{L}, \theta}[  \hat{g}_\mathcal{L}^A(\mathbf{x}, \{\theta\})])^2].
\ee

So far the calculation for ensembles is almost identical to that of a single estimator. However, since the aggregate estimator is a sum of estimators, its variance implicitly depends on the correlations between the individual estimators in the 
ensemble. Using the definition of the aggregate estimator Eq.~\eqref{Ens:def_aggregate} and the definitions in Eq.~\eqref{Ens:cor_aggregate}, we see that
\begin{widetext}
\bea
Var(\mathbf{x}) &=& \mbbE_{\mathcal{L}, \theta}[ ( \hat{g}_\mathcal{L}^A(\mathbf{x}, \{\theta\})-\mbbE_{\mathcal{L}, \theta}[  \hat{g}_\mathcal{L}^A(\mathbf{x}, \{\theta\})])^2] \nonumber \\
&=& {1 \over M^2} \left[ \sum_{m,m^\prime} \mbbE_{\mathcal{L}, \theta}[ \hat{g}_\mathcal{L}(\mathbf{x}, \theta_m) \hat{g}_\mathcal{L}(\mathbf{x}, \theta_{m^\prime})]- M^2\sum_i  [\mu_{\mathcal{L}, \theta}(\mathbf{x}) ]^2 \right] \nonumber\\
&=& \rho(\mathbf{x}) \sigma_{\mathcal{L}, \theta}^2+ {1-\rho(\mathbf{x}) \over M} \sigma_{\mathcal{L}, \theta}^2.
\label{Ens:var_form}
\eea
\end{widetext}

This last formula is the key to understanding the power of random ensembles. Notice that by using large ensembles ($M \rightarrow \infty$), we can significantly reduce the variance, and for completely random ensembles
where the models are uncorrelated ($\rho(\mathbf{x})=0$), maximally suppresses the variance! Thus, using the aggregate predictor beats down fluctuations due to finite-sample effects. The key, as the formula indicates, is to decorrelate the models as much as possible while still using a very large ensemble. One can be worried that this comes at the expense of a very large bias. This turns out not to be the case. When models in the ensemble are completely random, the bias of the aggregate predictor is just the expected
bias of a single model
\bea
Bias^2(\mathbf{x}) &=&  (f(\mathbf{x})- \mbbE_{\mathcal{L}, \theta}[  \hat{g}_\mathcal{L}^A(\mathbf{x}, \{\theta\})])^2 \nonumber \\
&=&(f(\mathbf{x})- {1 \over M} \sum_{m=1}^M \mbbE_{\mathcal{L}, \theta}[  \hat{g}_\mathcal{L}(\mathbf{x}, \theta_m)])^2  \\
&=&(f(\mathbf{x})-\mu_{\mathcal{L},\theta})^2.
\eea
Thus, for a random ensemble one can always add more models without increasing the bias. This observation lies behind the immense power of random forest methods discussed below. For other methods, such as bagging, we will see that the bootstrapping procedure actually does increase the bias. But in many cases, this increase in bias is negligible compared to the reduction in variance.

\subsubsection{Summarizing the Theory and Intuitions behind Ensembles}

\begin{figure}[t!]
	\includegraphics[width=1.0\columnwidth]{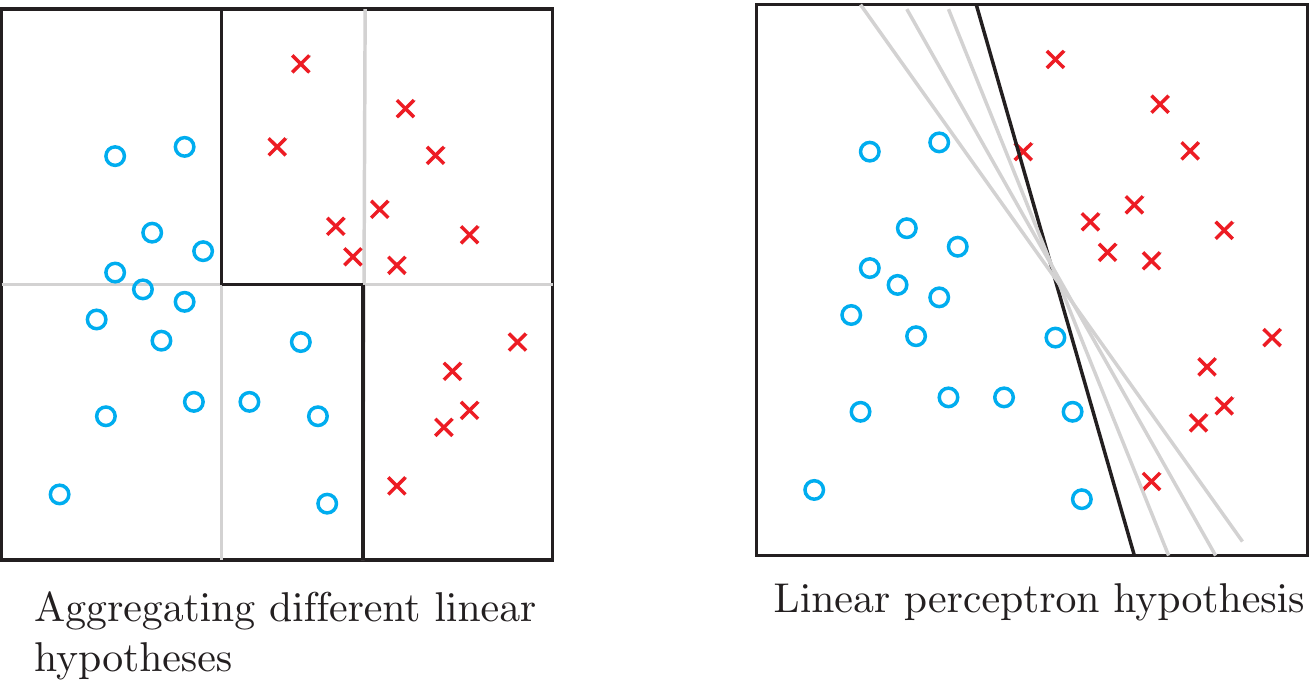}
	\caption{ Why combining models? On the left we show that by combining simple linear hypotheses (grey lines) one can achieve better and more flexible classifications (dark line), which is in stark contrast to the case in which one only uses a single perceptron hypothesis as shown on the right.}
	\label{fig:whyagg}
\end{figure}

Before discussing specific methods, let us briefly summarize why ensembles have proven so successful in many ML applications. Dietterich \cite{dietterich2000ensemble} identifies three distinct shortcomings that are fixed by ensemble methods: statistical, computational,
and representational. These are explained in the following discussion from Ref.~\cite{louppe2014understanding}:
\begin{quote}
The first reason is statistical. When the learning set is too small, a learning algorithm can typically find several models in the hypothesis space $\mathcal{H}$  that all give the same performance on the training data. Provided their predictions are uncorrelated, averaging several models reduces the risk of choosing the wrong hypothesis. The second reason is computational. Many learning algorithms rely on some greedy assumption or local search that may get stuck in local optima. As such, an ensemble made of individual models built from many different starting points may provide a better approximation of the true unknown function than any of the single models. Finally, the third reason is representational. In most cases, for a learning set of finite size, the true function cannot be represented by any of the candidate models in $\mathcal{H}$. By combining several models in an ensemble, it may be possible to expand the space of representable functions and to better model the true function.
\end{quote}

The increase in representational power of ensembles can be simply visualized. For example, the classification task shown in Fig.~\ref{fig:whyagg} reveals that it is more advantageous to combine a group of simple hypotheses (vertical or horizontal lines) than to utilize a single arbitrary linear classifier. This of course comes with the price of introducing more parameters to our learning procedure. But if the problem itself can never be learned through a simple hypothesis, then there is no reason to avoid applying a more complex model. Since ensemble methods reduce the variance and are often easier to train than a single complex model, they are a powerful way of increasing representational power (also called expressivity in the ML literature).

Our analysis also gives several intuitions for how we should construct ensembles. First, we should try to randomize ensemble construction as much as possible to reduce the correlations between predictors in the ensemble. This ensures that our variance will be reduced while minimizing an increase in bias due to correlated errors. Second, the ensembles will work best for procedures where the error of the predictor is dominated by the variance and not the bias. Thus, these methods are especially well suited for unstable procedures whose results are sensitive to small changes in the training dataset. 

Finally, we note that although the discussion above was derived in the context of continuous predictors such as regression, the basic intuition behind using ensembles applies equally well to classification tasks. Using an ensemble allows one to reduce the variance by averaging the result of many independent classifiers. As with regression, this procedure works best for unstable predictors for which errors are dominated by variance due to finite sampling rather than bias.  

\subsection{Bagging}
\label{subsec:bagging}

BAGGing, or Bootstrap AGGregation, first introduced by Leo Breiman, is one of the most widely employed and simplest ensemble-inspired methods~\cite{breiman1996bagging}. Imagine we have a very large dataset $\mathcal{L}$ that we could partition into
$M$ smaller data sets which we label $ \{ \mathcal{L}_1, \ldots, \mathcal{L}_M\}$. If each partition is sufficiently large to learn a predictor, we can create an ensemble aggregate predictor composed of predictors trained on each subset of the data. For continuous predictors like regression, this is just the average of all the individual predictors: 
\be
\hat{g}_\mathcal{L}^A (\mathbf{x}) = {1 \over M} \sum_{i=1}^M g_{\mathcal{L}_i}(\mathbf{x}).
\ee  
For classification tasks where each predictor predicts a class label $j \in \{1, \ldots, J \}$, this is just a majority vote of all the predictors,
\be
\hat{g}_\mathcal{L}^A (\mathbf{x}) = \argmax_j   \sum_{i=1}^M I[g_{\mathcal{L}_i}(\mathbf{x})=j],
\ee
where $I[g_{\mathcal{L}_i}(\mathbf{x})=j]$ is an indicator function that is equal to one if $g_{\mathcal{L}_i}(\mathbf{x})=j$ and zero otherwise. From the theoretical discussion above, we know that this can significantly reduce the variance without increasing the bias.

\begin{figure}[h!]
\includegraphics[width=0.9\columnwidth]{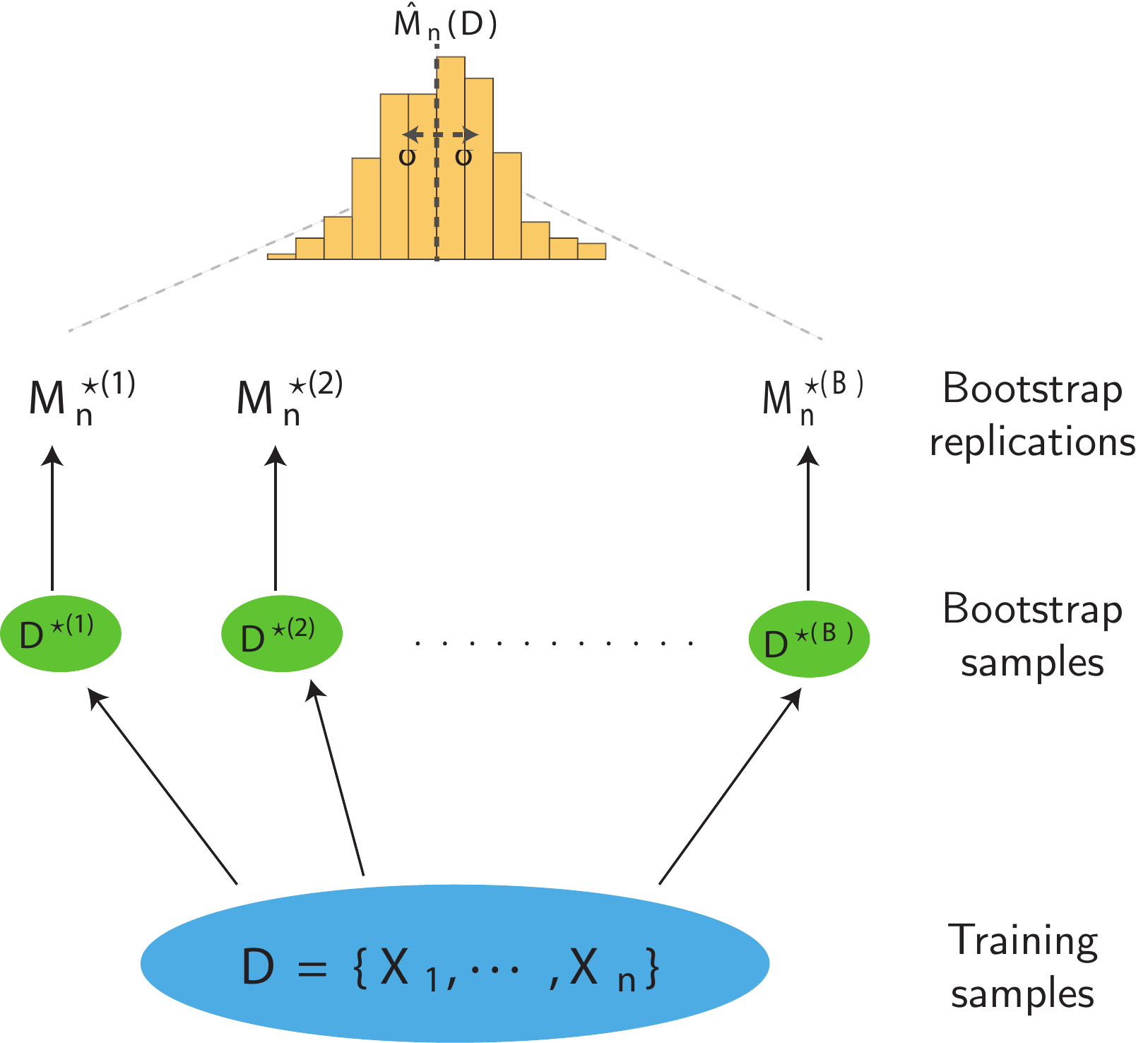}
\caption{\label{fig:bootstrap_illustration} Shown here is the procedure of empirical bootstrapping. The goal is to assess the accuracy of a statistical quantity of interest, which in the main text is illustrated as the sample median $\hat{M}_n(\mathcal{D})$. We start from a given dataset $\mathcal{D}$ and bootstrap $B$ size $n$ datasets $\mathcal{D}^{\star(1)}, \cdots, \mathcal{D}^{\star(B)}$ called the bootstrap samples. Then we compute the statistical quantity of interest on these bootstrap samples to get the median $M_n^{\star (k)}$, for $k=1,\cdots, B$. These are then used to evaluate the accuracy of $\hat{M}_n(\mathcal{D})$ (see also box on Bootstrapping in main text). It can be shown that in the $n\rightarrow\infty$ limit the distribution of $M_n^{\star (k)}$ would be a Gaussian centered around $\hat{M}_n(\mathcal{D})$ with variance $\sigma^2$ defined by Eq. \eqref{eq:bootstrapvar} scales as $1/n$. }
\end{figure}

While simple and intuitive, this form of aggregation clearly works only when we have enough data in each partitioned set $\mathcal{L}_i$. To see this, one can consider the extreme limit where $\mathcal{L}_i$ contains exactly one point. In this case, the base hypothesis $g_{\mathcal{L}_i}(\mathbf{x})$ (e.g.~linear regressor) becomes extremely poor and the procedure above fails. One way to circumvent this shortcoming is to resort to \textbf{empirical bootstrapping}, a resampling technique in statistics introduced by Efron \cite{efron1979bootstrap} (see accompanying box and  Fig.~\ref{fig:bootstrap_illustration}). The idea of empirical bootstrapping is to use sampling with replacement to create new ``bootstrapped'' datasets $\{L_1^{BS}, \ldots, L_M^{BS}\}$  from our original dataset $\mathcal{L}$.  These bootstrapped datasets share many points, but due to the sampling with replacement, are all somewhat different from each other. In the bagging procedure, we create an aggregate estimator by replacing the $M$ independent datasets by the $M$ bootstrapped estimators:
\be
\hat{g}_\mathcal{L}^{BS} (\mathbf{x}) = {1 \over M} \sum_{i=1}^M g_{\mathcal{L}_i^{BS}}(\mathbf{x}).
\ee  
and
\be
\hat{g}_\mathcal{L}^{BS} (\mathbf{x}) = \argmax_j   \sum_{i=1}^M I[g_{\mathcal{L}_i^{BS}}(\mathbf{x})=j].
\ee
This bootstrapping procedure allows us to construct an approximate ensemble and thus reduce the variance. For unstable predictors, this can significantly improve the predictive performance. The price we pay for using bootstrapped training datasets, as opposed to really partitioning the dataset, is an increase in the bias of our bagged estimators. To see this, note that  as the number of datasets $M$ goes to infinity, the expectation with respect to the bootstrapped samples converges to the empirical distribution describing the training data set $p_\mathcal{L}(\mathbf{\mathbf{x}})$ (e.g.~a delta function at each datapoint in $\mathcal{L}$) which in general is different from the true generative distribution for the data $p(\mathbf{x})$.

\begin{figure}[h!]
\includegraphics[width=0.9\columnwidth]{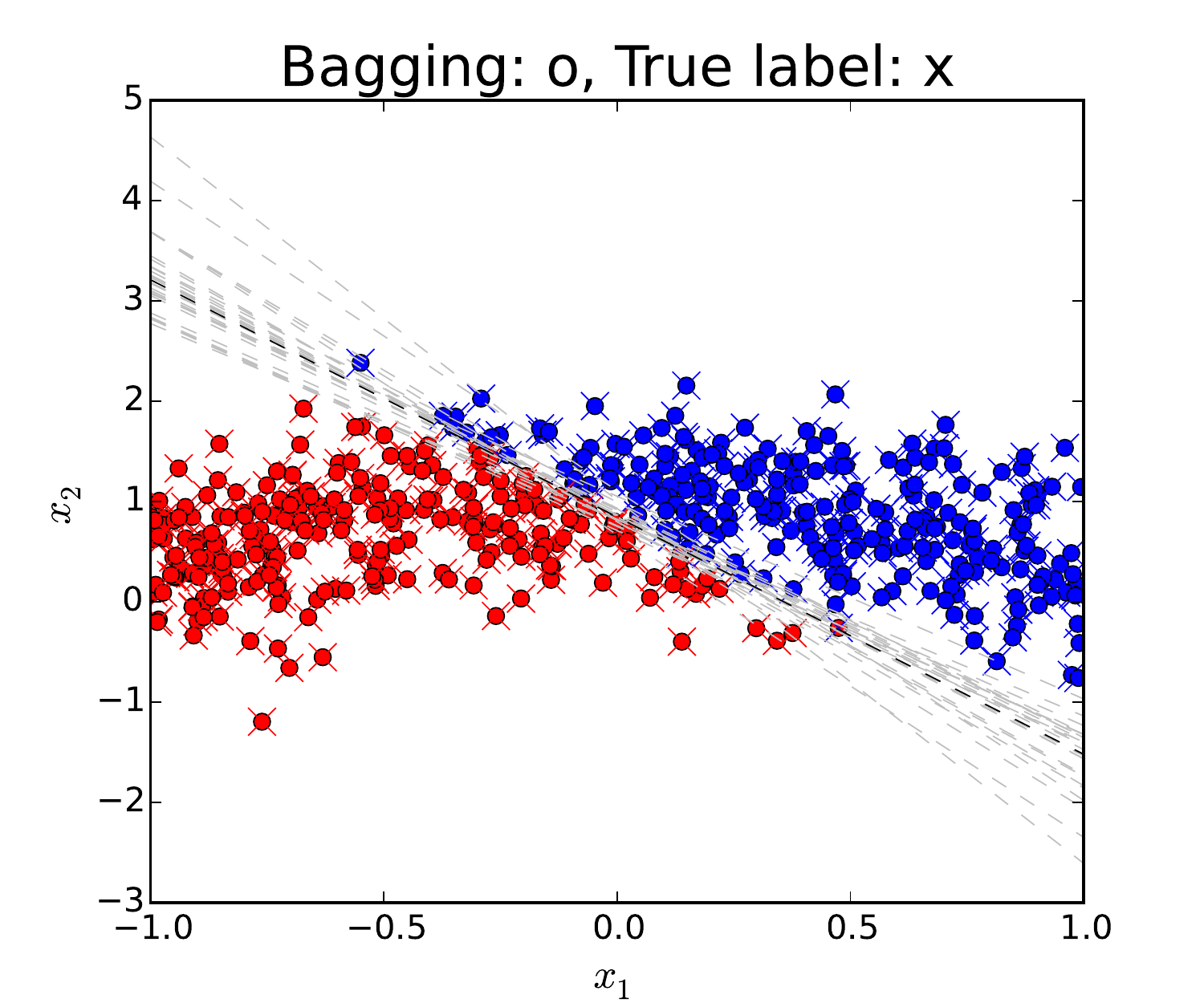}
\caption{ {\bf Bagging applied to the perceptron learning algorithm (PLA)}. Training data size $n=500$, number of bootstrap datasets $B=25$, each contains $50$ points. Colors corresponds to different classes while the marker indicates how these points are labelled: cross for true label and circle for that obtained by bagging. Each gray dashed line indicates the prediction made, based on every bootstrap set while the dark dashed black line is the average of these. }
\label{fig:baggingPLA_example}
\end{figure}

In Fig.~\ref{fig:baggingPLA_example} we demonstrate bagging with a perceptron  (linear classifier) as the base classifier that constitutes the elements of the ensemble. It is clear that, although each individual classifier in the ensemble performs poorly at classification, bagging these estimators yields reasonably good predictive performance. This raises questions like \emph{why bagging works} and \emph{how many bootstrap samples are needed}. As mentioned in the discussion above, bagging is effective on ``unstable'' learning algorithms where small changes in the training set result in large changes in predictions~\cite{breiman1996bagging}. When the procedure is unstable, the prediction error is dominated by the variance and one can exploit the aggregation component of bagging to reduce the prediction error. In contrast, for a stable procedure the accuracy is limited by the bias introduced by using bootstrapped datasets. This means that there is an instability-to-stability transition point beyond which bagging stops improving our prediction.

\fbox{
\begin{minipage}{22em}
\begin{center}
{\bf Brief Introduction to Bootstrapping}
\end{center}

Suppose we are given a finite set of $n$ data points $\mathcal{D}=\{X_1,\cdots, X_n\}$ as training samples and our job is to construct measures of confidence for our sample estimates (e.g.~the confidence interval or mean-squared error of sample median estimator). To do so, one first samples $n$ points \textbf{with replacement} from $\mathcal{D}$ to get a new set $\mathcal{D}^{\star(1)}=\{X^{\star(1)}_1,\cdots, X^{\star(1)}_n\}$, called a \textbf{bootstrap sample}, which possibly contains repetitive elements. Then we repeat the same procedure to get in total $B$ such sets: $\mathcal{D}^{\star(1)}, \cdots, \mathcal{D}^{\star(B)}$. The next step is to use these $B$ bootstrap sets to get the \textbf{bootstrap estimate} of the quantity of interest. For example, let $M_n^{\star (k)}=Median(\mathcal{D}^{\star (k)})$ be the sample median of bootstrap data $\mathcal{D}^{\star (k)}$. Then we can construct the variance of the distribution of bootstrap medians as :
\be\label{eq:bootstrapvar}
\widehat{Var}_B(M_n)=\frac{1}{B-1}\sum_{k=1}^B\left(M_n^{\star (k)} - \bar{M}_n^{\star}\right)^2,
\ee
where 
\be
\bar{M}_n^{\star}=\frac{1}{B}\sum_{k=1}^B M_n^{\star (k)}
\ee
is the mean of the median of all bootstrap samples. Specifically, Bickel and Freedman \cite{bickel1981some} and Singh \cite{singh1981asymptotic} showed that in the $n\rightarrow \infty$ limit, the distribution of the bootstrap estimate will be a Gaussian centered around $\hat{M}_n(\mathcal{D})=Median (X_1,\cdots, X_n)$  with standard deviation proportional to $1/\sqrt{n}$. This means that the bootstrap distribution $\hat{M}^{\star}_n -\hat{M}_n$ approximates fairly well the sampling distribution $\hat{M}_n - M$ from which we obtain the training data $\mathcal{D}$. Note that $M$ is the median based on which the true distribution $\mathcal{D}$ is generated. In other words, if we plot the histogram of $\{M_n^{\star (k)}\}_{k=1}^B$, we will see that in the large $n$ limit it can be well fitted by a Gaussian which sharp peaks at $\hat{M}_n(\mathcal{D})$ and vanishing variance whose definition is given by Eq.~\eqref{eq:bootstrapvar} (see Fig.~\ref{fig:bootstrap_illustration}). 
\end{minipage}}

\subsection{Boosting}
\label{subsec:boosting}

Another powerful and widely used ensemble method is \emph{Boosting}. In bagging, the contribution of all predictors is weighted equally in the bagged (aggregate) predictor. However, in principle, there are myriad ways to combine different predictors. In some problems one might prefer to use an autocratic approach that emphasizes the best predictors, while in others it might be better to opt for more `democratic' ways as is done in bagging. In all cases, the idea is to build a strong predictor by combining many weaker classifiers. 

In boosting, an ensemble of weak classifiers $\{ g_k(\mathbf{x})\}$ is combined into an aggregate, boosted classifier. However, unlike bagging, each classifier is associated with a weight  $\alpha_k$ that indicates how much it contributes to the aggregate classifier
\be\label{eq:DT_summary}
g_A(\mathbf{x})= \sum_{K=1}^M \alpha_k g_k(\mathbf{x}),
\ee
where $\sum_k \alpha_k=1$.  For the reasons outlined above, boosting, like all ensemble methods, works best when we combine simple, high-variance classifiers into a more complex whole.

Here, we focus on ``adaptive boosting'' or AdaBoost, first proposed by Freund and Schapire in the mid 1990s  \cite{freund1995desicion, freund1999short, schapire2012boosting}. The basic idea behind AdaBoost, is to form the aggregate classifier in an iterative process. Importantly, at each iteration we reweight the error function to "highlight" data points where the aggregate classifier performs poorly (so that in the next round the procedure put more emphasis on making those right.) In this way, we can successively ensure that our classifier has good performance over the whole dataset. 

We now discuss the AdaBoost procedure in greater detail. Suppose that we are given a data set $\mathcal{L}=\{(\bd{x}_i,y_i),\,i=1,\cdots, N\}$ where $\bd{x}_i\in \mathcal{X}$\; and $y_i\in\mathcal{Y}=\{+1,-1\}$. Our objective is to find an optimal hypothesis/classifier $g:\mathcal{X}\rightarrow\mathcal{Y}$ to classify the data. Let $\mathcal{H}=\{g:\mathcal{X}\rightarrow\mathcal{Y}\}$ be the family of classifiers available in our ensemble. In the AdaBoost setting, we are concerned with the classifiers that perform somehow better than ``tossing a fair coin''. This means that for each classifier, the family $\mathcal{H}$ can predict $y_i$ correctly at least half of the time. 

We construct the boosted classifier as follows: 
\begin{itemize}
\item \textbf{Initialize} $w_{t=1}(\bd{x}_n)=1/N, n=1,\cdots,N$.
\item \textbf{For} $t=1\cdots,T$(desired termination step), \textbf{do}:
\begin{enumerate}
\item Select a hypothesis $g_t\in\mathcal{H}$ that minimizes the weighted error
\be\label{eq:bagging-hypselect}
\epsilon_t=\sum_{i=1}^N w_{t}(\bd{x}_i)\mathbb{1}(g_t(\bd{x}_i)\neq y_i)
\ee
\item Let $\alpha_t=\frac{1}{2}\ln\frac{1-\epsilon_t}{\epsilon_t}$, update the weight for each data $\bd{x}_n$ by
\[w_{t+1}(\bd{x}_n)\leftarrow w_t(\bd{x}_n)\frac{\exp[-\alpha_ty_ng_t(\bd{x}_n)]}{Z_t},\]
where $Z_t=\sum_{n=1}^Nw_t(\bd{x}_n)e^{-\alpha_ty_ng_t(\bd{x}_n)}$ ensures all weights add up to unity.
\end{enumerate}
\item \textbf{Output} $g_A(\bd{x})=\text{sign}\left(\sum_{t=1}^T\alpha_tg_t(\bd{x})\right)$
\end{itemize}
There are many theoretical and empirical studies on the performance of AdaBoost but they are beyond the scope of this review. We refer interested readers to the extensive literature on boosting \cite{freund1999short}.

\subsection{Random Forests}

\begin{figure}[h!]
\includegraphics[width=0.9\columnwidth]{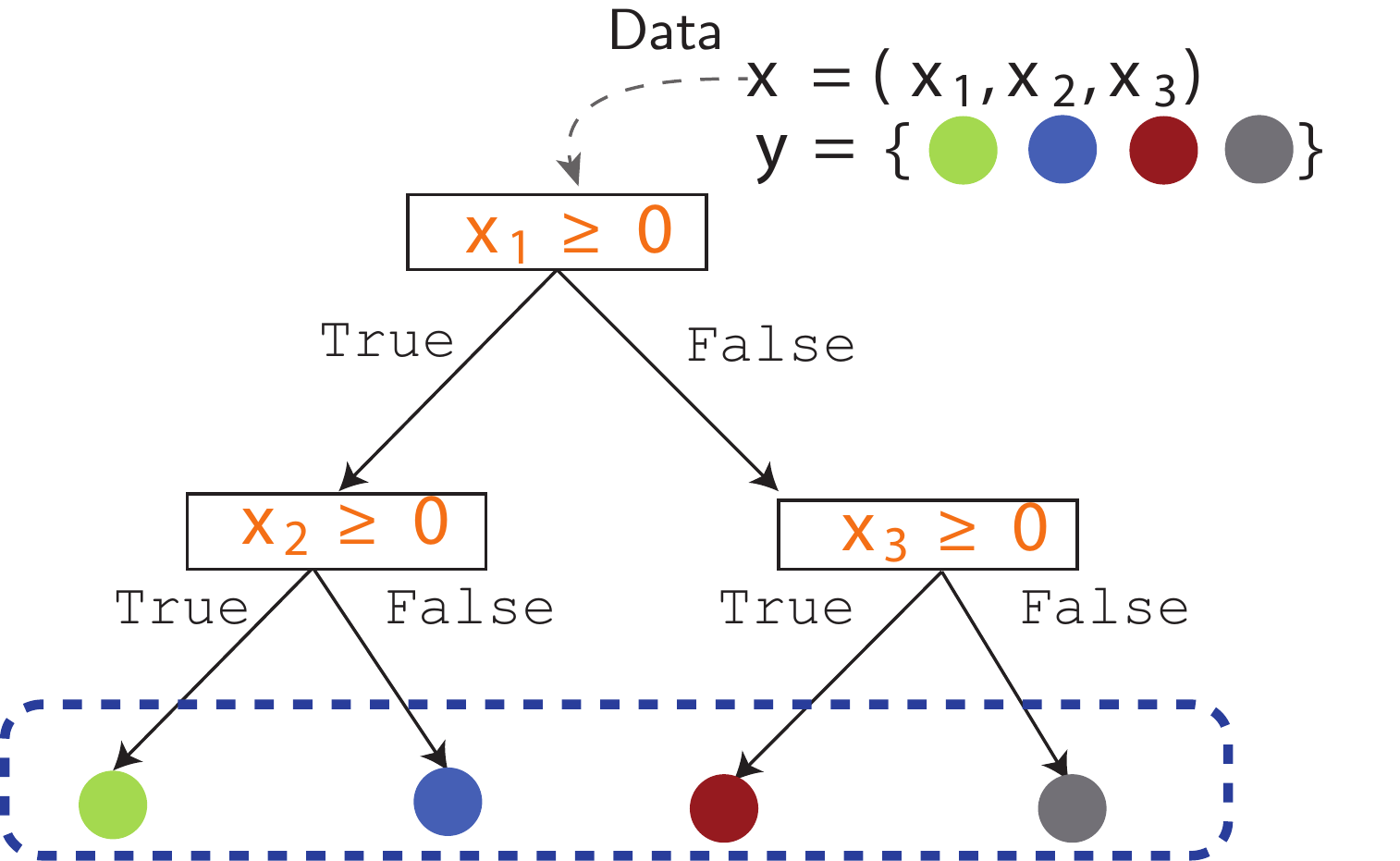}
\caption{\label{fig:DTschematics} Example of a decision tree. For an input observation $\bd{x}$, its label $y$ is predicted by traversing it from the root all the way down the leaves, following branches it satisfies.}
\end{figure}

\label{subsec:randomforest}
We now briefly review one of the most widely used  and versatile algorithms in data science and machine learning, \emph{Random Forests} (RF). Random Forests is an ensemble method widely deployed for complex classification tasks.  A random forest is composed of a family of (randomized) tree-based classifier decision trees (discussed below). Decision trees are high-variance, weak classifiers that can be easily randomized, and as such, are ideally suited for ensemble-based methods. Below, we give a brief high-level introduction to these ideas.

A decision tree uses a series of questions to  hierarchically partition the data.  Each branch of the decision tree consists of a question that splits the data into smaller subsets (e.g.~is some feature larger than a given number? See Fig.~\ref{fig:DTschematics}), with the leaves (end points) of the tree corresponding to the ultimate partitions of the data. When using decision trees for classification, the goal is to construct trees such that the partitions are informative about the class label (see Fig. \ref{fig:DTschematics}). It is clear that more complex decision trees lead to finer partitions that give improved  performance on the training set. However, this generally leads to over-fitting\footnote{One extreme limit is an $n-$node tree, with $n$ being the number of data point in the dataset given.}, limiting the out-of-sample performance. For this reason, in practice almost all decision trees use some form of regularization (e.g.~maximum depth for the tree) to control complexity and reduce overfitting.  Decision trees also have extremely high variance, and are often extremely sensitive to many details of the training data. This is not surprising since decision trees are learned by partitioning the training data. Therefore, individual decision trees are weak classifiers. However, these same properties make them ideal for incorporation in an ensemble method. 

In order to create an ensemble of decision trees, we must introduce a randomization procedure. As discussed above, the power of ensembles to reduce variance only manifests when randomness reduces correlations between the classifiers within the ensemble. Randomness is usually introduced into random forests in one of three distinct ways. The first is to use bagging and simply ``bag'' the decision trees by training each decision tree on a different bootstrapped dataset \cite{breiman2001random}. Strictly speaking, this procedure does not constitute a random forest but rather a bagged decision tree. The second procedure is to only use a different random subset of the features at each split in the tree. This ``feature bagging'' is the distinguishing characteristic of random forests~\cite{breiman2001random, ho1998random}. Using feature bagging reduces correlations between decision trees that can arise when only a few features are strongly predictive of the class label. Finally, extremized random forests (ERFs) combine ordinary and feature bagging with an extreme randomization procedure where splitting is done randomly instead of using optimality criteria (see for details Refs.~\cite{geurts2006extremely, louppe2014understanding}). Even though this reduces the predictive power of each individual decision tree, it still often improves the predictive power of the ensemble because it dramatically reduces correlations between members and prevents overfitting. 

\begin{figure}[t!]
\includegraphics[width=1.0\columnwidth]{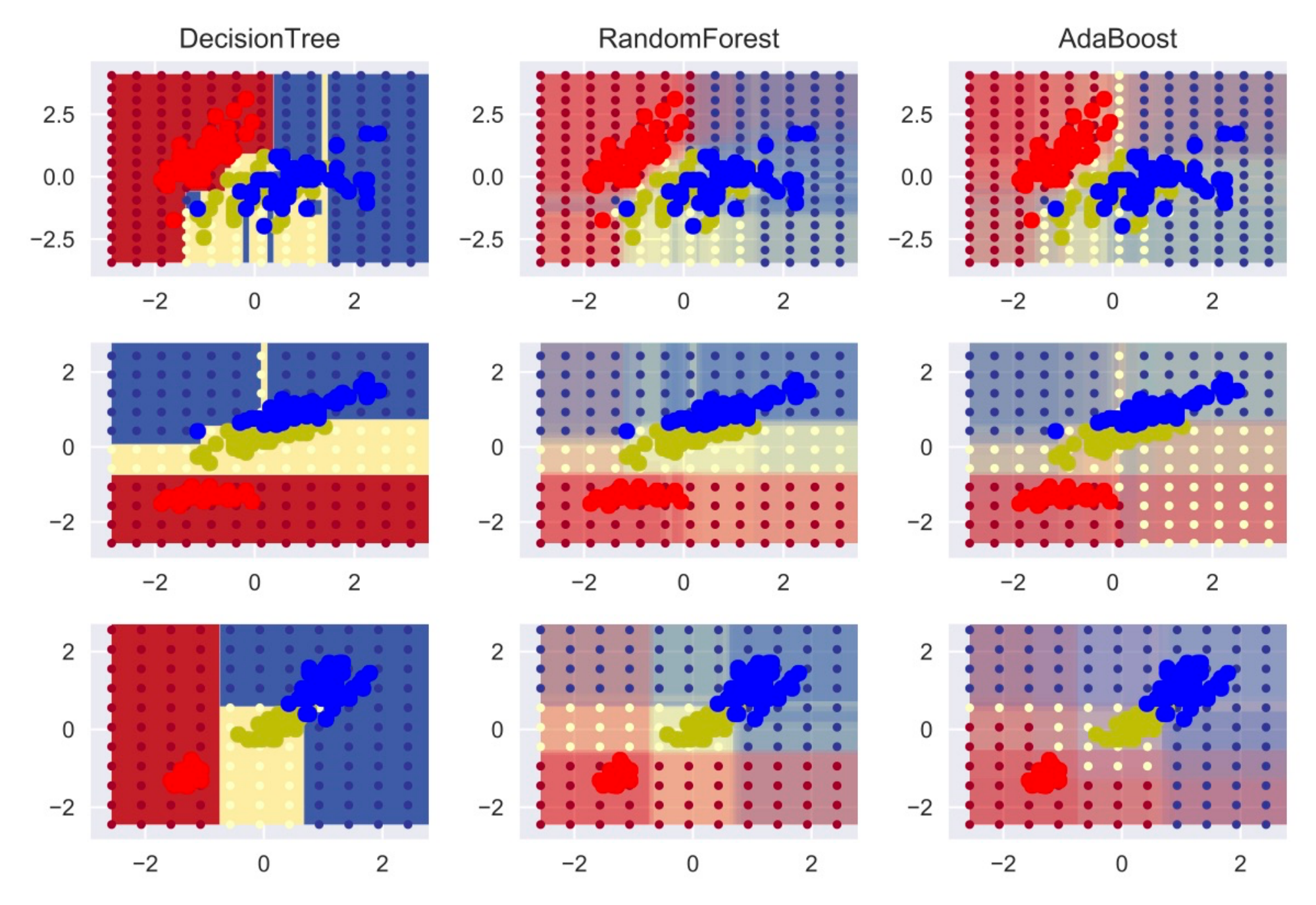}
\caption{ Classifying Iris dataset with aggregation models for scikit learn tutorial. This dataset seeks to classify iris flowers into three types (labeled in red, blue, or yellow)
based on a measurement of four features: septal length septal width, petal length, and petal width. To visualize the decision surface, we trained classifiers using only two of the four potential features (e..g septal length, septal width). Each row corresponds to a different subset of two features and the columns to a Decision Tree with 10-fold CV (first column), Random Forest with 30 trees and 10-fold CV (second column) and AdaBoost with 30 base hypotheses (third column). Decision surface learned is highlighted by color shades. See the corresponding \href{http://scikit-learn.org/stable/auto_examples/ensemble/plot_forest_iris.html}{tutorial} for more details \cite{scikit-learn} }
\label{fig:Iris_aggregation}
\end{figure}

Examples of the kind of decision surfaces found by decision trees, random forests, and Adaboost are shown in Fig.~\ref{fig:Iris_aggregation}. We invite the reader to check out the corresponding scikit-learn
~\href{http://scikit-learn.org/stable/auto_examples/ensemble/plot_forest_iris.html}{tutorial} for more details of how these are implemented in python \cite{scikit-learn}.

There are many different types of decision trees and training procedures. A full discussion of decision trees (and random forests) lies beyond the scope of this review and we refer readers to the extensive literature on these topics \cite{loh2011classification, lim2000comparison, louppe2014understanding}. Recently, decision trees were applied in high-energy physics to study to learn non-Higgsable gauge groups~\cite{wang2018learning}. 

\subsection{Gradient Boosted Trees and XGBoost}

Before we turn to applications of these techniques, we briefly discuss one final class of ensemble methods that has become increasingly popular in the last few years: \emph{Gradient-Boosted Trees} \cite{friedman2001greedy, chen2016xgboost}. The basic idea of gradient-boosted trees is to use intuition from boosting and gradient descent (in particular Newton's method, see Sec.~\ref{sec:gradient_descent}) to construct ensembles of decision trees. Like in boosting, the ensembles are created by iteratively adding new decision trees to the ensemble. In gradient boosted trees, one critical component is the a cost function that measures the performance of our ensemble.  At each step, we compute the gradient of the cost function with respect to the predicted value of the ensemble and add trees that move us in the direction of the gradient. Of course, this requires a clever way of mapping gradients to decision trees. We give a brief overview of how this is done within XGBoost (Extreme Gradient Boosting), which has recently been applied, to classify and rank transcription factor binding in DNA sequences~\cite{li2018quantum}. Below, we follow closely the~\href{https://xgboost.readthedocs.io/en/latest/model.html}{XGboost tutorial}.

Our starting point is a clever parametrization of decision trees. Here, we use notation where the decision tree makes continuous predictions (regression trees), though this can also easily be generalized to classification tasks. We parametrize a decision tree $j$, denoted as $g_j(\mathbf{x})$, with $T$ leaves by two quantities: a function $q(\mathbf{x})$
that maps each data point to one of the leaves of the tree,  $q: \mathbf{x} \in \mathbb{R}^d \rightarrow \{1,2 \ldots, T\}$ and a weight vector $\bd{w} \in  \mathbb{R}^T$ that assigns a predicted value to each leaf. In other words, the $j$-th decision tree's  prediction for the datapoint $\mathbf{x}_i$ is simply: $g_j(\mathbf{x}_i)=w_{q(\mathbf{x}_i)}$.

In addition to a parametrization of decision trees, we also have to specify a cost function which measures predictions. The prediction of our ensemble for a datapoint $(y_i,\mathbf{x}_i)$ is given by 
\be
\hat{y}_i= g_A(\mathbf{x}_i)= \sum_{j=1}^M g_j(\mathbf{x}_i),\quad g_j\in \mathcal{F}
\ee
where $g_j(\mathbf{x}_i)$ is the prediction of the $j$-th decision tree on datapoint $\mathbf{x}_i$, $M$ is the number of members of the ensemble, and $\mathcal{F}=\{g(\mathbf{x})=w_{q(\mathbf{x})}\}$ is the space of trees. As discussed in the context of random trees above, without regularization, decision trees tend to overfit the data by dividing it into smaller and smaller partitions. For this reason, our cost function is generally composed of two terms, a term that measures the goodness of predictions on each datapoint, $l_i(y_i, \hat{y}_i)$, which is assumed to be differentiable and convex, and for each tree in the ensemble, a regularization term $\Omega(g_j)$
that does not depend on the data:
\be
 \mathcal{C}( \boldsymbol{X}, g_A)= \sum_{i=1}^N  l(y_i, \hat{y}_i) + \sum_{j=1}^M \Omega(g_j),
\ee
where the index $i$ runs over data points and the index $j$ runs over decision trees in our ensemble. In XGBoost, the regularization function is chosen to be
\be\label{eq:XGBoost-regOmega}
\Omega(g) =\gamma T + {\lambda \over 2}||\bd{w}||_2^2,
\ee
with $\gamma$ and $\lambda$ regularization parameters that must be chosen appropriately. Notice that this regularization penalizes both large weights on the leaves (similar to $L^2$-regularization) and
having large partitions with many leaves. 

As in boosting, we form the ensemble iteratively. For this reason, we define a family of predictors $\hat{y}_i^{(t)}$ as
\be
\hat{y}_i^{(t)}= \sum_{j=1}^t g_j(\mathbf{x}_i)= \hat{y}_i^{(t-1)} + g_t(\mathbf{x}_i).
\ee
Note that by definition $y_i^{(M)}=g_A(\mathbf{x}_i)$. The central idea is that for large $t$, each decision tree is a small perturbation to the predictor (of order $1/T$) and hence we can perform a Taylor expansion on our loss function to second order:
\bea
 \mathcal{C}_t &=&\sum_{i=1}^N  l(y_i, \hat{y}_i^{(t-1)}+g_t(\mathbf{x}_i)) + \Omega(g_t) \nonumber \\
&\approx &  \mathcal{C}_{t-1} + \Delta \mathcal{C}_t,
\eea
with
\be
 \Delta \mathcal{C}_t = a_ig_t(\mathbf{x}_i) +{1 \over 2}b_i g_t(\mathbf{x}_i)^2 +\Omega(g_t),
\ee
where
\be
a_i= \partial_{\hat{y}_i^{(t-1)}}l(y_i, \hat{y}_i^{(t-1)}),
\ee
\be
b_i= \partial^2_{\hat{y}_i^{(t-1)}}l(y_i, \hat{y}_i^{(t-1)}).
\ee
We then choose the $t$-th decision tree $g_t$ to minimize $\Delta \mathcal{C}_t$. This is almost identical to how we derived the Newton method update in the section on gradient descent, see Sec.~\ref{sec:gradient_descent}.

We can actually derive an expression for the parameters of $g_t$ that minimize $\Delta \mathcal{C}_t$ analytically. To simplify notation, it is useful to define the set of points $\mbf{x}_i$ that get mapped to leaf $j$: $I_j =\{i:q_t(\mbf{x}_i)=j\}$ and the functions $B_j = \sum_{i \in I_j}b_i$ and $A_j=\sum_{i \in I_j}a_i$. Notice that in terms of these quantities, we can write
\be
\Delta \mathcal{C}_t= \sum_{j=1}^T [ A_j w_j+ {1 \over 2}(B_j+\lambda)w_j^2] +\gamma T,
\ee
where we made the $t$-dependence of all parameters implicit. Note that $\lambda$ comes from the regularization term, $\Omega(g_t)$, through Eq.\eqref{eq:XGBoost-regOmega}. To find the optimal $w_j$, just as in Newton's method we take the gradient of the above expression with respect to $w_j$ and set this equal to zero, to get
\be
w_j^{opt}=  -\frac{A_j}{B_j +\lambda}.
\ee
Plugging this expression into  $\Delta \mathcal{C}_t$ gives
\be
\Delta \mathcal{C}_t^{opt}= -{1 \over 2} \sum_{j=1}^T {A_j^2 \over B_j+ \lambda} + \gamma T.
\ee
It is clear that $\Delta \mathcal{C}_t^{opt}$ measures the in-sample performance of $g_t$ and we should find the decision tree that minimizes this value. In principle, one could enumerate all possible trees over the data and find the tree that minimizes $\Delta \mathcal{C}_t^{opt}$. However, in practice this is impossible. Instead, an approximate  greedy algorithm is run that optimizes one level of the tree at a time by trying to find optimal splits of the data. This leads to a tree that is a good local minimum of $\Delta \mathcal{C}_t^{opt}$ which is then added to the ensemble. We emphasize that this is only a very high level sketch of how the algorithm works. In practice, additional regularization such as shrinkage\cite{friedman2002stochastic} and feature subsampling\cite{breiman2001random, friedman2003importance} is also used. In addition, there are many numerical and technical tricks used for the approximate algorithm and how to find splits of the data that give good decision trees \cite{chen2016xgboost}.

\subsection{\label{subsec:ensemble_appl}Applications to the Ising model and Supersymmetry Datasets}

\begin{figure}[t!]
\includegraphics[width=1.0\columnwidth]{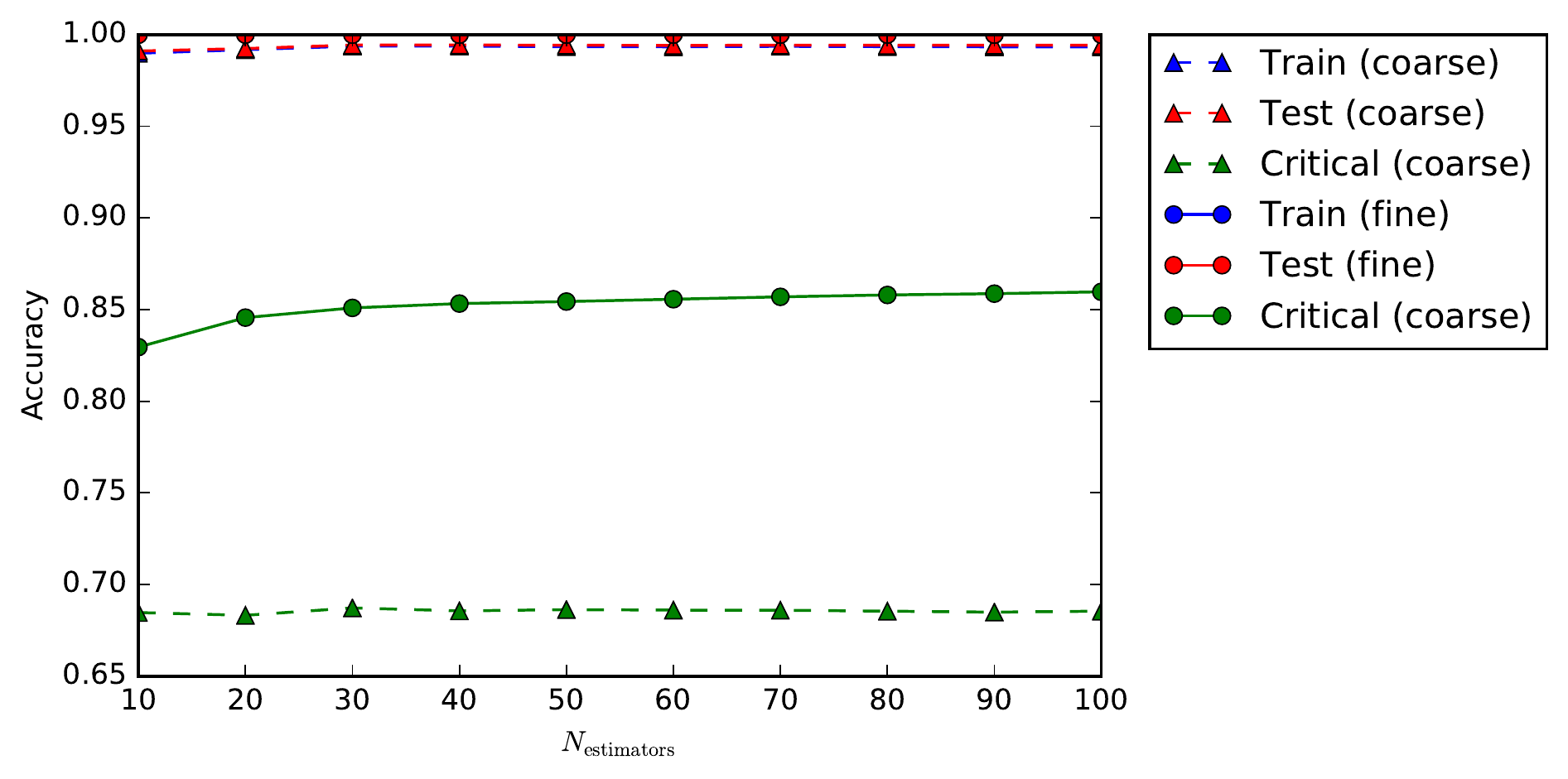}
\includegraphics[width=1.0\columnwidth]{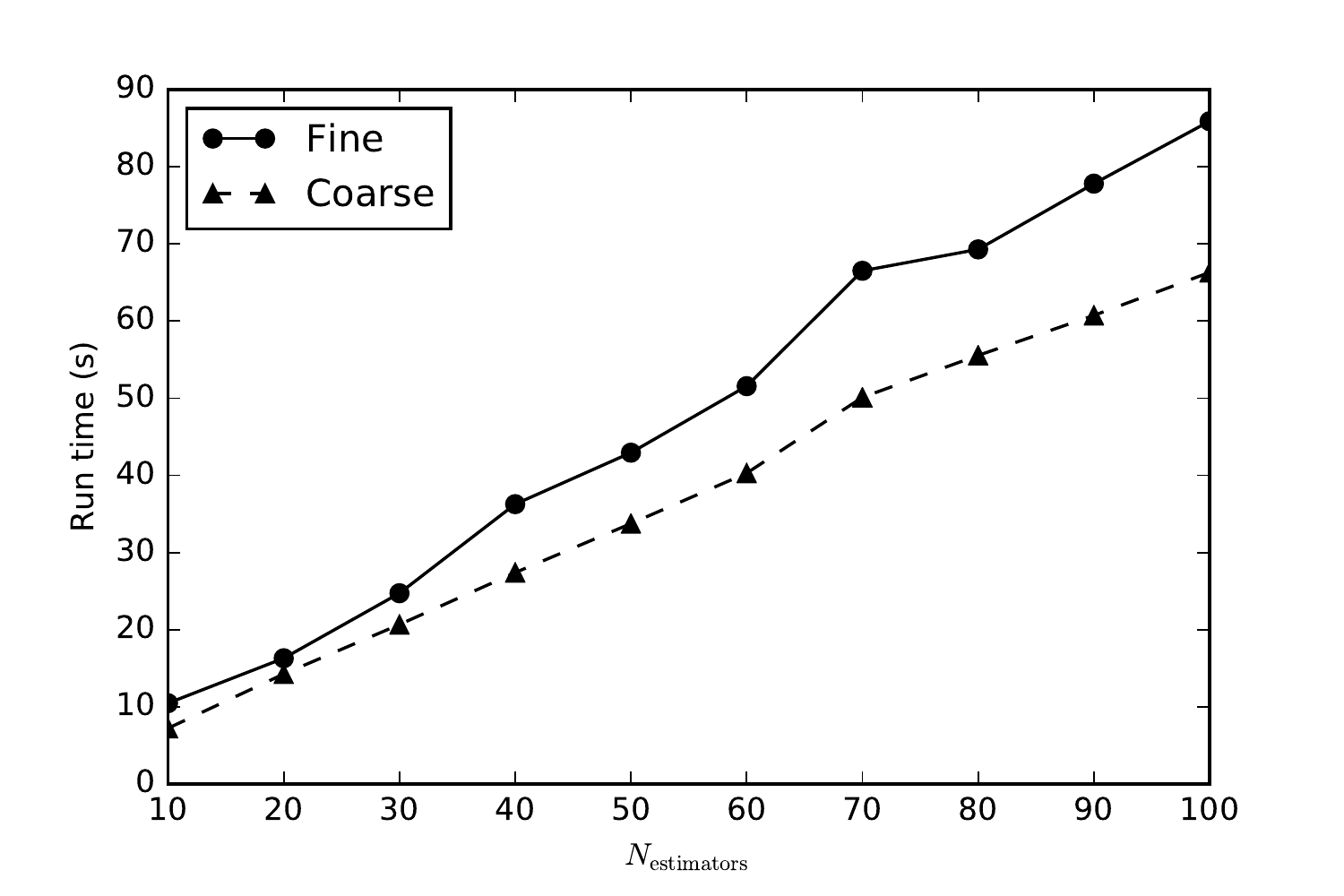}
\caption{Using Random Forests (RFs) to classify Ising Phases. (Top)  Accuracy of RFs for classifying the phase of samples from the Ising mode for the training set (blue), test set (red), and critical region (green) using coarse trees with a few leaves (triangles) and fine decision trees with many leaves (filled circles). RFs were trained on samples from ordered and disordered phases but were \emph{not} trained on samples from the critical region. (Bottom) The time it takes to train RFs scales linearly with the number of estimators in the ensemble. For the upper panel, note that the train case (blue) overlaps with the test case (red). Here `fine' and `coarse' refer to trees with 2 and 10,000 leaves, respectively. For implementation details, see \href{https://physics.bu.edu/~pankajm/ML-Notebooks/HTML/NB_CVIII-randomforests_ising.html}{Jupyter notebooks 9}}
\label{Fig:Ising-RF}
\end{figure}

We now illustrate some of these ideas using two examples drawn from physics: (i) classifying the phases of the spin configurations of the $2D$-Ising model above and below the critical temperature using random forests and (ii) classifying Monte-Carlo simulations of collision events in the SUSY dataset as supersymmetric or standard using an XGBoost implementation of gradient-boosted trees. Both examples were analyzed in Sec.~\ref{subsec:logreg_examples} using logistic regression. Here we show that on the Ising dataset, the RFs perform significantly better than logistic regression models whereas gradient boosted trees seem to yield an accuracy of about 80\%, comparable to published results. The two accompanying \href{https://physics.bu.edu/~pankajm/MLnotebooks.html}{Jupyter notebooks} discuss practical details of implementing these examples and the readers are encouraged to experiment with the notebooks.

The Ising dataset used for classification by RFs here is identical to that used to study logistic regression in Sec.~\ref{subsec:logreg_examples}. We assign a label to each state according to its phase: $0$ if the state is disordered, and $1$ if it is ordered. We divide the dataset into three categories according to the temperature at which samples are drawn: ordered ($T/J<2.0$), near-critical ($2.0\leq T/J\leq 2.5)$ and disordered ($T/J>2.5$) (see Figure \ref{fig:2D_Ising_states}). We use the ordered and disordered states to train a random forest and evaluate our learned model on a test set of unseen ordered and disordered states (test sets). We also ask how well our RF can predict the phase of samples drawn in the critical region (i.e. predict whether the temperature of a critical sample is above or below the critical temperature). Since our model is never trained on samples in the critical region, prediction in this region is a test of the algorithm's ability to generalize to new regions in phase space.

The results of fits using RFs to predict phases are shown in Figure \ref{Fig:Ising-RF}.  We used two types of RF classifiers, one where the ensemble consists of coarse decision trees with a few leaves and another with finer decision trees with many leaves (see corresponding notebook). RFs have extremely high accuracy on the training and test sets  (over 99\%) for both coarse and fine trees. However, notice that the RF consisting of coarse trees perform extremely poorly on samples from the critical region whereas the RF with fine trees classifies critical samples with an accuracy of nearly 85\%. Interestingly, and unlike with logistic regression, this performance in the critical region requires almost no parameter tuning. This is because, as discussed above, RFs are largely immune to overfitting problems even as the number of estimators in the ensemble becomes large. Increasing the number of estimators in the ensemble does increase performance but at a large cost in computational time (Fig. \ref{Fig:Ising-RF} bottom).

\begin{figure}[t!]
\includegraphics[width=1.0\columnwidth]{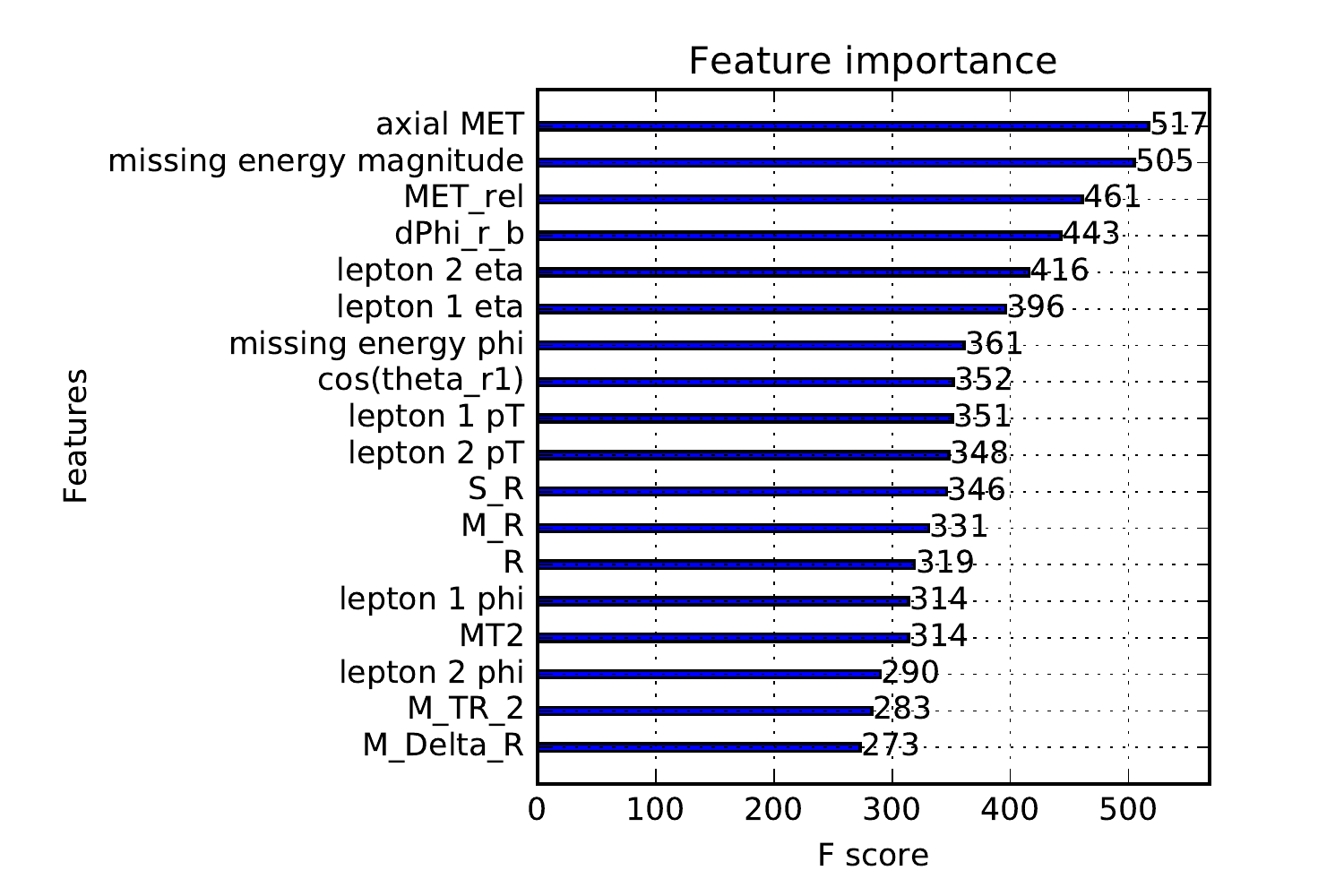}
\caption{Feature Importance Scores in SUSY dataset from applying XGBoost to $100,000$ samples. See \href{https://physics.bu.edu/~pankajm/MLnotebooks.html}{Notebook 10} for more details.}
\label{Fig:FscoresXGBoostSUSY}
\end{figure}

In the second application of ensemble methods to physics-related datasets, we used the XGBoost implementation of gradient boosted trees to classify Monte-Carlo collisions from the SUSY dataset. With default parameters using a small subset of the data ($100,000$ out of the full $5,000,000$ samples), we were able to achieve a classification accuracy of about $79$\%, which could be improved to nearly $80$\% after some fine-tuning (see accompanying \href{https://physics.bu.edu/~pankajm/MLnotebooks.html}{notebook}). This is comparable to published results  \cite{baldi2014searching} and those obtained using logistic regression in earlier chapters. One nice feature of ensemble methods such as XGBoost  is that they automatically allow us to calculate feature scores (Fscores) that rank the importance of various features for classification. The higher the Fscore, the more important the feature for classification. Figure \ref{Fig:FscoresXGBoostSUSY} shows the feature scores from our XGBoost algorithm for the production of electrically-charged supersymmetric particles ($\chi \pm$) which decay to $W$ bosons and an electrically neutral supersymmetric particle $\chi^0$, which is invisible to the detector. The features are a mix of eight directly measurable quantities from the detector, as well as ten hand crafted features chosen using physics knowledge. Consistent with the physics of these supersymmetric decays in the lepton channel, we find that the most informative features for classification are the missing transverse energy along the vector defined by the charged leptons (Axial MET) and the missing energy magnitude due to $\chi_0$.

%% file: sections/DNNs.tex
\section{An Introduction to Feed-Forward Deep Neural Networks (DNNs)}
\label{sec:DNNs}

Over the last decade, neural networks  have emerged as the one of most powerful and widely-used  supervised learning techniques. Deep Neural Networks (DNNs) have a long history \cite{bishop1995neural, schmidhuber2015deep}, but re-emerged to prominence after a rebranding as ``Deep Learning'' in the mid 2000s \cite{hinton2006fast, hinton2006reducing}. DNNs truly caught the attention of the wider machine learning community and industry in 2012 when Alex Krizhevsky, Ilya Sutskever, and Geoff Hinton used a GPU-based DNN model (AlexNet) to lower the error rate on the ImageNet Large Scale Visual Recognition Challenge (ILSVRC) by an incredible twelve percent from 28\% to 16\% \cite{krizhevsky2012imagenet}. Just three years later, a machine learning group from Microsoft achieved an error of 3.57\% using an ultra-deep residual neural network (ResNet) with 152 layers \cite{he2016deep}!  Since then, DNNs have become the workhorse technique for many image and speech recognition based machine learning tasks. The large-scale industrial deployment of DNNs has given rise to a number of high-level libraries and packages (Caffe, Keras, Pytorch, TensorFlow, etc.) that make it easy to quickly code and deploy DNNs.

Conceptually, it is helpful to divide neural networks into four categories:  (i) general purpose neural networks for supervised learning, (ii) neural networks designed specifically for image processing, the most prominent example of this class being Convolutional Neural Networks (CNNs), (iii) neural networks for sequential data such as Recurrent Neural Networks (RNNs), and (iv) neural networks for unsupervised learning such as Deep Boltzmann Machines. Here, we will limit our discussions to the first two categories (unsupervised learning is discussed later in the review). Though increasingly important for many applications such as audio and speech recognition, for the sake of brevity, we omit a discussion of sequential data and RNNs from this review. For an introduction to RNNs and LSTM networks see Chris Olah's blog,~\href{https://colah.github.io/posts/2015-08-Understanding-LSTMs/}{https://colah.github.io/posts/2015-08-Understanding-LSTMs/}, and Chapter 13 of \cite{bishop2006pattern} as well as the introduction to RNNs in Chapter 10 of \cite{Goodfellow-et-al-2016} for sequential data. 

Due to the number of recent books on deep learning (see for example Michael Nielsen's introductory online book~\cite{nielsen2015neural} and the more advanced \cite{Goodfellow-et-al-2016}), the goal of this section is to give a high-level introduction to the basic ideas behind DNNs, and provide some practical knowledge for coding simple neural nets for supervised learning tasks (see the accompanying \href{http://physics.bu.edu/~pankajm/MLnotebooks.html}{Notebooks}). This section assumes the reader is familiar with the basic concepts introduced in earlier sections on logistic and linear regression. Throughout, we strive to provide intuition behind the inner workings of DNNs, as well as highlight limitations of present-day algorithms. 

The influx of corporate and industrial interests has rapidly transformed the field in the last few years. This massive influx of money and researchers has given rise to new dogmas and best practices that change rapidly. As with most intellectual fields experiencing rapid expansion, many commonly accepted heuristics many turn out not to be as powerful as thought \cite{wilson2017marginal}, and widely held beliefs not as universal as once imagined \cite{zhang2016understanding,  lee2017first}. This is especially true in modern neural networks where results are largely empirical and heuristic and lack the firm footing of many earlier machine learning methods. For this reason, in this review we have chosen to emphasize tried and true fundamentals, while pointing out what, from our current vantage point, seem like promising new techniques. The field is rapidly evolving and readers are urged to read papers and to implement these algorithms themselves in order to gain a deep appreciation for the incredible power of modern neural networks, especially in the context of image, speech, and natural language processing, as well as limitations of the current methods.

In physics, DNNs and CNNs have already found numerous applications. In statistical physics, they have been applied to detect phase transitions in 2D Ising~\cite{tanaka2017detection} and Potts~\cite{li2017applications} models, lattice gauge theories~\cite{wetzel2017machine}, and different phases of polymers~\cite{wei2017identifying}. It has also been shown that deep neural networks can be used to learn free-energy landscapes~\cite{sidky2017learning}. At the same time, methods from statistical physics have been applied to the field of deep learning to study the thermodynamic efficiency of learning rules~\cite{goldt2017thermodynamic}, to explore the hypothesis space that DNNs span, make analogies between training DNNs and spin glasses~\cite{baldassi2017role,baity2018comparing}, and to characterize phase transitions with respect to network topology in terms of errors~\cite{li2017exploring}. In relativistic hydrodynamics, deep learning has been shown to capture features of non-linear evolution and has the potential to accelerate numerical simulations~\cite{huang2018applications}, while in mechanics CNNs have been used to predict eigenvalues of photonic crystals~\cite{finol2018deep}. Deep CNNs were employed in lensing reconstruction of the cosmic microwave background~\cite{caldeira2018deepcmb}. Recently, DNNs have been used to improve the efficiency of Monte-Carlo algorithms~\cite{shen2018self}.

Deep learning has also found interesting applications in quantum physics. Various quantum phase transitions~\cite{arai2017deep,broecker2017quantum,suchsland2018parameter,van2017learning2,iakovlev2018supervised} can be detected and studied using DNNs and CNNs, including the transverse-field Ising model~\cite{ohtsuki2017deep}, topological phases~\cite{yoshioka2017learning,zhang2017machine_top,zhang2017machine} and non-invasive topological quality control~\cite{caio2019machine}. DNNs found applications even in non-equilibrium many-body localization~\cite{schindler2017probing,venderley2017machine,van2017learning2,van2017learning}, and the characterization of photoexcited quantum states~\cite{shinjo2019characterization}. Experimentally, DNNs were recently employed in cold atoms to identify critical points~\cite{rem2018identifying}. Representing quantum states as DNNs~\cite{gao2017efficient,saito2017machine,levine2017deep,gao2017experimental} and
quantum state tomography~\cite{torlai2018neural} are among some of the impressive achievements to reveal the potential of deep learning to facilitate the study of quantum systems. Machine learning techniques involving neural networks were also used to study quantum and fault-tolerant error correction~\cite{baireuther2017machine,breuckmann2017scalable,krastanov2017deep,davaasuren2018general,maskara2018advantages,chamberland2018deep}, estimate rates of coherent and incoherent quantum processes~\cite{greplova2017quantum}, to obtain spectra of $1/f$-noise in spin-qubit devices~\cite{zhang2018spin}, and the recognition of state and charge configurations and auto-tuning in quantum dots~\cite{kalantre2017machine}. In quantum information theory, it has been shown that one can perform gate decompositions with the help of neural nets~\cite{swaddle2017generating}. In lattice quantum chromodynamics, DNNs have been used to learn action parameters in regions of parameter space where principal component analysis fails~\cite{shanahan2018machine}. CNNs were applied to data from a high-energy experiment to identify particle interactions in sampling calorimeters used commonly in neutrino physics~\cite{aurisano2016convolutional}. Last but not least, DNNs also found place in the study of quantum control~\cite{yang2017neural}, and in scattering theory to learn the $s$-wave scattering length~\cite{wu2018visualizing} of potentials.

\subsection{Neural Network Basics}

Neural networks (also called neural nets) are neural-inspired nonlinear models for supervised learning.  As we will see, neural nets can be viewed as natural, more powerful extensions of supervised learning methods such as linear and logistic regression and soft-max methods.

\subsubsection{The basic building block: neurons}

\begin{figure}[t!]
\includegraphics[width=0.9\columnwidth]{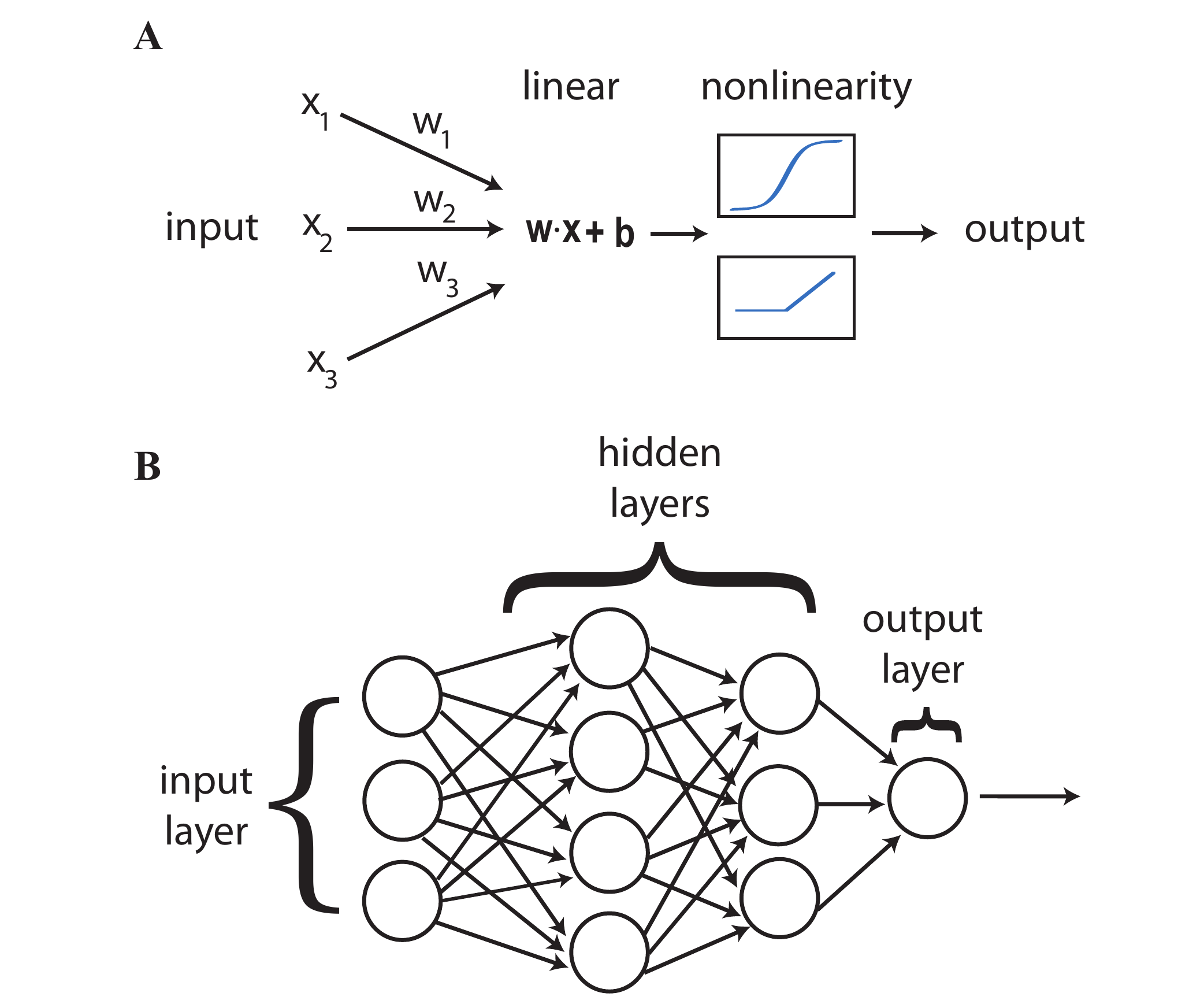}
\caption{{\bf Basic architecture of neural networks. } (A) The basic components of a neural network are stylized neurons consisting of a linear transformation that weights the importance of various inputs, followed by a non-linear activation function. (b) Neurons are arranged into layers with the output of one layer serving as the input to the next layer.}
\label{Fig:NNarchitecture}
\end{figure}

The basic unit of a neural net is a stylized ``neuron'' $i$ that takes a vector of $d$ input features $\boldsymbol{x}=(x_1, x_2, \ldots, x_d)$ and produces a scalar output $a_i(\boldsymbol{x})$. A neural network consists of many such neurons stacked into layers, with the output of one layer serving as the input for the next (see Figure \ref{Fig:NNarchitecture}). The first layer in the neural net is called the input layer, the middle layers are often called ``hidden layers'', and the final layer is called the output layer.

The exact function $a_i$ varies depending on the type of non-linearity used in the neural network. However, in essentially all cases $a_i$ can be decomposed into a linear operation that weights the relative importance of the various inputs, and a non-linear transformation $\sigma_i(z)$ which is usually the same for all neurons. The linear transformation in almost all neural networks takes the form of a dot product with a set of neuron-specific weights $\boldsymbol{w}^{(i)}=(w_1^{(i)}, w_2^{(i)}, \ldots, w_d^{(i)})$  followed by re-centering with a neuron-specific bias $b^{(i)}$: 
\be
 z^{(i)}=\boldsymbol{w}^{(i)}\cdot \boldsymbol{x}+b^{(i)} = \mathbf{x}^T\cdot\mathbf{w}^{(i)}, 
\ee
where $\mathbf{x} = (1,\boldsymbol{x})$ and $\mathbf{w}^{(i)} = (b^{(i)}, \boldsymbol{w}^{(i)} )$.
In terms of $z^{(i)}$ and the non-linear function $\sigma_i(z)$, we can write the full input-output function as
\begin{equation}
\label{eq:DNN_output_fn}
a_i(\mathbf{x})=\sigma_i(z^{(i)}),
\end{equation}
see Figure~\ref{Fig:NNarchitecture}.

\begin{figure*}[t!]
\includegraphics[width=1.8\columnwidth]{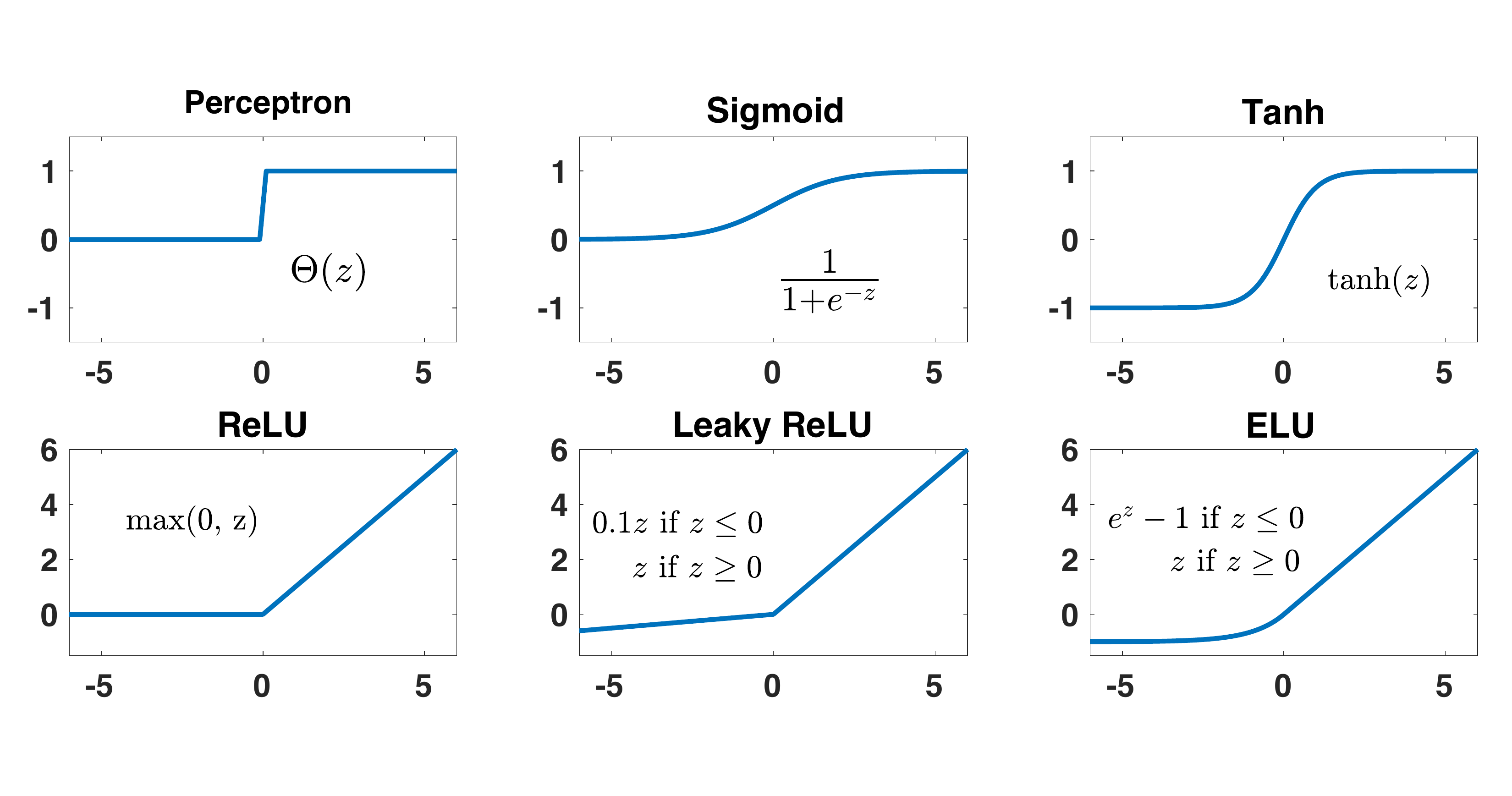}
\caption{\label{fig:nonlinear_activation} {\bf Possible non-linear activation functions for neurons.} In modern DNNs, it has become common to use non-linear functions that do not saturate for large inputs (bottom row) rather than saturating functions (top row). }
\end{figure*}

Historically in the neural network literature, common choices of nonlinearities included step-functions (perceptrons), sigmoids (i.e.~Fermi functions), and the hyperbolic tangent. More recently, it has become more common to use rectified linear units (ReLUs), leaky rectified linear units (leaky ReLUs), and  exponential linear units (ELUs) (see Figure \ref{fig:nonlinear_activation}). Different choices of non-linearities lead to different computational  and training properties for neurons. The underlying reason for this is that we train neural nets using gradient descent based methods, see Sec.~\ref{sec:gradient_descent}, that require us to take derivatives of the neural input-output function with respect to the weights $\boldsymbol{w}^{(i)}$ and the bias $b^{(i)}$. 

Notice that the derivatives of the aforementioned non-linearities $\sigma(z)$ have very different properties. The derivative of the perceptron is zero everywhere except where the input is zero. This discontinuous behavior makes it impossible to train perceptrons using gradient descent. For this reason, until recently the most popular choice of non-linearity was the tanh function or a sigmoid/Fermi function. However, this choice of non-linearity has a major drawback. When the input weights become large, as they often do in training, the activation function saturates and the derivative of the output with respect to the weights tends to zero since ${ \partial  \sigma / \partial z} \rightarrow 0$ for $z\gg 1$. Such ``vanishing gradients'' are a feature of any saturating activation function (top row of Fig.~\ref{fig:nonlinear_activation}), making it harder to train deep networks. In contrast, for a non-saturating activation function such as ReLUs or ELUs, the gradients stay finite even for large inputs.

\subsubsection{Layering neurons to build deep networks: network architecture.}

The basic idea of all neural networks is to layer neurons in a hierarchical fashion, the general structure of which is known as the network architecture (see Fig. \ref{Fig:NNarchitecture}). In the simplest feed-forward networks, each neuron in the \emph{input layer} of the neurons takes the inputs $\mathbf{x}$ and produces an output $a_i(\mathbf{x})$ that depends on its current weights, see Eq.~\eqref{eq:DNN_output_fn}. The outputs of the input layer are then treated as the inputs to the next \emph{hidden layer}. This is usually repeated several times until one reaches the top or \emph{output layer}. The output layer is almost always a simple classifier of the form discussed in earlier sections: a logistic regression or soft-max function in the case of categorical data (i.e.~discrete labels) or a linear regression layer in the case of continuous outputs. Thus, the whole neural network can be thought of as a complicated nonlinear transformation of the inputs $\mathbf{x}$ into an output $\hat{y}$ that depends on the weights and biases of all the neurons in the input, hidden, and output layers.

The use of hidden layers greatly expands the representational power of a neural net when compared with a simple soft-max or linear regression network. Perhaps, the most formal expression of the increased representational power of neural networks (also called the expressivity) is the universal approximation theorem which states that a neural network with a single hidden layer can approximate any continuous, multi-input/multi-output function with arbitrary accuracy. The reader is strongly urged to read the beautiful graphical proof of the theorem in \href{http://neuralnetworksanddeeplearning.com/chap4.html}{Chapter 4 of Nielsen's free online book} \cite{nielsen2015neural}. The basic idea behind the proof is that hidden neurons allow neural networks to generate step functions with arbitrary offsets and heights. These can then be added together to approximate arbitrary functions. The proof also makes clear that the more complicated a function, the more hidden units (and free parameters) are needed to approximate it. Hence, the applicability of the approximation theorem to practical situations should not be overemphasized. In condensed matter physics, a good analogy are matrix product states, which can approximate any quantum many-body state to an arbitrary accuracy, provided the bond dimension can be increased arbitrarily -- a severe requirement not met in any useful practical implementation of the theory.

Modern neural networks generally contain multiple hidden layers (hence the ``deep'' in deep learning). There are many ideas of why such deep architectures are favorable for learning.  Increasing the number of layers increases the number of parameters and hence the representational power of neural networks. Indeed, recent numerical experiments suggests that as long as the number of parameters is larger than the number of  data points, certain classes of neural networks can fit arbitrarily labeled random noise samples \cite{zhang2016understanding}. This suggests that large neural networks of the kind used in practice can express highly complex functions. Adding hidden layers is also thought to allow neural nets to learn more complex features from the data. Work on convolutional networks suggests that the first few layers of a neural network learn simple, ``low-level''  features that are then combined into higher-level, more abstract features in the deeper layers. Other works suggest that it is computationally and algorithmically easier to train deep networks rather than shallow, wider nets, though this is still an area of major controversy and active research \cite{mhaskar2016learning}.

Choosing the exact network architecture for a neural network remains an art that requires extensive numerical experimentation and intuition, and is often times problem-specific. Both the number of hidden layers and the number of neurons in each layer can affect the performance of a neural network. There seems to be no single recipe for the right architecture for a neural net that works best. However, a general rule of thumb that seems to be emerging is that the number of parameters in the neural net should be large enough to prevent under-fitting (see theoretical discussion in \cite{advani2017high}).

Empirically, the best architecture for a problem depends on the task, the amount and type of data that is available, and the computational resources at one's disposal. Certain architectures are easier to train, while others might be better at capturing complicated dependencies in the data and learning relevant input features. Finally, there have been numerous works that move beyond the simple deep, feed-forward neural network architectures discussed here.  For example, modern neural networks for image segmentation often incorporate ``skip connections'' that skip layers of the neural network \cite{he2016deep}. This allows information to directly propagate to a hidden or output layer, bypassing intermediate layers and often improving performance. 

\subsection{Training deep networks}

In the previous section, we introduced the basic architecture for neural networks. Here we discuss how to efficiently train large neural networks. Luckily, the basic procedure for training neural nets is the same as we used for training simpler supervised learning algorithms, such as logistic and linear regression: construct a cost/loss function and then use gradient descent to minimize the cost function and find the optimal weights and biases. Neural networks differ from these simpler supervised procedures in that generally they contain multiple hidden layers that make taking the gradient computationally more difficult. We will return to this in Sec.~\ref{subsec:backprop} which discusses the ``backpropagation'' algorithm for computing gradients.

Like all supervised learning procedures, the first thing one must do to train a neural network is to specify a loss function. Given a data point $(\mathbf{x}_i, y_i)$, $\mathbf{x}_i \in\mathbb{R}^{d+1}$, the neural network makes a prediction $\hat{y}_i (\mathbf{w})$, where $\mathbf{w}$ are the parameters of the neural network. Recall that in most cases, the top output layer of our neural net is either a continuous predictor or a classifier that makes discrete (categorical) predictions. Depending on whether one wants to make continuous or categorical predictions, one must utilize a different kind of loss function.  

For continuous data, the loss functions that are commonly used to train neural networks are identical to those used in linear regression, and include the mean squared error
\be
E(\mathbf{w})=  {1 \over n}\sum_{i=1}^n (y_i-\hat{y}_i(\mathbf{w}))^2,
\ee
where $n$ is the number of data points, and the mean-absolute error (i.e.~$L_1$ norm)
\be
E(\mathbf{w})=  {1 \over n}\sum_i |y_i-\hat{y}_i(\mathbf{w})|.
\ee
The full  cost function often includes additional terms that implement regularization (e.g.~$L_1$ or $L_2$ regularizers), see Sec.~\ref{sec:lin_reg}.

For categorical data, the most commonly used loss function is the cross-entropy (Eq.~\eqref{eq:cross_entropy} and Eq.~\eqref{eq:cross_entropy_softmax}), since the output layer is often taken to be a logistic classifier for binary data with two types of labels, or a soft-max classifier if there are more than two types of labels. The cross-entropy was already discussed extensively in earlier sections on logistic regression and soft-max classifiers, see Sec.~\ref{sec:log_reg}. Recall that for classification of binary data, the output of the top layer of the neural network is the probability $\hat{y}_i(\mathbf{w})=p(y_i=1| \mathbf{x}_i; \mathbf{w})$ that data point $i$ is predicted to be in category $1$. The  cross-entropy between the true labels $y_i \in \{ 0,1\}$ and the predictions is given by 
\be
\label{eq:cross_entropy_NN}
E(\mathbf{w})= -\sum_{i=1}^n y_i\log \hat{y}_i(\mathbf{w})+ (1-y_i)\log\left[1-\hat{y}_i(\mathbf{w})\right].\nonumber
\ee

More generally, for categorical data, $y$ can take on $M$ values so that $y \in \{0, 1, \ldots, M-1\}$. For each datapoint $i$, define a vector $y_{im}$ called a `one-hot' vector, such that
\be
y_{im}=
    \begin{cases}
      1, & \text{if}\ y_i=m \\	
      0, & \text{otherwise}.
    \end{cases}
\ee
We can also define the probability that the neural network assigns a datapoint to category $m$: $\hat{y}_{im}(\mathbf{w})= p(y_i=m| \mathbf{x}_i; \mathbf{w})$. Then, the categorical cross-entropy 
is defined as
\bea
\label{eq:cross_entropy_softmax_NN}
E(\mathbf{w}) &=& -\sum_{i=1}^n\sum_{m=0}^{M-1} y_{im}\log \hat{y}_{im}(\mathbf{w})\nonumber \\
&&+ (1-y_{im})\log\left[ 1-\hat{y}_{im}(\mathbf{w})\right] .
\eea
As in linear and logistic regression, this loss function is often supplemented by additional terms	that implement regularization.

Having defined an architecture and a cost function, we must now train the model. Similar to other supervised learning methods, we make use of gradient descent-based methods to optimize the cost function. Recall that the basic idea of gradient descent is to update the parameters $\mathbf{w}$ to move in the direction of the gradient of the cost function $\nabla_\mathbf{w} E(\mathbf{w})$. In Sec.~\ref{sec:gradient_descent}, we discussed numerous optimizers that implement variations of stochastic gradient descent (SGD, Nesterov, RMSProp, Adam, etc.) Most modern neural network packages, such as Keras, allow the user to specify which of these optimizers they would like to use in order to train the neural network. Depending on the architecture, data, and computational resources, different optimizers may work better on the problem, though vanilla SGD is a good first choice.

\begin{figure}[t!]
	\includegraphics[width=1.0\columnwidth]{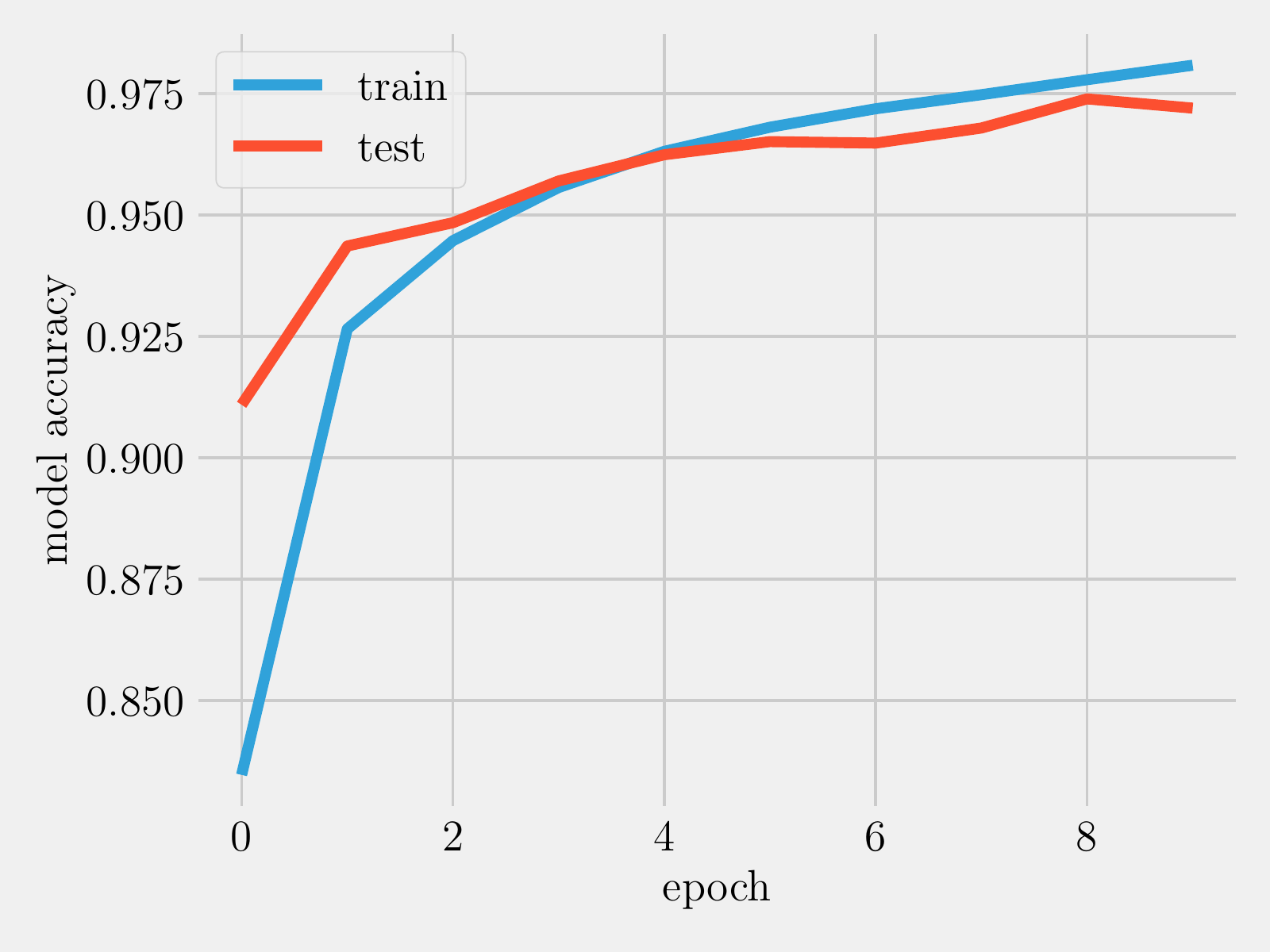}
	\caption{\label{fig:DNN_MNIST_accuracy} Model accuracy of a DNN to study the MNIST problem as a function of the training epochs (see \href{http://physics.bu.edu/~pankajm/MLnotebooks.html}{Notebook 11}). Besides the input and output layers, the DNN has four layers of size (100,400,400,50) with different nonlinearities $\sigma(z)$.}
\end{figure}

\begin{figure}[t!]
	\includegraphics[width=1.0\columnwidth]{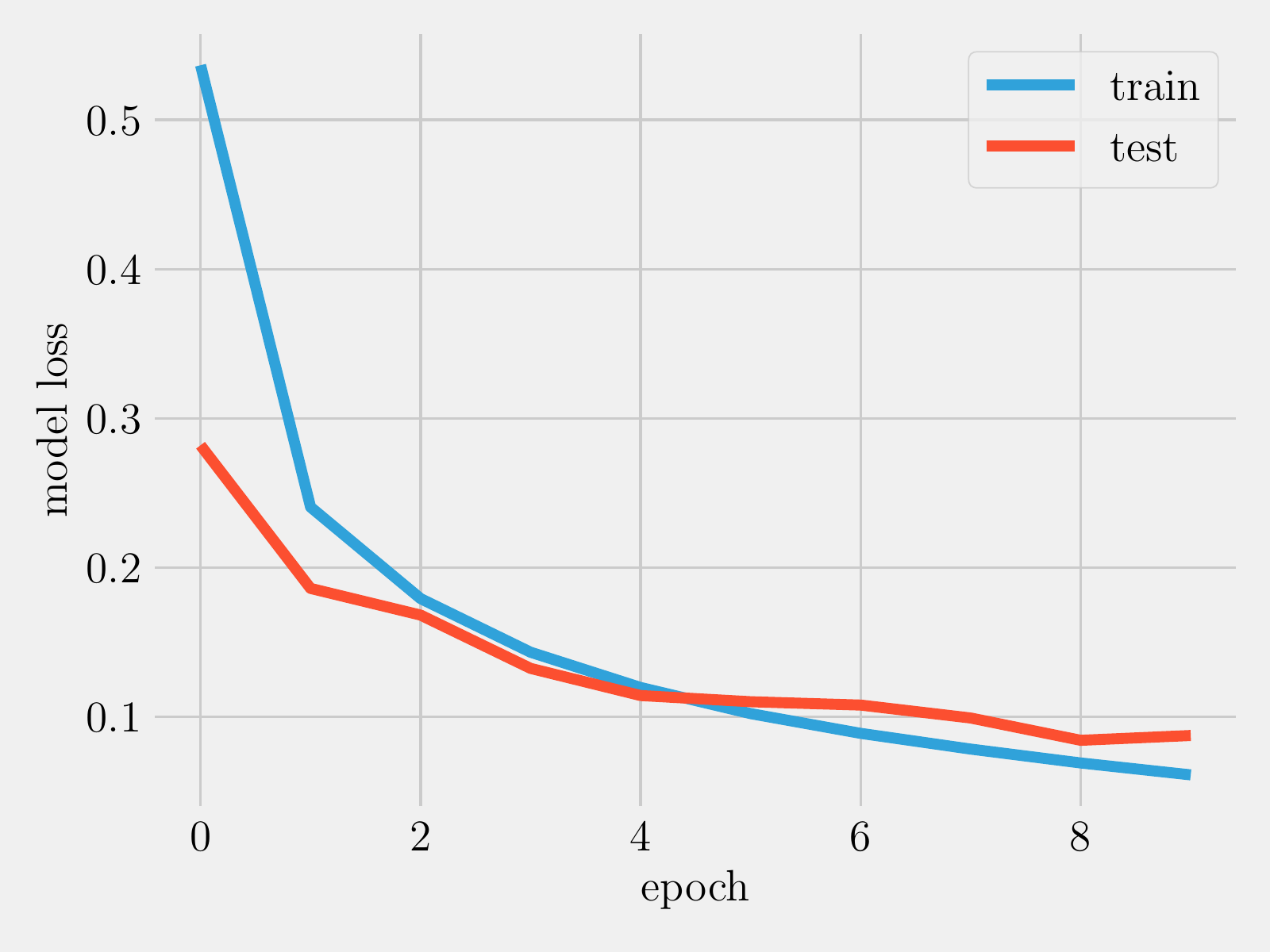}
	\caption{\label{fig:DNN_MNIST_loss} Model loss of a DNN to study the MNIST problem as a function of the training epochs (see \href{http://physics.bu.edu/~pankajm/MLnotebooks.html}{Notebook 11}). Besides the input and output layers, the DNN has four layers of size (100,400,400,50) with different nonlinearities $\sigma(z)$.}
\end{figure}

Finally, we note that unlike in linear and logistic regression,  calculating the gradients  for a neural network requires a specialized algorithm, called Backpropagation (often abbreviated backprop) which forms the heart of any neural network training procedure. Backpropagation has been discovered multiple times independently but was popularized for modern neural networks in 1985 \cite{rumelhart1985feature}. We will turn to the backpropagation algorithm in the next section. Before reading it, the reader is strongly encouraged to play with \href{http://physics.bu.edu/~pankajm/MLnotebooks.html}{Notebook 11} in order to gain some intuition about how to build a DNN in practice using the high-level Keras Python package.  \href{http://physics.bu.edu/~pankajm/MLnotebooks.html}{Notebook 11} discusses a simple example where we build a feed-forward deep neural network for classifying hand-written digits from the MNIST dataset. Figures~\ref{fig:DNN_MNIST_accuracy} and~\ref{fig:DNN_MNIST_loss} show the accuracy and the loss as a function of the training episodes.


\subsection{\label{subsec:backprop}The Backpropagation algorithm}

In the last section, we saw how to deploy a high-level package, Keras, to design and train a simple neural network. This training procedure requires us to be able to calculate the derivative of the cost function with respect to all the parameters of the neural network (the weights and biases of all the neurons in the input, hidden, and visible layers). A brute force calculation is out of the question since it requires us to calculate as many gradients as parameters at each step of the gradient descent. The backpropagation algorithm \cite{rumelhart1985feature} is a clever procedure that exploits the layered structure of neural networks to more efficiently compute gradients (for a more detailed discussion with Python code examples see Chapter 2 of \cite{nielsen2015neural}).

\subsubsection{Deriving and implementing the backpropagation equations}
At its core, backpropagation is simply the ordinary chain rule for partial differentiation, and can be summarized using four equations. In order to see this, we must first establish some useful notation.  We will assume that there are $L$ layers in our network with $l=1, \ldots, L$ indexing the layer. 
Denote by $w_{jk}^l$ the weight for the connection from the $k$-th neuron in layer $l-1$ to the $j$-th neuron in layer $l$. We denote the bias of this neuron by $b_j^l$. By construction, in a feed-forward neural network the activation $a_j^l$ of the $j$-th neuron in the $l$-th layer can be related to the activities of the neurons in the layer $l-1$ by the equation
\be
a_j^l=\sigma\left(\sum_k w_{jk}^l a_k^{l-1} + b_j^l\right)= \sigma(z_j^l),
\label{eq:activation_eq_NN}
\ee
where we have defined the linear weighted sum
\be
z_j^l =\sum_k w_{jk}^l a_k^{l-1} + b_j^l.
\label{eq:activation_eq_NN_z}
\ee

By definition, the cost function $E$ depends directly on the activities of the output layer $a_j^L$. It of course also \emph{indirectly} depends on all the activities of neurons in lower layers in the neural network through iteration of Eq.~\eqref{eq:activation_eq_NN}. Let us define the error $\Delta_j^L$ of the $j$-th neuron in the $L$-th layer as the change in cost function with respect to the weighted input $z_j^L$
\be
\Delta_j^L = {\partial{E} \over \partial z_j^L}.
\ee
This definition is the first of the four backpropagation equations.

We can analogously define the error of neuron $j$ in layer $l$,  $\Delta_j^l$, as the change in the cost function w.r.t.~the weighted input $z_j^l$:
\be
\tag{I}
\Delta_j^l= {\partial{E} \over \partial z_j^l}=  {\partial{E} \over \partial a_j^l} \sigma^\prime(z_j^l),
\label{eq:BP1}
\ee
where $\sigma^\prime(x)$ denotes the derivative of the non-linearity $\sigma(\cdot)$ with respect to its input evaluated at $x$. 
Notice that the error function $\Delta_j^l$ can also be interpreted as the partial derivative of the cost function with respect to the bias $b_j^l$, since
\be
\tag{II}
\Delta_j^l= {\partial{E} \over \partial z_j^l}={\partial{E} \over \partial b_j^l}{\partial b_j^l \over \partial z_j^l}= {\partial{E} \over \partial b_j^l},
\label{eq:BP2}
\ee
where in the last line we have used the fact that ${ \partial b_j^l / \partial z_j^l}=1$, cf.~Eq.~\eqref{eq:activation_eq_NN_z}. This is the second of the four backpropagation equations. 

We now derive the final two backpropagation equations using the chain rule. Since the error depends on neurons in layer $l$ only through the activation of neurons in the subsequent layer $l+1$, we can use the chain rule to write
\begin{align}
\Delta_j^l &= {\partial{E} \over \partial z_j^l}= \sum_{k}{\partial E \over \partial z_{k}^{l+1}} {\partial z_{k}^{l+1} \over \partial z_j^l} \nonumber\\
&= \sum_k \Delta_k^{l+1} {\partial z_k^{l+1} \over \partial z_j^l} \nonumber \\
	&= \left(\sum_k  \Delta_k^{l+1} w_{kj}^{l+1} \right)\sigma^\prime(z_j^l).\tag{III}
\label{eq:BP3}
\end{align}
This is the third backpropagation equation. The final equation can be derived by differentiating of the cost function with respect to the weight $w_{jk}^l$ as
\begin{equation}
{\partial E \over \partial w_{jk}^l} = {\partial E \over \partial z_{j}^l}{\partial z_{j}^l \over \partial w_{jk}^l} = \Delta_j^l a_{k}^{l-1} \tag{IV}
\label{eq:BP4}
\end{equation}

Together, Eqs.~\eqref{eq:BP1},~\eqref{eq:BP2},~\eqref{eq:BP3},  and~\eqref{eq:BP4} define the four backpropagation equations relating the gradients of the activations of various neurons $a_{j}^l$, the weighted inputs $z_{j}^l=\sum_k w_{jk}^l a_k^{l-1} + b_j^l $, and the errors $\Delta_j^l$. These equations can be combined into a simple, computationally efficient algorithm to calculate the gradient with respect to all parameters \cite{nielsen2015neural}.
{\center \bf The  Backpropagation Algorithm}
\begin{enumerate}
\item {\bf Activation at input layer:} calculate the activations $a_j^1$ of all the neurons in the input layer.
\item {\bf Feedforward:} starting with the first layer, exploit the feed-forward architecture through Eq.~\eqref{eq:activation_eq_NN} to compute $z^l$ and $a^l$ for each subsequent layer. 
\item {\bf Error at top layer:} calculate the error of the top layer using Eq.~\eqref{eq:BP1}. This requires to know the expression for the derivative of both the cost function $E(\mathbf{w}) = E(\boldsymbol{a}^L)$ and the activation function $\sigma(z)$.
\item {\bf ``Backpropagate'' the error:} use Eq.~\eqref{eq:BP3} to propagate the error backwards and calculate $\Delta_j^l$ for all layers.
\item{\bf Calculate gradient:} use Eqs.~\eqref{eq:BP2} and~\eqref{eq:BP4} to calculate ${\partial{E} \over \partial b_j^l}$ and ${\partial E \over \partial w_{jk}^l}$.
\end{enumerate}

We can now see where the name backpropagation comes from. The algorithm consists of a forward pass from the bottom layer to the top layer where one calculates the weighted inputs and activations of all the neurons. One then backpropagates the error starting with the top layer down to the input layer and uses these errors to calculate the desired gradients. This description makes clear the incredible utility and computational efficiency of the backpropagation algorithm. We can calculate all the derivatives using a single ``forward'' and ``backward'' pass of the neural network. This computational efficiency is crucial since we must calculate the gradient with respect to all parameters of the neural net at each step of gradient descent. These basic ideas also underly almost all modern automatic differentiation packages such as Autograd (Pytorch).

\subsubsection{Computing gradients in deep networks: what can go wrong with backprop?}

Armed with backpropagation and gradient descent, it seems like it should be straightforward to train any neural network. However, until fairly recently it was widely believed that training deep networks was an extremely difficult task. One reason for this was that even with backpropagation, gradient descent on large networks is extremely computationally expensive. However, the great advances in computational hardware (and the widespread use of GPUs) has made this a much less vexing problem than even a decade ago. It is hard to understate the impact these advances in computing have had on the practical utility of neural networks.

On a more technical and mathematical note, another problem that occurs in deep networks, which transmit information through many layers, is that gradients can vanish or explode. This is, appropriately, known as \emph{the problem of vanishing or exploding gradients}. This problem is especially pronounced in neural networks that try to capture long-range dependencies, such as Recurrent Neural Networks for sequential data. We can illustrate this problem by considering a simple network with one neuron in each layer. We further assume that all weights are equal, and denote them by $w$. The behavior of the backpropagation equations for such a network can be inferred from repeatedly using Eq.~\eqref{eq:BP3}:
\be
\Delta_j^1=  \Delta_j^L  \prod_{j=0}^{L-1} w\sigma^\prime(z_j) =  \Delta_j^L (w)^L   \prod_{j=0}^{L-1} \sigma^\prime(z_j),
\ee 
where $\Delta_j^L$ is the error in the $L$-th topmost layer, and $(w)^L$ is the weight to the power $L$.
Let us now also assume that the magnitude $\sigma^\prime(z_j)$ is fairly constant and we can approximate $\sigma^\prime(z_j) \approx \sigma^\prime_0$. In this case, notice that for large $L$, the error $\Delta_j^1$ has very different behavior depending on the value of $w  \sigma^\prime_0$. If $w \sigma^\prime_0>1$, the errors and the gradient blow up. On the other hand, if $w \sigma^\prime_0<1$ the errors and gradients vanish. Only when the weights satisfy $w \sigma^\prime_0\approx 1$ and the neurons are not saturated will the gradient stay well behaved for deep networks. 

This basic behavior holds true even in more complicated networks. Rather than considering a single weight, we can ask about the eigenvalues (or singular values) of the weight matrices $w_{jk}^l$. In order for the gradients to be finite for deep networks, we need these eigenvalues to stay near unity even after many gradient descent steps. In modern feedforward and ReLU neural networks, this is achieved by initializing the weights for the gradient descent in clever ways and using non-linearities that do not saturate, such as ReLUs (recall that for saturating functions, $\sigma^\prime  \rightarrow 0$, which will cause the gradient to vanish). Proper initialization and regularization schemes such as gradient clipping (cutting-off gradients with very large values), and batch normalization also help mitigate the vanishing and exploding gradient problem.

\subsection{Regularizing neural networks and other practical considerations}

DNNs, like all supervised learning algorithms, must navigate the bias-variance tradeoff. Regularization techniques play an important role in ensuring that DNNs generalize well to new data. The last five years have seen a wealth of new specialized regularization techniques for DNNs  beyond the simple $L_1$ and $L_2$ penalties discussed in the context of linear and logistic regression, see Secs.~\ref{sec:lin_reg} and~\ref{sec:log_reg}. These new techniques include Dropout and Batch Normalization. In addition to these specialized regularization techniques, large DNNs seem especially well-suited to implicit regularization that already takes place in the Stochastic Gradient Descent (SGD)~\cite{wilson2017marginal}, cf.~Sec.~\ref{sec:gradient_descent}. The implicit stochasticity and local nature of SGD often prevent overfitting of spurious correlations in the training data, especially when combined with techniques such as Early Stopping. In this section, we give a brief overview of these regularization techniques.

\subsubsection{Implicit regularization using SGD: initialization, hyper-parameter tuning, and Early Stopping}

The most commonly employed and effective optimizer for training neural networks is SGD (see Sec.~\ref{sec:gradient_descent} for other alternatives). SGD acts as an implicit regularizer by introducing stochasticity (from the use of mini-batches) that prevents overfitting. In order to achieve good performance, it is important that the weight initialization is chosen randomly, in order to break any leftover symmetries. One common choice is drawing the weights from a Gaussian centered around zero with some variance that scales inversely with number of inputs to the neuron \cite{sutskever2013importance, he2015delving}. Since SGD is a local procedure, as networks get deeper, \emph{choosing a good weight initialization becomes increasingly important} to ensure that the gradients are well behaved. Choosing an initialization with a variance that is too large or too small will cause gradients to vanish and the network to train poorly -- even a factor of $2$ can make a huge difference in practice \cite{he2015delving}. For this reason, it is important to experiment with different variances.

The second important thing is to appropriately choose the learning rate or step-size by searching over five logarithmic grid points  \cite{wilson2017marginal}.  If the best performance occurs at the edge of the grid,  repeat this procedure until the optimal learning rate is in the middle of the grid parameters. Finally, it is common to center or whiten the input data (just as we did for linear and logistic regression).

Another important form of regularization that is often employed in practice is Early Stopping. The idea of Early Stopping is to divide the training data into two portions, the dataset we train on, and a smaller \textit{validation set} that serves as a proxy for out-of-sample performance on the test set. As we train the model, we plot both the training error and the validation error. We expect the training error to continuously decrease during training. However, the validation error will eventually increase due to overfitting. The basic  idea of early stopping is to halt the training procedure when the validation error starts to rise.
This Early Stopping procedure ensures that we stop the training and avoid fitting sample specific features in the data. Early Stopping is a widely used essential tool in the deep learning regularization toolbox.

\subsubsection{Dropout}

Another important regularization schemed that has been widely adopted in the neural networks literature is Dropout \cite{srivastava2014dropout}. The basic idea of Dropout is to prevent overfitting by reducing spurious correlations between neurons within the network by introducing a randomization procedure similar to that underlying ensemble models such as Bagging. Recall that the basic idea behind ensemble methods is to train an ensemble of models that are created using a randomization procedure to ensure that the members of the ensemble are uncorrelated, see Sec.~\ref{sec:combining}. This reduces the variance of statistical predictions without creating too much additional bias. 

\begin{figure}
\includegraphics[width=1.0 \columnwidth]{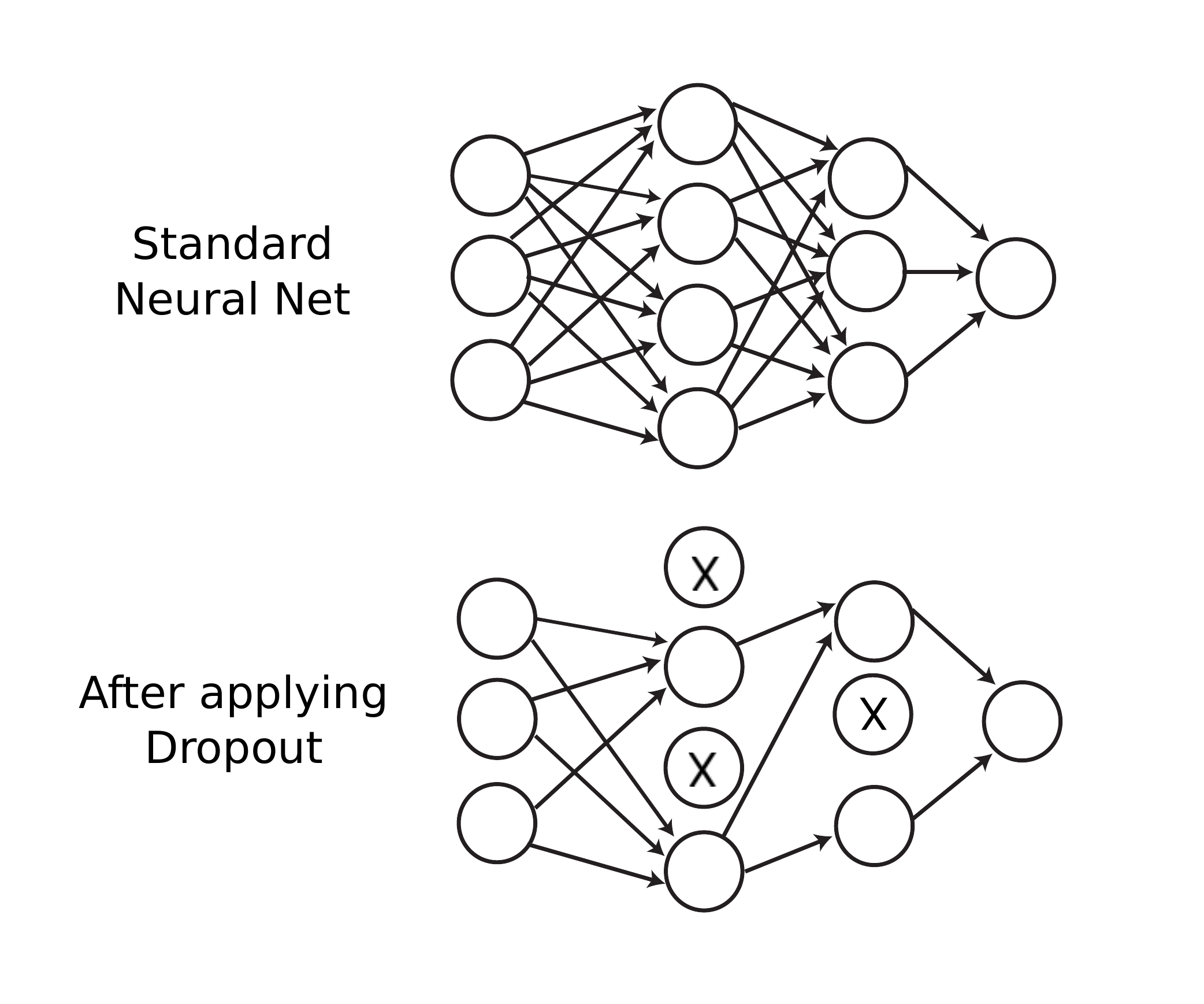}
\caption{{\bf Dropout } During the training procedure neurons are  randomly ``dropped out'' of the neural network with some probability $p$ giving rise to a thinned network. This prevents overfitting by reducing correlations among neurons and reducing the variance in a method similar in spirit to ensemble methods.}
\label{Fig:dropout}
\end{figure}

In the context of neural networks, it is extremely costly to train an ensemble of networks, both from the point of view of the amount of data needed, as well as computational resources and parameter tuning required. Dropout circumnavigates these problems by randomly dropping out neurons (along with their connections) from the neural network during each step of the training (see Figure \ref{Fig:dropout}). Typically, for each mini-batch in the gradient descent step, a neuron is dropped from the neural network with a probability $p$. The gradient descent step is then performed only on the weights of the ``thinned'' network of individual predictors.  

Since during training, on average weights are only present a fraction $p$ of the time, predictions are made by reweighing the weights by $p$: $\mathbf{w}_{\mathrm{test}}  = p \mathbf{w}_\mathrm{train}$.The learned weights can be viewed as some  ``average'' weight over all possible thinned neural network. This averaging of weights is similar in spirit to the Bagging procedure discussed in the context of ensemble models, see Sec.~\ref{sec:combining}.

\subsubsection{\label{subsec:DNNs_batchnorm} Batch Normalization}

Batch Normalization is a regularization scheme that has been quickly adopted by the neural network community since its introduction in 2015 \cite{ioffe2015batch}. The basic inspiration behind Batch Normalization is the long-known observation that training in neural networks works best when the inputs are centered around zero with respect to the bias. The reason for this is that it prevents neurons from saturating and gradients from vanishing in deep nets. In the absence of such centering, changes in parameters in lower layers can give rise to saturation effects in higher layers, and vanishing gradients. The idea of Batch Normalization is to introduce additional new ``BatchNorm'' layers that standardize the inputs by the mean and variance of the mini-batch.

Consider a layer $l$ with $d$ neurons whose inputs are $(z_1^l, \ldots, z_d^l)$. We standardize each dimension so that 
\be
{z}_k^l \longrightarrow \hat{z}_k^l= {z_k^l-\mathbb{E}[z_k^l] \over \sqrt{\mathrm{Var}[z_k^l]}},
\label{eq:batch_norm}
\ee
where the mean and variance are taken over all samples in the mini-batch. One problem with this procedure is that it may change the representational power of the neural network. For example, for $\tanh$ non-linearities, it may force the network to live purely in the linear regime around $z=0$. Since non-linearities are crucial to the representational power of DNNs, this could dramatically alter the power of the DNN. 

For this reason, one introduces two new parameters $\gamma_{k}^l$ and  $\beta_{k}^l$ for each neuron that can additionally shift and scale the normalized input 
\be
\hat{z}_k^l \longrightarrow \hat{\mathrm{z}}_k^l = \gamma_k^l \hat{z}_k^l + \beta_k^l.
\label{eq:batch_norm_rescale}
\ee
One can think of Eqs.~\eqref{eq:batch_norm} and~\eqref{eq:batch_norm_rescale} as adding new extra layers $\hat{\mathrm{z}}_k^l$ in the deep net architecture. Hence, the new parameters $\gamma_{k}^l$ and  $\beta_{k}^l$ can be learned just like the weights and biases using backpropagation (since this is just an extra layer for the chain rule). We initialize the neural network so that at the beginning of training the inputs are being standardized. Backpropagation then adjusts $\gamma$ and $\beta$ during training.

In practice, Batch Normalization considerably improves the learning speed by preventing gradients from vanishing. However, it also seems to serve as a powerful regularizer for reasons that are not fully understood. One plausible explanation is that in batch normalization, the gradient for a sample depends not only on the sample itself but also on all the properties of the mini-batch. Since a single sample can occur in different mini-batches, this introduces additional randomness into the training procedure which seems to help regularize training.

\subsection{\label{subsec:DN_examples}Deep neural networks in practice: examples}

Now that we have gained sufficient high-level background knowledge about deep neural nets, let us discuss how to use them in practice. 

\subsubsection{Deep learning packages}

In~\href{http://physics.bu.edu/~pankajm/MLnotebooks.html}{Notebook 11}, we demonstrated that the numerical implementation of DNNs is greatly facilitated by open source python packages, such as \href{https://keras.io/}{Keras}, \href{https://www.tensorflow.org/}{TensorFlow}, \href{http://pytorch.org/}{Pytorch}, and others. The complexity and learning curves for these packages differ, depending on the user's level of familiarity with Python. The reader should keep mind mind that there are DNN packages written in other languages, such as Caffe which uses \texttt{C++}, but we do not discuss them in this review for brevity.

Keras is a high-level framework which does not require any knowledge about the inner workings of the underlying deep learning algorithms. Coding DNNs in Keras is particularly simple, see~\href{http://physics.bu.edu/~pankajm/MLnotebooks.html}{Notebook 11}, and allows one to quickly grasp the big picture behind the theoretical concepts which we introduced above. However, for advanced applications, which may require more direct control over the operations in between the layers, Keras' high-level design may turn out insufficient. 

If one opens up the Keras black box, one will find that it wraps the functionality of another package -- TensorFlow\footnote{While Keras can also be used with a Theano backend, we do not discuss this here since Theano support has been discontinued.}. Over the last years, TensorFlow, which is supported by Google, has been gaining popularity and has become the preferred library for deep learning. It is frequently used in Kaggle competitions, university classes, and industry. In TensorFlow one constructs data flow graphs, the nodes of which represent mathematical operations, while the edges encode multidimensional tensors (data arrays). A deep neural net can then be thought of as a graph with a particular architecture and connectivity. One needs to understand this concept well before one can truly unleash TensorFlow's full potential. The learning curve can sometimes be rather steep for TensorFlow beginners, and requires a certain degree of perseverance and devoted time to internalize the underlying ideas. 

There are, however, many other open source packages which allow for control over the inter- and intra-layer operations, without the need to introduce computational graphs. Such an example is Pytorch, which offers libraries for automatic differentiation of tensors at GPU speed. As we discussed above, manipulating neural nets boils down to fast array multiplication and contraction operations and, therefore, the \texttt{torch.nn} library often does the job of providing enough access and controllability to manipulate the linear algebra operations underlying deep neural nets.

For the benefit of the reader, we have prepared Jupyter \href{https://physics.bu.edu/~pankajm/MLnotebooks.html}{notebooks} for DNNs using all three packages for the deep learning problems we discuss below. We invite the reader to carefully examine the differences in the code which should help them decide on which package they prefer to use.

\subsubsection{Approaching the learning problem}

Let us now analyze a typical procedure for using neural networks to solve supervised learning problems. As can be seen already from the code snippets in~\href{http://physics.bu.edu/~pankajm/MLnotebooks.html}{Notebook 11}, constructing a deep neural network to solve ML problems is a multiple-stage process. Generally, one can identify a set of key steps:
\begin{enumerate}
	\item {\bf Collect and pre-process the data.}
	\item {\bf Define the model and its architecture.}
	\item {\bf Choose the cost function and the optimizer.}
	\item {\bf Train the model.}
	\item {\bf Evaluate and \emph{study} the model performance on the test data.}
	\item {\bf Use the validation data to adjust the hyperparameters (and, if necessary, network architecture) to optimize performance for the specific dataset.}
\end{enumerate}

At this point a few remarks are in order. While we treat Step 1 above as consisting mainly of loading and reshaping a dataset prepared ahead of time, we emphasize that obtaining a sufficient amount of data is a typical challenge in many applications. Oftentimes insufficient data serves as a major bottleneck on the ultimate performance of DNNs. In such cases one can consider data augmentation, i.e.~distorting data samples from the existing dataset in some way to enhance size the dataset. Obviously, if one knows how to do this, one already has partial information about the important features in the data.

One of the first questions we are typically faced with is how to determine the sizes of the training and test data sets. The MNIST dataset, which has $10$ classification categories, uses $80\%$ of the available data for training and $20\%$ for testing. On the other hand, the \href{http://www.image-net.org/}{ImageNet} data which has $1000$ categories is split $50\%-50\%$. As a rule of thumb, the more classification categories there are in the task, the closer the sizes of the training and test datasets should be in order to prevent overfitting. Once the size of the training set is fixed, it is common to reserve $20\%$ of it for validation, which is used to fine-tune the hyperparameters of the model. 

Also related to data preprocessing is the standardization of the dataset. It has been found empirically that if the original values of the data differ by orders of magnitude, training can be slowed down or impeded. This can be traced back to the vanishing and exploding gradient problem in backprop discussed in Sec.~\ref{subsec:backprop}. To avoid such unwanted effects, one often resorts to two tricks: (i) all data should be mean-centered, i.e.~from every data point we subtract the mean of the entire dataset, and (ii) rescale the data, for which there are two ways: if the data is approximately normally distributed, one can rescale by the standard deviation. Otherwise, it is typically rescaled by the maximum absolute value so the rescaled data lies within the interval $[-1,1]$. Rescaling ensures that the weights of the DNN are of a similar order of magnitude (notice the similarity of this idea to Batch Normalization, cf.~Sec.~\ref{subsec:DNNs_batchnorm}).

The next issue is how to choose the right hyperparameters to start training the model. According to Bengio, the optimal learning rate is often an order of magnitude lower than the smallest learning rate that blows up the loss~\cite{bengio2012practical}. One should also keep in mind that, depending on how ambitious of a problem one is dealing with, training the model can take a considerable amount of time. This can severely slow down any progress on improving the model in Step 6. Therefore, it is usually a good idea to play with a small enough fraction of the training data to get a rough feeling about the correct hyperparameter regimes, the usefulness of the DNN architecture, and to debug the code. The size of this small `play set' should be such that training on it can be done fast and in real time to allow to quickly adjust the hyperparameters. A typical strategy of exploring the hyperparameter landscape is to use grid searches.

Whereas it is always possible to view Steps 1-5 as generic and independent of the particular problem we are trying to solve, it is only when these steps are put together in Step 6 that the real benefit of deep learning is revealed, compared to less sophisticated methods such as  regression or bagging, see Secs.~\ref{sec:lin_reg},~\ref{sec:log_reg}, and~\ref{sec:combining}. The optimal choice of network architecture, cost function, and optimizer is determined by the properties of the training and test datasets, which are usually revealed when we try to improve the model.

While there is no ``one-size-fits-them-all'' recipe to approach ML problems, we believe that the above list gives a good overview and can be a useful guideline to the layman. Furthermore, as it becomes clear, this `recipe' can be applied to generic supervised learning tasks, not just DNNs. We refer the reader to Sec.~\ref{sec:DNN-III} for more useful hints and tips on how to use the validation data during the training process.

\subsubsection{SUSY dataset}

As a first example from physics, we discuss a DNN approach to the SUSY dataset already introduced in the context of logistic regression in Sec.~\ref{subsubsec:susy_logreg}, and Bagging in Sec.~\ref{subsec:ensemble_appl}.  For a detailed description of the SUSY dataset and the corresponding classification problem, we refer the reader to Sec.~\ref{subsubsec:susy_logreg}. There is an interest in using deep learning methods to automate the discovery of collision features from data. Benchmark results using Bayesian Decision Trees from a standard physics package, and five-layer neural networks using Dropout were presented in the original paper~\cite{baldi2014searching}; they demonstrate the ability of deep learning to bypass the need of using hand-crafted high-level features. Our goal here is to study systematically the accuracy of a DNN classifier as a function of the learning rate and the dataset size.

\begin{figure}[t!]
	\includegraphics[width=1.0\columnwidth]{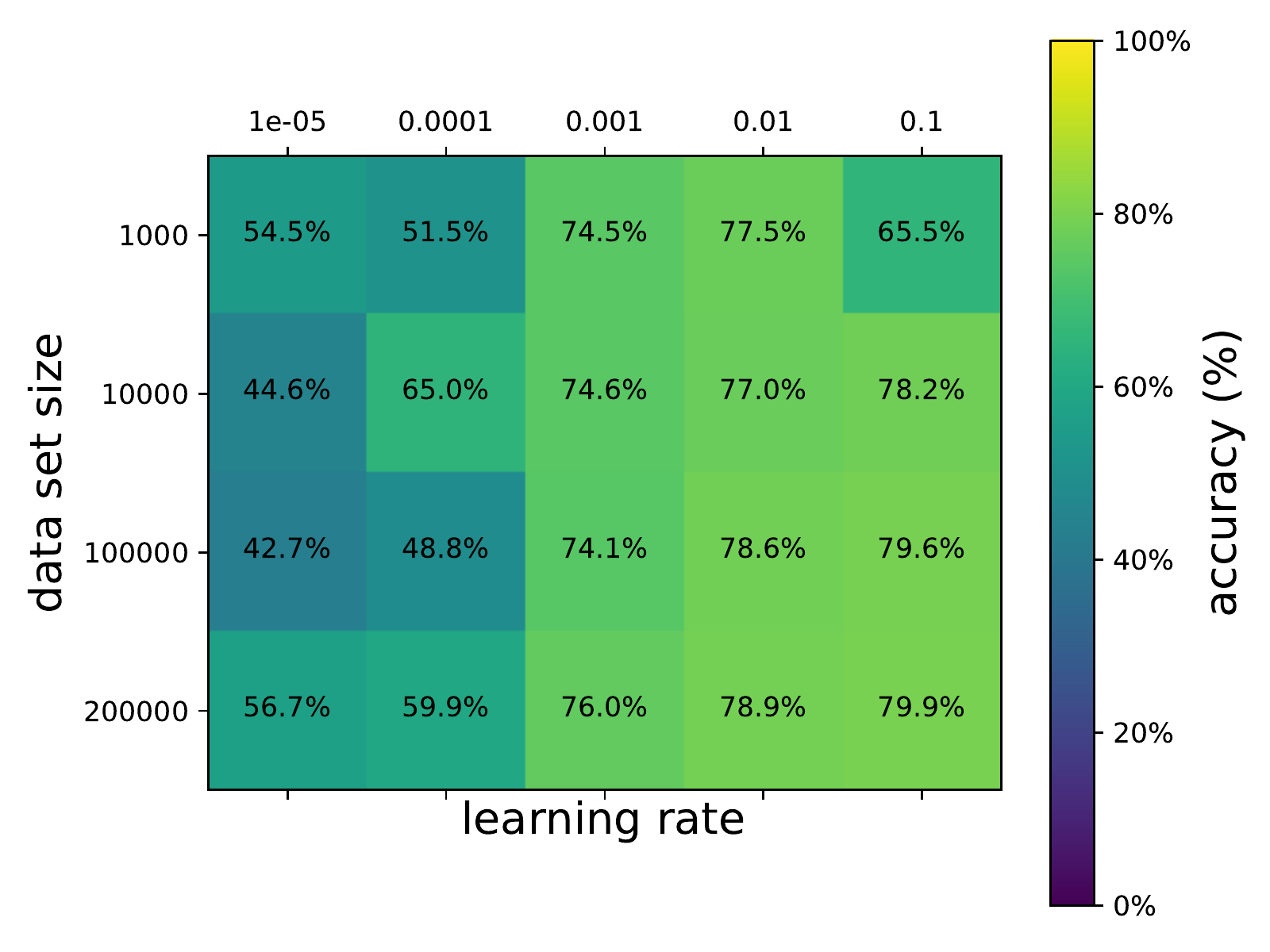}
	\caption{\label{fig:DNN_SUSY_accuracy_test} Grid search results for the test set accuracy of the DNN for the SUSY problem as a function of the learning rate and the size of the dataset. The data used includes all high- and low-level features.}
\end{figure} 

Unlike the MNIST example where we used Keras, here we use the opportunity to introduce the Pytorch package, see the corresponding \href{https://physics.bu.edu/~pankajm/MLnotebooks.html}{notebook}. We leave the discussion of the code-specific details for the accompanying notebook. 

To classify the SUSY collision events, we construct a DNN with two dense hidden layers of $200$ and $100$ neurons, respectively. We use ReLU activation between the input and the hidden layers, and a sofmax output layer. We apply dropout regularization on the weights of the DNN. Similar to MNIST, we use the cross-entropy as a cost function and minimize it using SGD with batches of size $10\%$ of the training data size. We train the DNN for $10$ epochs.

Figure~\ref{fig:DNN_SUSY_accuracy_test} shows the accuracy of our DNN on the test data as a function of the learning rate and the size of the dataset. It is considered good practice to start with a logarithmic scale to search through the hyperparameters, to get an overall idea for the order of magnitude of the optimal values.  In this example, the performance peaks at the largest size of the dataset and a learning rate of $0.1$, and is of the order of $80\%$. Since the optimal performance is obtained at the edge of the grid, we encourage the reader to extend the grid size to beat our result.
For comparison, in the original study ~\cite{baldi2014searching}, the authors achieved $\approx 89\%$ by using the entire dataset with $5,000,000$ points and a more sophisticated network architecture, trained using GPUs.

\subsubsection{Phases of the 2D Ising model}

As a second example from physics, we discuss a DNN approach to the Ising dataset introduced in Sec.~\ref{subsubsec:ising_phases_logreg}. We study the problem of classifying the states of the 2D Ising model with a DNN~\cite{tanaka2017detection}, focussing on the model performance as a function of both the number of hidden neurons and the learning rate. The discussion is accompanied by a \href{https://physics.bu.edu/~pankajm/MLnotebooks.html}{notebook} written in TensorFlow. As in the previous example, the interested reader can find the discussion of the code-specific details in the notebook. 

\begin{figure}[t!]
	\includegraphics[width=1.0\columnwidth]{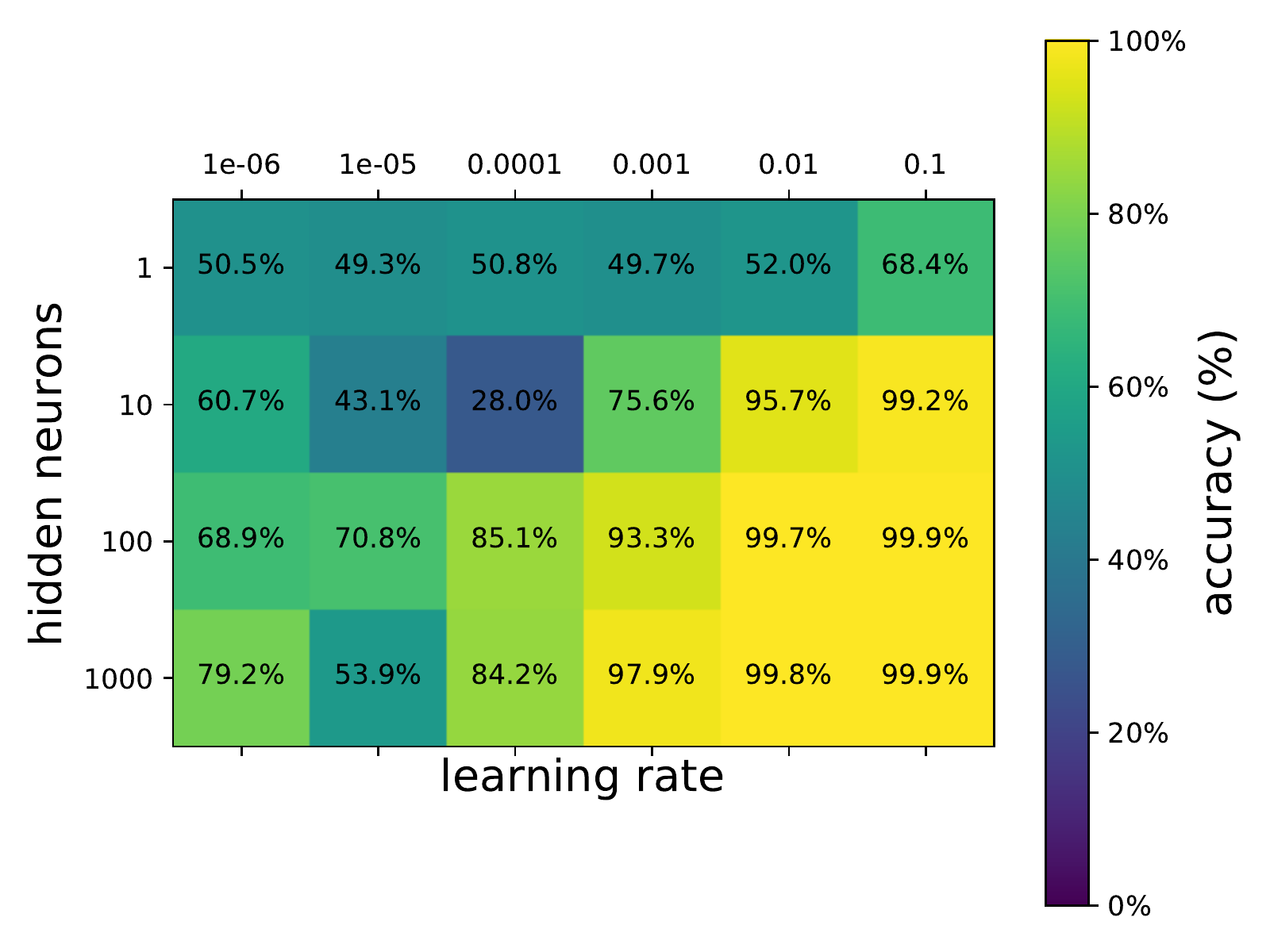}
	\includegraphics[width=1.0\columnwidth]{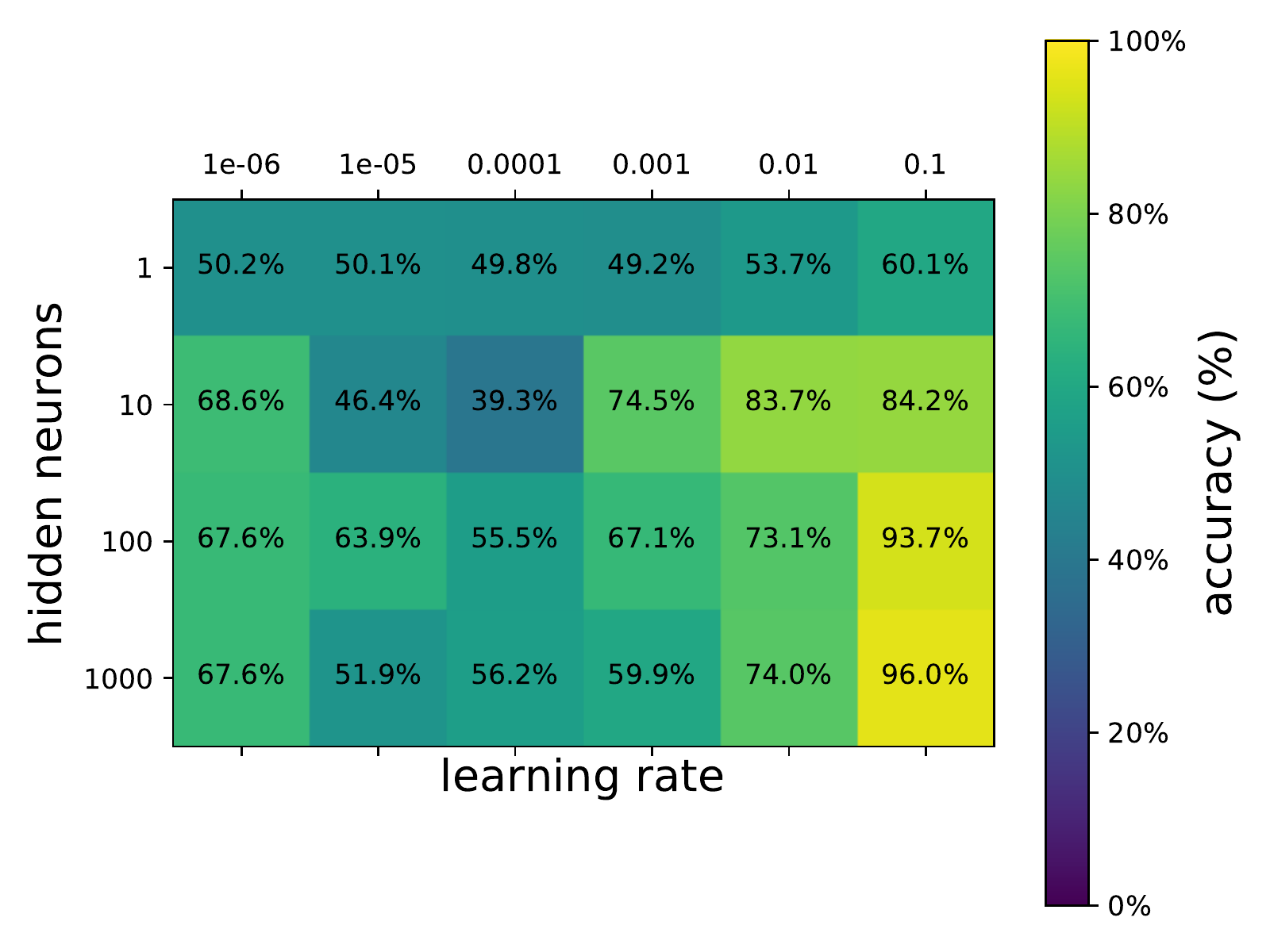}
	\caption{\label{fig:DNN_Ising_accuracy} Grid search results for the test set accuracy (top) and the critical set accuracy (bottom) of the DNN for the Ising classification problem as a function of the learning rate and the number of hidden neurons.}
\end{figure} 

To classify whether a given spin configuration is in the ordered or disordered phase, we construct a minimalistic model for a DNN with a single hidden layer containing a number of hidden neurons. The network architecture thus includes a ReLU-activated input layer, the hidden layer, and the softmax output layer. We pick the categorical cross-entropy as a cost function and minimize it using SGD with mini-batches of size $100$. We train the DNN for $100$ epochs.

Figure~\ref{fig:DNN_Ising_accuracy} shows the outcome of a grid search over a log-spaced learning rate and the number of neurons in the hidden layer. We see that about $10$ neurons are enough at a learning rate of $0.1$ to get to a very high accuracy on the test set. However, if we aim at capturing the physics close to criticality, clearly more neurons are required to reliably learn the more complex correlations in the Ising configurations.

\section{Convolutional Neural Networks (CNNs)}
\label{sec:CNNs}

\begin{figure*}[t!]
	\includegraphics[width=2.0\columnwidth]{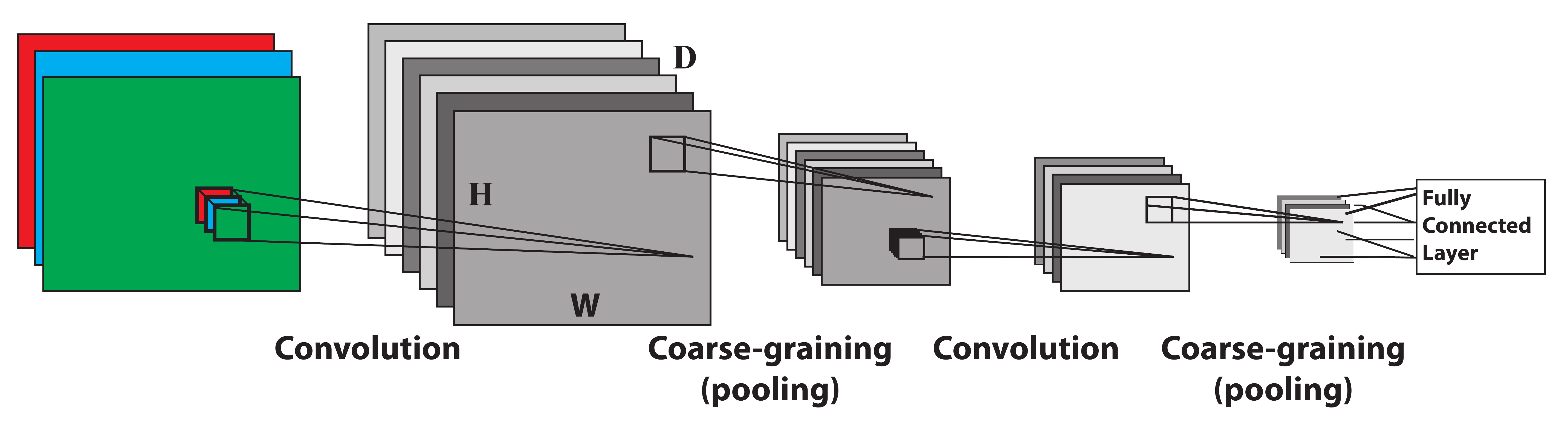}
	\caption{ {\bf Architecture of a Convolutional Neural Network (CNN).}  The neurons in a CNN are arranged in three dimensions: height ($H$), width ($W$), and depth ($D$). For the input layer, the depth corresponds to the number of channels (in this case 3 for RGB images). Neurons in the convolutional layers calculate the convolution of  the image with a local spatial filter (e.g.~$3\times 3$ pixel grid, times 3 channels for first layer) at a given location in the spatial $(W,H)$-plane. The depth $D$ of the convolutional layer corresponds to the number of filters used in the convolutional layer. Neurons at the same depth correspond to the same filter. Neurons in the convolutional layer mix inputs at different depths but preserve the spatial location. Pooling layers perform a spatial coarse graining (pooling step) at each depth to give a smaller height and width while preserving the depth. The convolutional and pooling layers are followed by a fully connected layer and classifier (not shown).}
\label{fig:CNN-architecture}
\end{figure*}

One of the core lessons of physics is that we should exploit symmetries and invariances when analyzing  physical systems. Properties such as locality and translational invariance are often built directly into the physical laws. Our statistical physics models often directly incorporate everything we know about the physical system being analyzed. For example, we know that in many cases it is sufficient to consider only local couplings in our Hamiltonians, or work directly in momentum space if the system is translationally invariant. This basic idea, tailoring our analysis to exploit additional structure, is a key feature of modern physical theories from general relativity, through gauge theories, to critical phenomena. 

Like physical systems, many datasets and supervised learning tasks also possess additional symmetries and structure. For instance, consider a supervised learning task where we want to label images from some dataset as being pictures of cats or not. Our statistical procedure must first learn features associated with cats. Because a cat is a physical object, we know that these features are likely to be local (groups of neighboring pixels in the two-dimensional image corresponding to whiskers, tails, eyes, etc).   We also know that the cat can be anywhere in the image. Thus, it does not really matter where in the picture these features occur (though relative positions of features likely do matter). This is a manifestation of translational invariance that is built into our supervised learning task. This example makes clear that, like many physical systems, many ML tasks  (especially in the context of computer vision and image processing) also possess additional structure, such as locality and translation invariance. 

The all-to-all coupled neural networks in the previous section fail to exploit this additional structure. For example, consider the image of the digit `four' from the MNIST dataset shown in Fig.~\ref{fig:MNIST_data_vis}. In the all-to-all coupled neural networks used there, the $28 \times 28$ image was considered a one-dimensional vector of size $28^2=796$. This clearly throws away lots of the spatial information contained in the image. Not surprisingly, the neural networks community realized these problems and designed a class of neural network architectures, convolutional neural networks or CNNs, that take advantage of this additional structure (locality and translational invariance) \cite{lecun1995convolutional}. Furthermore,  what is especially interesting from a physics perspective is the recent finding that these CNN architectures are intimately related to models such as tensor networks~\cite{stoudenmire2016supervised,stoudenmire2018learning} and, in particular, MERA-like architectures that are commonly used in physical models for quantum condensed matter systems \cite{levine2017deep}.

\subsection{The structure of convolutional neural networks}

A convolutional neural network is a translationally invariant neural network that respects locality of the input data. CNNs are the backbone of many modern deep learning applications and here we just give a high-level overview of CNNs that  should allow the reader to delve directly into the specialized literature. The reader is also strongly encouraged to consult the excellent, succinct notes for the Stanford CS231n Convolutional Neural Networks class developed by Andrej Karpathy and Fei-Fei Li (\url{https://cs231n.github.io/}). We have drawn heavily on the pedagogical style of these notes in crafting this section. 

There are two kinds of basic layers that make up a CNN: a convolution layer that computes the convolution of the input with a bank of filters (as a mathematical operation, see this practical guide to image kernels: \url{http://setosa.io/ev/image-kernels/}), and pooling layers that coarse-grain the input while maintaining locality and spatial structure, see Fig.~\ref{fig:CNN-architecture}. For two-dimensional data, a layer $l$ is characterized by three numbers: height $H_l$, width $W_l$, and depth $D_l$\footnote{The depth $D_l$ is often called ``number of channels'', to distinguish it from the depth of the neural network itself, i.e.~the total number of layers (which can be convolutional, pooling or fully-connected), cf.~Fig.~\ref{fig:CNN-architecture}.}. The height and width correspond to the sizes of the two-dimensional spatial $(W_l,H_l)$-plane (in neurons), and the depth $D_l$ (marked by the different colors in Fig.~\ref{fig:CNN-architecture}) -- to the number of filters in that layer. All neurons corresponding to a particular filter have the same parameters (i.e. shared weights and bias).

\begin{figure}[t!]
\includegraphics[width=1.0\columnwidth]{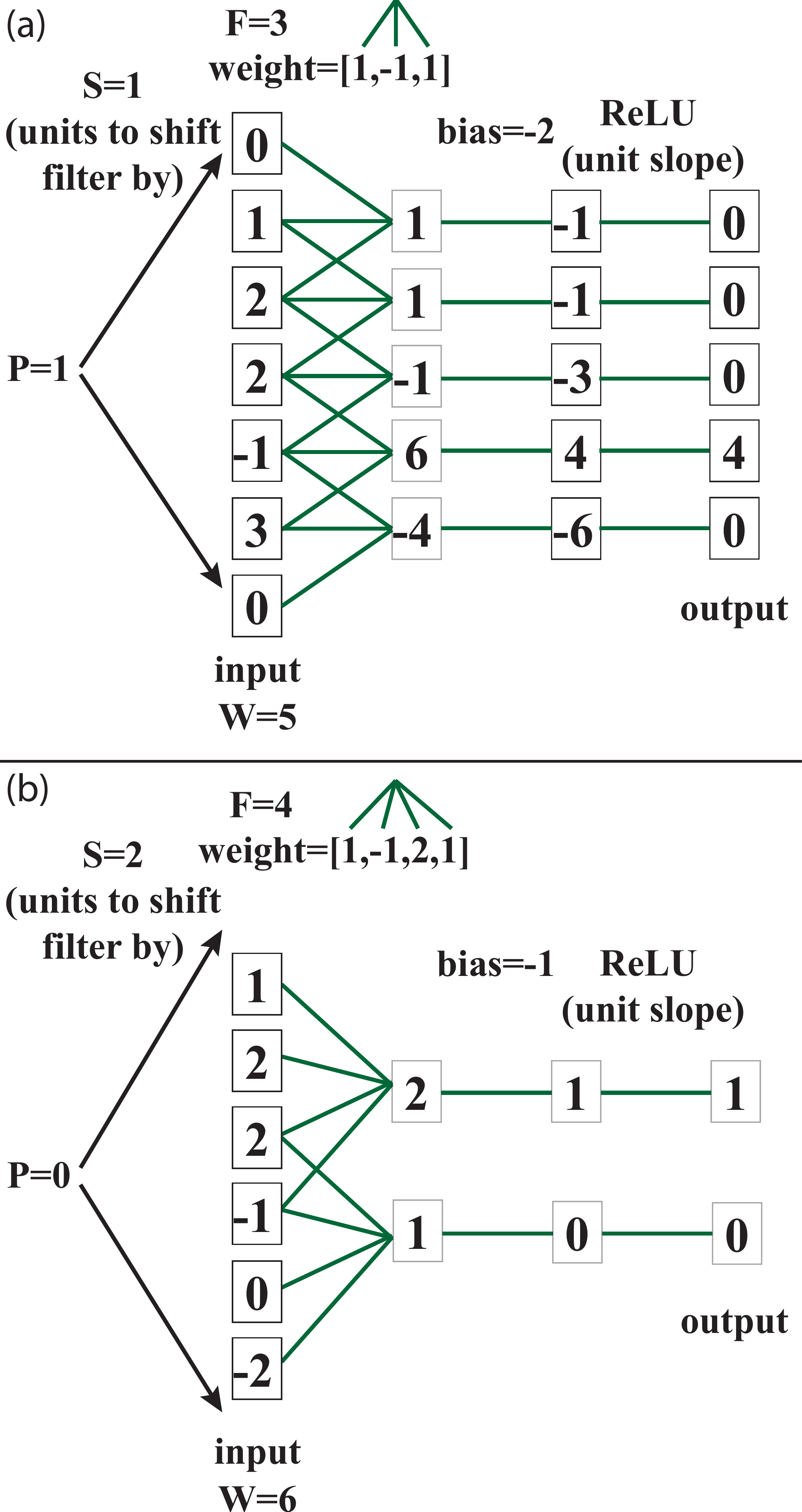}
\caption{ {\bf Two examples to illustrate a one-dimensional convolutional layer with ReLU nonlinearity}. Convolutional layer for a spatial filter of size $F$ for a one-dimensional input of width $W$ with stride $S$ and padding $P$ followed by a ReLU non-linearity. }
	\label{fig:1Dconv}
\end{figure}

In general, we will be concerned with local spatial filters (often called a receptive field in analogy with neuroscience) that take as inputs a small spatial patch of the previous layer at all depths. For instance, a square filter of size $F$ is a three-dimensional array of size $F \times F \times D_{l-1}$. The convolution consists of running this filter over all locations in the spatial plane. To demonstrate how this works in practice, let us a consider the simple example consisting of a one-dimensional input of depth $1$, shown in Fig.~\ref{fig:1Dconv}. In this case, a filter of size $F\times 1\times 1$ can be specified by a vector of weights $w$ of length $F$. The stride, $S$, encodes by how many neurons we translate the filter by when performing the convolution. In addition, it is common to pad the input with $P$ zeros (see Fig.~\ref{fig:1Dconv}). For an input of width $W$, the number of neurons (outputs) in the layer is given by $ (W-F+2P)/S+1$. We invite the reader to check out this visualization of the convolution procedure, \url{https://github.com/vdumoulin/conv_arithmetic/blob/master/README.md}, for a square input of unit depth. After computing the filter, the output is passed through a non-linearity, a ReLU in Fig. \ref{fig:1Dconv}. In practice, one often inserts a BatchNorm layer before the non-linearity, cf.~Sec.~\ref{subsec:DNNs_batchnorm}.

\begin{figure}[t!]
\includegraphics[width=1.0\columnwidth]{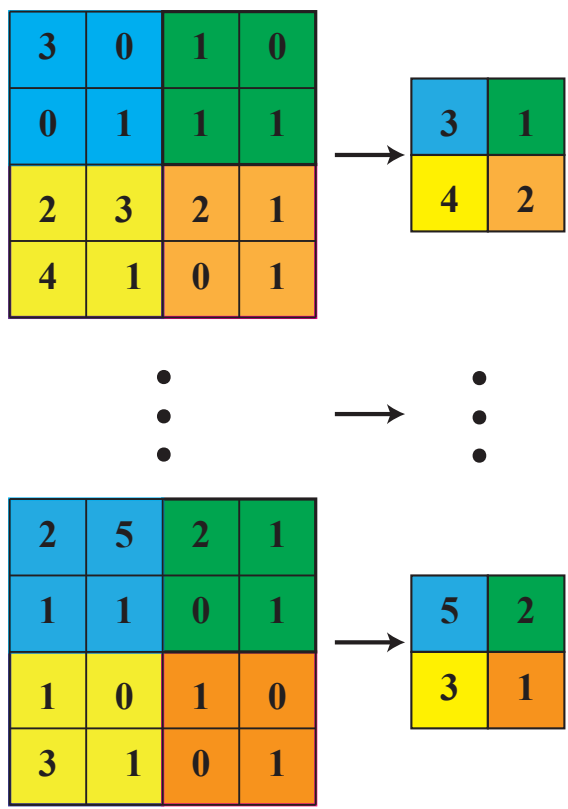}
\caption{{\bf Illustration of Max Pooling}.  Illustration of max-pooling over a $2 \times 2$ region. Notice that pooling is done at each depth (vertical axis) separately. The number of outputs is halved along each dimension due to this coarse-graining.}
	\label{fig:MaxPool}
\end{figure}

These convolutional layers are interspersed with pooling layers that coarse-grain spatial information by performing a subsampling at each depth. One common pooling operation is the max pool. In a max pool, the spatial dimensions are coarse-grained by replacing a small region (say $2 \times 2$ neurons) by a single neuron whose output is the maximum value of the output in the region. In physics, this pooling step is very similar to the decimation step of RG~\cite{mehta2014exact,lin2017does,koch2017mutual,iso2018scale}. This generally reduces the dimension of outputs. For example, if the region we pool over is $2 \times 2$, then both the height and the width of the output layer will be halved. Generally, pooling operations do not reduce the depth of the convolutional layers because pooling is performed separately at each depth. A simple example of a max-pooling operation is shown in Fig.~\ref{fig:MaxPool}. There are some studies suggesting that pooling might be unnecessary \cite{springenberg2014striving}, but pooling layers remain a staple of most CNNs.

In a CNN, the convolution and max-pool layers are generally followed by an all-to-all connected layer and a high-level classifier such as a soft-max. This allows us to train CNNs as usual using the backprop algorithm, cf.~Sec.~\ref{subsec:backprop}. From a backprop perspective, CNNs are almost identical to fully connected neural network architectures except with tied parameters.

Apart from introducing additional structure, such as translational invariance and locality, this convolutional structure also has important practical and computational benefits. All neurons at a given layer represent the same filter, and hence can all be described by a single set of weights and biases. This reduces the number of free parameters by a factor of $H \times W$ at each layer. For example, for a layer with $D=10^2$ and $H=W=10^2$, this gives a reduction in parameters of nearly $10^6$! This allows for the training of much larger models than would otherwise be possible with fully connected layers. We are familiar with similar phenomena in physics: e.g.~in translationally invariant systems we can parametrize all eigenmodes by specifying only their momentum (wave number) and functional form ($\sin$, $\cos$, etc.), while without translation invariance much more information is required.

\subsection{Example: CNNs for the 2D Ising model}

\begin{figure}[t!]
\includegraphics[width=1.0\columnwidth]{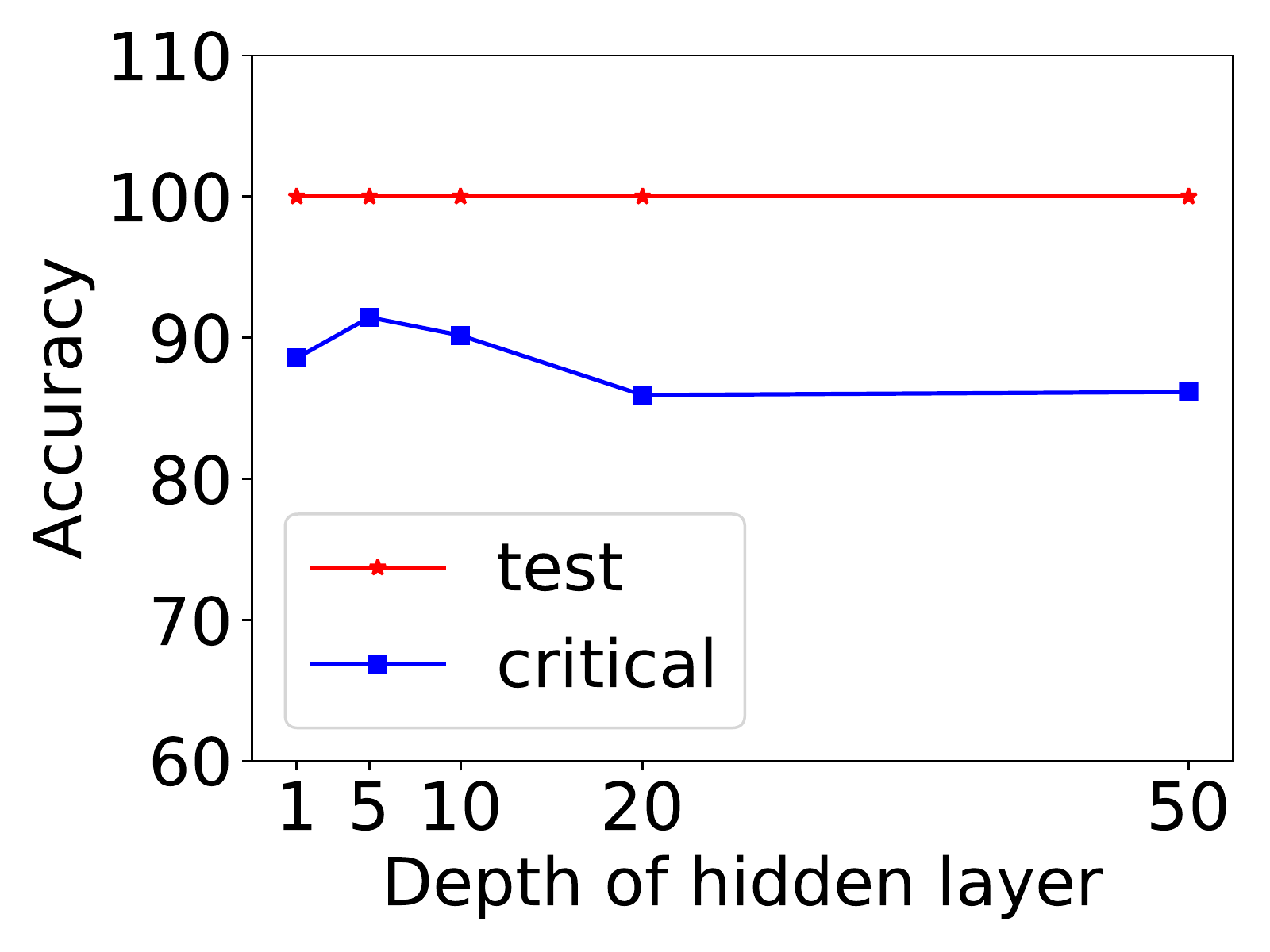}
\caption{{\bf  Single-layer convolutional  network for classifying phases in the Ising mode}. Accuracy on test set and critical samples for a convolutional neural network
with single layer of varying depth with filters of size 2, max-pool layer with receptive field of size $2$, followed by soft-max classifier. Notice that the test accuracy is $100\%$ even
for a CNN of depth one with a single set of weights. Accuracy on the near-critical dataset is significantly below that for the test set. }
	\label{fig:CNN-Ising}
\end{figure}

The inclusion of spatial structure in CNNs is an important feature that can be exploited when designing neural networks for studying physical systems. In the accompanying \href{https://physics.bu.edu/~pankajm/MLnotebooks.html}{notebook}, we used Pytorch to implement a simple CNN  composed of a single convolutional layer followed by a soft-max layer. Every input data point (i.e.~Ising configuration) is shaped as a two-dimensional array. We varied the output depth (i.e.~the number of output channels) of the convolutional layer from unity -- a single set of weights and one bias -- to an output depth of $50$ distinct weights and biases. The CNN was then trained using SGD for five epochs using a training set consisting of samples from far in the paramagnetic and ordered phases. The results are shown in Fig.~\ref{fig:CNN-Ising}. The CNN  achieved a 100\% accuracy on the test set for all architectures, even for a CNN with depth one. We also checked the performance of the CNN on samples drawn from the near-critical region for temperatures $T$ slightly above and below the critical temperature $T_c$. The CNN performed admirably even on these critical samples with an accuracy of between $80\%$ and $90\%$. As is the case with all ML and neural networks, the performance on parts of the data that are missing from the training set is considerably worse than on test data that is similar to the training data. This highlights the importance of properly constructing an accurate training dataset and the considerable obstacles of generalizing to novel situations. We encourage the interested reader to explore the corresponding notebook and design better CNN architectures with improved generalization performance on the near-critical set.

The reader may wish to check out the second part of the MNIST notebook for a discussion of CNNs applied to the digit recognition using the high-level Keras package. Regarding the SUSY dataset, we stress that the absence of spatial locality in the collision features renders applying CNNs to that problem inadequate.

\subsection{Pre-trained CNNs and transfer learning}

The immense success of CNNs for image recognition has resulted in the training of huge networks on enormous datasets, often by large industrial research teams from Google, Microsoft, Amazon, etc. Many of these models are known by name: AlexNet, GoogLeNet, ResNet, InceptionNet, VGGNet, etc. Most researchers and practitioners do not have the resources, data, or time to train networks on this scale. Luckily, the trained models have been released and are now available in standard packages such as the Torch Vision library in Pytorch or the Caffe framework. These models can be used directly as a basis for fine-tuning in different supervised image recognition tasks through a process called transfer learning.

The basic idea behind transfer learning is that the filters (receptive fields) learned by the convolution layers of these networks should be informative for most image recognition based tasks, not just the ones they were originally trained for. In other words, we expect that, since images reflect the natural world, the filters learned by these CNNs should transfer over to new tasks with only slight modifications and fine-tuning. In practice, this turns out to be true for many tasks one might be interested in.

There are three distinct ways one can take a pre-trained CNN and repurpose it for a new task. The following discussion draws heavily on the notes from the \href{https://cs231n.github.io/}{course CS231n} mentioned in the introduction to this section.
\begin{itemize}
\item {\bf Use CNN as fixed feature detector at top layer.}  If the new dataset we want to train on is small  and similar to the original dataset, we can simply use the CNN as a fixed feature detector and retrain our classifier. In other words, we  remove the classifier (soft-max) layer at the top of the CNN and replace it with a new classifier (linear support vector machine (SVM) or soft-max) relevant to our supervised learning problem. In this procedure,  the CNN serves as a fixed map from images to relevant features (the outputs of the top fully-connected layer right before the original classifier). This procedure prevents overfitting on small, similar datasets and is often a useful starting point for transfer learning. 

\item {\bf Use CNN as fixed feature detector at intermediate layer.} If the dataset is small and quite different from the dataset used to train the original image, the features at the top level might not be suitable for our dataset. In this case, one may want to instead use features in the middle of the CNN to train our new classifier. These features are thought to be less fine-tuned and more universal (e.g.~edge detectors). This is motivated by the idea that CNNs learn increasingly complex features the deeper one goes in the network (see discussion on representational learning in next section).

\item {\bf Fine-tune the CNN.} If the dataset is large, in addition to replacing and retraining the classifier in the top layer, we can also fine-tune the weights of the original CNN using backpropagation. One may choose to freeze some of the weights in the CNN during the procedure or retrain all of them simultaneously. 
\end{itemize}

All these procedures can be carried out easily by using packages such as Caffe or the Torch Vision library in PyTorch. PyTorch provides a nice python notebook that serves as tutorial on transfer learning. The reader is strongly urged to read the \href{http://pytorch.org/tutorials/}{Pytorch tutorials} carefully if interested in this topic.

\section{\label{sec:DNN-III}High-level Concepts in Deep Neural Networks}
\label{sec:DNN_concepts}

In the previous sections, we introduced deep neural networks and discussed how we can use these networks to perform supervised learning. Here, we take a step back and discuss some high-level questions about the practice and performance of neural networks. The first part of this section presents a deep learning workflow inspired by the bias-variance tradeoff. This workflow is especially relevant to industrial applications where one is often trying to employ neural networks to solve a particular problem. In the second part of this section, we shift gears and ask the question, \emph{why have neural networks been so successful?} We provide three different high-level explanations that reflect current dogmas. Finally, we end the section by discussing the limitations of supervised learning methods and current neural network architectures. 

\subsection{Organizing deep learning workflows using the bias-variance tradeoff}

\begin{figure}[t!]
\includegraphics[width=1.0\columnwidth]{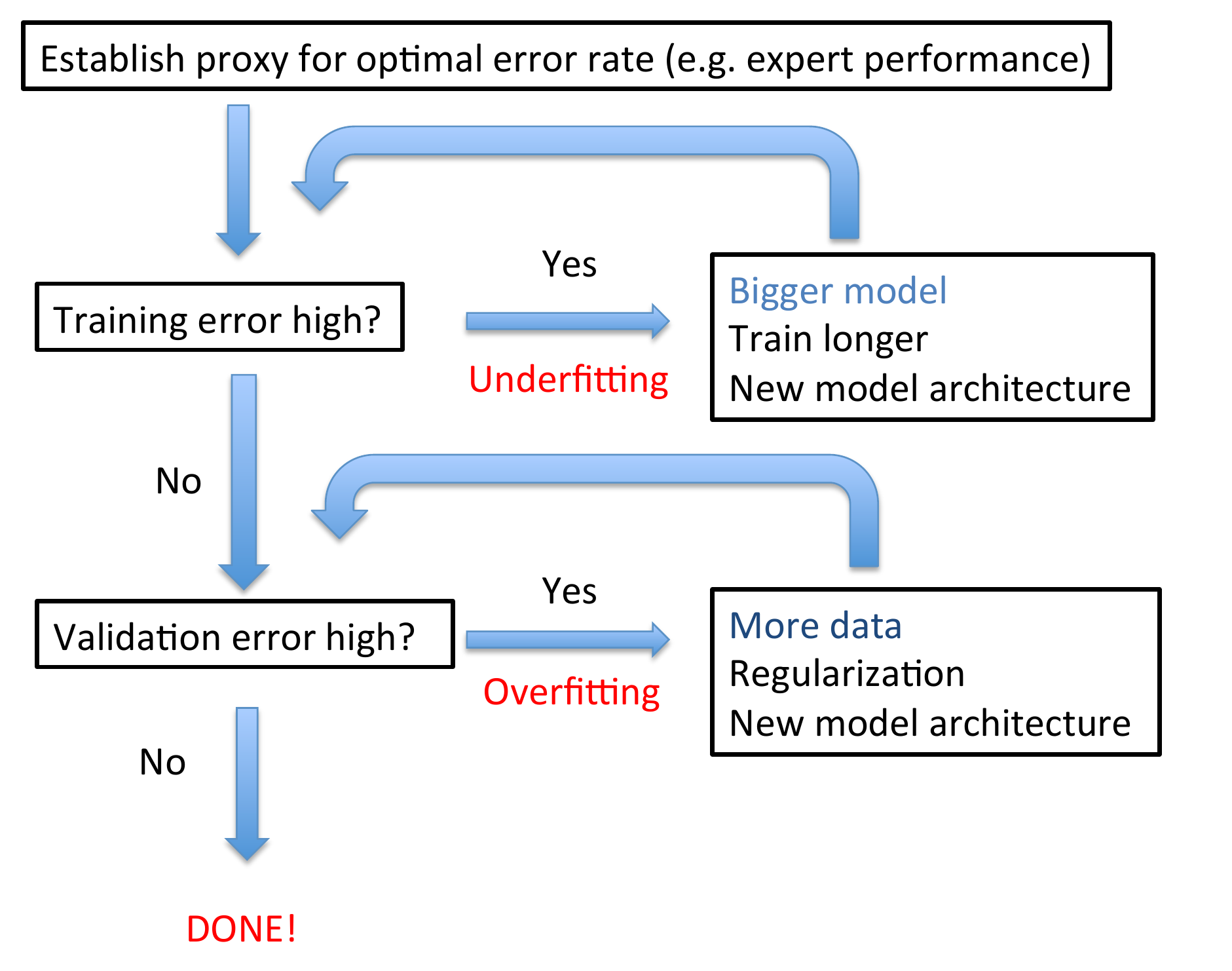}
\caption{{\bf Organizing a workflow for Deep Learning.} Schematic illustrating a deep learning workflow inspired by 
navigating the bias-variance tradeoff (Figure based on Andrew Ng's talk at the 2016 Deep Learning School available
at \url{https://www.youtube.com/watch?v=F1ka6a13S9I}.) In this diagram, we have assumed that there in no mismatch between the distributions the
training and test sets are drawn from.}
\label{Fig:NN_andrew_ng}
\end{figure}

Imagine that you are given some data and asked to design a neural network for learning how to perform a supervised learning task. What are the best practices for organizing a systematic workflow that allows us to efficiently do this? Here, we present a simple deep learning workflow inspired by thinking about the bias-variance tradeoff (see Figure \ref{Fig:NN_andrew_ng}). This section draws heavily on Andrew Ng's tutorial at the Deep Learning School (available online at  \href{https://www.youtube.com/watch?v=F1ka6a13S9I}{https://www.youtube.com/watch?v=F1ka6a13S9I}) which readers are strongly encouraged to watch.\\

The first thing we would like to do is divide the data into three parts. A training set, a validation or dev (development) set, and a test set. The test set is the data on which we want to make predictions. The dev set is a subset of the training data we use to check how well we are doing out-of-sample, after training the model on the training dataset. We use the validation error as a proxy for the test error in order to make tweaks to our model. It is crucial that we do not use any of the test data to train the algorithm. This is a cardinal sin in ML. We thus suggest the following workflow: \\

\noindent {\bf Estimate optimal error rate (Bayes rate).---}The first thing one should establish is the difficulty of the task and the best performance one can expect to achieve. No algorithm can do better than the ``signal'' in the dataset. For example, it is likely much easier to classify objects in high-resolution images than in very blurry, low-resolution images. Thus, one needs to establish a proxy or baseline for the optimal performance that can be expected from any algorithm. In the context of Bayesian statistics, this is often called the Bayes rate. Since we do not know this \emph{a priori},  we must get an estimate of this. For many tasks such as speech or object recognition, we can approximate this by the performance of humans on the task. For a more specialized task, we would like to ask how well experts, trained at the task, perform. This expert performance then serves as a proxy for our Bayes rate.\\

\noindent {\bf Minimize underfitting (bias) on training data set.---}After we have established the Bayes rate, we want to make sure that we are using a sufficiently complex model to avoid underfitting on the training dataset. In practice, this means comparing the training error rate to the Bayes rate. Since the training error does not care about generalization (variance), our model should approach the Bayes rate on the training set. If it does not, the bias of the DNN model is too large and one should try training the model longer and/or using a larger model. Finally, if none of these techniques work, it is likely that the model architecture is not well suited to the dataset, and one should modify the neural architecture in some way to better reflect the underlying structure of the data (symmetries, \emph{locality}, etc.).\\

\noindent{ \bf Make sure you are not overfitting.---} Next, we run our algorithm on the validation or dev set. If the error is similar to the training error rate and Bayes rate, we are done. If it is not, then we are overfitting the training data. Possible solutions include, regularization and, importantly, collecting more data. Finally, if none of these work, one likely has to change the DNN architecture.\\

If the validation and test sets are drawn from the same distributions, then good performance on the validation set should lead to similarly good performance on the test set. (Of course performance will typically be slightly worse on the test set because the hyperparameters were fit to the validation set.) However, sometimes the training data and test data differ in subtle ways because, for example, they are collected using slightly different methods, or because it is cheaper to collect data in one way versus another. In this case, there can be a mismatch between the training and test data. This can lead to the neural network overfitting these small differences between the test and training sets, and a poor performance on the test set despite having a good performance on the validation set. To rectify this, Andrew Ng suggests making two validation or dev sets, one constructed from the training data and one constructed from the test data. The difference between the performance of the algorithm on these two validation sets quantifies the train-test mismatch. This can serve as another important diagnostic when using DNNs for supervised learning.

\subsection{Why neural networks are so successful: three high-level perspectives on neural networks}

Having discussed the basics of neural networks, we conclude by giving three complementary perspectives on the success of DNNs and Deep Learning. This high-level discussion reflects various dogmas and intuitions about the success of DNNs and is in no way definitive or conclusive. As the reader was already warned in the introduction to DNNs, the field is rapidly expanding and many of these perspectives may turn out to be only partially true or even false. Nonetheless, we include them here as a guidepost for readers.

\subsubsection{Neural networks as representation learning}

One important and powerful aspect of the deep learning paradigm is the ability to learn relevant features of the data with relatively little domain knowledge, i.e. with minimal hand-crafting. Often, the power of deep learning stems from its ability to act like a black box that can take in a large stream of data and find good features that capture properties of the data we are interested in. This ability to learn good representations with very little hand-tuning is one of the most attractive properties of DNNs. Many of the other supervised learning algorithms discussed here (regression-based models, ensemble methods such as random forests or gradient-boosted trees) perform comparably or even better than neural networks but when using hand-crafted features with small-to-intermediate sized datasets.

The hierarchical structure of deep learning models is thought to be crucial to their ability to represent complex, abstract features.  For example, consider the use of CNNs for image classification tasks. The analysis of CNNs suggests that the lower-levels of the neural networks learn elementary features, such as edge detectors, which are then combined into higher levels of the networks into more abstract, higher-level features (e.g.~the famous example of a neuron that ``learned to respond to cats'') \cite{le2013building}. More recently, it has been shown that CNNs can be thought of as performing tensor decompositions on the data similar to those commonly used in numerical methods in modern quantum condensed matter \cite{cohen2016expressive}.

One of the interesting consequences of this line of thinking is the idea that one can train a CNN on one large dataset and the features it learns should also be useful for other supervised tasks. This results in the ability to learn important and salient features directly from the data and then transfer this knowledge to a new task. Indeed, this ability to learn important, higher-level, coarse-grained features is reminiscent of ideas like the renormalization group (RG) in physics where the RG flows separate out relevant and irrelevant directions, and certain unsupervised deep learning architectures have a natural interpretation in terms of variational RG schemes \cite{mehta2014exact}.

\subsubsection{Neural networks can exploit large amounts of data} 
With the advent of smartphones and the internet, there has been an explosion in the amount of data being generated. This data-rich environment favors supervised learning methods that can fully exploit this rich data world. One important reason for the success of DNNs is that they are able to exploit the additional signal in large datasets for difficult supervised learning tasks. Fundamentally, modern DNNs are unique in that they contain millions of parameters, yet can still be trained on existing hardwares. The complexity of  DNNs  (in terms of parameters) combined with their simple architecture (layer-wise connections) hit a sweet spot between expressivity (ability to represent very complicated functions) and trainability (ability to learn millions of parameters).  

\begin{figure}[t!]
\includegraphics[width=1.0 \columnwidth]{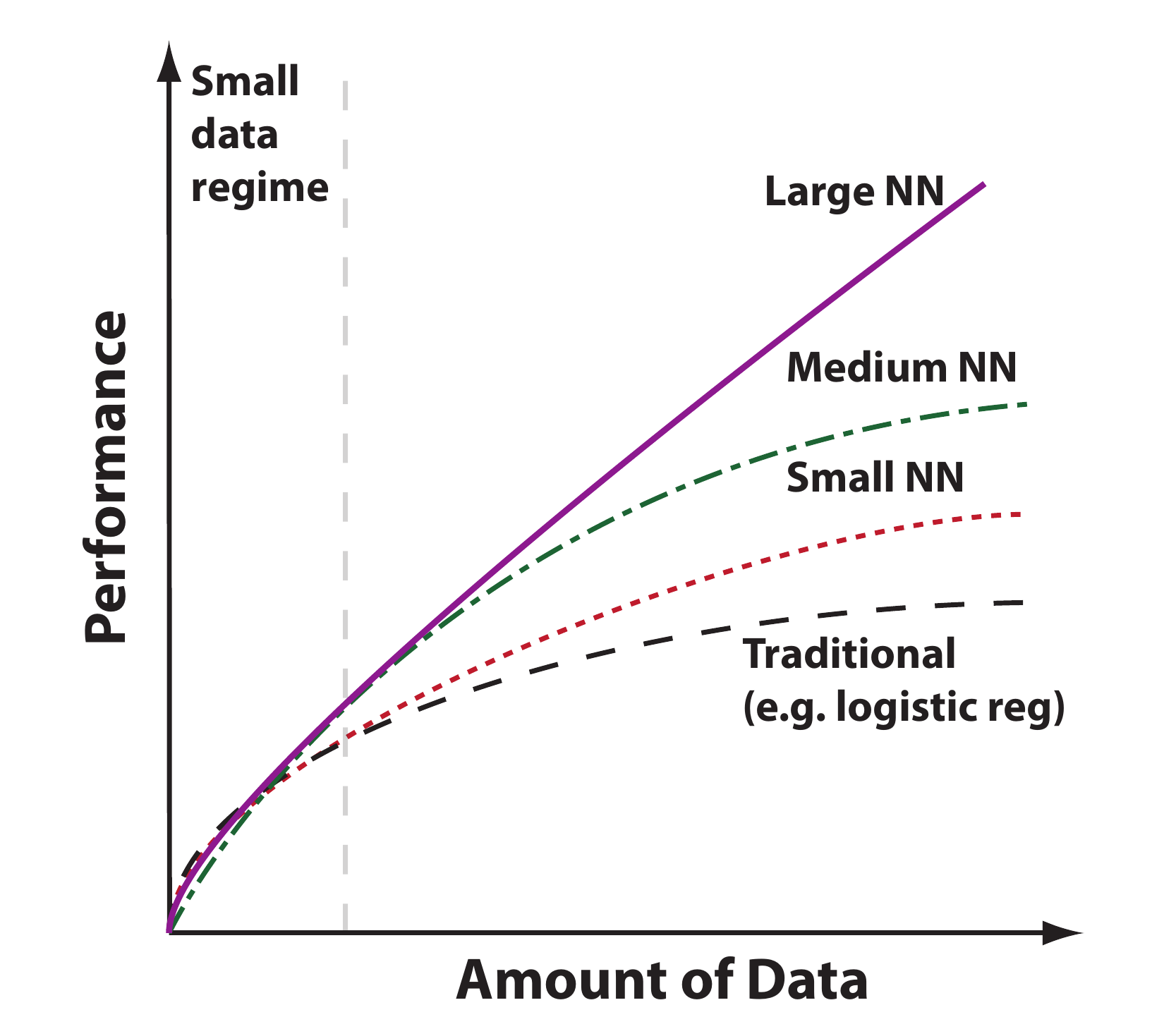}
\caption{{\bf Large neural networks can exploit the vast amount of data now available.} Schematic of how neural network performance depends on amount of available data (Figure based on Andrew Ng's talk at the 2016 Deep Learning School available
at \url{https://www.youtube.com/watch?v=F1ka6a13S9I}.)}
\label{Fig:NNdata}
\end{figure}

Indeed, the ability of large DNNs to exploit huge datasets is thought to differ from many other commonly employed supervised learning methods such as Support Vector Machines (SVMs). Figure \ref{Fig:NNdata} shows a schematic depicting the expected performance of DNNs of different sizes with the number of data samples and compares them to supervised learning algorithms such as SVMs or ensemble methods. When the amount of data is small, DNNs offer no substantial benefit over these other methods and often perform worse. However, large DNNs seem to be able to exploit additional data in a way other methods cannot. The fact that one does not have to hand engineer features makes the DNN even more well suited for handling large datasets. Recent theoretical results suggest that as long as a DNN is large enough, it should generalize well and not overfit \cite{advani2017high}. In the data-rich world we live in (at least in the context of images, videos, and natural language), this is a recipe for success. In other areas where data is more limited, deep learning architectures have (at least so far) been less successful.

\subsubsection{Neural networks scale up well computationally}

A final feature that is thought to underlie the success of modern neural networks is that they can harness the immense increase in computational capability that has occurred over the last few decades. The architecture of neural networks naturally lends itself to parallelization and the exploitation of fast but specialized processors such as graphical processing units (GPUs). Google and NVIDIA set on a course to develop TPUs (tensor processing units) which will be specifically designed for the mathematical operations underlying deep learning architectures. The layered architecture of neural networks also makes it easy to use modern techniques such as automatic differentiation that make it easy to quickly deploy them. Algorithms such as stochastic gradient descent and the use of mini-batches make it easy to parallelize code and train much larger DNNs than was thought possible fifteen years ago. Furthermore, many of these computational gains are quickly incorporated into modern packages with industrial resources. This makes it easy to perform numerical experiments on large datasets, leading to further engineering gains.

\subsection{Limitations of supervised learning with deep networks}

Like all statistical methods, supervised learning using neural networks has important limitations. This is especially important when one seeks to apply these methods, especially to physics problems. Like all tools, DNNs are not a universal solution. Often, the same or better performance on a task can  be achieved by using a few hand-engineered features (or even a collection of random features). This is especially important for hard physics problems where data (or Monte-Carlo samples) may be hard to come by.

Here we list some of the important limitations of supervised neural network based models. \begin{itemize}
\item{\bf Need labeled data.---}Like all supervised learning methods, DNNs for supervised learning require labeled data. Often, labeled data is harder to acquire than unlabeled data (e.g. one must pay for human experts to label images).
\item{\bf Supervised neural networks are extremely data intensive.---}DNNs are data hungry. They perform best when data is plentiful. This is doubly so for supervised methods where the data must also be labeled. The utility of DNNs is extremely limited if data is hard to acquire or the datasets are small (hundreds to a few thousand samples). In this case, the performance of other methods that utilize hand-engineered features can exceed that of DNNs.
\item{\bf Homogeneous data.---}Almost all DNNs deal with homogeneous data of one type. It is very hard to design architectures that mix and match data types (i.e.~some continuous variables, some discrete variables, some time series). In applications beyond images, video, and language, this is often what is required. In contrast, ensemble models like random forests or gradient-boosted trees have no difficulty handling mixed data types.
\item{\bf Many physics problems are not about prediction.---}In physics, we are often not interested in solving prediction tasks such as classification. Instead, we want to learn something about the underlying distribution that generates the data. In this case, it is often difficult to cast these ideas in a supervised learning setting. While the problems are related, it's possible to make good predictions with a ``wrong'' model. The model might or might not be useful for understanding the physics.
\end{itemize}

Some of these remarks are particular to DNNs, others are shared by all supervised learning methods. This motivates the use of unsupervised methods which in part circumnavigate these problems.

%% file: sections/dim_reduce.tex
Unsupervised learning is concerned with discovering structure in unlabeled data. In this section, we will begin our foray into unsupervised learning by way of data visualization. Data visualization methods are important for modeling as they can be used to identify correlated or redundant features along with irrelevant features (noise) from raw or processed data.  Conceivably, being able to identify and capture such characteristics in a dataset can help in designing better predictive models.  For data involving a relatively small number of features, studying pair-wise correlations (i.e.~pairwise scatter plots of all features) may suffice in performing a complete analysis. This rapidly becomes impractical for datasets involving a large number of measured featured (such as images). Thus, in practice, we often have to perform \emph{dimensional reduction}, namely, project or embed the data onto a lower dimensional space, which we refer to as the \emph{latent space}. As we will discuss, part of the complication of dimensional reduction lies in the fact that low-dimensional representations of high-dimensional data necessarily incurs information lost. Below, we introduce some common linear and nonlinear methods for performing dimensional reduction with applications in data visualization of high-dimensional data.

\subsection{Some of the challenges of high-dimensional data}
Before we begin exploring some specific dimensional reduction techniques, it is useful to highlight some of the generic difficulties encountered when dealing with high-dimensional data.

\paragraph{High-dimensional data lives near the edge of sample space.}
Geometry in high-dimensional space can be counterintuitive. One example that is pertinent to machine learning is the following. Consider data distributed uniformly at random in a $D$-dimensional hypercube $\mathcal{C}=[-e/2,e/2]^D$, where $e$ is the edge length. Consider also a $D$-dimensional hypersphere $\mathcal{S}$ of radius $e/2$ centered at the origin and contained within $\mathcal{C}$. The probability that a data point $\boldsymbol{x}$ drawn uniformly at random in $\mathcal{C}$ is contained within $\mathcal{S}$ is well approximated by the ratio of the volume of $\mathcal{S}$ to that of $\mathcal{C}$ : $p(\left\lVert\boldsymbol{x}\right\lVert_2< e/2) \sim (1/2)^D$. Thus, as the dimension of the feature space $D$ increases, $p$ goes to zero exponentially fast. In other words, most of the data will concentrate outside the hypersphere, in the corners of the hypercube. In physics, this basic observation underlies many properties of ideal gases such as the Maxwell distribution and the equipartition theorem (see Chapter 3 of \cite{sethna2006statistical} for instance). 

\paragraph{Real-world data vs.~uniform distribution.}
Fortunately, real-world data is not random or uniformly distributed! In fact, real data usually lives in a much lower dimensional space than the original space in which the features are being measured. This is sometimes referred to as the ``blessing of non-uniformity'' (in opposition to the curse of dimensionality). Data will typically be locally smooth, meaning that a local variation of the data will not incur a change in the target variable~\cite{bishop2006pattern}. This idea is central to statistical physics and field theories, where properties of systems with an astronomical number of degrees of freedom can be well characterized by low-dimensional ``order parameters'' or effective degrees of freedom. Another instantiation of this idea is manifest in the description of the bulk properties of a gas of weakly interacting particles, which can be simply described by the thermodynamic variables (temperature, pressure, etc.) that enter the equation of state rather than the enormous number of dynamical variables (i.e.~position and momentum) of each particle in the gas. 

\begin{figure}[t!]
\centering
\includegraphics[scale=0.35]{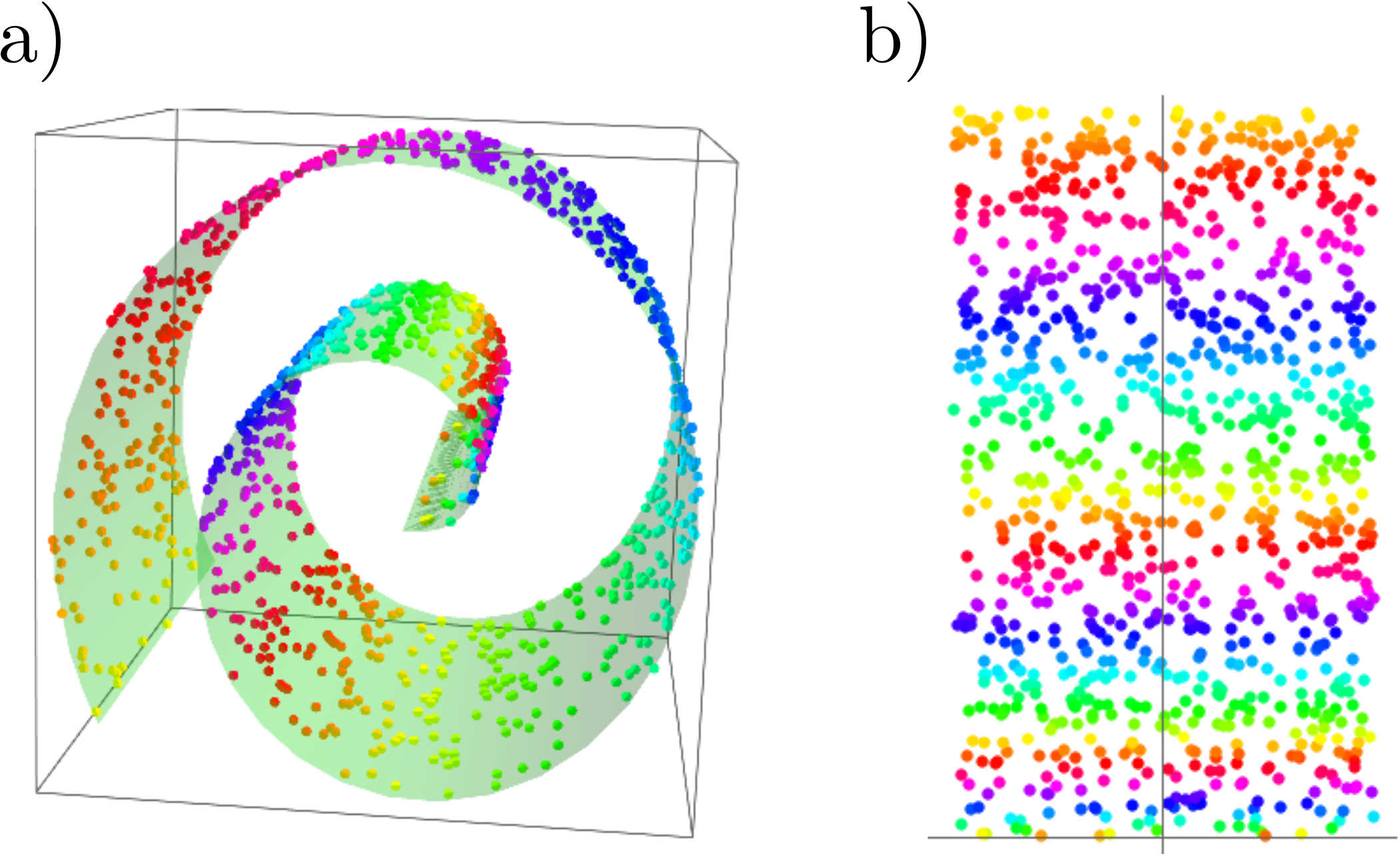}
\caption{The ``Swiss roll''. Data distributed in a three-dimensional space (a) that can effectively be described on a two-dimensional surface (b). A common goal of dimensional reduction techniques is to preserve ordination in the data: points that are close-by in the original space are also near-by in the mapped (latent) space. This is true of the mapping (a) to (b) as can be seen by inspecting the color gradient.
\label{spiral-manifold}
}
\end{figure}
\paragraph{Intrinsic dimensionality and the crowding problem.}
A recurrent objective of dimensional reduction techniques is to preserve the relative pairwise distances (or defined similarities) between data points from the original space to the latent space. This is a natural requirement, since we would like for nearby data points (as measured in the original space) to remain close-by after the corresponding mapping to the latent space.

\begin{figure}[t!]
\centering
\includegraphics[scale=0.85]{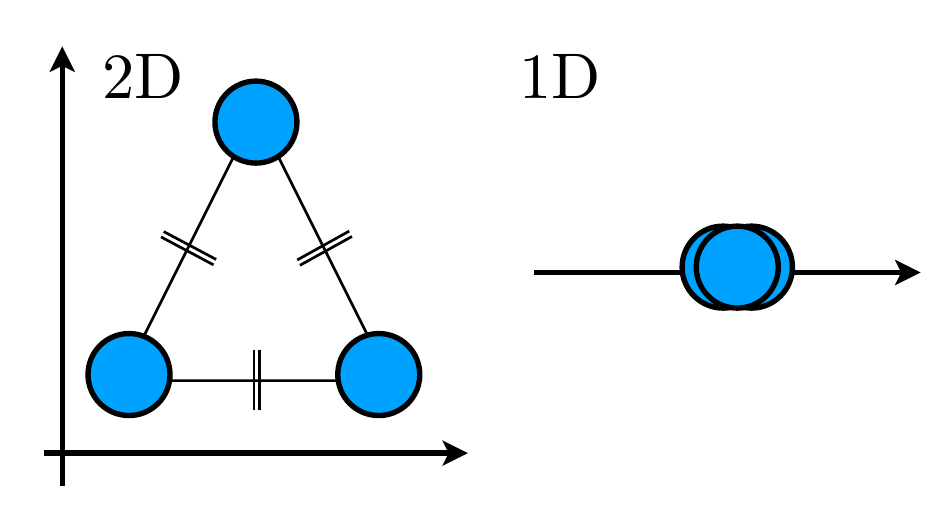}
\caption{Illustration of the crowding problem. (Left) A two-dimensional dataset $\boldsymbol{X}$ consisting of 3 equidistant points. (Right) Mapping $\boldsymbol{X}$ to a one-dimensional space while trying to preserve relative distances leads to a collapse of the mapped data points. 
\label{crowding}
}
\end{figure}

Consider the example of the ``Swiss roll'' presented in FIG. \ref{spiral-manifold}a. There, the relevant structure of the data corresponds to nearby points with similar colors and is encoded in the ``unrolled'' data in the latent space, see FIG. \ref{spiral-manifold}b. Clearly, in this example a two-dimensional space is sufficient to capture almost the entirety of the information in the data. 
A concept which stems from signal processing that is relevant to our current exposition is that of the intrinsic dimensionality of the data. Qualitatively, it refers to the minimum number of dimensions required to capture the signal in the data. In the case of the Swiss roll, it is 2 since the Swiss roll can effectively be parametrized using only two parameters, i.e.  $\boldsymbol{X}\in \{(x_1\sin(x_1), x_1\cos(x_1), x_2)\}$. 
The minimum number of parameters required for such a parametrization is the \emph{intrinsic} dimensionality of the data \cite{bennett1969intrinsic}. Attempting to represent data in a space of dimensionality lower than its intrinsic dimensionality can lead to a ``crowding'' problem \cite{maaten2008visualizing} (see schematic, FIG. \ref{crowding}). In short, because we are attempting to satisfy too many constraints (e.g.~preserve all relative distances of the original space), this results in a trivial solution for the latent space where all mapped data points collapse to the center of the map.

To alleviate this, one needs to weaken the constraints imposed on the visualization scheme. Powerful methods such as t-distributed stochastic embedding \cite{maaten2008visualizing} (in short, t-SNE, see section \ref{dimreduce:tsne}) and uniform manifold approximation and projection (UMAP) \cite{umap2018} have been devised to circumvent this issue in various ways.

\subsection{Principal component analysis (PCA)}
\label{subsec:PCA}

\begin{figure}[h]
\centering
\includegraphics[scale=0.5]{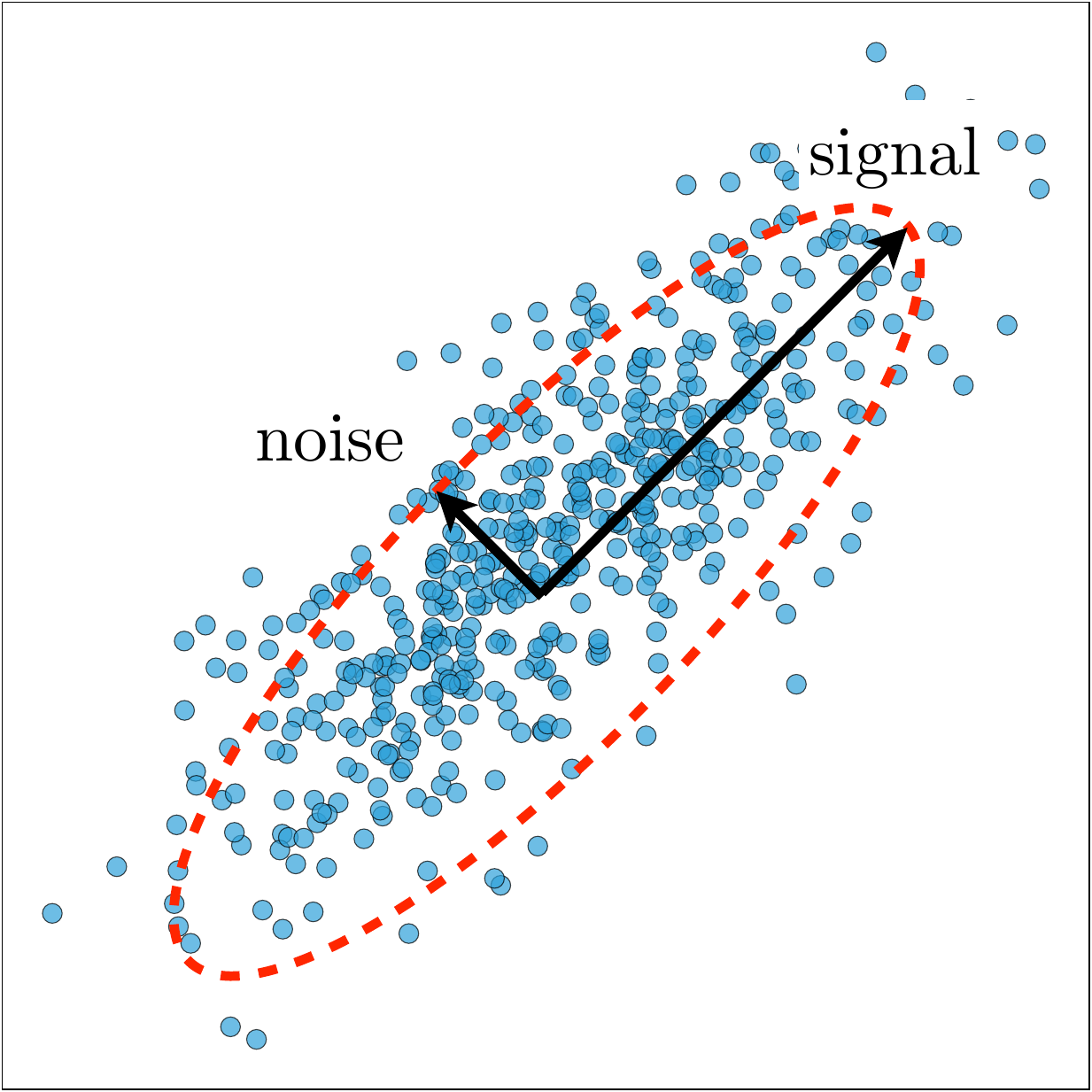}\\
\caption{PCA seeks to find the set of orthogonal directions with largest variance. This can be seen as ``fitting'' an ellipse to the data with the major axis corresponding to the first principal component (direction of largest variance). PCA assumes that directions with large variance correspond to the true signal in the data while directions with low variance correspond to noise.
\label{PCA:Ellipse}
}
\end{figure}

A ubiquitous method for dimensional reduction, data visualization and analysis is Principal Component Analysis (PCA). The goal of PCA is to perform an orthogonal transformation of the data in order to find high-variance directions. PCA is inspired by the observation that in many cases, the relevant information in a signal is contained in the directions with largest\footnote{This assumes that the features are measured and compared using the same units.} variance (see FIG. \ref{PCA:Ellipse}).  Directions with small variance are ascribed to ``noise'' and can potentially be removed or ignored.

\begin{figure}[h]
\centering
\includegraphics[scale=0.5]{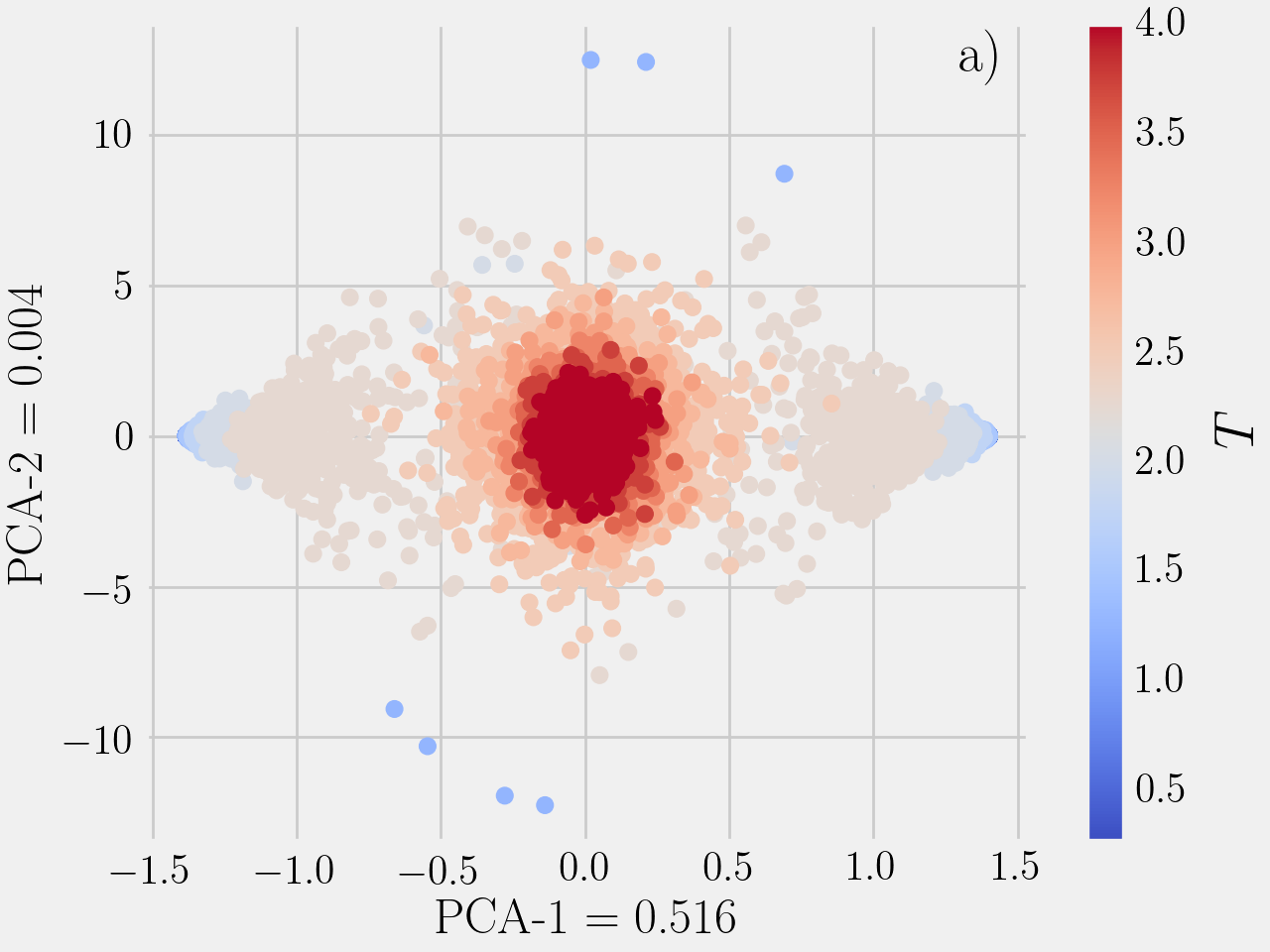}\\
\includegraphics[scale=0.5]{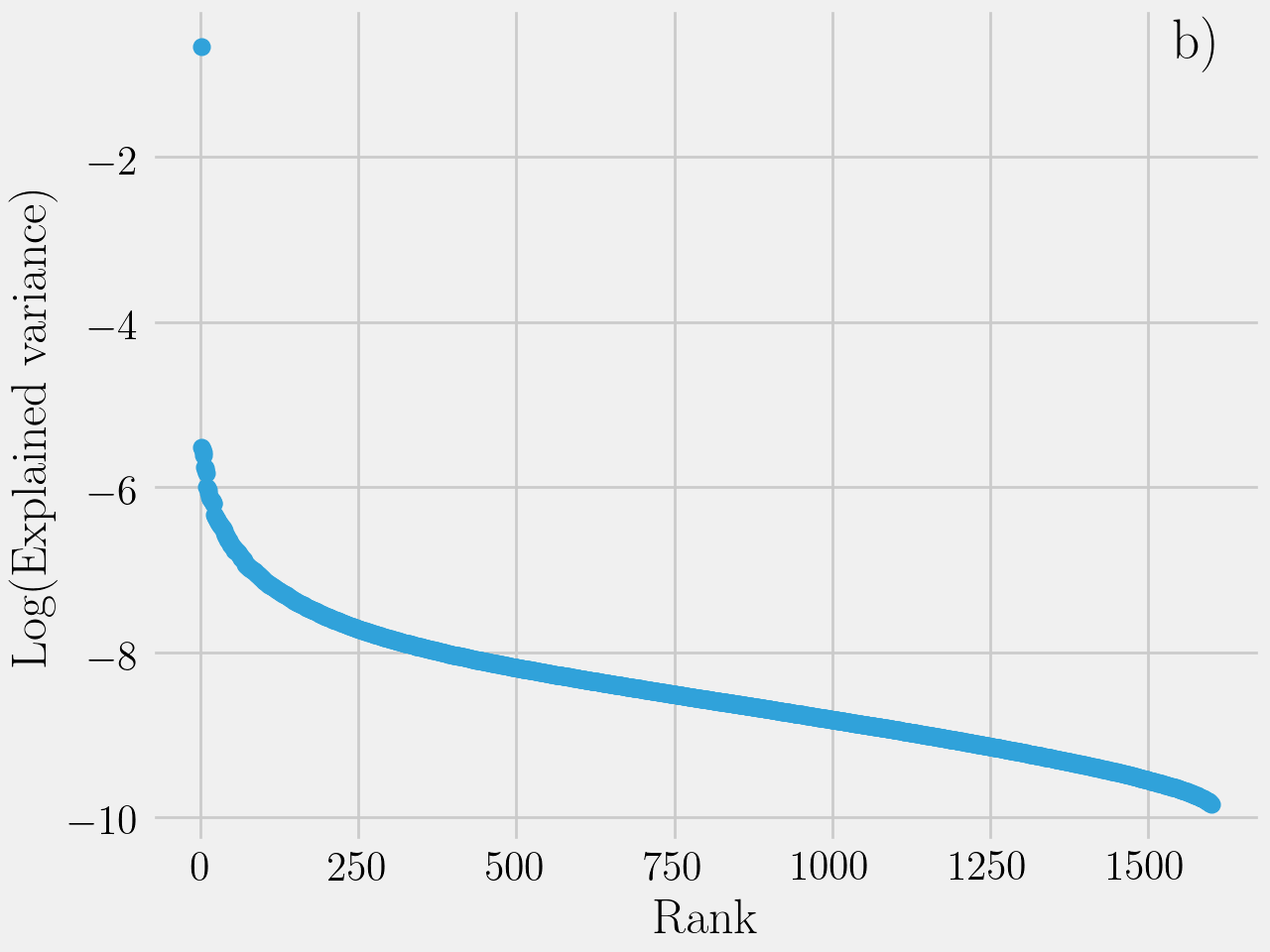}
\caption{(a) The first 2 principal component of the Ising dataset with temperature indicated by the coloring. PCA was performed on a joined dataset of 1000 samples taken at each temperatures $T=0.25,0.5,\cdots,4.0$. Almost all the variance is explained in the first component which corresponds to the magnetization order parameter (linear combination of the features with weights all roughly equal). The paramagnetic phase corresponds to the middle cluster and the left and right clusters correspond to the symmetry-related ferromagnetic phases (b) Log of the spectrum of the covariance matrix versus rank ordering. Only one dimension has high-variance. \label{PCA:Ising}
}
\end{figure}

Surprisingly, such PCA-based projections often capture a lot of the large scale structure of many datasets. For example, Figure \ref{PCA:Ising} shows the projection of samples drawn from the 2D Ising model at various temperatures on the first two principal components. Despite living in a $1600$ dimensional space (the samples are $40 \times 40$ spin configurations),  a single principal component (i.e.~a single direction in this 1600 dimensional space) can capture 50\% of the variability contained in our samples. In fact, one can verify that this direction weights all $1600$ spins nearly equally and thus corresponds to the magnetization order parameter. Thus, even without any prior physical knowledge, one can extract relevant order parameters using a simple PCA-based projection. Recently, a correspondence between PCA and Renormalization Group flows across the phase transition in the 2D Ising model~\cite{foreman2017rg} and in a more general setting~\cite{bradde2017pca} has been proposed. In statistical physics, PCA has also found application in detecting phase transitions~\cite{wetzel2017unsupervised}, e.g.~in the XY model on frustrated triangular and union jack lattices~\cite{wang2017unsupervised}. PCA was also used to classify dislocation patterns in crystals~\cite{papanikolaou2017learning,wang2018machine}, and to find correlations in the shear flow of athermal amorphous solids~\cite{ruscher2018correlations}. PCA is widely employed in biological physics when working with high-dimensional data. Physics has also inspired PCA-based algorithms to infer relevant features in unlabelled data~\cite{beny2018inferring}.
Concretely, consider $N$ data points, $\{ \boldsymbol{x}_1, \ldots \boldsymbol{x}_N\}$ that live in a $p$-dimensional feature space $\mathbb{R}^p$. Without loss of generality, we assume that the empirical mean $\bar{\boldsymbol{x}}=N^{-1} \sum_i \boldsymbol{x}_i$ of these data points is zero\footnote{We can always center around the mean: $\bar{\boldsymbol{x}}\leftarrow \boldsymbol{x}_i -\bar{\boldsymbol{x}}$}. Denote the $N \times p$ design matrix as $\boldsymbol{X}= [\boldsymbol{x}_1, \boldsymbol{x}_2, \ldots; \boldsymbol{x}_N]^T$ whose rows are the data points and columns correspond to different features. The $p \times p$ (symmetric) covariance matrix is therefore 
\be\label{eq:pca-cov}
\boldsymbol{\Sigma}(\bd{X}) ={1 \over N-1} \boldsymbol{X}^T\boldsymbol{X}.
\ee
Notice that the $j$-th diagonal entry of $\boldsymbol{\Sigma}(\bd{X})$ corresponds to the variance of the $j$-th feature and $\boldsymbol{\Sigma}(\bd{X})_{ij}$ measures the covariance (i.e. connected correlation in the language of physics) between feature $i$ and feature $j$.

We are interested in finding  a new basis for the data that emphasizes highly variable directions while reducing redundancy between basis vectors. In particular, we will look for a linear transformation  that reduces the covariance between different features. To do so, we first perform singular value decomposition (SVD) on the design matrix $\bd{X}$, namely, $\bd{X}=\bd{US}\bd{V}^T$, where $\bd{S}$ is a diagonal matrix of singular value $s_i$, the orthogonal matrix $\bd{U}$ contains (as its columns) the left singular vectors of $\bd{X}$, and similarly $\bd{V}$ contains (as its columns) the right singular vectors of $\bd{X}$. With this, one can rewrite the covariance matrix as
\bea
\bd{\Sigma}(\bd{X})  &=& \frac{1}{N-1}\bd{VS}\bd{U}^T \bd{US}\bd{V}^T\nonumber \\
&=& \bd{V}\left(\frac{\bd{S}^2}{N-1}\right)\bd{V}^T\nonumber\\
&\equiv&\bd{V}\bd{\Lambda}\bd{V}^T.
\eea
where $\boldsymbol{\Lambda}$ is a diagonal matrix with eigenvalues $\lambda_i$ in the decreasing order along the diagonal (i.e.~eigendecomposition). It is clear that the right singular vectors of $\bd{X}$ (i.e.~the columns of $\bd{V}$) are principal directions of $\boldsymbol{\Sigma}(\bd{X}) $, and the singular values of $\bd{X}$ are related to the eigenvalues of the covariance matrix $\boldsymbol{\Sigma}(\bd{X})$ via $\lambda_i=s_i^2/(N-1)$. To reduce the dimensionality of data from $p$ to $\tilde{p}<p$, we first construct the $p \times \tilde{p}$ projection matrix $\tilde{\boldsymbol{V}}_{p'}$ by selecting the singular components with the $\tilde{p}$ largest singular values. The projection of the data from $p$ to a $\tilde{p}$ dimensional space is simply $\tilde{\boldsymbol{Y}}= \boldsymbol{X}\tilde{\boldsymbol{V}}_{p'}$. The same idea is central to matrix-product-state-like techniques used to compress the number of components in quantum wavefunctions in studies of low-dimensional many-body lattice systems.

The singular vector with the largest singular value (i.e the largest variance) is referred to as the first principal component; the singular vector with the second largest singular value as the second principal component, and so on. An important quantity is the ratio $\lambda_i/\sum_{i=1}^{p} \lambda_i$ which is referred as the percentage of the explained variance contained in a principal component (see FIG. \ref{PCA:Ising}.b).

It is common in data visualization to present the data projected on the first few principal components.  This is valid as long as a large part of the variance is explained in those components. Low values of explained variance may imply that the intrinsic dimensionality of the data is high or simply that it cannot be captured by a linear representation. For a detailed introduction to PCA, see the tutorials by Shlens \cite{shlens2014tutorial} and Bishop~\cite{bishop2006pattern}.

\subsection{Multidimensional scaling}

Multidimensional scaling (MDS) is a non-linear dimensional reduction technique which preserves the pairwise distance or dissimilarity $d_{ij}$ between data points \cite{cox2000multidimensional}. Moving forward, we use the term ``distance'' and ``dissimilarity'' interchangeably. There are two types of MDS: metric and non-metric. In metric MDS, the distance is computed under a pre-defined metric and the latent coordinates $\boldsymbol{\tilde Y}$ are obtained by minimizing the difference between the distance measured in the original space ($d_{ij}(\boldsymbol{X})$) and that in the latent space ($d_{ij}(\boldsymbol{Y})$): 
\begin{equation}
\boldsymbol{\tilde Y} = \arg\min_{\boldsymbol{Y}} \sum_{i<j}w_{ij} |d_{ij}(\boldsymbol{X})-d_{ij}(\boldsymbol{Y})|,
\end{equation}
where $w_{ij}\geq 0$ are weight values. The weight matrix $w_{ij}$ is a set of free parameters that specify the level of confidence (or precision) in the value of $d_{ij}(\boldsymbol{X})$. 
If Euclidean metric is used, MDS gives the same result as PCA and is usually referred to as classical scaling \cite{torgerson1958theory}. Thus MDS is often considered as a generalization of PCA. 
In non-metric MDS, $d_{ij}$ can be any distance matrix. The objective function is then to preserve the ordination in the data, i.e.~if $d_{12}(\boldsymbol{X}) < d_{13}(\boldsymbol{X})$ in the original space, then in the latent space we should have $d_{12}(\boldsymbol{Y}) < d_{13}(\boldsymbol{Y})$. 

Both MDS and PCA can be implemented using standard Python packages such as~\href{http://scikit-learn.org/stable/modules/generated/sklearn.manifold.MDS.html}{Scikit}. MDS algorithms typically have a scaling of $\mathcal{O}(N^3)$ where $N$ corresponds to the number of data points, and are thus very limited in their application to large datasets. However, sample-based methods have been introduce to reduce this scaling to $\mathcal{O}(N\log N)$ \cite{yang2006fast}.  In the case of PCA, a complete decomposition has a scaling of $\mathcal{O}(Np^2+p^3)$, where $p$ is the number of features. Note that the first term $Np^2$ is due to the computation of covariance matrix Eq.\eqref{eq:pca-cov} while the second, $p^3$, stems from eigenvalue decomposition. Nothe that PCA can be improved to bear complexity $\mathcal{O}(Np^2+p)$ if only the first few principal components are desired (using iterative approaches). PCA and MDS are often among the first data visualization techniques one resorts to.

\subsection{t-SNE \label{dimreduce:tsne}}

It is often desirable to preserve local structures in high-dimensional datasets. However, when dealing with datasets having clusters delimitated by complicated surfaces or datasets with a large number of clusters, preserving local structures becomes difficult using linear techniques such as PCA. Many non-linear techniques such as non-classical MDS \cite{cox2000multidimensional}, self-organizing map \cite{kohonen1998self}, Isomap \cite{tenenbaum2000global} and Locally Linear Embedding \cite{roweis2000nonlinear} have been proposed and to address this class of problems. These techniques are generally good at preserving local structures in the data but typically fail to capture structures at the larger scale such as the clusters in which the data is organized \cite{maaten2008visualizing}. 

Recently, $t$-stochastic neighbor embedding (t-SNE) has emerged as one of the go-to methods for visualizing high-dimensional data. It has been shown to offer insightful visualization for many benchmark high-dimensional datasets \cite{maaten2008visualizing}. t-SNE is a non-parametric\footnote{It does not explicitly parametrize feature extraction required to compute the embedding coordinates. Thus it cannot be applied to find the coordinate of new data points.} method that constructs non-linear embeddings. Each high-dimensional training point is mapped to low-dimensional embedding coordinates, which are optimized in a way to preserve the local structure in the data. 

When used appropriately, t-SNE is a powerful technique for unraveling the hidden structure of high-dimensional datasets while at the same time preserving locality. In physics, t-SNE has recently been used to reduce the dimensionality and classify spin configurations, generated with the help of Monte Carlo simulations, for the Ising \cite{carrasquilla2017machine} and Fermi-Hubbard models at finite temperatures~\cite{ch2017unsupervised}. It was also applied to study clustering transitions in glass-like problems in the context of quantum control \cite{2018arXiv180310856D}.

The idea of stochastic neighbor embedding is to associate a probability distribution to the neighborhood of each data (note $x\in \mathbb{R}^p$, $p$ is the number of features):
\begin{equation}
p_{i|j} = \frac{\exp(-||x_i-x_j||^2/2\sigma_i^2)}{\sum_{k\neq i}\exp(-||x_i-x_k||^2/2\sigma_i^2)},\label{tSNE:gaussian}
\end{equation}
where $p_{i|j}$ can be interpreted as the likelihood that $x_j$ is $x_i$'s neighbor (thus we take $p_{i|i}=0$). $\sigma_i$ are free bandwidth parameters that are usually determined by fixing the local entropy $H(p_i)$ of each data point:
\begin{equation}
H(p_i)\equiv-\sum_j p_{j|i}\log_2 p_{j|i}.
\end{equation}
The local entropy is then set to equal a constant across \emph{all data points} $\Sigma=2^{H(p_i)}$, where $\Sigma$ is called the \emph{perplexity}. The perplexity constraint determines $\sigma_i~\forall~i$ and implies that points in regions of high-density will have smaller $\sigma_i$. 

\begin{figure}[h]
	\centering
	\includegraphics[scale=0.7]{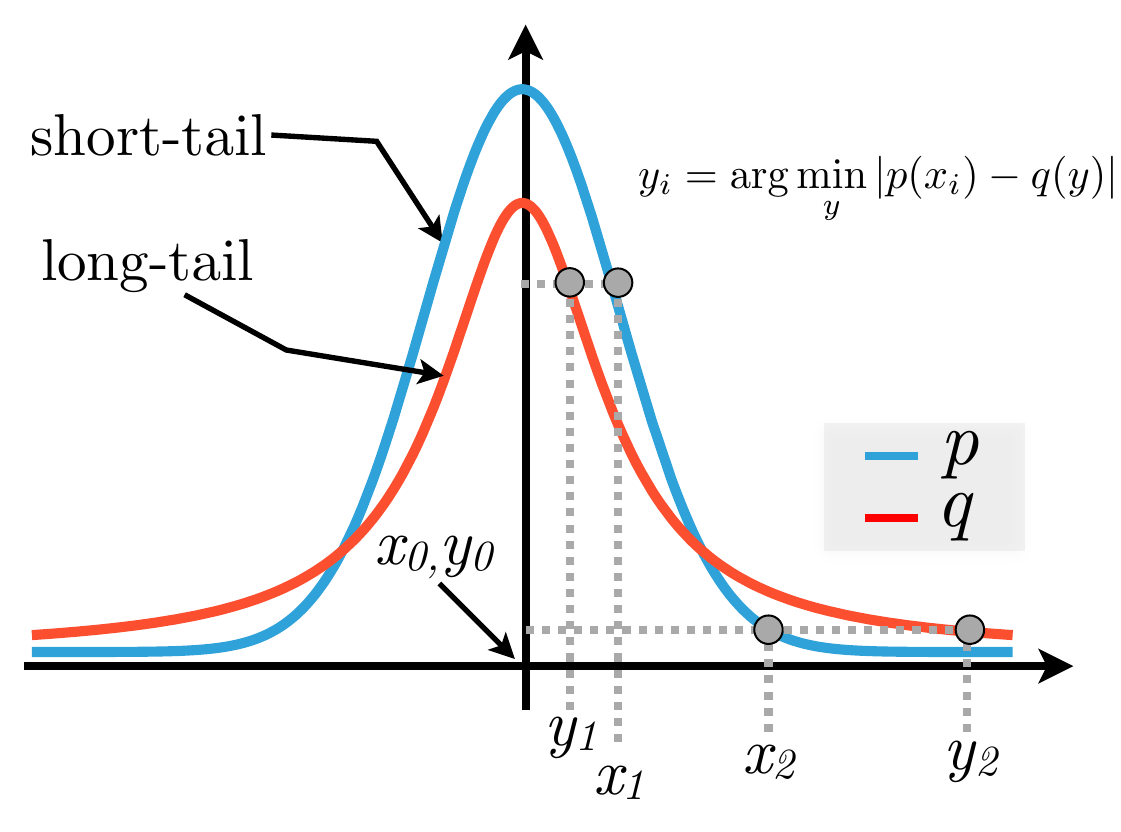}
	\caption{Illustration of the t-SNE embedding. $x_i$ points correspond to the original high-dimensional points while the $y_i$ points are the corresponding low-dimensional map points produced by t-SNE. Here we consider two points, $x_1$, $x_2$, that are respectively ``close'' and ``far'' from $x_0$. The high-dimensional Gaussian (short-tail) distribution $p(x)$ of $x_0$'s neighbors is shown in blue. The low-dimensional Cauchy (fat-tail) distribution $q(y)$ of $x_0$'s neighbors is shown in red. The map point $y_i$, are obtained by minimizing the difference $|q(y)-p(x_i)|$ (similar to minimizing the KL divergence). We see that point $x_1$ is mapped to short distances $|y_1-y_0|$. In contrast, far-away points such as $x_2$ are mapped to relatively large distances $|y_2-y_0|$. \label{t-SNE:cartoon}}
\end{figure}

Using Gaussian likelihoods in $p_{i|j}$ implies that only points that are nearby $x_i$ contribute to its probability distribution. 
While this ensures that the similarity for nearby points is well represented, this can be a problem for points that are far away from $x_i$ (i.e. outliers): they have exponentially vanishing contributions to the distribution, which in turn means that their embedding coordinates are ambiguous \cite{maaten2008visualizing}. One way around this is to define a symmetrized distribution $p_{ij} \equiv (p_{i|j}+p_{j|i})/(2N)$. This \emph{guarantees} that $\sum_j p_{ij} > 1/(2N)$ for all data points $x_i$, resulting in each data point $x_i$ making a significant contribution to the cost function to be defined below. 

t-SNE constructs a \emph{similar} probability distribution $q_{ij}$ in a low dimensional latent space (with coordinates $Y=\{y_i\},~y_i\in \mathbb{R}^{p'}$,  where $p' < p$  is the dimension of the latent space): 
\begin{equation}
q_{ij}=\frac{(1+||y_i-y_j||^2)^{-1}}{\sum_{k\neq i}(1+||y_i-y_k||^2)^{-1}}.
\end{equation}
The crucial point to note is that $q_{ij}$ is chosen to be a long tail distribution. This preserves short distance information (relative neighborhoods) while strongly repelling two points that are far apart in the original space (see FIG. \ref{t-SNE:cartoon}). In order to find the latent space coordinates $y_i$, t-SNE minimizes the Kullback-Leibler divergence between $q_{ij}$ and $p_{ij}$:
\begin{equation}
\mathcal{C}(Y)=D_{KL}(p||q)\equiv\sum_{ij} p_{ij} \log\left(\frac{p_{ij}}{q_{ij}}\right).
\label{KL:tsne}
\end{equation}
This minimization is done via gradient descent (see section \ref{sec:gradient_descent}). We can gain further insights on what the embedding cost-function $\mathcal{C}$ is capturing by computing the gradient of \eqref{KL:tsne} with respect to $y_i$ explicitly:
\begin{align}
  \partial_{y_i}\mathcal{C}&= \sum_{j\neq i} 4p_{ij}q_{ij}Z_i(y_i-y_j)-\sum_{j\neq i}4q_{ij}^2Z_i(y_i-y_j), \nonumber \\
                            &= F_{\mathrm{attractive},i}-F_{\mathrm{repulsive},i},
\label{gradient:tsne}
\end{align}
where $Z_i=1/(\sum_{k\neq i}(1+||y_k-y_i||^2)^{-1})$. We have separated the gradient of point $y_i$ into an attractive $F_{\mathrm{attractive}}$ and repulsive term $F_{\mathrm{repulsive}}$. Notice that $F_{\mathrm{attractive},i}$ induces a significant attractive force only between points that are nearby point $i$ in the \emph{original space} since it involves the $p_{ij}$ term. Finding the embedding coordinates $y_i$ is thus equivalent to finding the equilibrium configuration of particles interacting through the forces in \eqref{gradient:tsne}.

Below, we list some important properties that one should bear in mind when analyzing t-SNE plots.
\begin{itemize}
\item \emph{ t-SNE can rotate data}. The KL divergence is invariant under rotations in the latent space, since it only depends on the distance between points. For this reason, t-SNE plots that are rotations of each other should be considered equivalent.

\item \emph{t-SNE results are stochastic}.  In applying gradient descent the solution will depend on the initial seed. Thus, the map obtained may vary depending on the seed used and different t-SNE runs will give slightly different results.

\item \emph{t-SNE generally preserves short distance information}. As a rule of thumb, one should expect that nearby points on the t-SNE map are also closeby in the original space, i.e.~t-SNE tends to preserve ordination (but not actual distances). For a pictorial explanation of this, we refer the reader to Figure \ref{t-SNE:cartoon}.

\item \emph{Scales are deformed in t-SNE}. Since a scale-free distribution is used in the latent space, one should not put too much emphasis on the meaning of the size of any clusters observed in the latent space. 

\item \emph{t-SNE is computationally intensive}. Finally, a direct implementation of t-SNE has an algorithmic complexity of $\mathcal{O}(N^2)$ which is only applicable to small to medium data sets. Improved scaling of the form $\mathcal{O}(N\log N)$ can be achieved at the cost of approximating Eq.~\eqref{KL:tsne} by using the Barnes-Hut method \cite{van2014accelerating} for $N$-body simulations \cite{barnes1986hierarchical}. More recently extremely efficient t-SNE implementation making use of fast Fourier transforms for kernel summations in \eqref{gradient:tsne} have been made available on \url{https://github.com/KlugerLab/FIt-SNE} \cite{2017arXiv171209005L}.
\end{itemize}

\begin{figure}[h]
	\centering
	\includegraphics[scale=0.21]{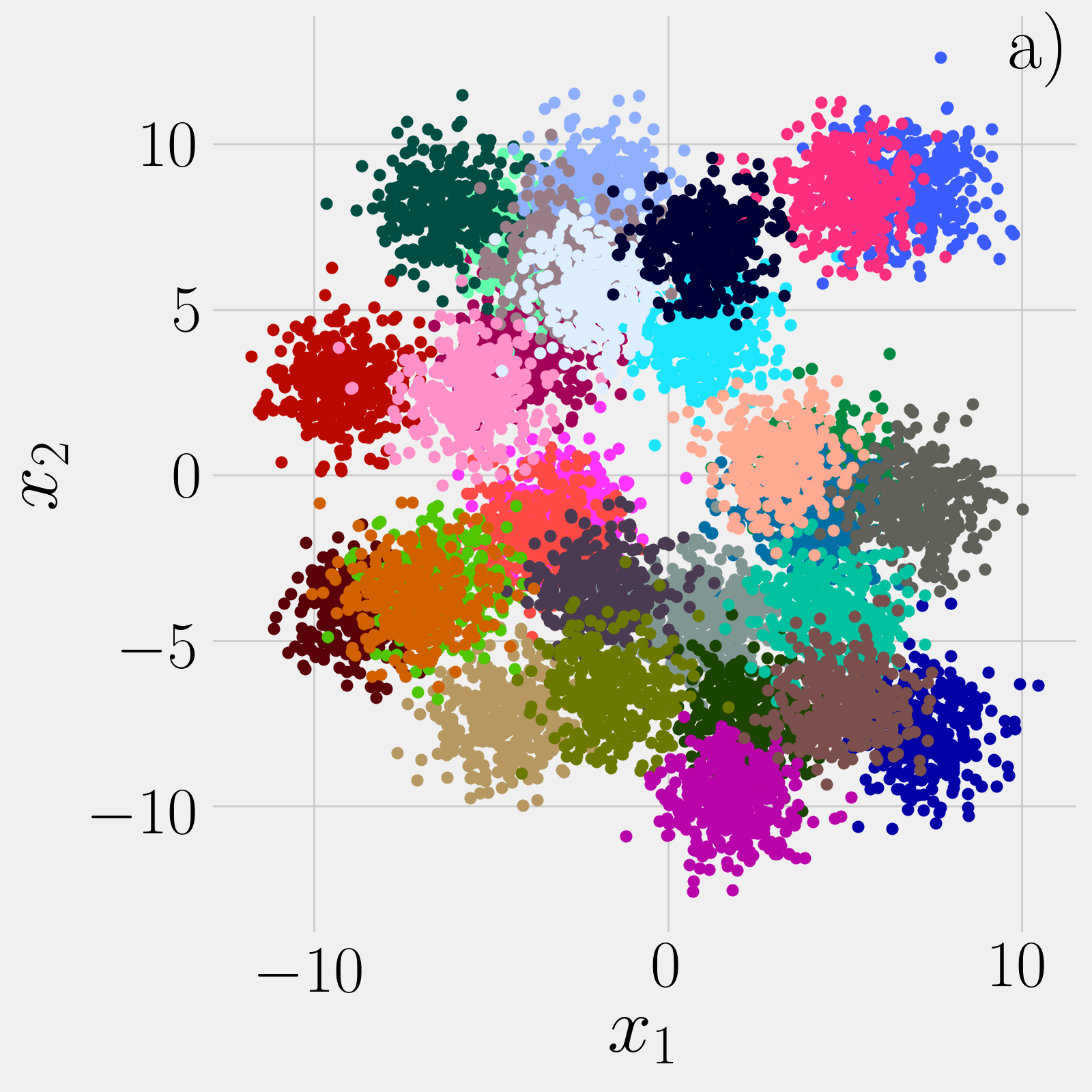}
	\includegraphics[scale=0.21]{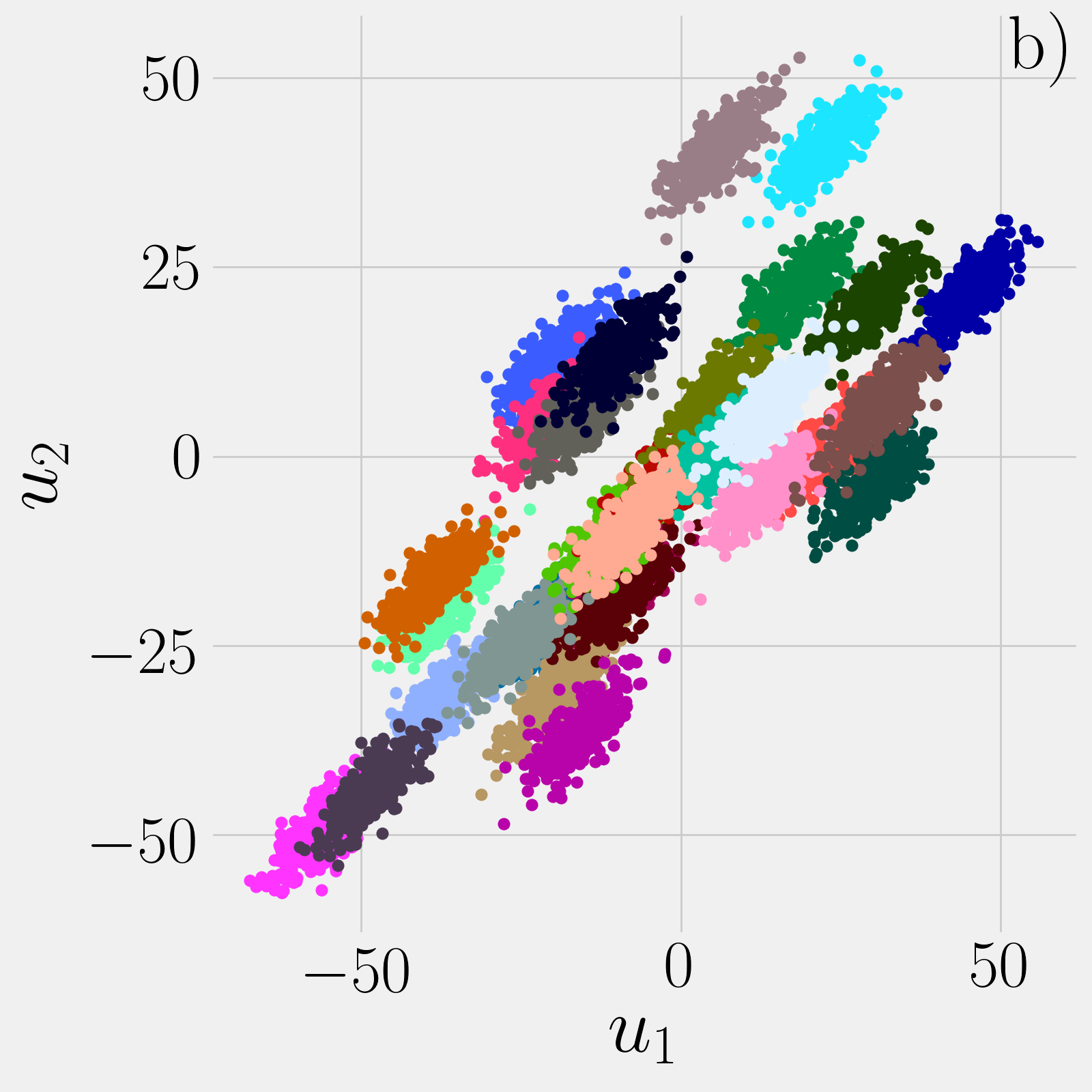}\\
	\includegraphics[scale=0.21]{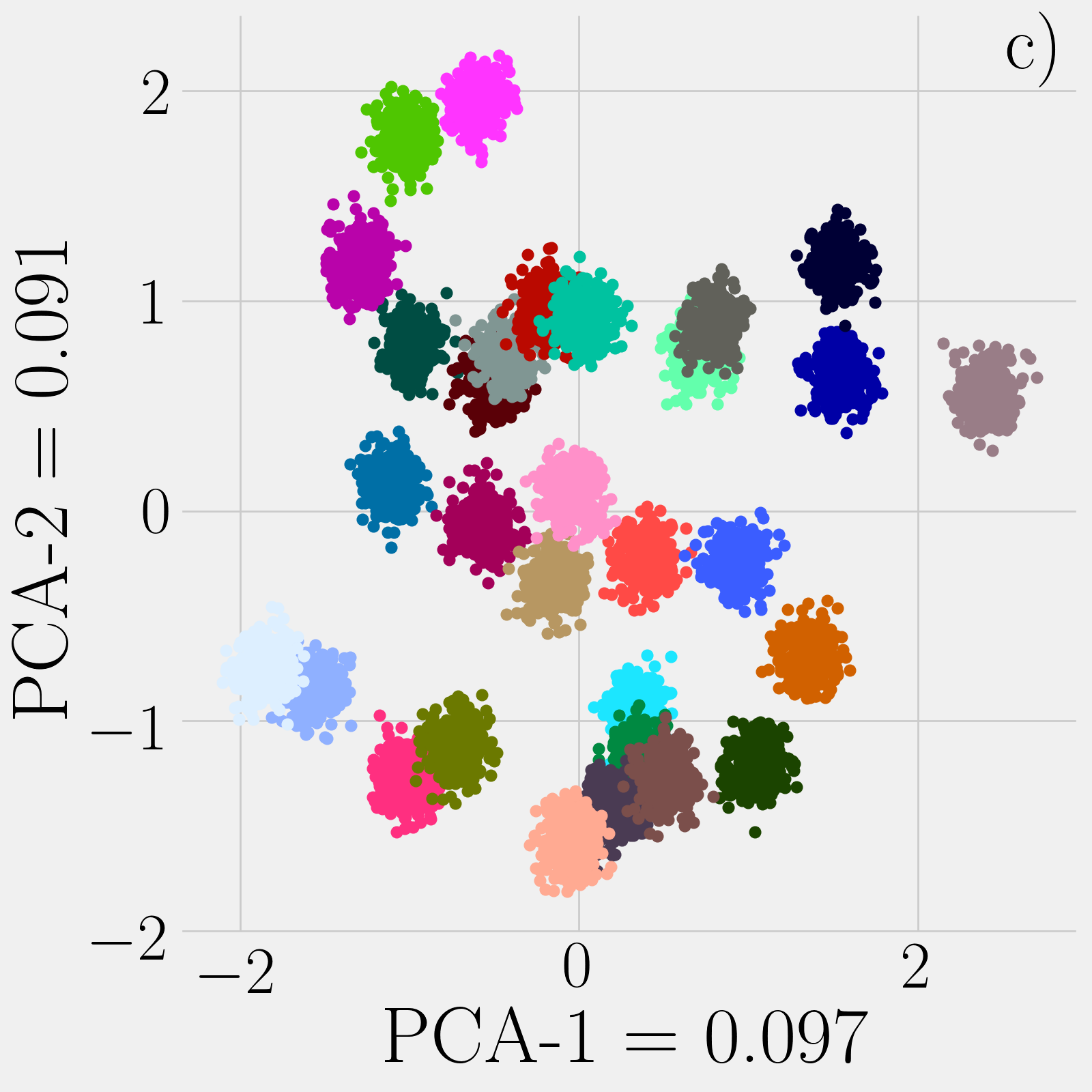}
	\includegraphics[scale=0.21]{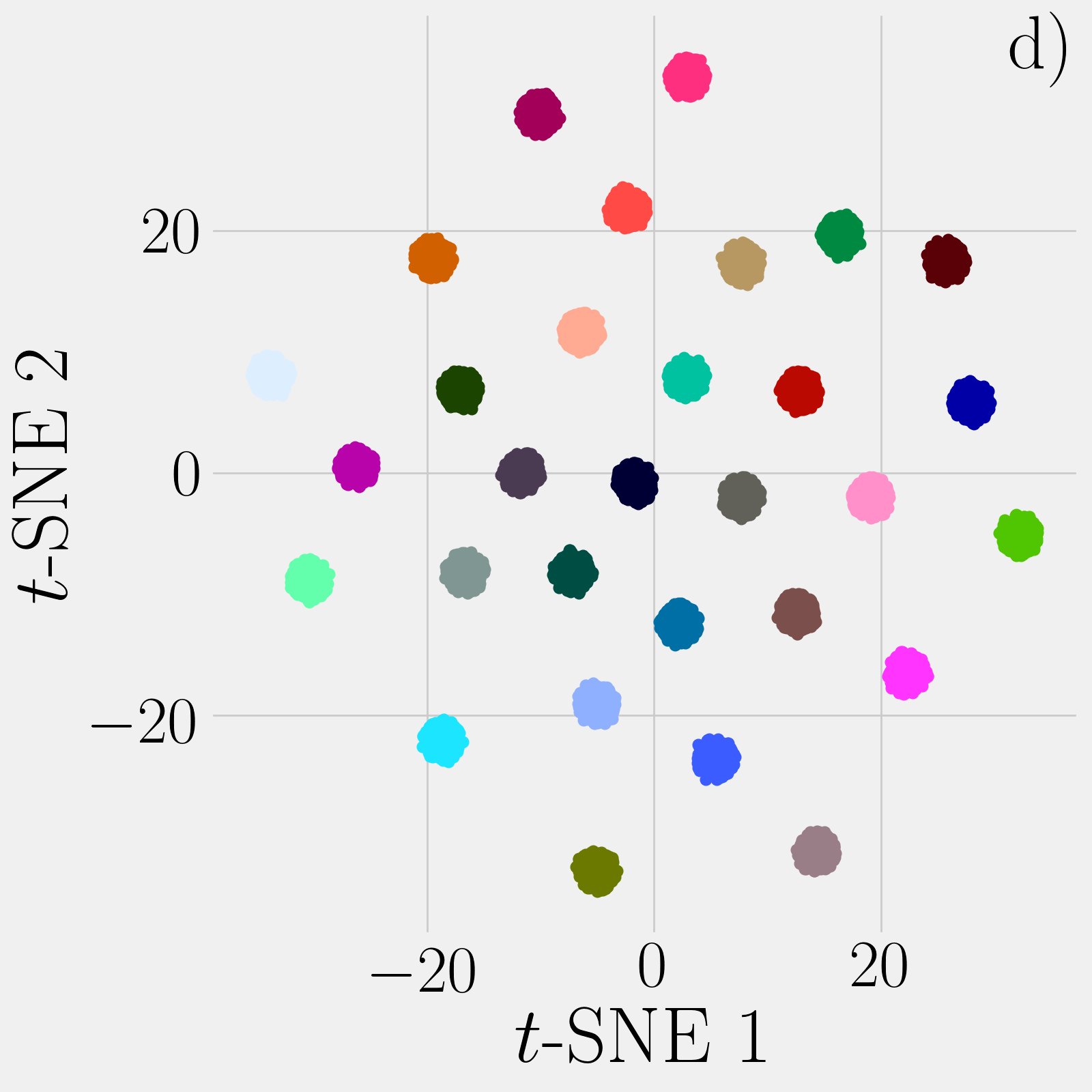}
	\caption{Different visualizations of a Gaussian mixture formed of $K=30$ mixtures in a $D=40$ dimensional space. The Gaussians have
		the same covariance but have means drawn uniformly at random in the space $[-10,10]^{40}$. (a) Plot of the first two coordinates. The labels of the
		different Gaussian is indicated by the different colors. Note that in a realistic setting, label information is of course not available, thus making it very hard to distinguish the different clusters. (b) Random projection of the data onto a 2 dimensional space. (c) 
		projection onto the first 2 principal components. Only a small fraction of the variance is explained by those components (the ratio is indicated along the axis). (d) t-SNE embedding (perplexity = 60, $\#$ iteration = 1000) in a 2 dimensional latent space. t-SNE captures correctly the local structure of the data. 
		\label{t-SNE:gaussian-mixture}}
\end{figure}

As an illustration, in Figure \ref{t-SNE:gaussian-mixture} we applied t-SNE to a Gaussian mixture model consisting of thirty Gaussians, whose means are uniformly distributed in forty-dimensional space. We compared the results to a random two-dimensional projection and PCA. It is clear that unlike more na{\"i}ve dimensional reduction techniques, both PCA and t-SNE can identify the presence of well-formed clusters. The t-SNE visualization cleanly separates all the clusters while certain clusters blend together in the PCA plot. This is a direct consequence of the fact that t-SNE keeps nearby points close together while repelling points that are far apart. 

\begin{figure}[h!]
\centering
\includegraphics[scale=0.21]{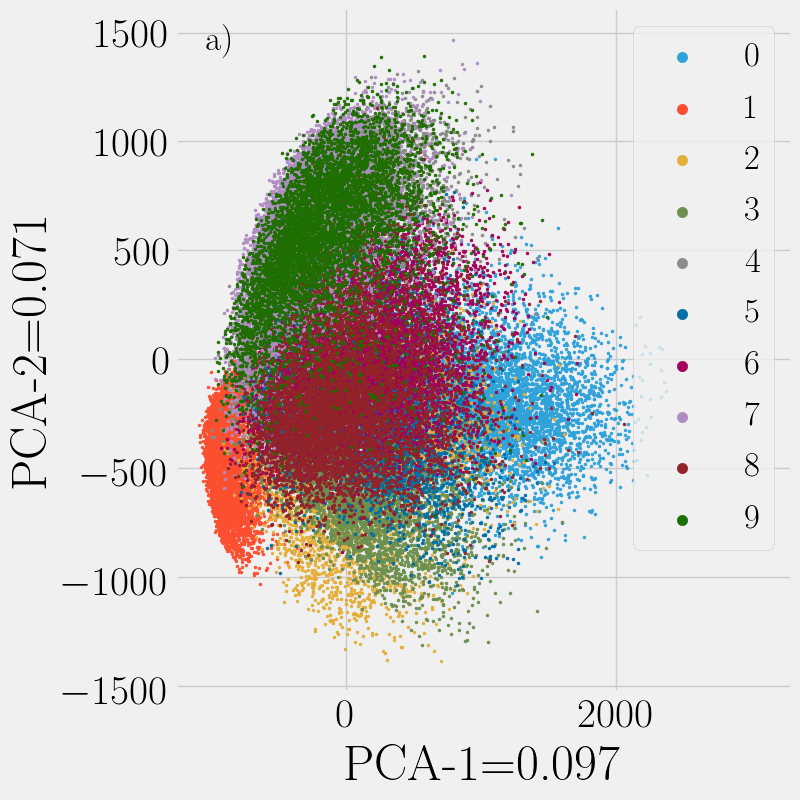}
\includegraphics[scale=0.21]{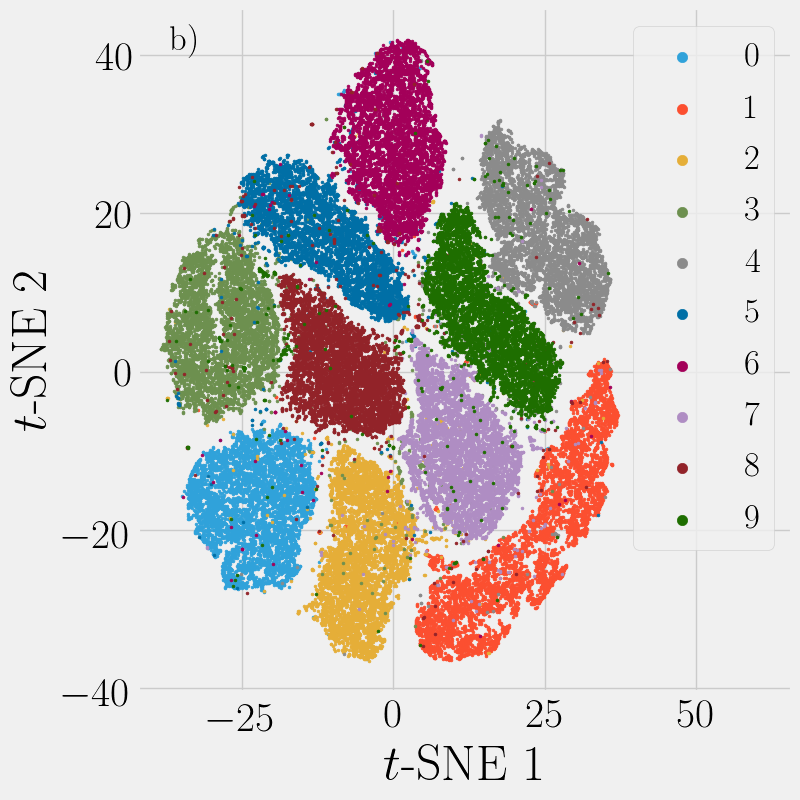}
\caption{Visualization of the MNIST handwritten digits training dataset (here $N=60000$). (a) First two principal components. (b) t-SNE applied with a perplexity of 30, a Barnes-Hut angle of 0.5 and 1000 gradient descent iterations. In order to reduce the noise and speed-up computation, PCA was first applied to the dataset to project it down to 40 dimensions. We used an open-source implementation to produce the results \cite{2017arXiv171209005L}, see \url{https://github.com/KlugerLab/FIt-SNE}.
  \label{dimreduce:PCAtSNE}
 }
\end{figure}

Figure~\ref{dimreduce:PCAtSNE} shows t-SNE and PCA plots for the MNIST dataset of ten handwritten numerical digits (0-9). It is clear that the non-linear nature of t-SNE makes it much better at capturing and visualizing the complicated correlations between digits, compared to PCA.

%% file: sections/Clustering.tex
In this section, we continue our discussion of unsupervised learning methods. Unsupervised learning is concerned with discovering structure in unlabeled data (for instance learning local structures for data visualization, see section \ref{sec:dim-red}). 
The lack of labels make unsupervised learning much more difficult and subtle than its supervised counterpart. What is somewhat surprising is that even without labels it is still possible to uncover and exploit the hidden structure in the data. 
Perhaps, the simplest example of unsupervised learning is clustering. The aim of clustering is to group unlabelled data into clusters according to some similarity or distance measure. Informally, a cluster is thought of as a set of points sharing some pattern or structure. 

Clustering finds many applications throughout data mining \cite{larsen1999fast}, data compression and signal processing \cite{gersho2012vector, mackay2003information}. Clustering can be used to identify coarse features or high level structures in an unlabelled dataset. The technique also finds many applications in physical sciences, ranging from detecting celestial emission sources in astronomical surveys \cite{sander1998density} to inferring groups of genes and proteins with similar functions in biology \cite{eisen1998cluster}, and building entanglement classifiers~\cite{lu2017separability}. Clustering is perhaps the simplest way to look for hidden structure in a dataset and for this reason, is among the most widely used and employed data analysis and machine learning techniques.

The field of clustering is vast and there exists a flurry of clustering methods suited for different purposes. Some common considerations one has to take into account when choosing a particular method is the distribution of the clusters (overlapping/noisy clusters vs.\@ well-separated clusters), the geometry of the data (flat vs.\@ non-flat), the cluster size distribution (multiple sizes vs.\@ uniform sizes), the dimensionality of the data (low vs.\@ high dimensional) and the computational efficiency of the desired method (small vs.\@ large dataset). 

We begin section \ref{clustering:1} with a focus on popular practical clustering methods such as $K$-means clustering, hierarchical clustering and density clustering. Our goal is to highlight the strength, weaknesses and differences between these techniques, while laying out some of the theoretical framework required for clustering analysis. There exist many more clustering methods beyond those discussed in this section\footnote{Our complementary  \href{https://physics.bu.edu/~pankajm/MLnotebooks.html}{Python notebook} introduces some of these other methods.}. The methods we discuss were chosen for their pedagogical value and/or their applicability to problems in physics. 
 
In section \ref{clustering:2} we discuss gaussian mixture models and the formulation of clustering through latent variable models. This section introduces many of the methods we will encounter when discussing other unsupervised learning methods later in the review. Finally, in section \ref{clustering:3} we discuss the problem of clustering in high-dimensional data and possible ways to tackle this difficult problem. The reader is also urged to experiment with various clustering methods using \href{https://physics.bu.edu/~pankajm/MLnotebooks.html}{Notebook 15}.

\subsection{Practical clustering methods\label{clustering:1}}

Throughout this section we focus on the Euclidean distance as a similarity measure. Other measures may be better suited for specific problems. We refer the enthusiast reader to \cite{rokach2005clustering} for a more in-depth discussion of the different possible similarity measures.

\subsubsection{$K$-means}
\begin{figure}[h!]
\centering
\includegraphics[scale=0.24]{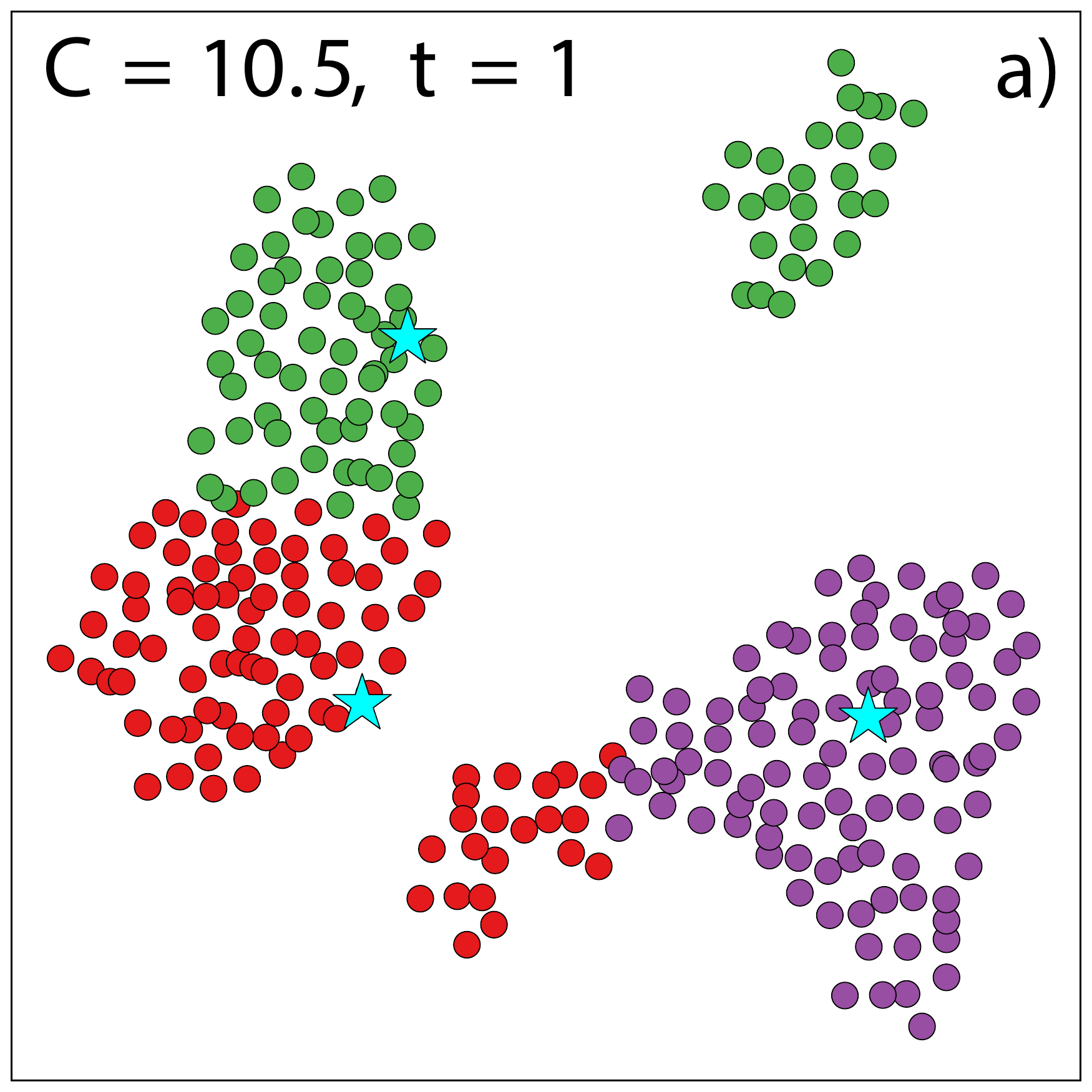}
\includegraphics[scale=0.24]{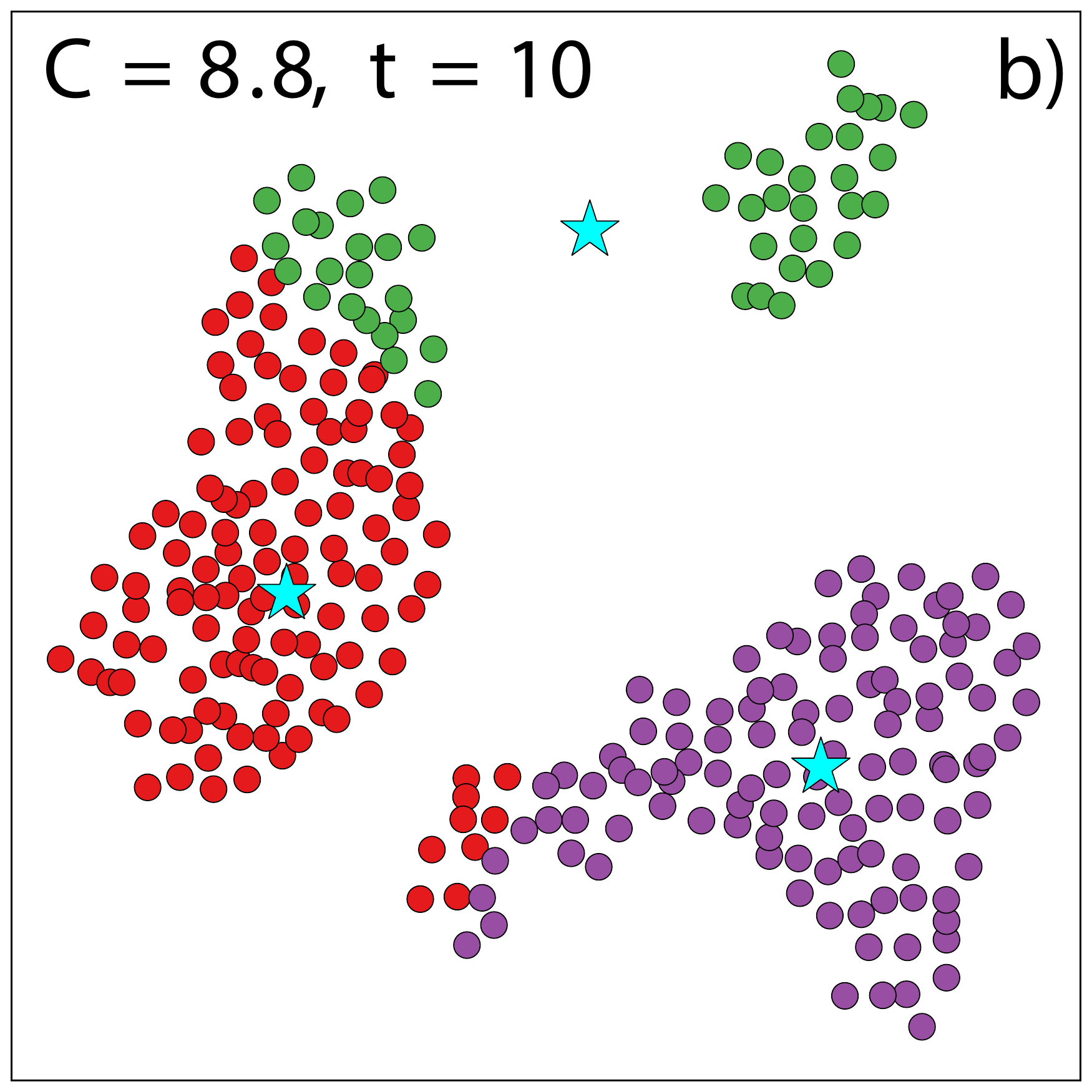}\\
\includegraphics[scale=0.24]{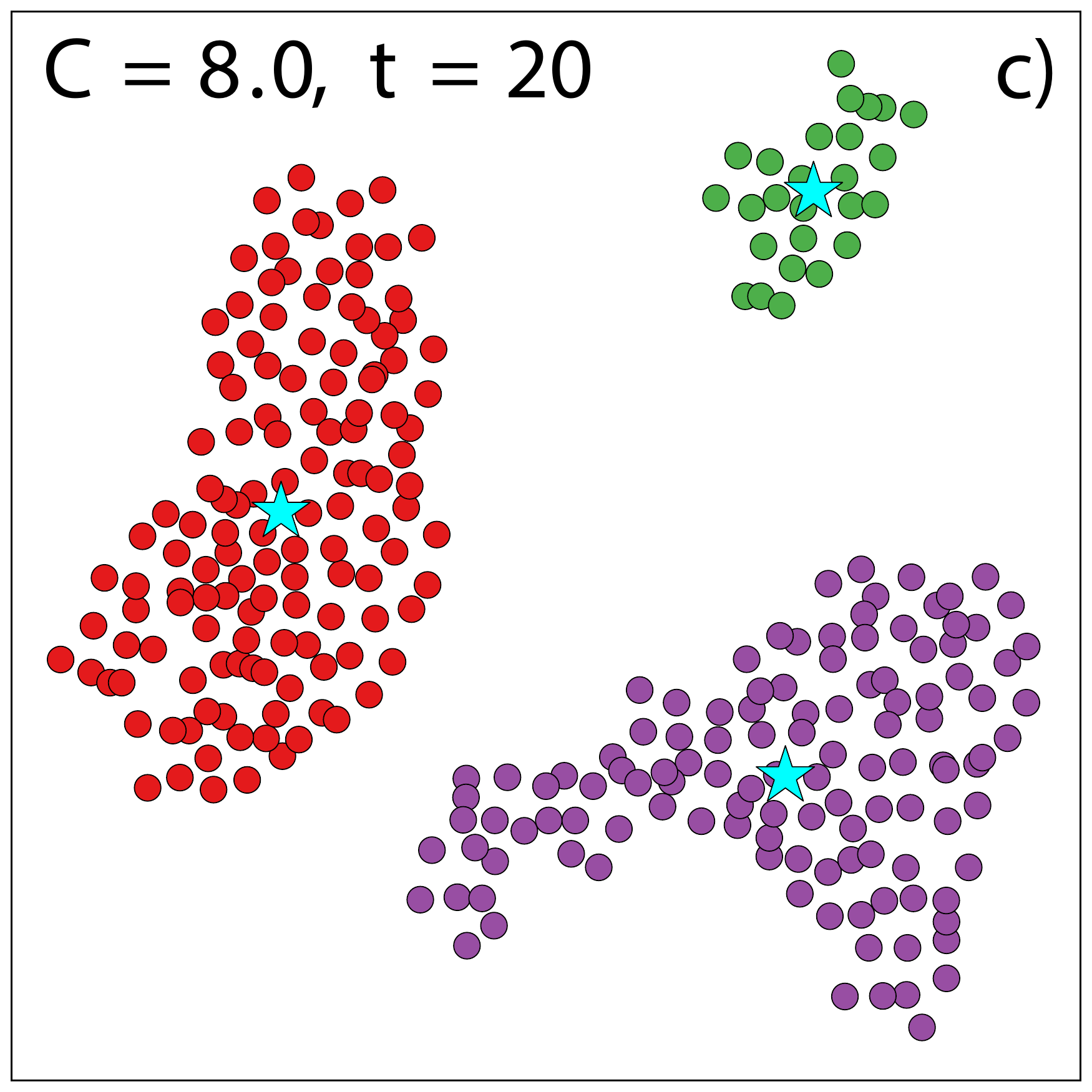}
\includegraphics[scale=0.24]{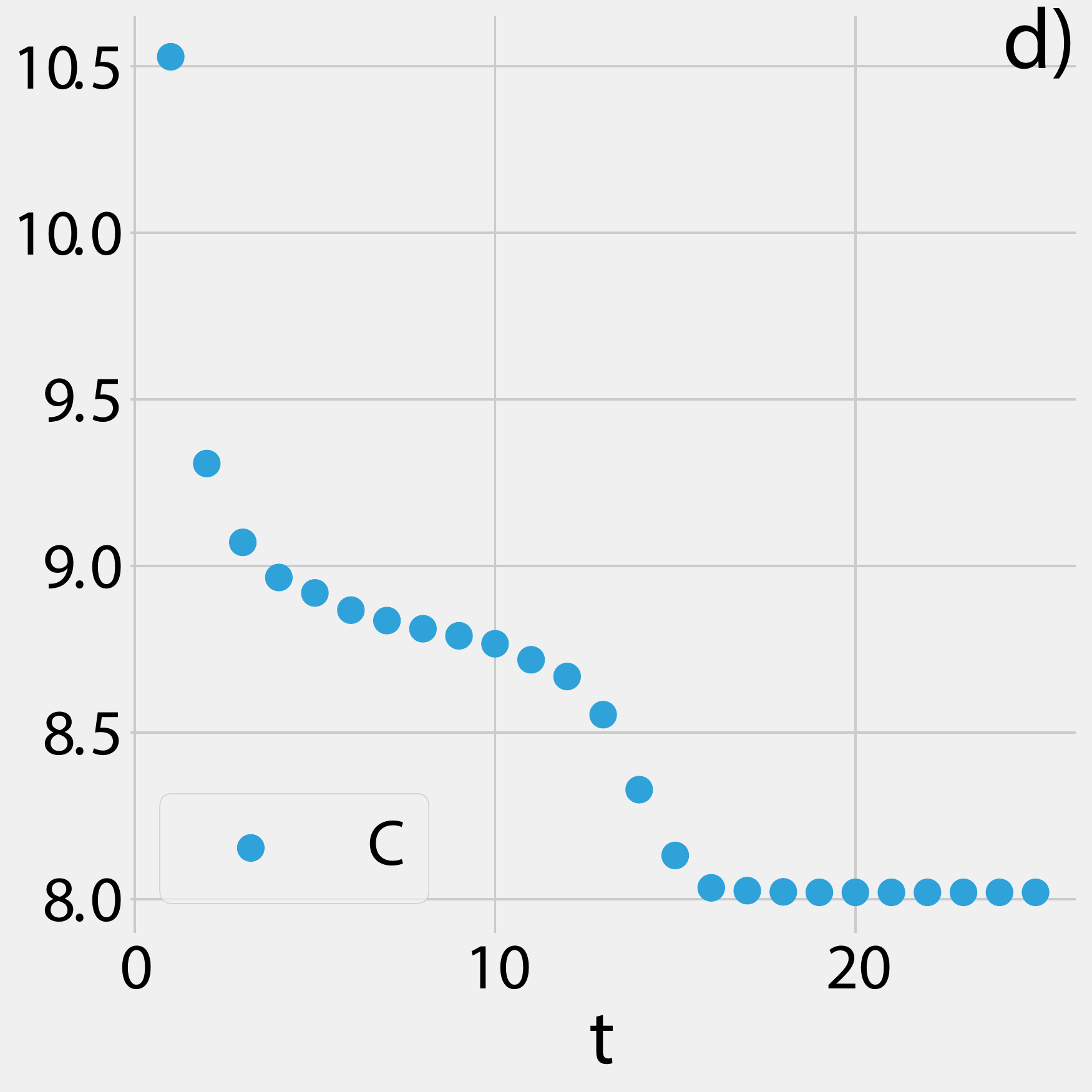}
\caption{$K$-means with $K=3$ applied to an artificial two-dimensional dataset. The cluster means at each iteration are indicated by cyan star markers. $t$ indicates the iteration number and $\mathcal{C}$ the value of the objective function. (a) The algorithm is initialized by randomly partitioning the space into 3 sectors to generate an initial assignment. (b)-(c) For well separated clusters, the algorithm converges rapidly to the true clusters. (d) The objective function as a function of the iteration. $\mathcal{C}$ converges after $t=18$ iterations for this choice of random seed (for center initialization).
\label{kmeans1}
}
\end{figure}

We begin our discussion with $K$-means clustering since this method is simple to implement and understand, and covers the core concepts of clustering. Consider a set of $N$ \emph{unlabelled} observations $\left\{\mathbf{x}_n\right\}_{n=1}^{N}$ where $\mathbf{x}_n\in \mathbb{R}^p$ and where $p$ is the number of features. Also consider a set of $K$ cluster centers called the cluster \emph{means}: $\left\{\bm{\mu}_k\right\}_{k=1}^K$, with $\bm{\mu}_k\in \mathbb{R}^p$, which we'll compute ``emperically" in the cluserting procedure. The cluster means can be thought of as the representatives of each cluster, to which data points are assigned (see FIG. \ref{kmeans1}). $K$-means clustering can be formulated as follows: given a fixed integer $K$, find the cluster means $\left\{\bm{\mu}\right\}$ and the data point assignments in order to minimize the following objective function:
\begin{equation}
\mathcal{C}(\left\{x,\bm{\mu}\right\})=\sum_{k=1}^K\sum_{n=1}^N r_{nk} (\mathbf{x}_n-\bm{\mu}_k)^2,
\label{KMeansCost}
\end{equation}
where $r_{nk}\in \left\{0,1\right\}$ is a binary variable called the assignment. The assignment $r_{nk}$ is 1 if $x_n$ is assigned to cluster $k$ and 0 otherwise. Notice that $\sum_k r_{nk} = 1~\forall~n$ and $\sum_n r_{nk} \equiv N_k$, where $N_k$ the number of points assigned to cluster $k$.
The minimization of this objective function can be understood as trying to find the best cluster means such that the variance within each cluster is minimized. In physical terms, $\mathcal{C}$ is equivalent to the sum of the moments of inertia of every cluster. Indeed, as we will see below, the cluster means $\bm{\mu}_k$ correspond to the centers of mass of their respective cluster. 
\paragraph*{\textbf{$K$-means algorithm.}}
The $K$-means algorithm alternates between two steps:
\begin{enumerate}
\item \emph{Expectation}: Given a set of assignments $\left\{r_{nk}\right\}$, minimize $\mathcal{C}$ with respect to $\bm{\mu}_k$. Taking a simple derivative and setting it to zero yields the following update rule:
\begin{equation}
\bm{\mu}_k=\frac{1}{N_k}\sum_n r_{nk} \mathbf{x}_n.
\end{equation}

\item \emph{Maximization}: Given a set of cluster means $\left\{\bm{\mu}_k\right\}$, find the assignments $\left\{r_{nk}\right\}$ which minimizes $\mathcal{C}$. Clearly, this is achieved by assigning each data point to their nearest cluster-mean:
\begin{equation}
r_{nk}=\begin{cases} 
      1 & \text{if } k = \argmin_{k'} (\mathbf{x}_n-\bm{\mu}_{k'})^2 \\  0& \text{otherwise}
   \end{cases}
\end{equation}
\end{enumerate}
$K$-means clustering consists in alternating between these two steps until some convergence criterion is met. Practically, the algorithm should terminate when the change in the objective function from one iteration to another becomes smaller than a pre-specified threshold. A simple example of $K$-means is presented in FIG. \ref{kmeans1}.

A nice property of the $K$-means algorithm is that it is guaranteed to converge. To see this, one can verify explicitly (by taking second-order derivatives) that the expectation step always decreases $\mathcal{C}$. This is also true for the assignment step. Thus, since $\mathcal{C}$ is bounded from below, the two-step iteration of $K$-means \emph{always} converges to a local minimum of $\mathcal{C}$. Since $\mathcal{C}$ is generally a non-convex function, in practice one usually needs to run the algorithm with different initial random cluster center initializations and post-select the best local minimum. A simple implementation of $K$-means has an average computational complexity which scales linearly in the size of the data set (more specifically the complexity is $\mathcal{O}(KN)$ per iteration) and is thus scalable to very large datasets.

As we will see in section \ref{clustering:2}, $K$-means is a hard-assignment limit of the Gaussian mixture model where all cluster variances are assumed to be the same. This highlights a common drawback of $K$-means: if the true clusters have very different variances (spreads), $K$-means can lead to spurious results since the underlying assumption is that the latent model has uniform variances.

\subsubsection{Hierarchical clustering: Agglomerative methods}

\begin{figure}[h!]
\centering
\includegraphics[scale=0.6]{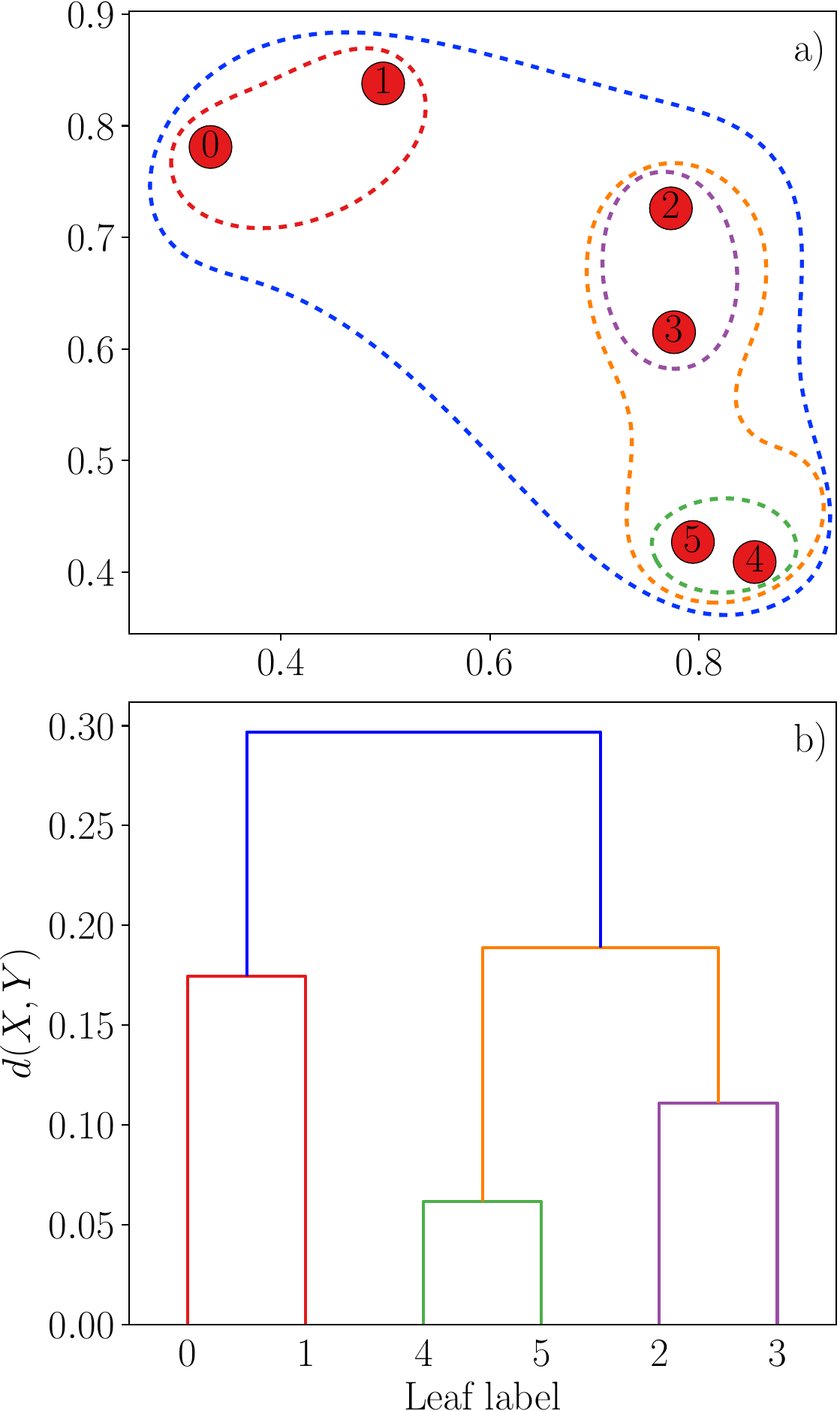}
\caption{Hierarchical clustering example with single linkage. (a) The data points are successively grouped as denoted by the colored dotted lines. (b) Dendrogram representation of the hierarchical decomposition. Each node of the tree represents a cluster. One has to specify a scale cut-off for the distance measure $d(X,Y)$ (corresponding to a horizontal cut in the dendrogram) in order to obtain a set of clusters. \label{hierarchical:1}}
\end{figure}

Agglomerative clustering is a bottom up approach that starts from small initial clusters which are then progressively merged to form larger clusters. The merging process generates a hierarchy of clusters that can be visualized in the form of a dendrogram (see FIG. \ref{hierarchical:1}). This hierarchy can be useful to analyze the relation between clusters and the subcomponents of individual clusters. Agglomerative methods are usually specified by defining a distance measure between clusters\footnote{Note that this measure need not be a metric.}. We denote the distance between clusters $X$ and $Y$ by $d(X,Y)\in \mathbb{R}$. Different choices of distance result in different clustering algorithms. At each step, the two clusters that are the closest with respect to the distance measure are merged until a single cluster is left.

\paragraph*{\textbf{Agglomerative clustering algorithm}}
Agglomerative clustering algorithms can thus be summarized as follows:
\begin{enumerate}
\item Initialize each point to its own cluster.
\item Given a set of $K$ clusters $X_1,X_2,\cdots,X_K$, merge clusters until one cluster is left ($K=1$):
\begin{enumerate}
\item Find the closest pair of clusters $(X_i, X_j)$: $(i,j)=\argmin_{(i',j')}d(X_{i'},X_{j'})$
\item Merge the pair. Update: $K\leftarrow K-1$ 
\end{enumerate}
\end{enumerate}

Here we list a few of the most popular distances used in agglomerative methods, often called linkage methods in the clustering literature.
\begin{enumerate}
\item Single-linkage: the distance between clusters $i$ and $j$ is defined as the minimum distance between two elements of the different clusters
\begin{equation}
d(X_i, X_j)=\min_{\mathbf{x}_i \in X_i,\mathbf{x}_j \in X_j}||\mathbf{x}_i-\mathbf{x}_j||_2.
\end{equation}
\item Complete linkage: the distance between clusters $i$ and $j$ is defined as the maximum distance between two elements of the different clusters.
\begin{equation}
d(X_i, X_j)=\max_{\mathbf{x}_i \in X_i,\mathbf{x}_j \in X_j}||\mathbf{x}_i-\mathbf{x}_j||_2
\end{equation}
\item Average linkage: average distance between points of different clusters
\begin{equation}
d(X_i, X_j)=\frac{1}{|X_i|\cdot |X_j|}\sum_{\mathbf{x}_i \in X_i, \mathbf{x}_j \in X_j}||\mathbf{x}_i-\mathbf{x}_j||_2
\end{equation}
\item Ward's linkage: This distance measure is analogous to the $K$-means method as it seeks to minimize the total inertia. The distance measure is the ``error squared'' before and after merging which simplifies to:
\begin{equation}
d(X_i, X_j)=\frac{|X_i||X_j|}{|X_i\cup X_j|}(\bm{\mu}_i-\bm{\mu}_j)^2,
\end{equation}
where $\bm{\mu}_j$ is the center of cluster $j$.
\end{enumerate}

A common drawback of hierarchical methods is that they do not scale well: at every step, a distance matrix between all clusters must be updated/computed. Efficient implementations achieve a typical computational complexity of $\mathcal{O}(N^2)$ \cite{mullner2011modern}, making the method suitable for small to medium-size datasets. A simple but major speed-up for the method is to initialize the clusters with $K$-means using a large $K$ (but still a small fraction of $N$) and then proceed with hierarchical clustering. This has the advantage of preserving the large-scale structure of the hierarchy while making use of the linear scaling of $K$-means. In this way, hierarchical clustering may be applied to very large datasets.

\subsubsection{Density-based (DB) clustering}

Density clustering makes the intuitive assumption that clusters are defined by regions of space with higher density of data points. Data points that constitute noise or that are outliers are expected to form regions of low density. Density clustering has the advantage of being able to consider clusters of multiple shapes and sizes while identifying outliers. The method is also suitable for large-scale applications.

The core assumption of DB clustering is that a \emph{relative} local density estimation of the data is possible. In other words, it is possible to order points according to their densities. Density estimates are usually accurate for low-dimensional data but become unreliable for high-dimensional data due to large sampling noise. Here, for brevity,  we confine our discussion to one of the most widely used density clustering algorithms, DBSCAN. We have also had great success with another recently introduced variant of DB clustering \cite{rodriguez2014clustering} that is similar in spirit which the reader is urged to consult. One of the authors (A.~D.) has also created a Python package, \url{https://pypi.org/project/fdc/}, which makes use of accurate density estimates via kernel methods combined with agglomerative clustering to produce fast and accurate density clustering (see also \href{https://github.com/alexandreday/fast_density_clustering}{GitHub repository}).

\paragraph*{\textbf{DBSCAN algorithm.}}
Here we describe the most prominent DB clustering algorithm: DBSCAN, or density-based spatial clustering of applications with noise \cite{ester1996density}. Consider once again a set of $N$ data points $X\equiv \left\{\mathbf{x}_n\right\}_{n=1}^{N}$.

\begin{figure}[h!]
\centering
\includegraphics[scale=0.8]{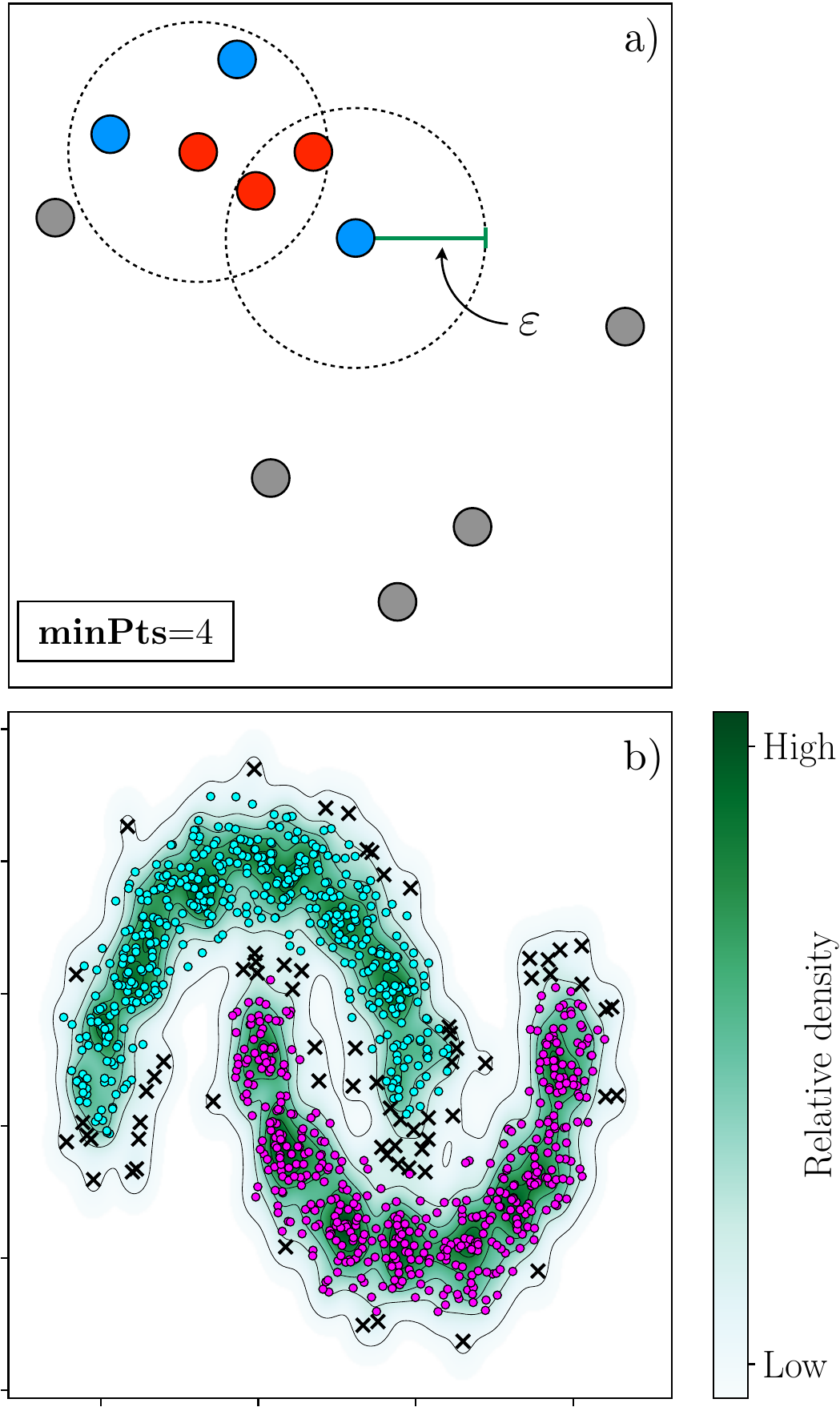}
\caption{(a) Illustration of DBSCAN algorithm with \textbf{minPts}$=4$. Two $\varepsilon$-neighborhood are represented as dashed circles of radius $\varepsilon$. Red points are the core points and blue points are density-reachable point that are not core points.  Outliers are gray colored. (b) Application of DBSCAN (\textbf{minPts}=40) to a noisy dataset with two non-convex clusters. Density profile is shown for clarity. Outliers are indicated by black crosses.
\label{db:1}}
\end{figure}

We start by defining the $\varepsilon$-neighborhood of point $\mathbf{x}_n$ as follows:
\begin{equation}
N_\varepsilon(\mathbf{x}_n) = \left\{ \mathbf{x} \in X | d(\mathbf{x},\mathbf{x}_n) < \varepsilon\right\}.
\end{equation}
$N_\varepsilon(\mathbf{x}_n)$ are the data points that are at a distance smaller than $\varepsilon$ from $\mathbf{x}_n$. As before, we consider $d(\cdot,\cdot)$ to be the Euclidean metric (which yields spherical neighborhoods, see Figure \ref{db:1}) but other metrics may be better suited depending on the specific data. $N_\varepsilon(\mathbf{x}_n)$ can be seen as a crude estimate of local density. $\mathbf{x_n}$ is considered to be a \emph{core-point} if at least $\textbf{minPts}$ are in its $\varepsilon$-neighborhood. $\textbf{minPts}$ is a free parameter of the algorithm that sets the scale of the size of the smallest cluster one should expect. Finally, a point $\mathbf{x}_i$ is said to be \emph{density-reachable} if it is in the $\varepsilon$-neighborhood of a \emph{core-point}. From these definitions, the algorithm can be simply formulated (see also Figure \ref{db:1}):
\begin{itemize}
\item[$\rightarrow$] Until all points in $X$ have been visited; \textbf{do}
\begin{itemize}
\item[$-$] Pick a point $\mathbf{x}_i$ that has not been visited
\item[$-$] Mark $\mathbf{x}_i$ as a visited point
\item[$-$] If $\mathbf{x}_i$ is a core point; \textbf{then}
\begin{itemize}
\item[$\cdot$] Find the set $\mathcal{C}$ of all points that are \emph{density reachable} from $\mathbf{x}_i$.
\item[$\cdot$] $\mathcal{C}$ now forms a cluster. Mark all points within that cluster as being visited.
\end{itemize}
\end{itemize}
\item[$\rightarrow$] Return the cluster assignments $\mathcal{C}_1, \cdots, \mathcal{C}_k$, with $k$ the number of clusters. Points that have not been assigned to a cluster are considered noise or outliers. 
\end{itemize}
Note that DBSCAN does not require the user to specify the number of clusters but only $\varepsilon$ and $\textbf{minPts}$. While, it is common to heuristically fix these parameters, methods such as cross-validation can be used for their determination. Finally, we note that DBSCAN is very efficient since efficient implementations have a computational cost of $\mathcal{O}(N\log N)$.

\subsection{Clustering and Latent Variables via the Gaussian Mixture Models \label{clustering:2}}

In the previous section, we introduced several practical methods for clustering. In this section, we will approach clustering from a more abstract vantage point, and in the process, introduce many of the core ideas underlying unsupervised learning. A central concept in many unsupervised learning techniques is the idea of a latent or hidden variable. Even though latent variables are not directly observable, they still influence the visible structure of the data. For example, in the context of clustering we can think of the cluster identity of each datapoint (i.e. which cluster does a datapoint belong to) as a latent variable. And even though we cannot see the cluster label explicitly, we know that points in the same cluster tend to be closer together. The latent variables in our data (cluster identity) are a way of representing and abstracting the correlations between datapoints.

In this language,  we can think of clustering as an algorithm to learn the most probable value of a latent variable (cluster identity) associated with each datapoint. Calculating this latent variable requires additional assumptions about the structure of our dataset. Like all unsupervised learning algorithms, in clustering we must make an assumption about the underlying probability distribution from which the data was generated. Our model for how the data is generated is called the \emph{generative model}. In clustering, we assume that data points are assigned a cluster, with each cluster characterized by some cluster-specific probability distribution (e.g.~a Gaussian with some mean and variance that characterizes the cluster). We then specify a procedure for finding the value of the latent variable. This is often done by choosing the values of the latent variable that minimize some cost function.

One common choice for a class of cost functions for many unsupervised learning problems is Maximum Likelihood Estimation (MLE), see Secs.~\ref{sec:bayesian_inference} and~\ref{sec:lin_reg}. In MLE, we choose the values of the latent variables that maximize the likelihood of the observed data under our generative model (i.e.~maximize the probability of getting the observed dataset under our generative model). Such MLE equations often give rise to the kind of Expectation Maximization (EM) equations that we first encountered in the last section in the context of $K$-means clustering.

Gaussian Mixtures models (GMM) are a generative model often used in the context of clustering. In GMM, points are drawn from one of  $K$ Gaussians, each with its own mean $\boldsymbol{\mu}_k$ and covariance matrix $\Sigma_k$,
\be
\mathcal{N}(\boldsymbol{x}|\boldsymbol{\mu}, \boldsymbol{\Sigma})\sim \exp\left[-\frac{1}{2}(\boldsymbol{x}-\boldsymbol{\mu})\boldsymbol{\Sigma}^{-1}(\boldsymbol{x}-\boldsymbol{\mu})^T\right].
\ee
Let us denote the probability that a point is drawn from mixture $k$ by $\pi_k$. Then, the probability of generating a point $\boldsymbol{x}$ in a GMM is given by
\be
p(\boldsymbol{x}| \{ \boldsymbol{\mu}_k,   \boldsymbol{\Sigma}_k, \pi_k \})= \sum_{k=1}^K \mathcal{N}(\boldsymbol{x}|\boldsymbol{\mu}_k,  \boldsymbol{\Sigma}_k) \pi_k.
\ee
Given a dataset $\boldsymbol{X}=\left\{\boldsymbol{x}_1,\cdots, \boldsymbol{x}_N \right\}$, we can write the likelihood of the dataset as
\be
p(\boldsymbol{X}| \{ \boldsymbol{\mu}_k,   \boldsymbol{\Sigma}_k, \pi_k \})=\prod_{i=1}^N p(\boldsymbol{x}_i | \{ \boldsymbol{\mu}_k,   \boldsymbol{\Sigma}_k, \pi_k \})
\label{GMM:likelihood}
\ee
For future reference, let us denote the set of parameters  (of $K$ Gaussians in the model) $\{ \boldsymbol{\mu}_k,   \boldsymbol{\Sigma}_k, \pi_k \}$ by $\boldsymbol{\theta}$.

To see how we can use GMM and MLE to perform clustering, we introduce discrete binary $K$-dimensional latent variables $\boldsymbol{z}$  for each data point $\boldsymbol{x}$ 
whose $k$-th component is $1$ if point $x$ was generated from the $k$-th Gaussian and zero otherwise (these are often called ``one-hot variables''). For instance if we were considering a Gaussian mixture with $K=3$, we would have three possible values for $\boldsymbol{z}\equiv(z_1, z_2, z_3)$ : $(1,0,0), (0,1,0)$ and $(0,0,1)$.   We cannot directly observe the variable $\boldsymbol{z}$. It is a latent variable that encodes the cluster identity of point $\boldsymbol{x}$. Let us also denote all the $N$ latent variables corresponding to a dataset $\boldsymbol{X}$ by $\boldsymbol{Z}$.

Viewing the GMM as a generative model, we can write the probability $p(\boldsymbol{x}| \boldsymbol{z})$ of observing a data point $\boldsymbol{x}$ given $\boldsymbol{z}$ as
\be
p(\boldsymbol{x}|\boldsymbol{z}; \{ \boldsymbol{\mu}_k, \boldsymbol{\Sigma}_k\} )=\prod_{k=1}^K \mathcal{N}(\boldsymbol{x}|\mu_k,\Sigma_k)^{z_k}\label{GMM:xgivenz}
\ee
as well as the probability of observing a given value of latent variable
\be
p(\boldsymbol{z}| \{ \pi_k \})= \prod_{k=1}^K \pi_k^{z_k}.
\ee
Using Bayes' rule, we can write the joint probability of a clustering assignment $\boldsymbol{z}$ and a data point $\boldsymbol{x}$ given the GMM parameters 
as
\be
p(\boldsymbol{x},\boldsymbol{z}; \bm{\theta}) = p(\boldsymbol{x}|\boldsymbol{z}; \{ \boldsymbol{\mu}_k, \boldsymbol{\Sigma}_k\} )p(\boldsymbol{z}| \{ \pi_k \}).
\ee

We can also use Bayes rule to rearrange this expression to give the conditional probability of the data point $\boldsymbol{x}$ being in the $k$-th cluster, $\gamma(z_k)$, given model parameters $\theta$ as
\begin{equation}
\gamma(z_k)\equiv p(z_k=1|\boldsymbol{x}; \theta)= \frac{\pi_k\mathcal{N}(\boldsymbol{x}|\mu_k,\Sigma_k)}{\sum_{j=1}^K\pi_j\mathcal{N}(x|\mu_j,\Sigma_j)}.
\label{GMM:responsibility}
\end{equation}
The  $\gamma(z_k)$ are often referred to as the ``responsibility'' that mixture $k$ takes for explaining $\boldsymbol{x}$. Just like in our discussion of soft-max classifiers, this can be made into a ``hard-assignment'' by assigning each point to the cluster with the largest probability: $\arg \max_{k}\gamma(z_k)$ over the responsibilities.

The complication is of course that we do not know the parameters $\bm{\theta}$ of the underlying GMM but instead must also learn them from the dataset $\boldsymbol{X}$. As discussed above, ideally we could do this by choosing the parameters that maximize the likelihood (or equivalently the
log-likelihood) of the data
\be
\hat{\bm{\theta}} = \argmax_{\bm{\theta}} \log{ p(\boldsymbol{X}| \bm{\theta})} 
\ee
where $\bm{\theta} = \{ \boldsymbol{\mu}_k,   \boldsymbol{\Sigma}_k, \pi_k \}$. Once we know the MLEs $\hat{\bm{\theta}}$, we could use Eq.~\eqref{GMM:responsibility} to calculate the optimal hard cluster assignment $\arg \max_{k}\hat{\gamma}(z_k)$ where $\hat{\gamma}(z_k)=p(z_k=1|\boldsymbol{x}; \hat{\bm{\theta}})$. 

In practice, due to the complexity of Eq.~\eqref{GMM:likelihood}, it is almost impossible to find the global maximum of the likelihood function. Instead, we must settle for a local maximum. One approach to finding a local maximum of the likelihood is to use a method like stochastic gradient descent on the negative log-likelihood, cf.~Sec~\ref{sec:gradient_descent}. Here, we introduce an alternative, powerful approach for finding local minima in latent variable models using an iterative procedure called Expectation Maximization (EM). Given an initial guess for the parameters $\theta^{(0)}$, the EM algorithm iteratively generates new estimates for the parameters $\theta^{(1)},\theta^{(2)}, \ldots$. Importantly, the likelihood is guaranteed to be non-decreasing under these iterations and hence EM converges to a local maximum of the likelihood~\cite{neal1998view}.

The central observation underlying EM is that it is often much easier to calculate the conditional likelihoods of the latent variables $\tilde{p}^{(t)}(\mathbf{Z})=p(\mathbf{Z}|\mathbf{X}; \theta^{(t)})$ given some choice of parameters, and the maximum of the expected log likelihood given an assignment of the latent variables: $ \theta^{(t+1)}=\argmax_\theta  E_{p(\mathbf{Z}|\mathbf{X}; \theta^{(t)})}[\log{p(\mathbf{X}, \mathbf{Z}; \theta)]}$. To get an intuition for this later quantity notice that we can write
\begin{widetext}
	\be
	\mathbb{E}_{\tilde{p}^{(t)}}[\log{p(\mathbf{X}, \mathbf{Z}; \theta)}]=\sum_{i=1}^N \sum_{k=1}^K  \gamma_{ik}^{(t)} \left[ \log{\mathcal{N}(\boldsymbol{x}_i|\boldsymbol{\mu}_k},  \boldsymbol{\Sigma}_k)+\log{\pi_k} \right],
	\ee
\end{widetext}
where we have used the shorthand $\gamma_{ik}^{(t)}=p({z}_{ik} |\mathbf{X}; \theta^{(t)})$ with $z_{ik}$ the $k$-th component of $\mathbf{z}_i$. Taking the derivative of this equation with respect to $\boldsymbol{\mu}_k$, $\boldsymbol{\Sigma}_k$, and $\pi_k$ (subject to the constraint $\sum_k  \pi_k=1$) and setting this to zero yields the intuitive equations
\begin{align} 
\boldsymbol{\mu}_{k}^{(t+1)}&={\sum_{i}^N \gamma_{ik}^{(t)} x_{i} \over \sum_i \gamma_{ik}^{(t)}} \nonumber \\
\boldsymbol{\Sigma_k}^{(t+1)}&={\sum_i^N \gamma_{ik}^{(t)} (\boldsymbol{x}_{i}-\boldsymbol{\mu}_k)(\mathbf{x}_i-\boldsymbol{\mu}_k)^T \over  \sum_i \gamma_{ik}^{(t)}} \nonumber \\
\boldsymbol{\pi_k}^{(t+1)}&=\frac{1}{N} \sum_k \gamma_{ik}^{(t)}
\label{GMM:M-step}
\end{align}
These are just the usual estimates for the mean and variance, with each data point weighed according to our current best guess for the probability that it belongs to cluster $k$. We can then use our new estimate $\theta^{(t+1)}$ to calculate responsibility $\gamma_{ik}^{(t+1)}$ and repeat the process. This is essentially the $K$-Means algorithm discussed in the first section.

This discussion of the Gaussian mixture model introduces several concepts that we will return to repeatedly in the context of unsupervised learning. First, it is often useful to think of the visible correlations between features in the data as resulting from hidden or latent variables. Second, we will often posit a generative model that encodes the structure we think exists in the data and then find parameters that maximize the likelihood of the observed data. Third, often we will not be able to directly estimate the MLE, and will have to instead look for a computationally efficient way to find a  local minimum of the likelihood. 

\begin{figure}[h]
	\centering
	\includegraphics[scale=0.5]{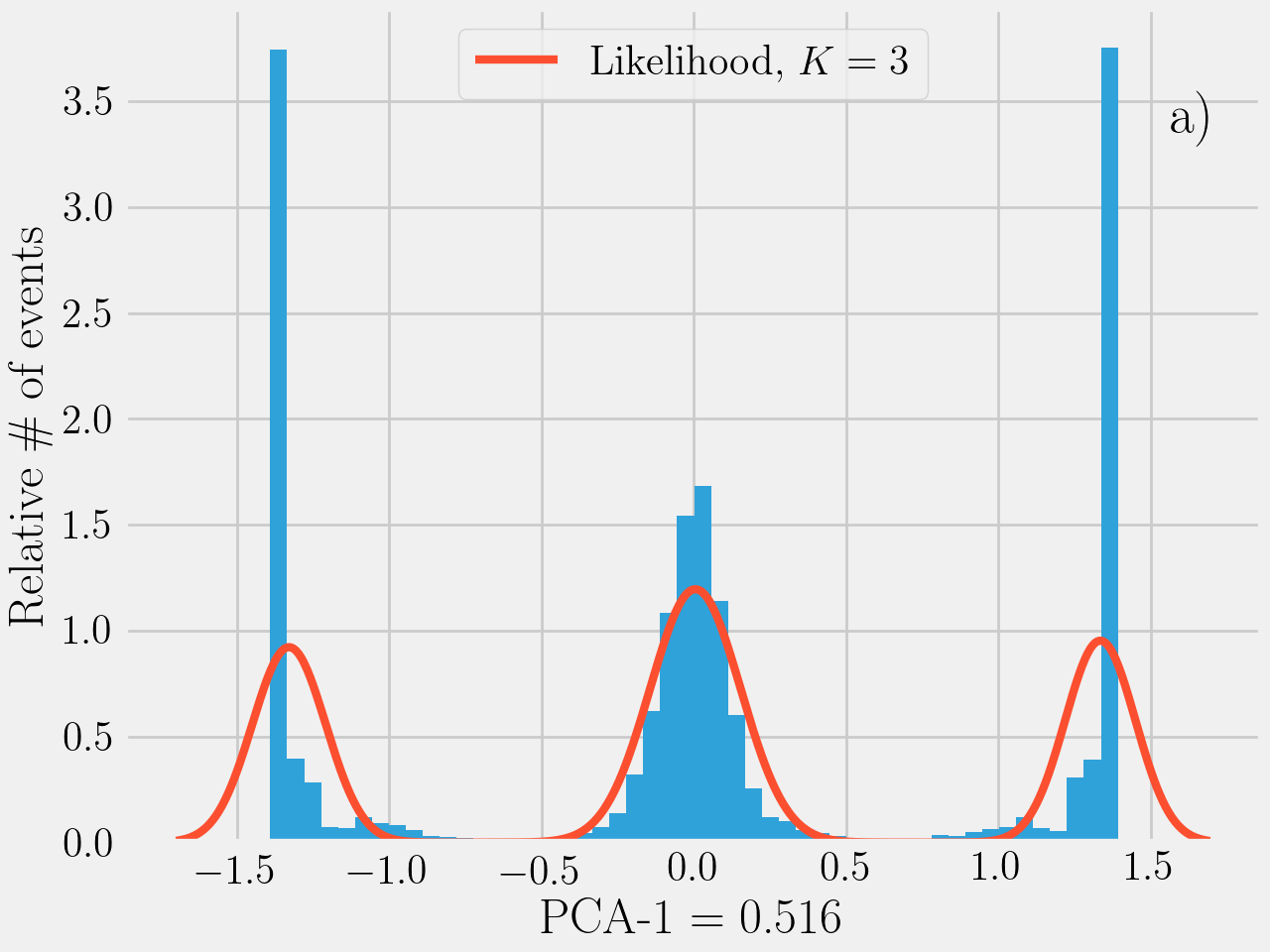}
	\includegraphics[scale=0.5]{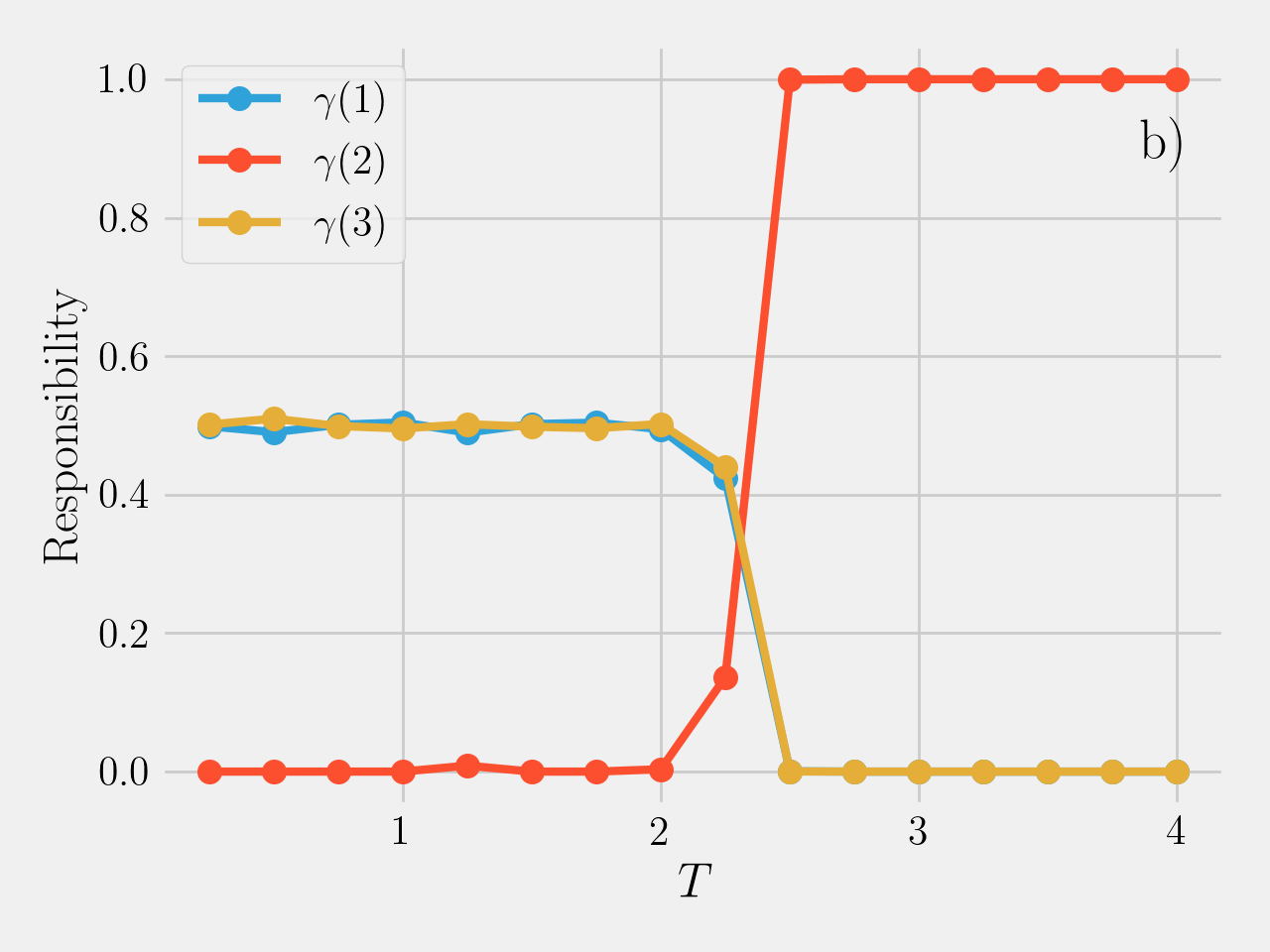}
	\caption{(a) Application of gaussian mixture modelling to the Ising dataset. The normalized histogram corresponds to the first principal component distribution of the dataset (or equivalently the magnetization in this case). The 1D data is fitted with a $K=3$-component gaussian mixture. The likehood of the fitted gaussian mixture is represented in red and is obtained via the expectation-maximization algorithm (b) The gaussian mixture model can be used to compute posterior probability (responsibilities), i.e. the probability of being in one of the phases. Note that the point where $\gamma(1)=\gamma(2)=\gamma(3)$ can be interpreted as the critical point. Indeed the crossing occurs at $T\approx 2.26$.
	}
\end{figure}

\subsection{Clustering in high dimensions \label{clustering:3}}
Clustering data in high-dimension can be very challenging. One major problem that is aggravated in high-dimensions is the generic accumulation of noise due to random measurement error for each feature. This in turn leads to increased errors for pairwise similarity and distance measures and thus tends to ``blur'' distances between data points \cite{kriegel2009clustering, domingos2012few, zimek2012survey}. Many clustering algorithms rely on the explicit use of a similarity measure or distance metrics that weigh all features equally. For this reason, one must be careful when using an off-the-shelf method in high dimensions. 

In order to perform clustering on high-dimensional data, it is often useful to denoise the data before proceeding with using a standard clustering method such as $K$-means \cite{kriegel2009clustering}. Figure \ref{dimreduce:PCAtSNE} illustrates an application of denoising to high-dimensional data.  PCA (section \ref{subsec:PCA}) was used to denoise the MNIST dataset by projecting the 784 original dimensions onto the 40 dimensions with the largest principal components. The resulting features were then used to construct a Euclidean distance matrix which was used by $t$-SNE to compute the two-dimensional embedding that is presented. Using $t$-SNE directly on original data leads to a
``blurring'' of the clusters (the reader is encouraged to test this themselves). 

However, simple feature selection or feature denoising (using PCA for instance) can sometimes be insufficient for learning clusters due to the presence of large variations in the signal and noise of the features that are relevant for identifying the underlying clusters \cite{kriegel2009clustering}. Recent promising work suggests that one way to overcome these limitations is to learn the latent space \emph{and} the cluster labels \emph{at the same time} \cite{xie2016unsupervised}.

Finally we end the clustering section with a short discussion on clustering validation, which can be particularly difficult for high-dimensional data. Often clustering validation, i.e.~verifying whether the obtained labels are ``valid'' is done by direct visual inspection. That is, the data is represented in a low-dimensional space and the cluster labels obtained are visually inspected to make sure that  different labels organize into distinct ``blobs''. For high-dimensional data, this is done by performing dimensional reduction (section \ref{sec:dim-red}). However, this can lead to the appearance of spurious clusters since dimensional reduction inevitably loses information about the original data. Thus, these methods should be used with care when trying to validate clusters [see \cite{wattenberg2016use} for an interactive discussion on how $t$-SNE can sometime be misleading and how to effectively use it].

A lot of work has been done  to devise ways of validating clusters based on various metrics and measures~\cite{kriegel2009clustering}. Perhaps one of the most intuitive way of defining a good clustering is by measuring how well clusters generalize. Clustering methods based on leveraging powerful classifiers to measure the generalization errors of the clusters have been developed by some of the authors \cite{agrday2018VAC}, see \url{https://pypi.org/project/hal-x/}. We believe this represents an especially promising research direction in high-dimensional clustering. Finally, we emphasize that this discussion is far from exhaustive and we refer the reader to \cite{rokach2005clustering}, Chapter 15, for an in-depth survey of the various validation techniques.

%% file: sections/Variational.tex
A common thread in many unsupervised learning tasks is accurately representing the underlying probability distribution from which a dataset is drawn. 
Unsupervised learning of high-dimensional, complex distributions presents a new set of technical and computational challenges that are different from those we encountered in a  supervised learning setting. When dealing with complicated probability distributions, it is often much easier to learn the \emph{relative weights} of different states or data points (ratio of probabilities), than \emph{absolute} probabilities. In physics, this is the familiar statement that the weights of a Boltzmann distribution are much easier to calculate than the partition function. The relative probability of two configurations, $\bd{x}_1$ and $\bd{x}_2$, are proportional to the difference between their Boltzmann weights
\be
{p(\mathbf{x}_1) \over p(\mathbf{x}_2)} =\mathrm  e^{-\beta \left(E(\mathbf{x}_1)-E(\mathbf{x}_2)\right)},
\ee
where as is usual in statistical mechanics  $\beta$ is the inverse temperature and $E(\mathbf{x}; \theta)$ is the energy of state $\mathbf{x}$ given some parameters (couplings) $\theta$ . However, calculating the absolute weight of a configuration requires knowledge of the partition function 
\be
Z_p =\mathrm{Tr}_{\mathbf{x} } \mathrm  e^{-\beta E(\mathbf{x})},
\ee
(where the trace is taken over all possible configurations $\mathbf{x}$) since 
\be
p(\mathbf{x}) ={ \mathrm e^{-\beta E(\mathbf{x})} \over Z_p}.
\ee
In general, calculating the partition function $Z_p$ is analytically and computationally intractable. 

For example, for the Ising model with $N$ binary spins, the trace involves calculating a sum over $2^N$ terms, which is a difficult task for most energy functions. For this reason, physicists (and machine learning scientists) have developed various numerical and computational methods for evaluating such partition functions. One approach is to use Monte-Carlo based methods to draw samples from the underlying distribution (this can be done knowing only the relative probabilities) and then use these samples to numerically estimate the partition function. This is the philosophy behind powerful methods such as Markov Chain Monte Carlo (MCMC) \cite{andrieu2003introduction} and annealed importance sampling \cite{neal1998view} which are widely used in both the statistical physics and machine learning communities. An alternative approach -- which we focus on here  -- is to approximate the the probability distribution $p(\mathbf{x})$ and partition function using a ``variational distribution'' $q(\mathbf{x}; \theta_q)$ whose partition function we can calculate exactly. The variational parameters $\theta_q$ are chosen to make the variational distribution as close to the true distribution as possible (how this is done is the focus of much of this section). 

One of the most-widely applied examples of a variational method in statistical physics is Mean-Field Theory (MFT). MFT can be naturally understood as a procedure for approximating the true distribution of the system by a factorized distribution. The deep connection between MFT and variational methods is discussed below.  These variational MFT methods have been extended to understand more complicated spin models (also called graphical models in the ML literature) and form the basis of powerful set of techniques that go under the name of Belief Propagation and Survey Propagation  \cite{yedidia2003understanding, mackay2003information, wainwright2008graphical}. 

Variational methods are also widely used in ML to approximate complex probabilistic models. For example, below we show how the Expectation Maximization (EM) procedure, which we discussed in the context of Gaussian Mixture Models for clustering, is actually a general method that can be derived for any latent (hidden) variable model using a variational procedure \cite{neal1998view}. This section serves as an introduction to this powerful class of variational techniques.  For readers interested in an in-depth discussion on variational inference for probabilistic graphical models, we recommend the great treatise written by Michael I. Jordan and others\cite{jordan1999introduction}, the more physics oriented discussion in \cite{yedidia2001idiosyncratic, yedidia2003understanding}, as well as David MacKay's outstanding book \cite{mackay2003information}.

\subsection{\label{subsec:MFT_Ising}Variational mean-field theory for the Ising model}

Ising models are a major paradigm in statistical physics. Historically introduced to study magnetism, it was quickly realized that their predictive power applies to a variety of interacting many-particle systems. Ising models are now understood to serve as minimal models for complex phenomena such as certain classes of phase transitions. In the Ising model, degrees of freedom called spins assume discrete, binary values, e.g.~$s_i=\pm 1$. Each spin variable $s_i$ lives on a lattice (or, in general, a graph), the sites of which are labeled by $i=1,2\dots, N$. Despite the extreme simplicity relative to real-world systems, Ising models exhibit a high level of intrinsic complexity, and the degrees of freedom can become correlated in sophisticated ways. Often, spins interact spatially locally, and respond to externally applied magnetic fields. 

A spin configuration $\bd{s}$ specifies the values $s_i$ of the spins at every lattice site. We can assign an ``energy'' to every such configuration
\begin{equation}
E(\bd{s},\bd{J}) = -\frac{1}{2}\sum_{i,j}J_{ij}s_i s_j - \sum_i h_i s_i,
\end{equation}
where $h_i$ is a local magnetic field applied to the spin $s_i$, and $J_{ij}$ is the interaction strength between the spins $s_i$ and $s_j$. In textbook examples, the coupling parameters $\bd{J}=(J,h)$ are typically uniform or, in studies of disordered systems, $(J_i,h_i)$ are drawn from some probability distribution (i.e. quenched disorder).  

The probability of finding the system in a given spin configuration at temperature $\beta^{-1}$ is given by
\begin{eqnarray}
p(\bd{s}|\beta,\bd{J}) &=& \frac{1}{Z_p(\bd{J})}\mathrm e^{-\beta E(\bd{s},\bd{J})},\nonumber\\
Z_p(\beta,\bd{J})&=&\sum_{\{s_i=\pm 1\}} \mathrm e^{-\beta E(\bd{s},\bd{J})},
\end{eqnarray}
with $\sum_{\{s_i=\pm 1\}}$ denoting the sum over all possible configurations of the spin variables. We write $Z_p$ to emphasize that this is the partition function corresponding to the probability distribution $p(\bd{s}|\beta,\bd{J})$, which will become important later. For a fixed number of lattice sites $N$, there are $2^N$ possible configurations, a number that grows exponentially with the system size. Therefore, it is not in general feasible to evaluate the partition function $Z_p(\beta,\bd{J})$ in closed form. This represents a major practical obstacle for extracting predictions from physical theories since the partition function is directly related to the free-energy through the expression
\begin{eqnarray}
\label{eq:exact_F}
\beta F_p(\bd{J})&=&-\log Z_p(\beta,\bd{J})= \beta \langle E(\bd{s},\bd{J})\rangle_p - H_p,
\end{eqnarray}
with 
\begin{equation}
\label{eq:entropy}
H_p = -\sum_{\{s_i=\pm 1\}}p(\bd{s}|\beta,\bd{J})\log p(\bd{s}|\beta,\bd{J})
\end{equation}
the entropy of the probability distribution $p(\bd{s}|\beta,\bd{J})$.

Even though the true probability distribution $p(\bd{s}|\beta,\bd{J})$ may be a very complicated object, we can still make progress by approximating $p(\bd{s}|\beta,\bd{J})$ by a \emph{variational  distribution} $q(\bd{s},\bd{\theta})$ which captures the essential features of interest, with $\theta$ some parameters that define our variational ansatz. The name variational distribution comes from the fact that we are going to vary the parameters $\bd{\theta}$ to make $q(\bd{s},\bd{\theta})$ as close to $p(\bd{s}|\beta,\bd{J})$ as possible. The functional form of $q(\bd{s},\bd{\theta})$ is based on an ``educated guess'', which oftentimes comes from our intuition about the problem. We can also define a variational free-energy
\be
\beta F_q(\bd{\theta}, \bd{J}) = \beta \langle E(\bd{s},\bd{J})\rangle_q - H_q,
\label{eq:main_var_eq}
\ee
where $ \langle E(\bd{s},\bd{J})\rangle_q$ is the expectation value of the energy $E(\bd{s},\bd{J})$ with respect to the distribution $q(\bd{s},\bd{\theta})$, and $H_q$ is the entropy of $q(\bd{s},\bd{\theta})$.

Before proceeding further, it is helpful to introduce a new quantity: the Kullback-Leibler divergence (KL-divergence or relative entropy) between two distributions $p(\mathbf{x})$ and $q(\mathbf{x})$. The KL-divergence measures the dissimilarity between the two distributions and is given by
 \be
D_{KL}(q\| p)=\mathrm{Tr}_{\mathbf{x}} q(\mathbf{x})\log{q(\mathbf{x}) \over p(\mathbf{x})},
\ee
which is the expectation w.r.t.~$q$ of the logarithmic difference between the two distributions $p$ and $q$. The trace $\mathrm{Tr}_{\mathbf{x}}$ denotes a sum over all possible configurations $\mathbf{x}$. Two important properties of the KL-divergence are (i) positivity: $D_{KL}(p\| q)\geq 0$ with equality if and only if $p=q$ (in the sense of probability distributions), and (ii) $D_{KL}(p\| q)\neq D_{KL}(q\| p)$, that is the KL-divergence is not symmetric in its arguments.

Variational mean-field theory is a systematic way for constructing such an approximate distribution $q(\bd{s},\bd{\theta})$.
The main idea is to choose parameters that minimize the difference between the variational free-energy $F_q(\bd{J},\bd{\theta})$ and the true free-energy $F_p(\bd{J}|\beta)$. We will show in Section \ref{subsubsec:EM} below that the difference between these two free-energies is actually the KL-divergence:
\be
F_q(\bd{J},\bd{\theta})=  F_p(\bd{J},\beta) + D_{KL}(q\|p).
\ee
This equality, when combined with the non-negativity of the KL-divergence has important consequences. First, it shows that the variational free-energy is always larger than the true free-energy, $F_q(\bd{J},\bd{\theta})\geq F_p(\bd{J})$, with equality if and only if  $q=p$ (the latter inequality is found in many physics textbooks and is known as the Gibbs inequality). Second, finding the best variational free-energy is equivalent to minimizing the KL divergence $D_{KL}(q\| p)$.

Armed with these observations, let us now derive a MFT of the Ising model using variational methods. In the simplest MFT of the Ising model, the variational distribution is chosen so that all spins are independent:
\begin{equation}
q(\bd{s},\bd{\theta}) = \frac{1}{Z_q}\exp\left(\sum_i \theta_i s_i\right) = \prod_i \frac{\mathrm e^{\theta_i s_i}}{2\cosh \theta_i}.
\end{equation}
In other words,  we have chosen a distribution $q$ which factorizes on every lattice site. An important property of this functional form is that we can analytically find a closed-form expression for  the variational partition function $Z_q$. This simplicity also comes at a cost:  ignoring correlations between spins. These correlations become less and less important in higher dimensions and the MFT ansatz becomes more accurate.

To evaluate the variational free-energy, we make use of Eq.~\eqref{eq:main_var_eq}. First, we need the entropy $H_q$ of the distribution $q$. Since $q$ factorizes over the lattice sites, the entropy separates into a sum of one-body terms
\begin{eqnarray}
H_q(\bd{\theta}) &=& -\sum_{\{s_i=\pm 1\}}q(\bd{s},\bd{\theta})\log q(\bd{s},\bd{\theta})\nonumber\\
&=& -\sum_i q_i\log q_i +(1-q_i)\log(1-q_i), 
\end{eqnarray}
where $q_i = \frac{\mathrm e^{\theta_i}}{2\cosh \theta_i}$ is the probability that spin $s_i$ is in the $+1$ state. Next, we need to evaluate the average of the Ising energy $E(\bd{s},\bd{J})$ with respect to the variational distribution $q$. Although the energy contains bilinear terms, we can still evaluate this average easily, because the spins are independent (uncorrelated) in the $q$ distribution. The mean value of spin $s_i$ in the $q$ distribution, also known as the on-site magnetization, is given by 
\begin{equation}
\label{eq:MF_self_consistency}
m_i = \langle s_i\rangle_q = \sum_{s_i=\pm 1 }s_i \frac{\mathrm e^{\theta_i s_i}}{2\cosh \theta_i} = \tanh(\theta_i).
\end{equation}
Since the spins are independent, we have
\begin{equation}
\langle E(\bd{s},\bd{J})\rangle_q =  -\frac{1}{2}\sum_{i,j}J_{ij}m_i m_j - \sum_i h_i m_i.
\end{equation}
The total variational free-energy is
\begin{equation*}
\beta F_q(\bd{J},\bd{\theta})  = \beta \langle E(\bd{s},\bd{J})\rangle_q - H_q,
\end{equation*}
and minimizing with respect to the variational parameters ${\bd \theta}$, we obtain
\begin{equation}
\frac{\partial}{\partial \theta_i} \beta F_q({\bd{J},\bd{\theta}}) = 2\frac{d q_i}{d\theta_i}\left(-\beta\left[\sum_j J_{ij} m_j + h_i\right] + \theta_i \right).
\end{equation}
Setting this equation to zero, we arrive at 
\begin{equation}
\label{eq:MF_Ising}
\theta_i = \beta\sum_j J_{ij}m_j(\theta_j) + h_i. 
\end{equation}
For the special case of a uniform field $h_i = h$ and uniform nearest neighbor couplings $J_{ij} = J$, by symmetry the variational parameters for all the spins are identical, with $\theta_i = \theta$ for all $i$. Then, the mean-field equations reduce to their familiar textbook form~\cite{sethna2006statistical}, $m = \tanh(\theta)$ and $\theta = \beta(z J m(\theta) + h)$, where $z$ is the coordination number of the lattice (i.e. the number of nearest neighbors). 

Equations~\eqref{eq:MF_self_consistency} and~\eqref{eq:MF_Ising} form a closed system, known as the mean-field equations for the Ising model. To find a solution to these equations, one method is to iterate through and update each $\theta_i$, once at a time, in an asynchronous fashion. Once can see the emerging relationship of this approach to solving the MFT equations to Expectation Maximization (EM) procedure first introduced in the context of the $K$-means algorithm in Sec.~\ref{clustering:1}. To make this explicit, let us spell out the iterative procedure to find the solutions to Eq.~\eqref{eq:MF_Ising}. We start by initializing our variational parameters to some $\bd{\theta}^{(0)}$ and repeat the following two steps until convergence:
\begin{enumerate}
\item \emph{Expectation}: Given a set of assignments at iteration $t$, $\bd{\theta}^{(t)}$, calculate the corresponding magnetizations $\bd{m}^{(t)}$ using Eq.~\eqref{eq:MF_self_consistency}
\item \emph{Maximization}: Given a set of magnetizations $m_t$, find new assignments $\theta^{(t+1)}$ which minimize the variational free energy $F_q$. From, Eq.~\eqref{eq:MF_Ising}
this is just
\be
\theta_i^{(t+1)}= \beta\sum_j J_{ij}m_j^{(t)} + h_i. 
\ee
\end{enumerate}
From these equations, it is clear that we can think of the MFT of the Ising model as an EM-like procedure similar to the one we used for $K$-means clustering and Gaussian Mixture Models in Sec.~\ref{sec:clustering}.

As is well known in statistical physics, even though MFT is not exact, it can often yield qualitatively and even quantitatively precise predictions (especially in high dimensions). The discrepancy between the true physics and MFT predictions stems from the fact that the variational distribution $q$ we chose cannot capture correlations between the spins. For instance, it predicts the wrong value for the critical temperature for the two-dimensional Ising model. It even erroneously predicts the existence of a phase transition in one dimension at a non-zero temperature. We refer the interested reader to standard textbooks on statistical physics for a detailed analysis of applicability of MFT to the Ising model. However, we emphasize that the failure of any particular variational ansatz does not compromise the usefulness of the approach. In some cases, one can consider changing the variational ansatz to improve the predictive properties of the corresponding variational MFT \cite{yedidia2001idiosyncratic, yedidia2003understanding}. The take-home message is that variational MFT is a powerful tool but one that must be applied and interpreted with care.

\subsection{\label{subsubsec:EM}Expectation Maximization (EM)}
Ideas along the lines of variational MFT have been independently developed in statistics and imported into machine learning to perform maximum likelihood (ML) estimates. In this section, we explicitly derive the Expectation Maximization (EM) algorithm and demonstrate further its close relation to variational MFT \cite{neal1998view}. We will focus on latent variable models where some of the variables are hidden and cannot be directly observed. This often makes maximum likelihood estimation difficult to implement. EM gets around this difficulty by using an iterative two-step procedure, closely related to variational  free-energy based approximation schemes in statistical physics. 

To set the stage for the following discussion, let $\bd{x}$ be the set of visible variables we can directly observe and $\bd{z}$ be the set of latent or hidden variables that we cannot directly observe. Denote the underlying probability distribution from which $\bd{x}$ and $\bd{z}$ are drawn by $p(\bd{z},\bd{x}|\bd{\theta})$, with $\bd{\theta}$ representing all relevant parameters. Given a dataset $\bd{x}$, we wish to find the maximum likelihood estimate of the parameters $\bd{\theta}$ that maximizes the probability of the observed data. 

As in variational MFT, we view $\bd{\theta}$ as variational parameters chosen to maximize the log-likelihood $L(\bd{\theta})= \langle \log p(\bd{x}|\bd{\theta})\rangle_{P_{\mathbf{x}}}$, where the expectation is taken with respect to the marginal distributions of $\bd{x}$. Algorithmically, this can be done by iterating the variational parameters $\bd{\theta}^{(t)}$ in a series of steps ($t=1,2, \dots$) starting from some arbitrary initial value $\bd{\theta}^{(0)}$:
\begin{enumerate}
	\item {\bf Expectation step (E step): }{ Given the known values of observed variable $\bd{x}$ and the current estimate of parameter $\bd\theta_{t-1}$, find the probability distribution of the latent variable $\bd{z}$:}
	\begin{equation}\label{eq:Estep}
	q_{t-1}(\bd{z})=p(\bd{z}|\bd{\theta}^{(t-1)},\bd{x})
	\end{equation}
	\item {\bf Maximization step (M step):} { Re-estimate the parameter $\bd{\theta}^{(t)}$ to be those with maximum likelihood, assuming $q_{t-1}(\bd z)$ found in the previous step is the true distribution of hidden variable $\bd z$:}
	\begin{equation}\label{eq:Mstep}
	\bd{\theta}_{t}=\argmax_{\bd{\theta}}\langle \log p(\bd{z},\bd{x}|\bd{\theta})\rangle_{q_{t-1}}
	\end{equation}
\end{enumerate}
It was shown \cite{dempster1977maximum} that {each EM iteration increases the true log-likelihood $L(\bd{\theta})$}, or at worst leaves it unchanged. { In most models, this iteration procedure converges to a  \emph{local maximum} of $L(\bd \theta)$.

\begin{figure}[h!]
	\includegraphics[width=1.0\columnwidth]{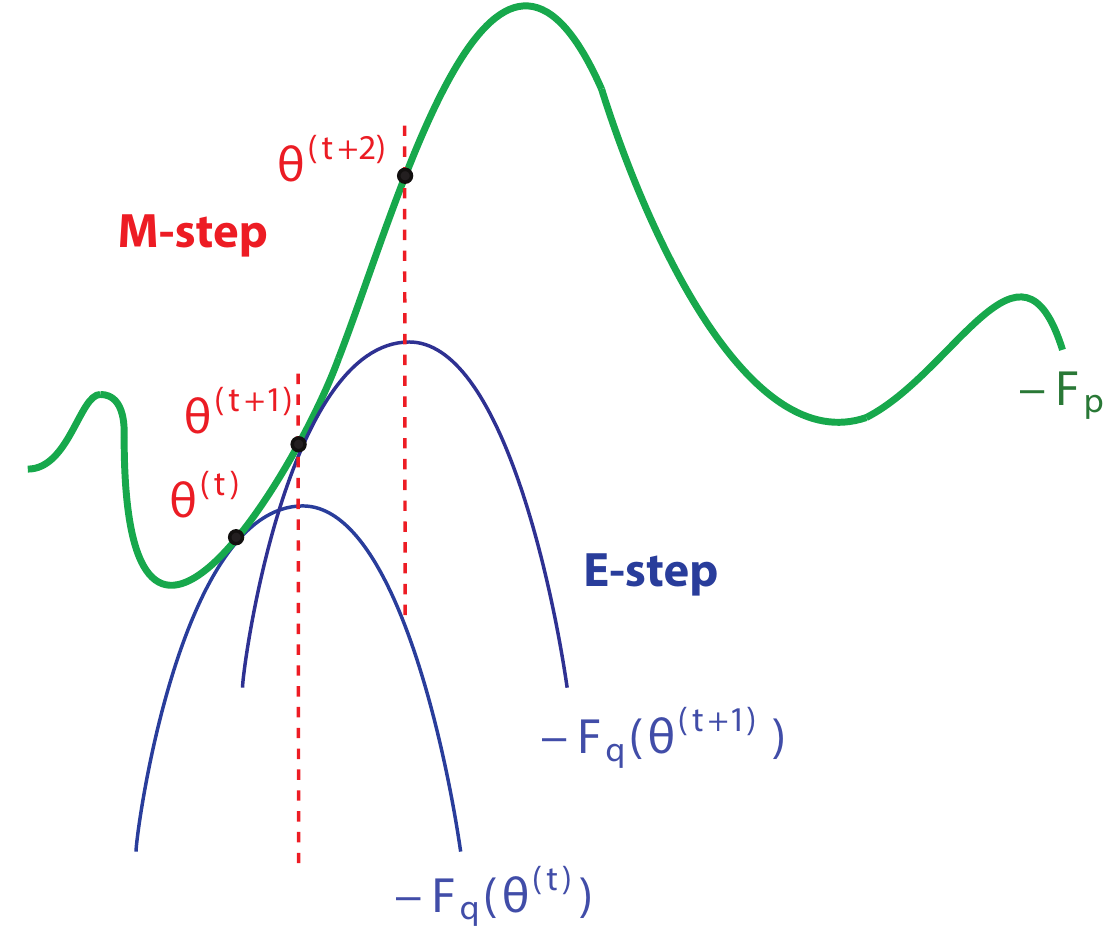}
	\caption{Convergence of EM algorithm. Starting from $\bd{\theta}^{(t)}$, E-step (blue) establishes $-F_{q}(\bd{\theta}^{(t)})$ which is always a lower bound of $-F_p:=\langle \log p(\bd x|\bd \theta)\rangle_{P_{\mathbf{x}}}$ (green). M-step (red) is then applied to update the parameter, yielding $\bd{\theta}^{(t+1)}$. The updated parameter $\bd{\theta}^{(t+1)}$ is then used to construct $-F_{q}(\bd{\theta}^{(t+1)})$  in the subsequent E-step. M-step is performed again to update the parameter, etc. }
	\label{fig:EM}
\end{figure}

To see how EM is actually performed and related to variational MFT, we make use of KL-divergence between two distributions introduced in the last section. Recall that our goal is to maximize the log-likelihood $L(\bd{\theta})$. With data $\bd{z}$ missing, we surely cannot just maximize $L(\bd{\theta})$ directly since parameter $\bd\theta$ might couple both $\bd z$ and $\bd x$. EM circumvents this by optimizing another objective function, $F_{q}(\bd{\theta})$, constructed based on estimates of the hidden variable distribution $q(\bd{z}|\bd{x})$. Indeed, the function optimized is none other than the {\it variational free energy} we encountered in the previous section:
\begin{eqnarray}
F_{q}(\bd\theta) := -  \langle \log p(\bd{z},\bd{x}|\bd{\theta})\rangle_{q, P_{\mathbf{x}}} -\langle H_q\rangle_{P_{\mathbf{x}}} \label{eq:Fptilde},
\end{eqnarray}
where $H_q$ is the Shannon entropy (defined in Eq.~\eqref{eq:entropy}) of $q(\bd z|\bd x)$. One can define the true free-energy $F_p(\bd\theta)$ as the negative log-likelihood of the observed data:
\be
-F_p(\bd\theta) = L(\bd{\theta}) = \langle\log p(\bd{x}|\bd{\theta})\rangle_{P_{\mathbf{x}}}.
\ee
In the language of statistical physics, $F_p(\bd \theta)$ is the {\it true} free-energy while $F_{q}(\bd\theta)$ is the variational free-energy we would like to minimize (see Table \ref{tbl:vMFT_EM}). Note that we have chosen to employ a physics sign convention here of defining the free-energy as minus log of the partition function. In the ML literature, this minus sign is often omitted \cite{neal1998view} and this can lead to some confusion. 
Our goal is to  choose $\bd{\theta}$ so that our variational free-energy  $F_{q}(\bd{\theta})$ is as close to the true free-energy  $F_p(\bd{\theta})$ as possible. The difference between these free-energies can be written as
\be
F_{q}(\bd\theta) - F_{p}(\bd\theta) = \langle f_q(\bd{x},\bd{\theta}) -  f_p(\bd{x},\bd{\theta})\rangle_{P_{\mathbf{x}}},
\ee
where 
\begin{eqnarray}
&& f_q(\bd{x},\bd{\theta}) -  f_p(\bd{x},\bd{\theta}) \nonumber\\
&=& \log p(\bd{x}|\bd{\theta}) -\sum_{\bd z}q(\bd{z}|\bd{x}) \log  p(\bd{z},\bd{x}|\bd{\theta}) \nonumber \\
+ &&\sum_{\bd z}q(\bd{z}|\bd{x}) \log q(\bd{z}|\bd{x}) \nonumber\\
&=&\sum_{\bd z}q(\bd{z}|\bd{x})  \log p(\bd{x}|\bd{\theta}) -\sum_{\bd z}q(\bd{z}|\bd{x}) \log  p(\bd{z},\bd{x}|\bd{\theta}) \nonumber\\
&& + \sum_{\bd z}q(\bd{z}|\bd{x}) \log q(\bd{z}|\bd{x}) \nonumber\\
&=&\! -\sum_{\bd z}q(\bd{z}|\bd{x}) \log \frac{p(\bd{z},\bd{x}|\bd{\theta})}{p(\bd{x}|\bd{\theta})}
\!+\!\sum_{\bd z}q(\bd{z}|\bd{x}) \log\tilde  p(\bd{z}) \nonumber\\
&=& \sum_{\bd z}q(\bd{z}|\bd{x}) \log \frac{q(\bd{z}|\bd{x})}{ p(\bd{z}|\bd{x},\bd{\theta})} \nonumber \\
&=& D_{KL}(q(\bd{z}|\bd{x}) \| p(\bd{z}|\bd{x},\bd{\theta}))\ge 0 \nonumber
\end{eqnarray} 
where we have used Bayes' theorem $ p(\bd z| \bd{x},\bd{\theta}) =  p (\bd z, \bd x|\bd \theta)/ p (\bd x|\bd \theta)$. Since the KL-divergence is always positive, this shows that the variational free-energy {$F_{q}$ is always an upper bound of the true free-energy $F_p$}. In physics, this result is known as Gibbs' inequality. 

From Eq.~\eqref{eq:Fptilde} and the fact that the the entropy term in Eq.~\eqref{eq:Fptilde} does not depend on $\bd\theta$, we can immediately see that the maximization step (M-step) in Eq.~\eqref{eq:Mstep} is equivalent to minimizing  the variational free-energy $F_{q}(\bd\theta)$. Surprisingly, the expectation step (E-step) can also viewed as the optimization of this variational free-energy. Concretely, one can show that the  distribution of hidden variables $\bd{z}$ given the observed variable $\bd x$ and the current estimate of parameter $\bd\theta$, Eq.~(\ref{eq:Estep}), is the \emph{unique} probability $q(\bd z)$ that minimizes $F_{q}(\bd\theta)$ (now seen as a functional of $q$). This can be proved by taking the functional derivative of Eq.~\eqref{eq:Fptilde}, plus a Lagrange multiplier that encodes $\sum_{\bd{z}} q(\bd z)=1$, with respect to $q(\bd z)$. Summing things up, we can re-write EM in the following form \cite{neal1998view}:
\begin{enumerate}
	\item \emph{Expectation step:} Construct the approximating probability distribution {of unobserved $\bd z$ given the values of observed variable $\bd x$ and parameter estimate $\bd\theta^{(t-1)}$}:
	\begin{equation}\label{eq:EM_Estep}
	q_{t-1}(\bd{z})=\argmin_{q}F_{q}(\bd{\theta}^{(t-1)})
	\end{equation}
	\item \emph{Maximization step:} {Fix $q$, }update the variational parameters:
	\begin{equation}\label{eq:EM_Mstep}
	\bd{\theta}^{(t)}=\argmax_{\bd{\theta}} -F_{q_{t-1}}(\bd{\theta}).
	\end{equation}
\end{enumerate}

To recapitulate, EM  implements ML estimation even with missing or hidden variables through optimizing a lower bound of the true log-likelihood. In statistical physics, this is reminiscent of optimizing a variational free-energy which is a lower bound of true free-energy due to Gibbs inequality. In Fig.~\ref{fig:EM}, we show pictorially how EM works. The E-step can be seen as representing the unobserved variable $\bd z$ by a probability distribution $q(\bd z)$. This probability is used to construct an alternative objective function $-F_{q}(\bd\theta)$, which is then maximized with respect to $\bd\theta$ in the M-step. By construction, maximizing the negative variational free-energy is equivalent to doing ML estimation on the joint data (i.e.~both observed and unobserved). The name ``M-step'' is intuitive since the parameters $\bd\theta$ are found by maximizing $-F_{q}(\bd\theta)$. The name ``E-step'' comes from the fact that one usually doesn't need to construct the probability of missing datas explicitly, but rather need only compute the ``expected" sufficient statistics over these data, cf.~Fig.~\ref{fig:EM}.

On the practical side, EM has been demonstrated to be extremely useful in parameter estimation, particularly in hidden Markov models and Bayesian networks (see, for example, \cite{wainwright2008graphical,barber2012bayesian}). Some of the authors have used EM in biophysics, to design algorithms which establish the equivalence of niche theory and the Minimum Environmental Perturbation Principle~\cite{marsland2019minimum}. One of the striking advantages of EM is that it is conceptually simple and easy to implement (see \href{https://physics.bu.edu/~pankajm/MLnotebooks.html}{Notebook 16}). In many cases, implementation of EM is guaranteed to increase the likelihood monotonically, which could be a perk during debugging. For readers interested in an overview on applications of EM, we recommend \cite{do2008expectation}.

Finally for advanced readers familiar with the physics of disordered systems, we note that it is possible to construct a one-to-one dictionary between EM for latent variable models and the MFT of spin systems with quenched disorder.  In a disordered spin systems, the Ising couplings $\bd{J}$ are commonly taken to be quenched random variables drawn from some underlying probability distribution. In the EM procedure, the quenched disorder is provided by the observed data points $\bd{x}$ which are drawn from some underlying probability distribution that characterizes the data. The spins $\bd{s}$ are like the hidden or latent variables $\bd{z}$. Similar analogues can be found for all the variational MFT quantities (see Table \ref{tbl:vMFT_EM}). This striking correspondence offers a glimpse into the deep connection between statistical mechanics and unsupervised latent variable models  -- a connection that we will repeatedly exploit to gain more intuition for the energy-based unsupervised models considered in the next few chapters.

\begin{table}[t!]
	\begin{tabular}{*{3}{|p{0.5\columnwidth}}}
		\hline
		\emph{\bf{statistical physics}}	&  \emph{\bf{Variational EM}} \\
		\hline
		spins/d.o.f.: $\bd{s}$	& hidden/latent variables $\bd{z}$  \\
		\hline
		couplings /quenched disorder: $\bd{J}$	& data observations: $\bd{x}$ \\
		\hline
		 Boltzmann factor $e^{-\beta E(\bd{s},\bd{J})}$	& Complete probability: $p({\bd x},{\bd z}|\bd{\theta})$ \\
		\hline
		partition function: $Z(\bd{J})$  &  marginal likelihood $p(\bd{x}|\theta)$ \\
		\hline
		energy: $\beta E(\bd{s},\bd{J})$  & negative log-complete data likelihood: $-\log p({\bd x},{\bd z}|\bd{\theta},m)$ \\
		\hline
		free energy: $\beta F_p(\bd{J}|\beta)$  & negative log-marginal likelihood:  $-\log p(\bd{x}|m)$ \\
		\hline

		variational distribution: $q(\bd{s})$  & variational distribution: $q(\bd{z}|\bd{x})$ \\
		\hline
		variational free-energy: $F_q(\bd{J}, \bd{\theta}) $  & variational free-energy:
		$F_{q}( \bd\theta)$ \\
		\hline
	\end{tabular}
	\caption{Analogy between quantities in statistical physics and variational EM.}
	\label{tbl:vMFT_EM}
\end{table}

%% file: sections/energyPM.tex
\section{Energy Based Models: Maximum Entropy (MaxEnt) Principle, Generative models, and Boltzmann Learning}
\label{sec:EnergyI}

Most of the models discussed in the previous sections (e.g.~linear and logistic regression, ensemble models, and supervised neural networks) are \emph{discriminative} -- they are designed to perceive differences between groups or categories of data. For example, recognizing differences between images of cats and images of dogs allows a discriminative model to label an image as ``cat'' or ``dog''. Discriminative models form the core techniques of most supervised learning methods. However, discriminative methods have several limitations. First, like all supervised learning methods, they require labeled data. Second, there are tasks that discriminative approaches simply cannot accomplish, such as drawing new examples from an unknown probability distribution. A model that can learn to represent and sample from a probability distribution is called \emph{generative}. For example, a generative model for images would learn to draw new examples of cats and dogs given a dataset of images of cats and dogs. Similarly, given samples generated from one phase of an Ising model we may want to generate new samples from that phase. Such tasks are clearly beyond the scope of discriminative models like the ensemble models and DNNs discussed so far in the review. Instead, we must turn to a new class of machine learning methods.

The goal of this section is to introduce the reader to \emph{energy-based} generative models. As we will see, energy-based models are closely related to the kinds of models commonly encountered in statistical physics. We will draw upon many techniques that have their origin in statistical mechanics (e.g.~Monte-Carlo methods). The section starts with a brief overview of generative models, highlighting the similarities and differences with the supervised learning methods encountered in earlier sections. Next, we introduce perhaps the simplest kind of generative models -- Maximum Entropy (MaxEnt) models. MaxEnt models have no latent (or hidden) variables, making them ideal for introducing the key concepts and tools that underlie energy-based generative models. We then present an extended discussion of how to train energy-based models. Much of this discussion will also be applicable to more complicated energy-based models such as Restricted Boltzmann Machines (RBMs) and the deep models discussed in the next section.
\subsection{An overview of energy-based generative models}

\begin{figure}[t]
\includegraphics[width=0.9\columnwidth]{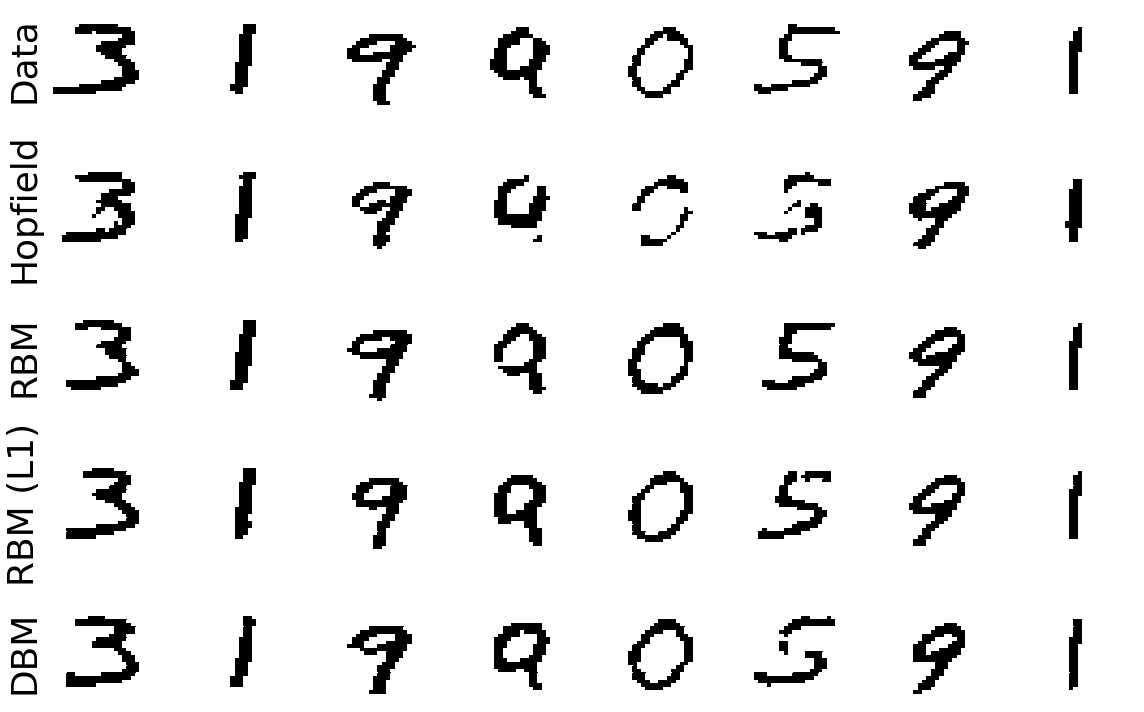}
\caption{Examples of handwritten digits (``reconstructions'') generated using various energy-based models using the powerful \emph{Paysage} package for unsupervised learning. Examples from top to bottom are: the original MNIST database,   an RBM with Gaussian units which is equivalent to a Hopfield Model, a Restricted Boltzmann Machine (RBM), a RBM with an $L_1$ penalty for regularization, and a Deep Boltzmann Machine (DBM) with $3$ layers. All models have $200$ hidden units. See Sec. \ref{sec:EnergyII} and corresponding notebook for details}
\label{fig:MNIST-generative}
\end{figure}

Generative models are a machine learning technique that allows to learn how to generate new examples similar to those found in a training dataset. The core idea of most generative models is to learn a parametric model for the probability distribution from which the data was drawn. Once we have learned a model, we can generate new examples by sampling from the learned generative model (see Fig.~\ref{fig:MNIST-generative}). As in statistical physics, this sampling is often done using Markov Chain Monte Carlo (MCMC) methods. A review of MCMC methods is beyond the scope of this discussion: for a concise and beautiful introduction to MCMC-inspired methods that bridges both statistical physics and ML the reader is encouraged to consult Chapters 29-32 of David MacKay's book \cite{mackay2003information} as well as the review by Michael I.~Jordan and collaborators~\cite{andrieu2003introduction}.

The added complexity of learning models directly from samples introduces many of the same fundamental tensions we encountered when discussing discriminative models. The ability to generate new examples requires models to be able to ``generalize'' beyond the examples they have been trained on, that is to generate new samples that are not samples of the training set. The models must be expressive enough to capture the complex correlations present in the underlying data distribution, but the amount of data we have is finite which can give rise to overfitting.

In practice, most generative models that are used in machine learning are flexible enough that, with a sufficient number of parameters, they can approximate any probability distribution. For this reason, there are three axes on which we can differentiate classes of generative models:
\begin{itemize}
\item The first axis is how easy the model is to train -- both in terms of computational time and the complexity of writing code for the algorithm.
\item  The second axis is how well the model generalizes from the training set to the test set. 
\item The third axis is which characteristics of the data distribution the model is capable of and focuses on capturing. 
\end{itemize}
All generative models must balance these competing requirements and generative models differ in the tradeoffs they choose. Simpler models capture less structure about the underlying distributions but are often easier to train. More complicated models can capture this structure but may overfit to the training data.

One of the fundamental reasons that energy-based models have been less widely-employed than their discriminative counterparts is that the training procedure for these models differs significantly from those for supervised neural networks models. Though both employ gradient-descent based procedures for minimizing a cost function (one common choice for generative models is the negative log-likelihood function), energy-based models do not use backpropagation (see Sec.~\ref{subsec:backprop}) and automatic differentiation for computing gradients. Rather, one must turn to ideas inspired by MCMC based methods in physics and statistics that sometimes go under the name ``Boltzmann Learning'' (discussed below). As a result, training energy-based models requires additional tools that are not immediately available in packages such as PyTorch and TensorFlow.

The open-source package -- {\it Paysage} -- that is built on top of PyTorch bridges this gap by providing the toolset for training energy-based models (Paysage is maintained by Unlearn.AI -- a company affiliated with two of the authors (CKF and PM)). Paysage makes it easy to quickly code and deploy energy-based models such as Restricted Boltzmann Machines (RBMs) and Stacked RBMs -- a ``deep'' unsupervised model. The package includes unpublished training methods that significantly improve the training performance, can be applied with various datatypes, and can be employed on GPUs. We make use of this package extensively in the next two sections and the accompanying Python notebooks. For example, Fig.~\ref{fig:MNIST-generative} (and the accompanying \href{https://physics.bu.edu/~pankajm/MLnotebooks.html}{Notebook 17}) show how the Paysage package can be used to quickly code and train a variety of energy-based models on the MNIST handwritten digit dataset.

Finally, we note that generative models at their most basic level are complex parametrizations of the probability distribution the data is drawn from. For this reason, generative models can do much more than just generate new examples. They can be used to perform a multitude of other tasks that require sampling from a complex probability distribution including ``de-noising'', filling in missing data, and even discrimination \cite{hinton2012practical}. The versatility of generative models is one of the major appeals of these unsupervised learning methods.

\subsection{Maximum entropy models: the simplest energy-based generative models}

Maximum Entropy (MaxEnt) models are one of the simplest classes of energy-based generative models.  MaxEnt models have their origin in a series of beautiful papers by Jaynes that reformulated statistical mechanics in information theoretic terms \cite{jaynes1957information, jaynes1957informationII}. Recently, the flood of new, large scale datasets has resulted in a resurgence of interest in MaxEnt models in many fields including physics (especially biological physics), computational neuroscience, and ecology \cite{schneidman2006weak, weigt2009identification, elith2011statistical}. MaxEnt models are often presented as the class of generative models that make the least assumptions about the underlying data. However, as we have tried to emphasize throughout the review, all ML and statistical models require assumptions, and MaxEnt models are no different. Overlooking this can sometimes lead to misleading conclusions, and it is important to be cognizant of these implicit assumptions \cite{schwab2014zipf, aitchison2016zipf}.

\subsubsection{MaxEnt models in statistical mechanics}

MaxEnt models were introduced by E.~T.~Jaynes in a two-part paper in 1957 entitled ``Information theory and statistical mechanics'' \cite{jaynes1957information, jaynes1957informationII}. In these incredible papers, Jaynes showed that it was possible to re-derive the Boltzmann distribution (and the idea of generalized ensembles) entirely from information theoretic arguments. Quoting from the abstract, Jaynes considered ``statistical mechanics as a form of statistical inference rather than as a physical theory'' (portending the close connection between statistical physics and machine learning). Jaynes showed that the Boltzmann distribution could be viewed as resulting from a statistical inference procedure for learning probability distributions describing physical systems where one only has partial information about the system (usually the average energy).

The key quantity in MaxEnt models is the information theoretic, or Shannon, entropy, a concept introduced by Shannon in his landmark treatise on information theory \cite{shannon1949communication}. The Shannon entropy quantifies the statistical uncertainty one has about the value of a random variable ${\bf x}$  drawn from a probability distribution $p({\bf x})$. The Shannon entropy of the distribution is defined as 
\be
S_p = - \mathrm{Tr}_{\bf x} p({\bf x}) \log p({\bf x})
\ee
where the trace is a sum/integral over all possible values a variable can take. Jaynes showed that the Boltzmann distribution follows from the Principle of Maximum Entropy. A physical system should be described by the probability distribution with the largest entropy subject to certain constraints (often provided by measuring the average value of conserved, extensive quantities such as the energy, particle number, etc.) The principle uniquely specifies a procedure for parametrizing the functional form of the probability distribution. Once we have specified and learned this form we can, of course, generate new examples by sampling this distribution.

Let us illustrate how this works in more detail. Suppose that we have chosen a set of functions $\{f_i({\bf x})\}$ whose average value we want to fix to some observed values $\langle f_i \rangle_\mathrm{obs}$. The Principle of Maximum Entropy states that we should choose the distribution $p({\bf x})$ with the largest uncertainty (i.e.~largest Shannon entropy $S_p$), subject to the constraints that the model averages match the observed averages:
\be
\langle f_i \rangle_\mathrm{model} = \int \mathrm{d} {\bf x} f_i({\bf x}) p({\bf x}) = \langle f_i \rangle_\mathrm{obs}.
\ee

We can formulate the Principle of Maximum Entropy as an optimization problem using the method of Lagrange multipliers by minimizing:
\begin{align}
\mathcal{L}[p] &=  -S_p + \sum_i \lambda_i \left(\langle f_i \rangle_\mathrm{obs} - \int \mathrm{d} {\bf x} f_i({\bf x}) p({\bf x})\right) \nonumber \\
&+ \gamma \left(1 - \int \mathrm{d} {\bf x} p({\bf x}) \right) \nonumber,
\end{align}
where the first set of constraints enforce the requirement for the averages and the last constraint enforces the normalization that the trace over the probability distribution equals one. We can solve for $p({\bf x})$ by taking the functional derivative and setting it to zero
\begin{align}
0=\frac{\delta \mathcal{L}}{ \delta p} = (\log p({\bf x}) + 1) -\sum_i \lambda_i f_i({\bf x}) - \gamma \nonumber.
\end{align}
The general form of the maximum entropy distribution is then given by
\begin{equation}
p({\bf x}) = \frac{1}{Z}\;  \mathrm e^{\;\sum_i \lambda_i f_i({\bf x})}
\label{eq:boltzmann-dist}
\end{equation}
where $Z({\lambda_i}) = \int \mathrm{d} {\bf x}\mathrm\;  e^{\sum_i \lambda_i f_i({\bf x})}$ is the partition function. 

The maximum entropy distribution is clearly just the usual Boltzmann distribution with energy $E({\bf x}) = - \sum_i \lambda_i f_i({\bf x})$.  The values of the Lagrange multipliers are chosen to match the observed averages for the set of functions  $\{f_i({\bf x})\}$ whose average value is being fixed:
\begin{equation}
\langle f_i \rangle_\mathrm{model} =  \int \mathrm{d} {\bf x} p({\bf x}) f_i({\bf x})= \frac{\partial \log Z}{\partial \lambda_i} =  \langle f_i \rangle_\mathrm{obs}.
\label{eq:moment-constraint}
\end{equation}
In other words, the parameters of the distribution can be chosen such that 
\be
\partial_{\lambda_i} \log Z = \langle f_i \rangle_\mathrm{data}.
\ee 

To gain more intuition for the MaxEnt distribution, it is helpful to relate the Lagrange multipliers to the familiar thermodynamic quantities we use to describe physical systems  \cite{jaynes1957information}.  Our $\mathbf{x}$ denotes the microscopic state of the system, i.e.~the MaxEnt distribution is a probability distribution over microscopic states. However, in thermodynamics we only have access to average quantities. If we know only the average energy $ \langle E(\mathbf{x}) \rangle_\mathrm{obs}$, the MaxEnt procedure tells us to maximize the entropy subject to the average energy constraint. This yields
\be
p({\mathbf x})=  \frac{1}{Z}\; \mathrm e^{- \beta  E(\mathbf x)},
\ee
where we have identified the Lagrange multiplier conjugate to the energy $\lambda_1=-\beta= 1/k_BT$ with the (negative) inverse temperature. Now, suppose we also constrain the particle number $\langle N(x) \rangle_\mathrm{obs}$. Then, an almost identical calculation yields a MaxEnt distribution of the functional form
\be
p({\mathbf x})=  \frac{1}{Z}\; \mathrm e^{-\beta( E({\mathbf x}) -\mu N(\mathbf{x}))},
\ee
where we have rewritten our Lagrange multipliers in the familiar thermodynamic notation $\lambda_1= -\beta$ and $\lambda_2 =  \mu/\beta$. Since this is just the Boltzmann distribution, we can also relate the partition function in our MaxEnt model to the thermodynamic free-energy via $F= -\beta^{-1} \log Z$. The choice of which quantities to constrain is equivalent to working in different thermodynamic ensembles.

\subsubsection{From statistical mechanics to machine learning}

The MaxEnt idea also provides a general procedure for learning a generative model from data. The key difference between MaxEnt models in (theoretical) physics and ML is that in ML we have no direct access to observed values $\langle f_i \rangle_\mathrm{obs}$. Instead, these averages must be directly estimated from data (samples). To denote this difference, we will call empirical averages calculated from data as $\langle f_i \rangle_\mathrm{data}$. We can think of MaxEnt as a statistical inference procedure simply by replacing $\langle f_i \rangle_\mathrm{obs}$ by  $\langle f_i \rangle_\mathrm{data}$ above. 

This subtle change has important implications for training MaxEnt models. First, since we do not know these averages exactly, but must estimate them from the data, our training procedures must be careful not to overfit to the observations (our samples might not be reflective of the true values of these statistics). Second, the averages of certain functions $f_i$ are easier to estimate from limited data than others. This is often an important consideration when formulating which MaxEnt model to fit to the data. Finally, we note that unlike in physics where conservation laws often suggest the functions $f_i$  whose averages we hold fix, ML offers no comparable guide for how to choose the $f_i$ we care about. For these reasons, choosing the $\{ f_i \}$ is often far from straightforward. As a final point, we note that here we have presented a physics-based perspective for justifying the MaxEnt procedure. We mention in passing that the MaxEnt in ML is also closely related to ideas from Bayesian inference \cite{jaynes1968prior, jaynes2003probability} and this latter point of view is more common in discussions of MaxEnt in the statistics and ML literature.

\subsubsection{Generalized Ising Models from MaxEnt}

The form of a MaxEnt model is completely specified once we choose the averages $\{f_i\}$ we wish to constrain. One common choice often used in MaxEnt modeling is to constrain the first two moments of a distribution. When our random variables $\bf{x}$ are continuous, the corresponding MaxEnt distribution is a multi-dimensional Gaussian. If the  $\bf{x}$ are binary (discrete), then the corresponding MaxEnt distribution is a generalized Ising (Potts) model with all-to-all couplings.

To see this, consider a random variable $\bf{x}$ with first and second moments $\langle x_i \rangle_\mathrm{data}$ and $\langle x_i x_j \rangle_\mathrm{data}$, respectively. According to the Principle of Maximum Entropy, we should choose to model this variable using a Boltzmann distribution with constraints on the first and second moments. Let $a_i$ be the Lagrange multiplier associated with $\langle x_i \rangle_\mathrm{data}$ and $J_{ij} / 2$ be the Lagrange multiplier associated with $\langle x_i x_j \rangle_\mathrm{data}$. Using Eq.~\eqref{eq:moment-constraint}, it is easy to verify that the energy function
\begin{equation}
E({\bf x}) = -\sum_i a_i x_i - \frac{1}{2} \sum_{ij} J_{ij} x_i x_j
\label{eq:pairwise-maxent}
\end{equation}
satisfies the above constraints.

Partition functions for maximum entropy models are often intractable to compute. Therefore, it is helpful to consider two special cases where ${\bf x}$ has different support (different kinds of data). First, consider the case that the random variables ${\bf x} \in \mathbb{R}^n$ are real numbers. In this case we can compute the partition function directly:
\begin{align}
Z 
&= \int \mathrm{d} {\bf x}\; \mathrm e^{{\bf a}^T {\bf x} + \frac{1}{2} {\bf x}^T J {\bf x} }
= \sqrt{(2 \pi)^n \text{det} J^{-1}} \mathrm e^{-\frac{1}{2} {\bf a}^T J^{-1} {\bf a}}.
\end{align}
The resulting probability density function is,
\begin{align}
p({\bf x}) &= Z^{-1} \mathrm e^{ -E({\bf x})} \nonumber \\
&=  \frac{1}{ \sqrt{(2 \pi)^n \text{det} J^{-1}}} \mathrm e^{\frac{1}{2} {\bf a}^T J^{-1} {\bf a} + {\bf a}^T {\bf x} + \frac{1}{2} {\bf x}^T J {\bf x} } \nonumber \\
&=  \frac{1}{ \sqrt{(2 \pi)^n \text{det} \Sigma }} \mathrm e^{-\frac{1}{2} ({\bf x} - {\bf \mu})^T \Sigma^{-1} ( {\bf x} - {\bf \mu}) },
\label{eq:gaussian-pdf}
\end{align}
where ${\bf \mu} = -J^{-1} {\bf a}$ and $\Sigma = -J^{-1}$. This, of course, is the normalized, multi-dimensional Gaussian distribution. 

Second, consider the case that the random variable ${\bf x}$ is binary with $x_i \in \{-1, +1\}$. The energy function takes the same form as Eq.~\eqref{eq:pairwise-maxent}, but the partition function can no longer be computed in a closed form. This model is known as the Ising model in the physics literature, and is often called a Markov Random Field in the machine learning literature. It is well known to physicists that calculating the partition function for the Ising Model is intractable. For this reason, the best we can do is estimate it using numerical techniques such MCMC methods or approximate methods like variational MFT methods, see Sec.~\ref{sec:varl_MFT}. Finally, we note that in ML it is common to use  binary variables which take on values in $x_i \in \{0,1\}$ rather than $\{\pm 1\}$. This can sometimes be a source of confusion when translating between ML and physics literatures and can lead to confusion when using ML packages for physics problems. 

\subsection{Cost functions for training energy-based models}

The MaxEnt procedure gives us a way of parametrizing an energy-based generative model. For any energy-based generative model, the energy function $E({\bf x}, \{\theta_i\})$ depends on some parameters $\theta_i$ -- couplings in the language of statistical physics -- that must be inferred directly from the data. For example, for the MaxEnt models the $\{ \theta_i\}$ are just the Lagrange multipliers $\{ \lambda_i\}$ introduced in the last section. The goal of the training procedure is to use the available training data to fit these parameters.

Like in many other ML techniques, we will fit these couplings by minimizing a cost function using stochastic gradient descent (cf.~Sec.~\ref{sec:gradient_descent}). Such a procedure naturally separates into two parts: choosing an appropriate cost function, and calculating the gradient of the cost function with respect to the model parameters. Formulating a cost function for generative models is a little bit trickier than for supervised, discriminative models. The objective of discriminative models is straightforward -- predict the label from the features. However, what we mean by a ``good'' generative model is much harder to define using a cost function. We would like the model to generate examples similar to those we find in the training dataset. However, we would also like the model to be able to generalize -- we do not want the model to reproduce ``spurious details'' that are particular to the training dataset. Unlike for discriminative models, there is no straightforward idea like cross-validation on the data labels that neatly addresses this issue. For this reason, formulating cost functions for generative models is subtle and represents an important and interesting open area of research. 

Calculating the gradients of energy-based models also turns out to be different than for discriminative models, such as deep neural networks. Rather than relying on automatic differentiation techniques and backpropagation (see Sec.~\ref{subsec:backprop}), calculating the gradient requires drawing on intuitions from MCMC-based methods. Below, we provide an in-depth discussion of Boltzmann learning for energy-based generative models, focusing on MaxEnt models. We put the emphasis on training procedures that generalize to more complicated generative models with latent variables such as RBMs discussed in the next section. Therefore, we largely ignore the incredibly rich physics-based literature on fitting Ising-like MaxEnt models (see the recent reviews  \cite{nguyen2017inverse,baldassi2018inverse} and references therein).

\subsubsection{Maximum likelihood }

By far the most common approach used for training a generative model is to maximize the log-likelihood of the training data set. Recall, that the log-likelihood characterizes the log-probability of generating the observed data using our generative model. By choosing the negative log-likelihood as the cost function, the learning procedure tries to find parameters that maximize the probability of the data. This cost function is intuitive and has been the work-horse of most generative modeling. However, we note that the Maximum Likelihood estimation (MLE) procedure has some important limitations that we will return to in Sec.~\ref{sec:vae}. 

In what follows, we employ a general notation that is applicable to all energy-based models, not just the MaxEnt models introduced above. The reason for this is that much of this discussion does not rely on the specific form of the energy function but only on the fact that our generative model takes a Boltzmann form. We denote the generative model by the probability distribution $p_\theta( {\bf x})$ and its corresponding partition function by $\log{Z(\{\theta_i\})}$. In MLE, the parameters of the model are fit by maximizing the log-likelihood:
\bea
\mathcal{L}(\{ \theta_i \}) &=& \langle \log \left( p_\theta( {\bf x}) \right) \rangle_\mathrm{data}  \nonumber \\
&=& -\langle E({\bf x}; \{ \theta_i\}) \rangle_\mathrm{data} - \log{Z(\{\theta_i\})},
\label{Eq:E1_likelihood}
\eea
where we have set $\beta=1$. In writing this expression we made use of two facts: (i) our generative distribution is of the Boltzmann form, and (ii) the partition function does not depend on the data:
\be
 \langle \log Z(\{\theta_i\}) \rangle_\mathrm{data}=\log{Z(\{\theta_i\})}.
\ee

\subsubsection{Regularization}

Just as for discriminative models like linear and logistic regression, it is common to supplement the log-likelihood with additional regularization terms (see Secs.~\ref{sec:lin_reg} and \ref{sec:log_reg}). Instead of minimizing the negative log-likelihood, one minimizes a cost function of the form
\be
-\mathcal{L}(\{ \theta_i \}) + {E}_\mathrm{reg}(\{\theta_i\}),
\ee
where ${E}_\mathrm{reg}(\{\theta_i\})$ is an additional regularization term that prevents overfitting. From a Bayesian perspective, this new term can be viewed as encoding a (negative) log-prior on model parameters and performing a maximum-a-posteriori (MAP) estimate instead of a MLE (see corresponding discussion in Sec.~\ref{sec:lin_reg}).

As we saw by studying linear regression, different forms of regularization give rise to different kinds of properties. A common choice for the regularization function are the sums of the $L_1$ or $L_2$ norms of the parameters
\be
E_\mathrm{reg}(\{\theta_i\})= \Lambda \sum_i |\theta_i|^\alpha, \, \alpha=1,2 
\ee
with $\Lambda$ controlling the regularization strength. For $\Lambda=0$, there is no regularization and we are simply performing MLE. In contrast, a choice of large $\Lambda$ will force many parameters to be close to or exactly zero. Just as in regression, an L1 penalty enforces sparsity, with many of the $\theta_i$ set to zero, and L2 regularization shrinks the size of the parameters towards zero.

One challenge of generative models is that it is often difficult to choose the regularization strength $\Lambda$.  Recall that, for linear and logistic regression, $\Lambda$ is chosen to maximize the out-of-sample performance on a validation dataset (i.e.~cross-validation). However, for generative models our data are usually unlabeled. Therefore, choosing a regularization strength is more subtle and there exists no universal procedure for choosing $\Lambda$. One common strategy is to divide the data into a training set and a validation set and monitor a summary statistic such as the log-likelihood, energy distance \cite{szekely2003statistics}, or variational free-energy of the generative model on the training and validation sets  (the variational free-energy was discussed extensively in Sec.~\ref{sec:varl_MFT} ) \cite{hinton2012practical}. If the gap between the training and validation datasets starts growing, one is probably overfitting the model even if the log-likelihood of the training dataset is still increasing. This also gives a procedure for ``early stopping'' -- a regularization procedure we introduced in the context of discriminative models. In practice, when using such regularizers it is important to try many different values of $\Lambda$ and then try to use a proxy statistic for overfitting to evaluate the optimal choice of $\Lambda$.

\subsection{Computing gradients }
\label{subsec:energy_gradients}

We still need to specify a procedure for minimizing the cost function. One powerful and common choice that is widely employed when training energy-based models is stochastic gradient descent (SGD)  (see Sec.~\ref{sec:gradient_descent}). Performing MLE using SGD requires calculating the gradient of the log-likelihood Eq.~\eqref{Eq:E1_likelihood}  with respect to the parameters $\theta_i$. To simplify notation and gain intuition, it is helpful to define ``operators'' $O_i(\mathbf{x})$, conjugate to the parameters $\theta_i$ 
\be
O_i(\mathbf{x}) = {\partial  E({\bf x}; \theta_i)  \over \partial \theta_i}.
\ee
Since the partition function is just the cumulant generating function for the Boltzmann distribution, we know that the usual statistical mechanics relationships between
expectation values and derivatives of the log-partition function hold:
\be
 \langle O_i (\mathbf{x}) \rangle_\mathrm{model}= \mathrm{Tr}_{\mathbf x} p_\theta(\mathbf{x}) O_i (\mathbf{x}) = -{ \partial \log{Z(\{\theta_i\})} \over \partial \theta_i}.
\ee
In terms of the operators $\{ O_i(\mathbf{x})\}$, the gradient of Eq.~\eqref{Eq:E1_likelihood}  takes the form \cite{ackley1987learning}
\bea
-{\partial \mathcal{L}(\{\theta_i \}) \over \partial \theta_i}&=& \Big\langle { \partial E({\bf x}; \theta_i) \over \partial \theta_i} \Big\rangle_\mathrm{data}+ {\partial \log{Z(\{\theta_i\})} \over \partial \theta_i} \nonumber \\
&=&  \langle O_i (\mathbf{x})  \rangle_\mathrm{data}-\langle O_i (\mathbf{x})  \rangle_\mathrm{model}.
\label{eq:Pos-neg-energy}
\eea

These equations have a simple and beautiful interpretation. The gradient of the log-likelihood with respect to a model parameter is a difference of moments -- one calculated directly from the data and one calculated from our model using the current model parameters. The data-dependent term is known as the \emph{positive phase} of the gradient and the model-dependent term is known as the \emph{negative phase} of the gradient. This derivation also gives an intuitive explanation for likelihood-based training procedures.  The gradient acts on the model to lower the energy of configurations that are near observed data points while raising the energy of configurations that are far from observed data points. Finally, we note that all information about the data only enters the training procedure through the expectations $\langle O_i (\mathbf{x})  \rangle_\mathrm{data}$ and our generative model is blind to information beyond what is contained in these expectations.

To use SGD, we must still calculate the expectation values that appear in Eq.~\eqref{eq:Pos-neg-energy}. The positive phase of the gradient -- the expectation values with respect to the data -- can be easily calculated using samples from the training dataset. However, the negative phase  -- the expectation values with respect to the model -- is generally much more difficult to compute. We will see that in almost all cases, we will have to resort to either numerical or approximate methods. The fundamental reason for this is that it is impossible to calculate the partition function exactly for most interesting models in both physics and ML.

There are exceptional cases in which we can calculate expectation values analytically. When this happens, the generative model is said to have a \emph{Tractable Likelihood}. One example of a generative model with a Tractable Likelihood is the Gaussian MaxEnt model for real valued data discussed in Eq.~\eqref{eq:gaussian-pdf}. The parameters/Lagrange multipliers for this model are the local fields $\mathbf{a}$ and the pairwise coupling matrix $J$. In this case, the usual manipulations involving Gaussian integrals allow us to exactly find the parameters $\mu=-J^{-1}\mathbf{a}$ and $\Sigma=-J^{-1}$, yielding the familiar expressions  ${\bf \mu} = \langle {\bf x} \rangle_\mathrm{data}$ and $\Sigma = \langle ({\bf x} -  \langle {\bf x} \rangle_\mathrm{data}) ({\bf x} -  \langle {\bf x} \rangle_\mathrm{data})^T \rangle_\mathrm{data}$. These are the standard estimates for the sample mean and covariance matrix. Converting back to the Lagrange multipliers yields 
\be
J = - \langle ({\bf x} -  \langle {\bf x} \rangle_\mathrm{data}) ({\bf x} -  \langle {\bf x} \rangle_\mathrm{data})^T \rangle_\mathrm{data}^{-1}.
\ee 

Returning to the generic case where most energy-based models have \emph{intractable likelihoods}, we must estimate expectation values numerically. One way to do this is draw  samples  $\mathcal{S}_\mathrm{model}=\{ \mathbf{x}_i ' \}$ from the model $p_\theta(\mathbf{x})$ and evaluate arbitrary expectation values using these samples:
\be
\langle f(\mathbf{x}) \rangle_\mathrm{model} = \int \mathrm{d}\mathbf{x} p_\theta(\mathbf{x}) f(\mathbf{x}) \approx \sum_{\mathbf{x}_i '\in \mathcal{S}_\mathrm{model}} f(\mathbf{x}_i').
\ee
The samples from the model $\mathbf{x}_i ' \in \mathcal{S}_\mathrm{model}$ are often referred to as \emph{fantasy particles} in the ML literature and can be generated using simple MCMC algorithms such as Metropolis-Hasting which are covered in most modern statistical physics classes. However, if the reader is unfamiliar with MCMC methods or wants a quick refresher, we recommend the concise and beautiful discussion of MCMC methods from both the physics and ML point-of-view in Chapters 29-32 of David MacKay's masterful book \cite{mackay2003information}.

Finally, we note that once we have the fantasy particles from the model, we can also easily calculate the gradient of any expectation value $\langle f(\mathbf{x}) \rangle_\mathrm{model}$ using what is commonly called the ``log-derivative trick'' in ML \cite{kleijnen1996optimization,fu2006gradient}:
\bea
{\partial \over \partial \theta_i}\langle f(\mathbf{x}) \rangle_\mathrm{model} &=& \int \mathrm{d}\mathbf{x} { \partial p_\theta(\mathbf{x}) \over \partial \theta_i} f(\mathbf{x}) \nonumber\\
&=& \Big \langle {\partial \log{ p_\theta(\mathbf{x})} \over \partial \theta_i}  f(\mathbf{x}) \Big \rangle_\mathrm{model} \nonumber \\
&=& \langle O_i(\mathbf{x})f(\mathbf{x})\rangle_\mathrm{model}\nonumber \\
&\approx& \sum_{\mathbf{x}_j' \in \mathcal{S}_\mathrm{model}} O_i(\mathbf{x}_j)f(\mathbf{x}_j').
\eea
This expression allows us to take gradients of more complex cost functions beyond the MLE procedure discussed here.

\subsection{Summary of the training procedure}

We now summarize the discussion above and present a general procedure for training an energy based model using SGD on the cost function (see Sec.~\ref{sec:gradient_descent}). Our goal is to fit the parameters of a model $p_{{\bf \lambda}}( \{\theta_i\} ) = Z^{-1} \mathrm e^{-E({\bf x}, \{\theta_i\})}$. Training the model involves the following steps:
\begin{enumerate}
\item Read a minibatch of data, $\{ {\bf x} \}$.
\item Generate fantasy particles $\{ {\bf x}' \}  \sim p_{{\bf \lambda}}$ using an MCMC algorithm (e.g., Metropolis-Hastings).  
\item Compute the gradient of log-likelihood using these samples and Eq.~\eqref{eq:Pos-neg-energy}, where the averages are taken over the minibatch of data and the fantasy particles from the model, respectively. 
\item Use the gradient as input to one of the gradient based optimizers discussed in section Sec.~\ref{sec:gradient_descent}.
\end{enumerate}
In practice, it is helpful to supplement this basic procedure with some tricks that help training. As with discriminative neural networks, it is important to initialize the parameters properly and print summary statistics during the training procedure on the training and validation sets to prevent overfitting. These and many other ``cheap tricks'' have been nicely summarized in a short note from the Hinton group \cite{hinton2012practical}.

A major computational and practical limitation of these methods is that it is often hard to draw samples from generative models. MCMC methods often have long mixing-times (the time one has to run the Markov chain to get uncorrelated samples) and this can result in biased sampling. Luckily, we often do not need to know the gradients exactly for training ML models (recall that noisy gradient estimates often help the convergence of gradient descent algorithms), and we can significantly reduce the computational expense by running MCMC for a reasonable time window. We will exploit this observation extensively in the next section when we discuss how to train more complex energy-based models with hidden variables.

\section{Deep Generative Models: Hidden Variables and Restricted Boltzmann Machines (RBMs)}
\label{sec:EnergyII}

The last section introduced many of the core ideas behind energy-based generative models.  Here, we extend this discussion to energy-based models that include latent or hidden variables. 

Including latent variables in generative models greatly enhances their expressive power -- allowing the model to represent sophisticated correlations between visible features without sacrificing trainability. By having multiple layers of latent variables, we can even construct powerful deep generative models that possess many of the same desirable properties as deep, discriminative neural networks.

We begin with a discussion that tries to provide a simple intuition for why latent variables are such a powerful tool for generative models. Next, we introduce a powerful class of latent variable models called Restricted Boltzmann Machines (RBMs) and discuss techniques for training these models. After that, we introduce Deep Boltzmann Machines (DBMs), which have multiple layers of latent variables. We then introduce the new Paysage package for training energy-based models and demonstrate how to use it on the MNIST dataset and samples from the Ising model. We conclude by discussing recent physics literature related to energy-based generative models.

\subsection{Why hidden (latent) variables?}

Latent or hidden variables are a powerful yet elegant way to encode sophisticated correlations between observable features. The underlying reason for this is that marginalizing over a subset of variables  -- ``integrating out'' degrees of freedom in the language of physics -- induces complex interactions between the remaining variables. The idea that integrating out variables can lead to complex correlations is a familiar component of many physical theories. For example, when considering free electrons living on a lattice, integrating out phonons gives rise to higher-order electron-electron interactions (e.g.~superconducting or magnetic correlations). More generally, in the Wilsonian renormalization group paradigm, all effective field theories can be thought of as arising from integrating out high-energy degrees of freedom \cite{wilson1974renormalization}.

Generative models with latent variables run this logic in reverse -- encode complex interactions between visible variables by introducing additional, hidden variables that interact with visible degrees of freedom in a simple manner, yet still reproduce the complex correlations between visible degrees in the data once marginalized over (integrated out). This allows us to encode complex higher-order interactions between the visible variables using simpler interactions at the cost of introducing new latent variables/degrees of freedom. This trick is also widely exploited in physics  (e.g.~in the Hubbard-Stratonovich transformation \cite{hubbard1959calculation,stratonovich1957method} or the introduction of ghost fields in gauge theory \cite{faddeev1967feynman}).

To make these ideas more concrete, let us revisit the pairwise Ising model introduced in the discussion of MaxEnt models, see Eq.~\eqref{eq:pairwise-maxent}. The model is described
by a Boltzmann distribution with energy 
\be
E({\bf v}) = - \sum_i a_i v_i - \frac{1}{2}  \sum_{ij} v_i J_{ij}v_j,
\ee
where $J_{ij}$ is a symmetric coupling matrix that encodes the pairwise constraints and $a_i$ enforce the single-variable constraint.

Our goal is to replace the complicated interactions between the visible variables $v_i$ encoded by $J_{ij}$, by interactions with a new set of latent variables $h_\mu$. In order to do this, it is helpful to rewrite the coupling matrix in a slightly different form. Using SVD, we can always express the coupling matrix in the form $J_{ij} = \sum_{\mu=1}^N  W_{i \mu} W_{j \mu}$, where $\{W_{i\mu}\}_i$ are appropriately normalized singular vectors. In terms of $W_{i \mu}$, the energy takes the form
\begin{equation}
E_\mathrm{Hop}({\bf v}) =  - \sum_i a_i v_i - \frac{1}{2}  \sum_{ij\mu} v_i W_{i \mu} W_{j \mu} v_j.
\label{eq:hopfield-energy}
\end{equation}

We note that in the special case when both $v_i \in \{-1, +1\}$ and $W_{i \mu} \in \{-1, +1 \}$ are binary variables, a model with this form of the energy function is known as the \emph{Hopfield model} \cite{hopfield1982neural, amit1985spin}. The Hopfield model has played an extremely important role in statistical physics, computational neuroscience, and machine learning, and a full discussion of its properties is well beyond the scope of this review [see \cite{amit1992modeling} for a beautiful discussion that combines all these perspectives]. Therefore, here we refer to all energy functions of the form Eq.~\eqref{eq:hopfield-energy} as (generalized) Hopfield models, even for the case when the $W_{i \mu}$ are continuous variables.

We now ``decouple'' the visible variables $v_i$ by introducing a set of normally, distributed continuous latent variables $h_{\mu}$ (in condensed matter language we perform a Hubbard-Stratonovich transformation). Using the usual identity for Gaussian integrals, we can rewrite the Boltzmann distribution for the generalized Hopfield model as 
\bea
p(\mathbf{v}) &=& {\mathrm e^{ \sum_i a_i v_i  + \frac{1}{2} \sum_{ij\mu} v_i W_{i \mu} W_{j \mu} v_j} \over Z} \nonumber \\
&=& { \mathrm e^{\sum_{i} a_i v_i }\prod_\mu \int \mathrm{d}h_\mu \mathrm e^{- \frac{1}{2} \sum_{\mu} h_{\mu}^2 - \sum_{i} v_i W_{i \mu} h_{\mu}} \over Z} \nonumber \\
&= &\int \mathrm{d}\mathbf{h}\; \mathrm e^{-E({\mathbf v}, {\mathbf h})} \over Z
\label{eq:HS-hopfield}
\eea
where $E( {\bf v}, {\bf h} )$ is a joint energy functional of both the latent and visible variables of the form
\be
E( {\bf v}, {\bf h} ) = - \sum_{i} a_i v_i + \frac{1}{2} \sum_{\mu} h_{\mu}^2 - \sum_{i \mu} v_i W_{i \mu} h_{\mu}.
\label{eq:hopfield-rbm}
\ee
We can also use the energy function $E({\mathbf v}, {\mathbf h})$ to define a new energy-based model $p({\bf v}, {\bf h})$ on both the latent and visible variables
\be
p({\bf v}, {\bf h})={\mathrm e^{-E({\mathbf v}, {\mathbf h})} \over Z'}.
\ee
Marginalizing over latent variables of course gives us back the generalized Hopfield model \cite{barra2012equivalence}
\be
p(\mathbf{v})= \int d{\bf h} p({\bf v}, {\bf h})={\mathrm e^{-E_\mathrm{Hop}({\bf v})} \over Z}.
\ee

Notice that $E( {\bf v}, {\bf h} )$ contains no direct interactions between visible degrees of freedom (or between hidden degree of freedom). Instead, the complex correlations between the $v_i$ are encoded in the interaction between the visible $v_i$ and latent variables $h_\mu$. It turns out that the model presented here is a special case of a more general class of powerful energy-based models called Restricted Boltzmann Machines (RBMs).

\subsection{Restricted Boltzmann Machines (RBMs)}
\label{sec:RBMs}

\begin{figure}[t]
\includegraphics[width=1.0\columnwidth]{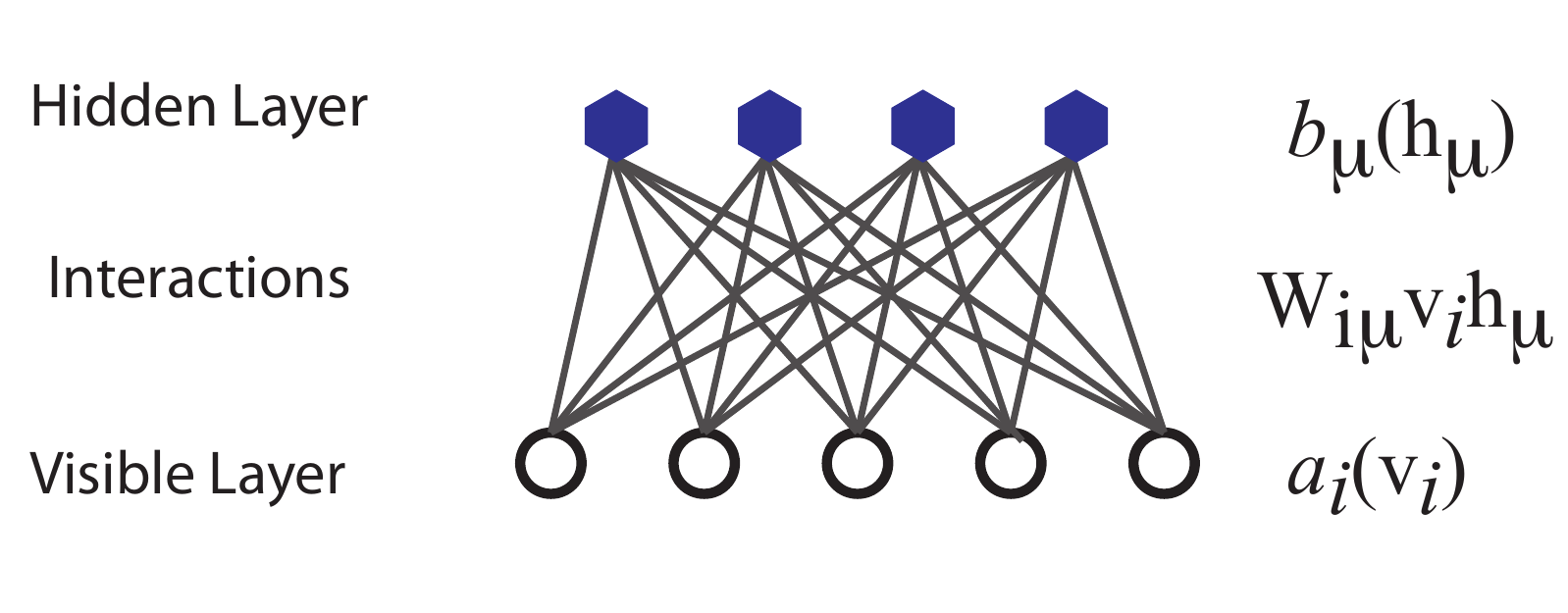}
\caption{A Restricted Boltzmann Machine (RBM) consists of visible units $v_i$ and hidden units $h_\mu$ that interact
with each other through interactions of the form $W_{i\mu}v_i h_\mu$. Importantly, there are no interactions between visible units themselves or hidden units themselves. }
\label{fig:RBM-architecture}
\end{figure}

A Restricted Boltzmann Machine (RBM) is an energy-based model with both visible and hidden units where the visible and hidden units interact with each other but do not interact among themselves. The energy function of an RBM takes the general functional form
\begin{equation}
E({\bf v}, {\bf h}) = -\sum_i a_i( v_i) - \sum_{\mu} b_{\mu}(h_{\mu}) - \sum_{i \mu} W_{i \mu} v_i h_{\mu},
\label{eq:rbm}
\end{equation}
where $a_i(\cdot)$ and $b_{\mu}(\cdot)$ are functions that we are free to choose. The most common choice is:
\[
    a_i( v_i)= 
\begin{cases}
    a_i v_i,& \text{if } v_i \in \{ 0, 1\} \text{ is binary} \\
   {v_i^2 \over 2 \sigma_i^2}, & \text{if } v_i \in \mathbb{R} \text{ is continuous},
\end{cases}
\]
and
\[
    b_\mu( h_\mu)= 
\begin{cases}
    b_\mu h_\mu ,& \text{if } h_\mu \in \{ 0, 1\} \text{ is binary} \\
   {h_\mu^2 \over 2 \sigma_\mu^2}, & \text{if } h_\mu \in \mathbb{R} \text{ is continuous}.
\end{cases}
\]
For this choice of $a_i(\cdot)$ and $b_{\mu}(\cdot)$, layers consisting of discrete binary units are often called Bernoulli layers, and layers consisting of continuous variables are often called Gaussian layers. The basic bipartite structure of an RBM -- i.e., a visible and hidden layer that interact with each other but not among themselves -- is often depicted using a graph of the form shown in Fig. \ref{fig:RBM-architecture}.

An RBM can have different properties depending on whether the hidden and visible layers are taken to be Bernoulli or Gaussian. The most common choice is to have both the visible and hidden units be Bernoulli. This is what is typically meant by an RBM. However, other combinations are also possible and used in the ML literature. When all the units are continuous, the RBM reduces to a multi-dimensional Gaussian with a very particular correlation structure. When the hidden units are continuous and the visible units are discrete, the RBM is equivalent to a generalized Hopfield model (see discussion above). When the the visible units are continuous and the hidden units are discrete, the RBM is often called a Gaussian Bernoulli Restricted Boltzmann Machine \cite{dahl2010phone, hinton2006reducing}. It is even possible to perform multi-modal learning with a mixture of continuous and discrete variables. For all these architectures, the important point is that all interactions occur only between the  visible and hidden units and there are no interactions between units within the hidden or visible layers, see Fig.~\ref{fig:RBM-architecture}. This is analogous to Quantum Electrodynamics, where a free fermion and a free photon interact with one another but not among themselves.

Specifying a generative model with this bipartite interaction structure has two major advantages: (i) it enables capturing both pairwise {\it and higher-order} correlations between the visible units and (ii) it makes it easier to sample from the model using an MCMC method known as block Gibbs sampling, which in turn makes the model easier to train.  
 
Before discussing training, it is worth better understanding the kind of correlations that can be captured using an RBM. To do so, we can marginalize over the hidden units and ask about the resulting distribution over just the visible units 
\be
p({\bf v})=\int \mathrm{d}{\bf h} p(\mathbf{v}, \mathbf{h}) =\int  \mathrm{d}{\bf h} {\mathrm e^{-E({\bf v}, {\bf h})} \over Z} 
\ee
where the integral should be replaced by a trace in all expressions for discrete units. 

We can also define a marginal energy using the expression
\be
p({\bf v})={\mathrm e^{-E({\bf v})} \over Z}.
\ee
Combining these equations,
\begin{align}
&E({\bf v}) = - \log \int \mathrm d {\bf h} \mathrm e^{-E({\bf v}, {\bf h})} \nonumber \\
&= - \sum_i a_i (v_i) - \sum_{\mu} \log \int \mathrm d h_{\mu} \mathrm e^{ b_{\mu} (h_{\mu}) + \sum_i v_i W_{i \mu} h_{\mu}} \nonumber 
\end{align}
To understand what correlations are captured by $p({\bf v})$ it is helpful to  introduce the distribution 
\be
q_{\mu}(h_{\mu}) = { \mathrm e^{ b_{\mu}(h_{\mu})} \over Z}
\ee
of hidden units $h_{\mu}$, ignoring the interactions between ${\bf v}$ and ${\bf h}$, and the cumulant generating function
\begin{equation}
K_{\mu}(t) = \log \int \mathrm d h_{\mu} q_{\mu}(h_{\mu}) \mathrm e^{t h_{\mu}} = \sum_{n} \kappa_{\mu}^{(n)} \frac{t^n}{n!}.
\end{equation}
$K_{\mu}(t)$ is defined such that the $n^{th}$ cumulant is $\kappa_{\mu}^{(n)} = \partial_t^{n} K_{\mu}\vert_{t=0}$. 

The cumulant generating function appears in the marginal free-energy of the visible units, which can be rewritten (up to a constant term) as:
\begin{eqnarray}
E({\bf v}) &=& - \sum_i a_i (v_i) - \sum_{\mu} K_{\mu}\left(\sum_i W_{i \mu} v_i\right) \nonumber \\
&=& - \sum_i a_i (v_i) - \sum_{\mu} \sum_{n} \kappa_{\mu}^{(n)} \frac{(\sum_i W_{i \mu} v_i)^n}{n!} \nonumber \\
&=& - \sum_i a_i (v_i) - \sum_i \left( \sum_{\mu} \kappa_{\mu}^{(1)}  W_{i \mu} \right) v_i \nonumber \\
&&- \frac{1}{2}  \sum_{ij}  \left(\sum_{\mu} \kappa_{\mu}^{(2)}W_{i \mu} W_{j \mu} \right) v_i v_j + \ldots
\label{eq:cumulant-rbm}
\end{eqnarray}
We see that the marginal energy includes all orders of interactions between the visible units, with the $n$-th order cumulants of $q_{\mu}(h_{\mu})$ weighting the $n$-th order interactions between the visible units. In the case of the Hopfield model we discussed previously, $q_{\mu}(h_{\mu})$ is a standard Gaussian distribution where the mean is $\kappa_{\mu}^{(1)} = 0$, the variance is $\kappa_{\mu}^{(2)} = 1$, and all higher-order cumulants are zero. Plugging these cumulants into Eq.~\eqref{eq:cumulant-rbm} recovers Eq.~\eqref{eq:hopfield-rbm}.

These calculations make clear the underlying reason for the incredible representational power of RBMs with a Bernoulli hidden layer. Each hidden unit can encode interactions of arbitrarily high order. By combining many different hidden units, we can encode very complex interactions at all orders. Moreover, we can learn which order of correlations/interactions are important directly from the data instead of having to specify them ahead of time as we did in the MaxEnt models. This highlights the power of  generative models with even the simplest interactions between visible and latent variables to encode, learn, and represent complex correlations present in the data.

\subsection{Training RBMs}

RBMs are a special class of energy-based generative models, which can be trained using the Maximum Likelihood Estimation (MLE) procedure described in detail in Sec.~\ref{sec:EnergyI}. To briefly recap, first, we must choose a cost function -- for MLE this is just the negative log-likelihood with or without an additional regularization term to prevent overfitting. We then minimize this cost function using one of the Stochastic Gradient Descent (SGD) methods described in Sec.~\ref{sec:gradient_descent}.

The gradient itself can be calculated using Eq.~\eqref{eq:Pos-neg-energy}. For example, for the Bernoulli-Bernoulli RBM in Eq.~\eqref{eq:rbm} we have
\bea
{\partial \mathcal{L}(\{W_{i \mu}, a_i, b_\mu \}) \over \partial  W_{i \mu}}  &=&  \langle v_i h_{\mu} \rangle_\mathrm{data} - \langle v_i h_{\mu} \rangle_\mathrm{model} \nonumber \\ 
{\partial \mathcal{L}(\{W_{i \mu}, a_i, b_\mu \}) \over \partial  a_i}  &=& \langle v_i \rangle_\mathrm{data} - \langle v_i  \rangle_\mathrm{model} \nonumber \\
{\partial \mathcal{L}(\{W_{i \mu}, a_i, b_\mu \}) \over \partial  b_\mu}  &=& \langle h_\mu \rangle_\mathrm{data} - \langle h_\mu  \rangle_\mathrm{model},
\eea
where the positive expectation with respect to the data is understood to mean sampling from the model while clamping the visible units to their observed values in the data. As before, calculating the negative phase of the gradient (i.e.~the expectation value with respect to the model) requires that we draw samples from the model. Luckily, the bipartite form of the interactions in RBMs were specifically chosen with this in mind.

\subsubsection{Gibbs sampling and contrastive divergence (CD)}

\begin{figure}[t]
\includegraphics[width=1.0\columnwidth]{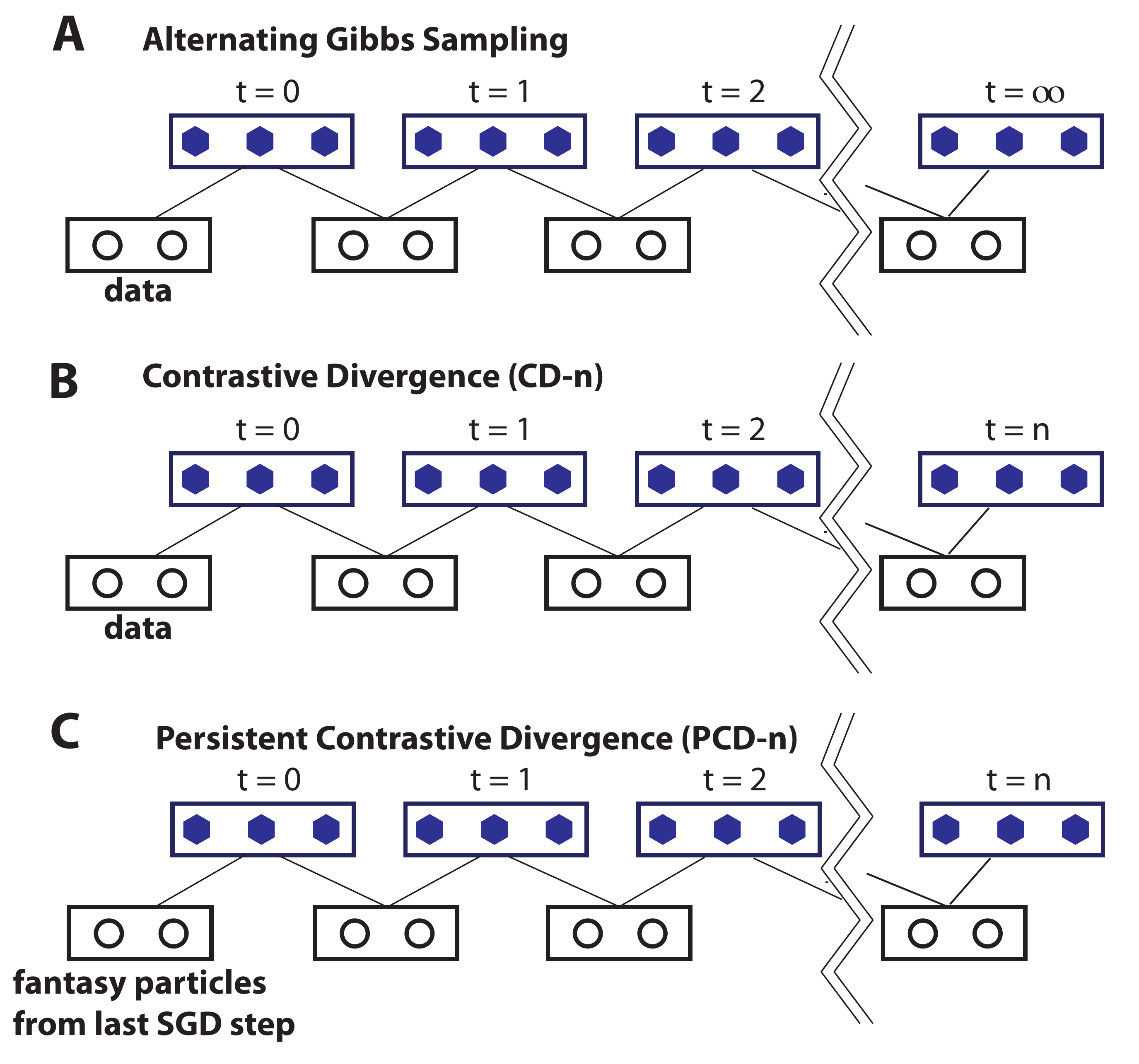}
\caption{ (Top) To draw fantasy particles (samples from the model) we can perform alternating (block) Gibbs sampling between the visible and hidden layers starting
with a sample from the data using the marginal distributions $p(\mathbf{h}|\mathbf{v})$ and $p(\mathbf{v}| \mathbf{h})$. The ``time'' $t$ corresponds to the time in the
Markov chain for the Monte Carlo and measures the number of passes between the visible and hidden states. (Middle) In Contrastive Divergence (CD), we approximately
sample the model by terminating the Gibbs sampling after $n$ steps (CD-$n$) starting from the data. (C) In Persistent Contrastive Divergence (PCD), instead of restarting
the Gibbs sampler from the data, we initialize the sampler with the fantasy particles calculated from the model at the last SGD step.}
\label{fig:CD-PCD}
\end{figure}

The bipartite interaction structure of an RBM makes it possible to calculate expectation values using a  Markov Chain Monte Carlo (MCMC) method known as Gibbs sampling. The key reason for this is that since there are no interactions of visible units with themselves or hidden units with themselves, the visible and hidden units of an RBM are conditionally independent:
 \begin{align}
 p({\bf v} | {\bf h}) &=  \prod_i p(v_i |{\bf h}) \nonumber \\
p({\bf h} | {\bf v}) &=  \prod_\mu p(h_\mu|{\bf v}),
 \end{align}
with
\begin{eqnarray}
\label{eq:RBM-conditional}
p(v_i = 1 | {\bf h}) &=& \sigma(a_i + \sum_{\mu} W_{i \mu} h_{\mu}) \\
p(h_{\mu} = 1 | {\bf v}) &=& \sigma(b_{\mu}+ \sum_{i} W_{i \mu} v_{i}) \nonumber
\end{eqnarray}
and where $\sigma(z) = 1 / (1 + \mathrm e^{-z})$ is the sigmoid function. 

Using these expressions it is easy to compute expectation values with respect to the data. The input to gradient descent is a minibatch of observed data. For each sample in the minibatch, we simply clamp the visible units to the observed values and apply Eq.~\eqref{eq:RBM-conditional} using the probability for the hidden variables. We then average over all samples in the minibatch to calculate expectation values with respect to the data. To calculate expectation values with respect to the model, we use (block) Gibbs sampling. The idea behind (block) Gibbs sampling is to iteratively sample from the conditional distributions ${\bf h}_{t+1} \sim p({\bf h} | {\bf v}_{t})$ and ${\bf v}_{t+1} \sim p({\bf v} | {\bf h}_{t+1})$ (see Figure \ref{fig:CD-PCD}, top). Since the units are conditionally independent, each step of this iteration can be performed by simply drawing random numbers.  The samples are guaranteed to converge to the equilibrium distribution of the model in the limit that $t \rightarrow \infty$. At the end of the Gibbs sampling procedure, one ends up with a minibatch of samples (fantasy particles).

One drawback of Gibbs sampling is that it may take many back and forth iterations to draw an independent sample. For this reason, the Hinton group introduced an approximate Gibbs sampling technique called Contrastive Divergence (CD) \cite{hinton2002training, hinton2006fast}. In CD-$n$, we just perform $n$ iterations of (block) Gibbs sampling, with $n$ often taken to be as small as 1 (see Figure \ref{fig:CD-PCD})! The price for this truncation is, of course, that we are not drawing samples from the true model distribution. But for our purpose -- using the expectations to estimate the gradient for SGD -- CD-$n$ has proven to work reasonably well. As long as the approximate gradients are reasonably correlated with the true gradient, SGD will move in a reasonable direction. CD-$n$ of course does come at a price.  Truncating the Gibbs sampler prevents sampling far away from the starting point, which for CD-$n$ are the data points in the minibatch.  Therefore, our generative model will be much more accurate around regions of feature space close to our training data. Thus, as is often the case in ML, CD-$n$ sacrifices the ability to generalize to some extent in order to make the model easier to train. 

Some of these undesirable features can be tempered by using a slightly different variant of CD called Persistent Contrastive Divergence (PCD) \cite{tieleman2009using}. In PCD, rather than restarting the Gibbs sampler from the data at each gradient descent step, we start the Gibbs sampling at the fantasy particles in the last gradient descent step (see Fig.~\ref{fig:CD-PCD}). Since parameters change slowly compared to the Gibbs sampling, samples that are high probability at one step of the SGD are also likely to be high probability at the next step. This ensures that PCD does not introduce large errors in the estimation of the gradients. The advantage of using fantasy particles to initialize the Gibbs sampler is to allow PCD to explore parts of the feature space that are much further from the training dataset than one could reach with ordinary CD. 

We note that, in applications using RBMs as a variational ansatz for quantum states, Gibbs sampling is not necessarily the best option for training, and in practice parallel tempering or other Metropolis schemes can outperform Gibbs sampling. In fact, Gibbs sampling is not even feasible with complex-valued weights required for quantum wavefucntions, whereas Metropolis schemes might be feasible~\cite{carleo_private}.

\subsubsection{Practical Considerations}
\label{subsubsec:energy_practical}

The previous section gave an overview of how to train RBMs. However, there are many ``tricks of the trade'' that are missing from this discussion. Luckily, a succinct summary of these has been compiled by Geoff Hinton and published as a note that readers interested in training RBMs are urged to consult  \cite{hinton2012practical}. 

For completeness, we briefly list some of the important points here:
\begin{itemize}
\item {\bf Initialization.---}The model must be initialized. Hinton suggests taking the weights $W_{i\mu}$ from a Gaussian with mean zero and standard deviation $\sigma= 0.01$ \cite{hinton2012practical}. An alternative initialization scheme proposed by Glorot and Bengio instead chooses the standard deviation to scale with the size of the layers: $\sigma = 2/\sqrt{N_v +N_h}$ where $N_v$ and $N_h$ are number of visible and hidden units respectively \cite{glorot2010understanding}. The bias of the hidden units is initialized to zero while the bias of the visible units is typically taken to be inversely proportional to the mean activation, $a_i= \langle v_i \rangle_\mathrm{data}^{-1}$.
\item {\bf Regularization.}---One can of course use an L1 or L2 penalty, typically only on the weight parameters, not the biases. Alternatively, Dropout has been shown to decrease overfitting when training with CD and PCD, which results in more interpretable learned features.
\item{\bf Learning Rates.}---Typically, it is helpful to reduce the learning rate in later stages of training.
\item{\bf Updates for CD and PCD.}---There are several computational tricks one can use for speeding up the alternating updates in CD and PCD, see Section 3 in~\cite{hinton2012practical}.
\end{itemize}


\subsection{Deep Boltzmann Machine}

\begin{figure}[t!]
\includegraphics[width=1.0\columnwidth]{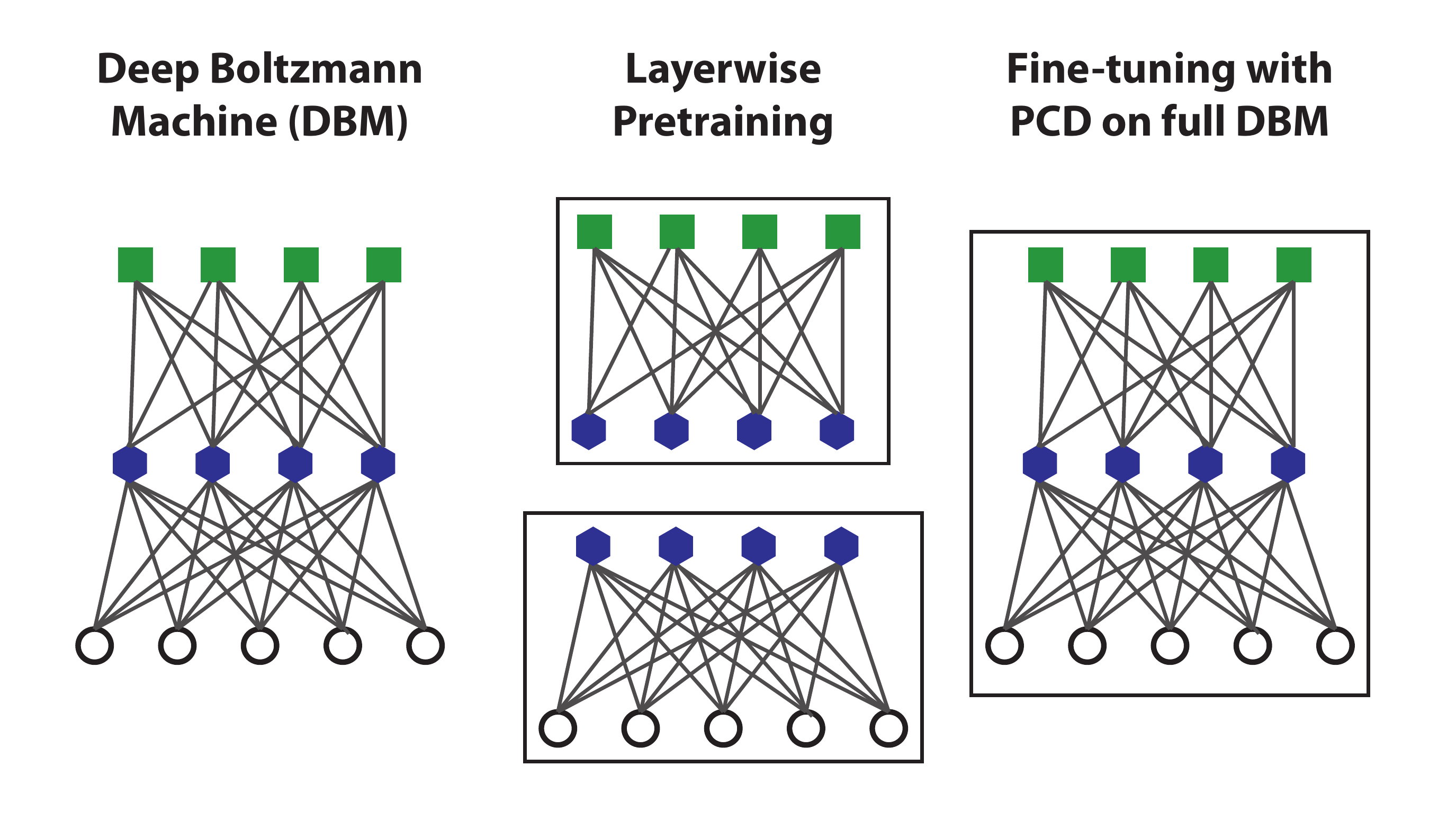}
\caption{Deep Boltzmann Machine contain multiple hidden layers. To train deep networks, first we perform layerwise training where each two layers are treated as a RBM. This
can be followed by fine-tuning using gradient descent and persistent contrastive divergence (PCD).}
\label{fig:DBM}
\end{figure}

In this section, we introduce Deep Boltzmann Machines (DBMs). Unlike RBMs, DBMs possess multiple hidden layers and were the first  models rebranded as ``deep learning'' \cite{hinton2006fast,hinton2006reducing} \footnote{Technically, these were Deep Belief Networks where only the top layer was undirected}. Many of the advantages that are thought to stem from having deep layers were already discussed in Sec.~\ref{sec:DNN-III} in the context of discriminative DNNs. Here, we revisit many of the same themes with emphasis on energy-based models.

An RBM is composed of two layers of neurons that are connected via an undirected graph, see Fig.~\ref{fig:RBM-architecture}. As a result, it is possible to perform sampling ${\bf v} \sim p({\bf v} | {\bf h})$ and inference ${\bf h} \sim p({\bf h} |{\bf v})$ with the same model. As with the Hopfield model, we can view each of the hidden units as representative of a pattern, or feature, that could be present in the data. \footnote{In general, one should instead think of activity patterns of hidden units representing features in the data.} The inference step involves assigning a probability to each of these features that expresses the degree to which each feature is present in a given data sample. In an RBM, hidden units do not influence each other during the inference step, i.e.~hidden units are conditionally independent given the visible units. There are a number of reasons why this is unsatisfactory. One reason is the desire for sparse distributed representations, where each observed visible vector will strongly activate a few (i.e.~more than one but only a very small fraction) of the hidden units. In the brain, this is thought to be achieved by inhibitory lateral connections between neurons. However, adding lateral intra-layer connections  between the hidden units  makes the distribution difficult to sample from, so we need to come up with another way of creating connections between the hidden units. 

With the Hopfield model, we saw that pairwise linear connections between neurons can be mediated through another layer. Therefore, a simple way to allow for effective connections between the hidden units is to add another layer of hidden units. Rather than just having two layers, one visible and one hidden, we can add additional layers of latent variables to account for the correlations between hidden units. Ideally, as one adds more and more layers, one might hope that the correlations between hidden variables become smaller and smaller deeper into the network. This basic logic is reminiscent of renormalization procedures that seek to decorrelate layers at each step \cite{mehta2014exact, vidal2007entanglement, li2018neural}. The price of adding additional layers is that the models become harder to train.

Training DBMs is more subtle than RBMs due to the difficulty of propagating information from visible to hidden units.  However, Hinton and collaborators realized that some of these problems could be alleviated via a layerwise procedure. Rather than attempting to the train the whole DBM at once, we can think of the DBM as a stack of RBMs (see Fig.~\ref{fig:DBM}). One first trains the bottom two layers of the DBM  -- treating it as if it is a stand-alone RBM. Once this bottom RBM is trained, we can  generate ``samples'' from the hidden layer and use these samples as an input to the next RBM (consisting of the first and second hidden layer -- purple hexagons and green squares in Fig. \ref{fig:DBM}). This procedure can then be repeated to pretrain all layers of the DBM.

This pretraining initializes the weights so that SGD can be used effectively when the network is trained in a supervised fashion. In particular, the pretraining helps the gradients to stay well behaved rather than vanish or blow up -- a problem that we discussed extensively in the earlier sections on DNNs. It is worth noting that once pretrained, we can use the usual Boltzmann learning rules in Eq.~\eqref{eq:Pos-neg-energy} to fine-tune the weights and improve the performance of the DBM \cite{ hinton2006fast,hinton2006reducing}.  As we demonstrate in the next section, the Paysage package presented here can be used to both construct and train DBMs using such a pretraining procedure.

\subsection{Generative models in practice: examples}


\subsubsection{MNIST}

First, we apply the open source package Paysage (French for \emph{landscape}) for training unsupervised energy-based models on the MNIST dataset.  

\begin{figure}[t!]
	\includegraphics[width=1.0\columnwidth]{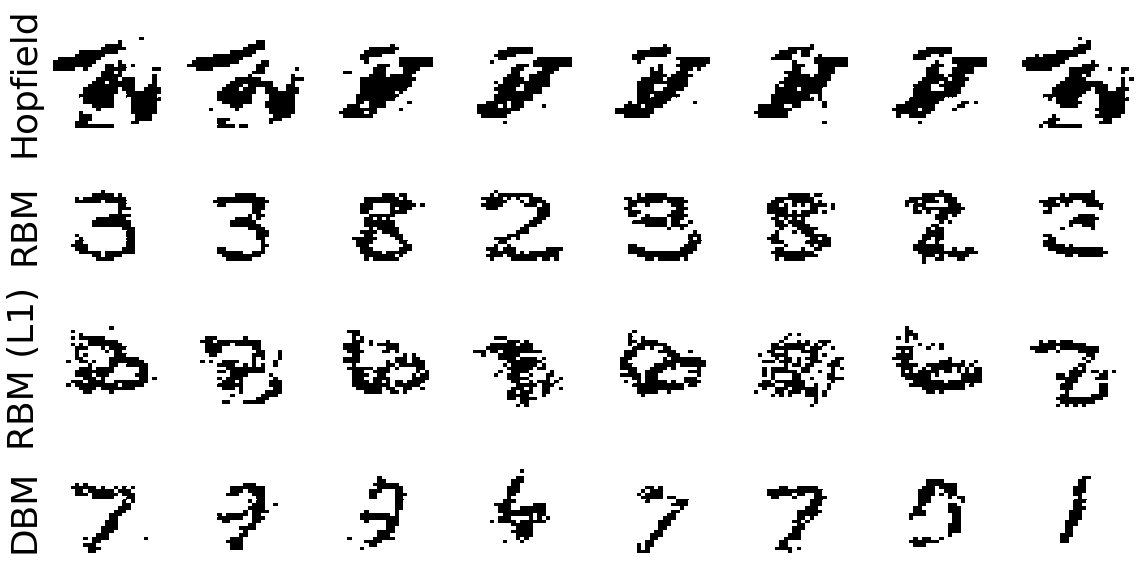}
	\caption{Fantasy particles (samples) generated using the indicated model trained on the MNIST dataset. Samples were generated by running (alternating) layerwise Gibbs sampling for $100$ steps. This allows the final sample to be very far away from the starting point in our feature space. Notice that the generated samples look much less like hand-written reconstructions than in Fig.~\ref{fig:MNIST-generative} which uses a single max-probability iteration of the Gibbs sampler, indicating that training is much less effective when exploring regions of probability space faraway from the training data. In the Sec.~\ref{sec:vae}, we will argue that this is likely a generic feature of Likelihood-based training.}
	\label{fig:MNISTfantasy}
\end{figure}

In \href{https://physics.bu.edu/~pankajm/MLnotebooks.html}{Notebook 17}, we explicitly demonstrate how to build and train four different kinds of models: (i) a ``Hopfield'' type RBM with Gaussian hidden units and Bernoulli (binary) visible units, (ii) a conventional RBM where both the visible and hidden units are Bernoulli, (iii) a conventional RBM with an additional L1-penalty that enforces sparsity, and (iv) a Deep Boltzmann Machine (DBM) with three Bernoulli layers with L1 penalty each. We refer the reader to the Notebook for the details of the code. In the following, we show and briefly discuss the results. 
	
After training the model, we compute reconstructions and fantasy particles from the validation data set. Recall that a reconstruction ${\bf v'}$ of a given data point ${\bf x}$ is computed in two steps: (i) we fix the visible layer ${\bf v}={\bf x}$ to be the data, and use MCMC sampling to find the state of the hidden layer ${\bf h}$ which maximizes the probability distribution $p({\bf h}\vert{\bf v})$. (ii) fixing the same obtained state ${\bf h}$, we find the reconstruction ${\bf v'}$ of the original data point which maximizes the probability $p({\bf v'}\vert{\bf h})$. In the case of a DBM, the forward pass continues until we reach the last of the hidden layers, and the backward pass goes in reverse. Figure~\ref{fig:MNIST-generative} shows the result.
	
We also used MCMC to draw samples from the learned probability distributions, the so-called fantasy particles. To this end, we did layer-wise Gibbs sampling for a total of a fixed number of equilibration steps. The result is shown in Figure~\ref{fig:MNISTfantasy}. 	
	
\begin{figure}[t]
	\includegraphics[width=1.0\columnwidth]{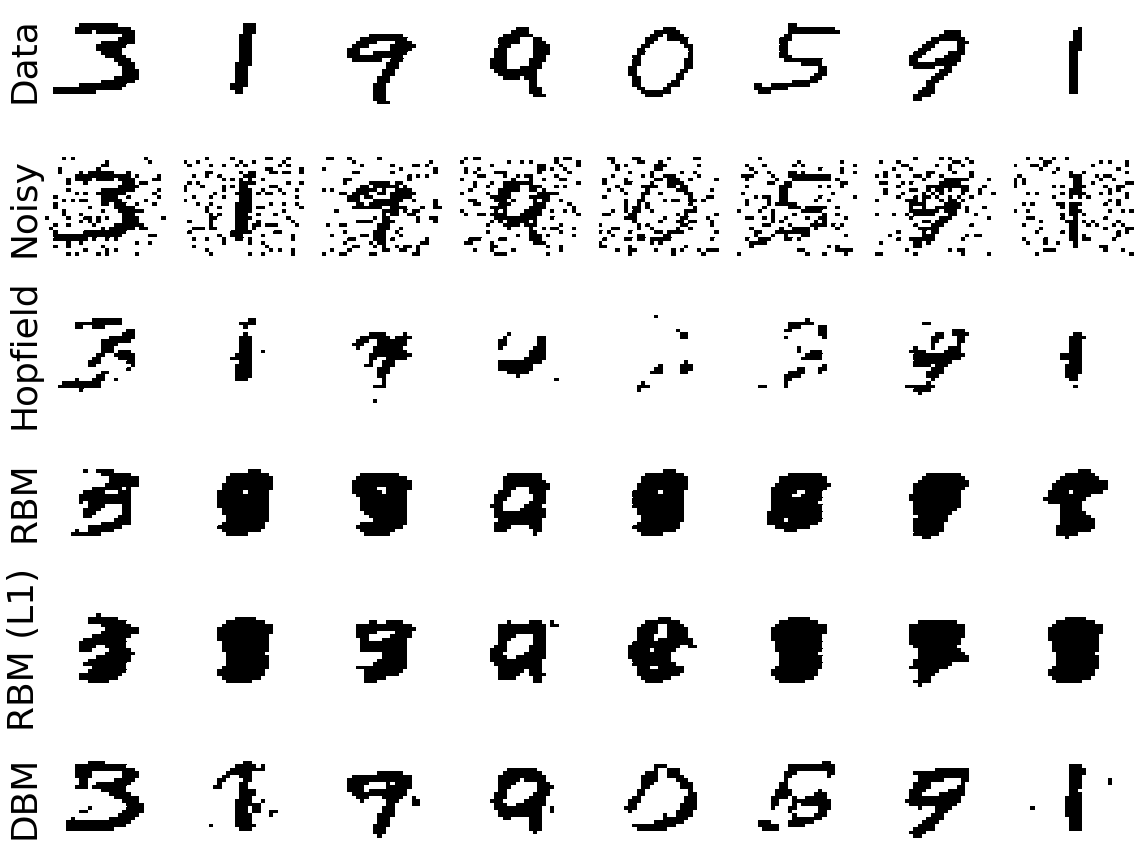}
	\caption{ Images from MNIST were randomly corrupted by adding noise. These noisy images were used as inputs to the visible layer of the
		generative model. The denoised images are obtained by a single ``deterministic'' (max probability) iteration $\mathbf{v} \rightarrow {\bf h} \rightarrow{\bf v'}$. }
	\label{fig:denoiseRBM}
	
\end{figure}

Finally, one can use generative models to reduce the noise in images (de-noising). We randomly flipped a fraction of the black\&white bits in the validation data, and use the models defined above to reconstruct (de-noise) the digit images. Figure~\ref{fig:denoiseRBM} shows the result. 

The full Paysage code used to generate Figs.~\ref{fig:MNIST-generative},~\ref{fig:MNISTfantasy} and~\ref{fig:denoiseRBM} is available in \href{https://physics.bu.edu/~pankajm/MLnotebooks.html}{Notebook 17}. The package was developed by one of the authors (CKF)  along with his colleagues at Unlearn.AI and makes it easy to build, train, and deploy energy-based generative models with different architectures. Paysage's documentation is available on GitHub under \href{https://github.com/drckf/paysage/tree/master/docs}{https://github.com/drckf/paysage/tree/master/docs}.

\begin{figure*}[t]
	
	\includegraphics[width=1.0\textwidth]{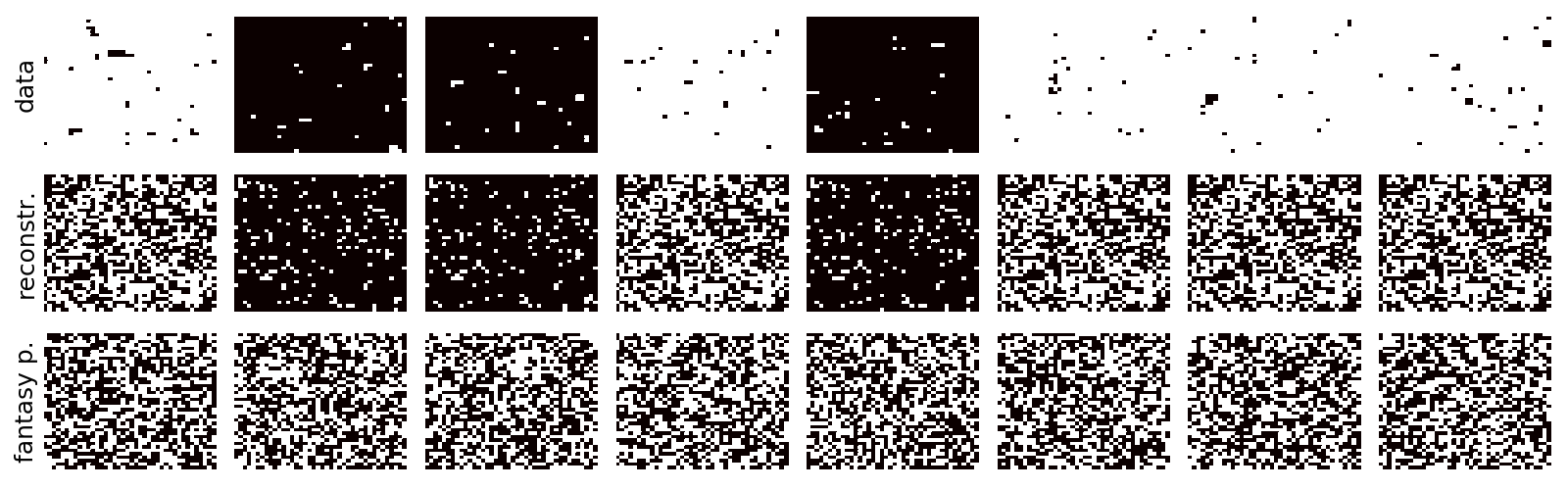}
	\caption{MC samples, their reconstructions and fantasy particles generated by a Deep Boltzmann Machine in the {\bf ordered phase} of the the 2D Ising data set at $T/J=1.75$. We used two hidden layers of $1000$ and $100$ layers, respectively.}
	\label{fig:RBM_Ising_T=1_75}
\end{figure*}

\begin{figure*}[t]
	
	\includegraphics[width=1.0\textwidth]{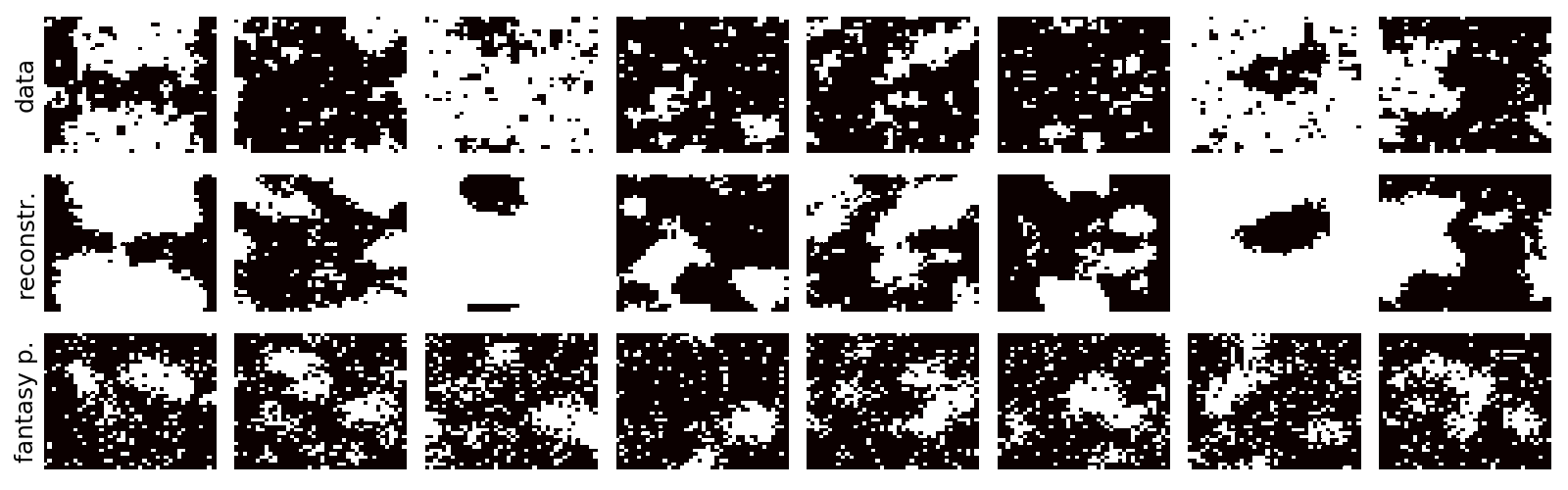}
	\caption{MC samples, their reconstructions and fantasy particles generated by a Deep Boltzmann Machine in the {\bf critical regime} of the the 2D Ising data set at $T/J=2.25$. We used two hidden layers of $1000$ and $100$ layers, respectively.}
	
	\label{fig:RBM_Ising_T=2_25}
\end{figure*}

\begin{figure*}[t]
	
	\includegraphics[width=1.0\textwidth]{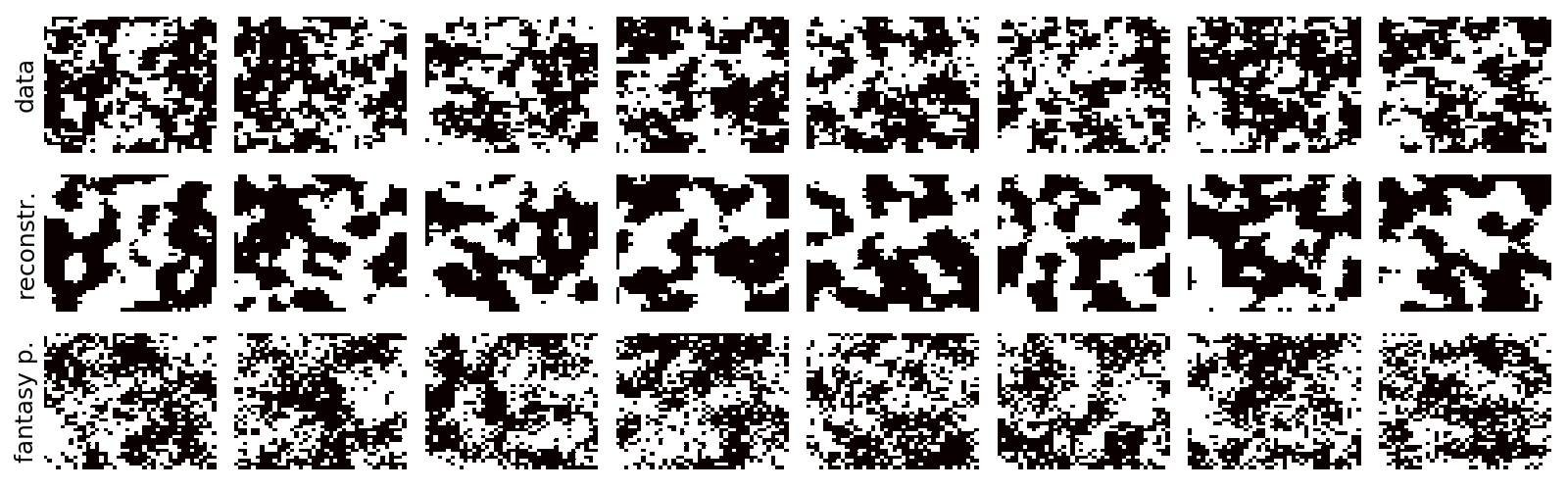}
	\caption{MC samples, their reconstructions and fantasy particles generated by a Deep Boltzmann Machine in the {\bf disordered phase} of the the 2D Ising data set at $T/J=2.75$. We used two hidden layers of $1000$ and $100$ layers, respectively.}
	
	\label{fig:RBM_Ising_T=2_75}
\end{figure*}

\subsubsection{Example: 2D Ising Model}

We can also analyze the 2D Ising data set. In previous sections, we used our knowledge of the critical point at $T_c/J\approx 2.26$ (see Onsager's solution) to label the spin configurations and study the problem of classifying the states according to their phase of matter. However, in more complicated models, where the precise position of $T_c$ is not known, one cannot label the states with such an accuracy, if at all. 

As we explained, generative models can be used to learn a variational approximation for the probability distribution that generated the data points. By using only the 2D spin configurations, we now attempt to train a Bernoulli RBM, the fantasy particles of which are thermal Ising configurations. Unlike in previous studies of the Ising dataset, here we perform the analysis at a fixed temperature $T$. We can then apply our model at three different values $T=1.75,2.25,2.75$ in the ordered, near-critical and disordered regions, respectively.

We define a Deep Boltzmann machine with two hidden layers of $N_\mathrm{hidden}$ and $N_\mathrm{hidden}/10$ units, respectively, and apply $L_1$ regularization to all weights. As in the MNIST problem above, we use layer-wise pre-training, and deploy Persistent Contrastive Divergence to train the DBM using ADAM.

One of the lessons from this problem is that this task is computationally intensive, see \href{https://physics.bu.edu/~pankajm/MLnotebooks.html}{Notebook 17}. The training time on present-day laptops easily exceeds that of previous studies from this review. Thus, we encourage the interested reader to try GPU-based training and study the resulting speed-up.

Figures~\ref{fig:RBM_Ising_T=1_75},~\ref{fig:RBM_Ising_T=2_25} and~\ref{fig:RBM_Ising_T=2_75} show the results of the numerical experiment at $T/J=1.75,2.25,2.75$ respectively, for a DBM with $N_\mathrm{hidden}=800$. Looking at the reconstructions and the fantasy particles, we see that our DBM works well in the disordered and critical regions. However, the chosen layer architecture is not optimal for $T=1.75$ in the ordered phase, presumably due to effects related to symmetry breaking.

\subsection{Generative models in physics}

Generative models have been studied and used extensively in the context of physics. For instance, in Biophysics, dynamic Boltzmann distributions have been used as effective models in chemical kinetics~\cite{ernst2018learning}. In Statistical Physics, they were used to identify criticality in the Ising model~\cite{morningstar2017deep}. In parallel, tools from Statistical Physics have been applied to analyze the learning ability of RBMs~\cite{huang2017statistical,decelle2018thermodynamics}, characterizing the sparsity of the weights, the effective temperature, the nonlinearities in the activation functions of hidden units, and the adaptation of fields maintaining the activity in the visible layer~\cite{tubiana2017emergence}. Spin glass theory motivated a deterministic framework for the training, evaluation, and use of RBMs~\cite{tramel2017deterministic}; it was demonstrated that the training process in RBMs itself exhibits phase transitions~\cite{barra2016phase,barra2017phase}; learning in RBMs was studied in the context of equilibrium~\cite{funai2018thermodynamics,cossu2018machine} and nonequilibrium~\cite{salazar2017nonequilibrium} thermodynamics, and spectral dynamics~\cite{decelle2017spectral}; mean-field theory found application in analyzing DBMs~\cite{huang2017mean}. Another interesting direction of research is the use of generative models to improve Monte Carlo algorithms~\cite{cristoforetti2017towards,tanaka2017towards,wang2017can,nagai2017self}. Ideas from quantum mechanics have been put forward to introduce improved speed-up in certain parts of the learning algorithms for Helmholtz machines~\cite{benedetti2017quantum,benedetti2016quantum}.

At the same time, generative models have applications in the study of quantum systems too. Most notably, RBM-inspired variational ansatzes were used to learn both complex-valued wavefunctions and the real-valued probability distribution associated with the absolute square of a quantum state~\cite{nomura2017restricted,carleo2017solving,carleo2018constructing,freitas2018neural,torlai2018neural} and, in this context, RBMs are sometimes called Born machines~\cite{cheng2017information}, including quantum state tomorgraphy~\cite{torlai2017neural,torlai2018neural,carrasquilla2018reconstructing}. Further applications include the detection of order in low-energy product states~\cite{rao2017identifying}, and learning Einstein-Podolsky-Rosen correlations on an RBM~\cite{weinstein2017learning}. Inspired by the success of tensor networks in physics, the latter have been used as a basis for both generative and discriminative learning~\cite{huggins2018towards}: RBMs~\cite{chen2018equivalence} were used to extract the spatial geometry from entanglement~\cite{you2017machine}, and generative models based on matrix product states have been developed~\cite{han2017unsupervised}. Last but not least, Quantum entanglement was studied using RBM-encoded states~\cite{deng2017quantum} and tensor product based generative models have been used to understand MNIST and other ML datasets \cite{stoudenmire2016supervised}.

%% file: sections/VAE_PM.tex
In the previous two sections, we considered energy-based generative models. Here, we  extend our discussion to  two new generative model frameworks that have gained wide appeal in the the last  few years: generative adversarial networks (GANs) \cite{goodfellow2014generative, radford2015unsupervised, goodfellow2016nips} and variational autoencoders (VAEs) \cite{kingma2013auto}. Unlike energy-based models, both these generative modeling frameworks are based on differentiable neural networks and consequently can be trained using backpropagation-based methods. VAEs, in particular, can be easily implemented and trained using high-level packages such as Keras making them an easy-to-deploy generative framework. These models also differ from the energy-based models in that they do not directly seek to maximize likelihood. GANs, for example,  employ a novel cost function based on adversarial learning (a concept we motivate and explain below). Finally we note that VAEs and GANs are already starting to make their way into physics~\cite{wetzel2017unsupervised, liu2017simulating, rocchetto2018learning,heimel2018qcd} and astronomy \cite{ravanbakhsh2017enabling}, and methods from physics may prove useful for furthering our understanding of these methods~\cite{alemi2017exponential}. More generally, GANs have found important applications in many artistic and image manipulation tasks (see references in \cite{goodfellow2016nips}).

The section is organized as follows. We start by motivating adversarial learning by discussing the limitations of maximum likelihood based approaches. We then give a high-level introduction to the main idea behind generative adversarial networks  and discuss how they overcome some of these limitations, simultaneously highlighting both the power of GANs and some of the difficulties. We then show how VAEs integrate the variational methods introduced in Sec.~\ref{sec:varl_MFT} with deep, differentiable neural networks to build more powerful generative models that move beyond the Expectation Maximization (EM). We then briefly discuss VAEs from an information theoretic perspective, before discussing practical tips for implementing and training VAEs. We conclude by using VAEs on examples using the Ising and MNIST datasets (see also \href{https://physics.bu.edu/~pankajm/MLnotebooks.html}{Notebooks 19 and 20}).

\subsection{The limitations of maximizing Likelihood}

\begin{figure}[t]
\includegraphics[width=0.9\columnwidth]{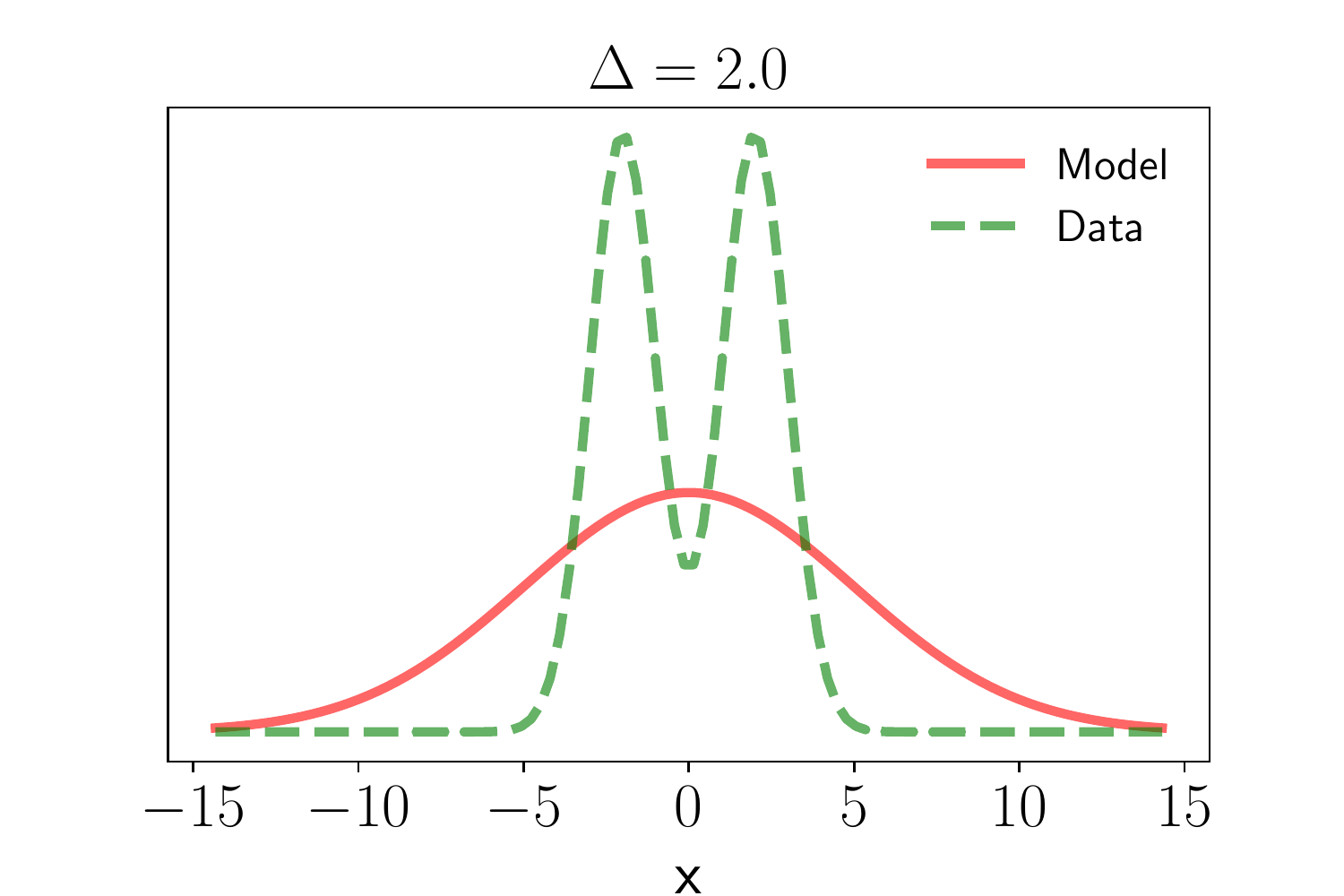}
\includegraphics[width=0.9\columnwidth]{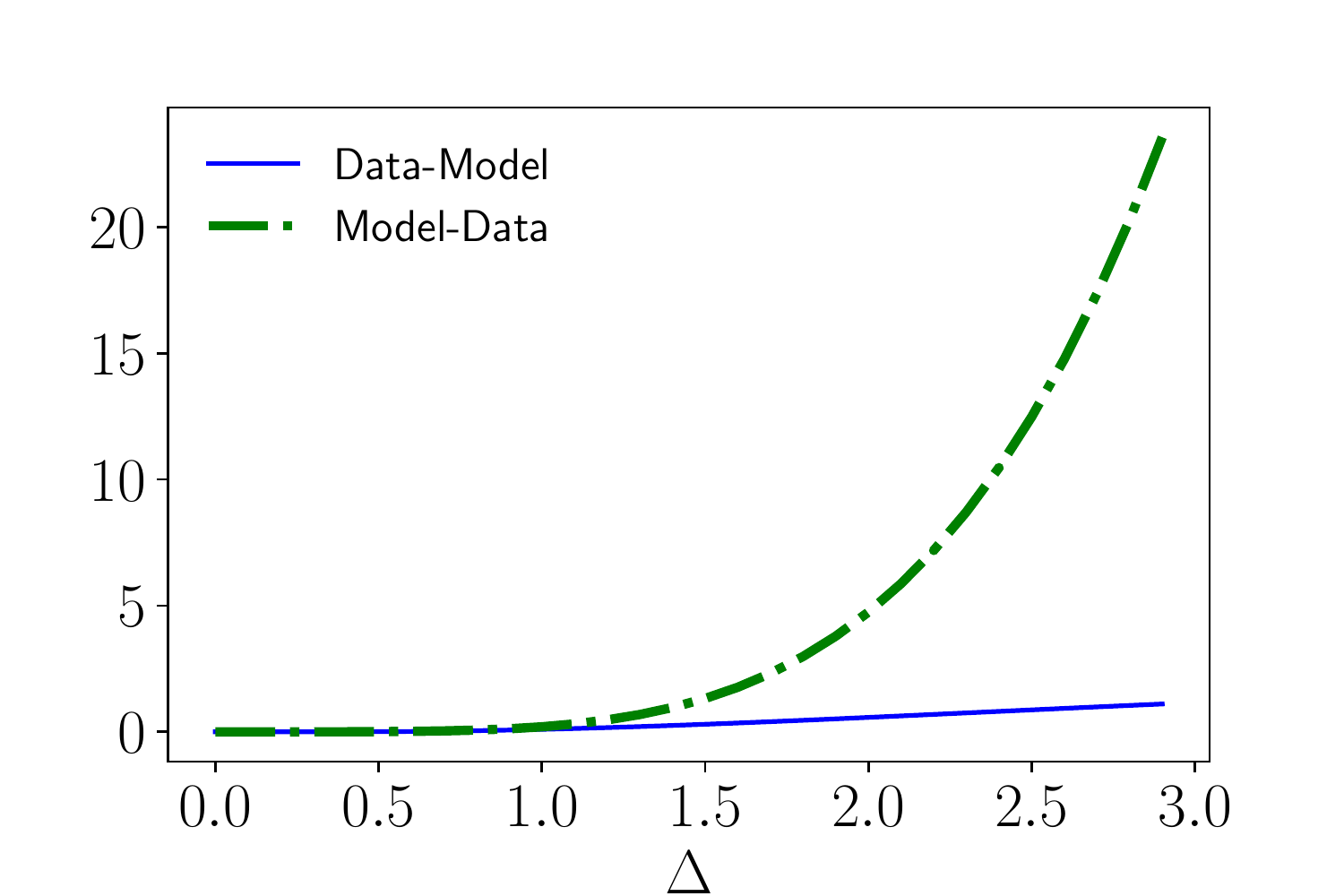}
\caption{ KL-divergences between the data distribution $p_\mathrm{data}$ and the model $p_\theta$. Data is drawn from a bimodal Gaussian distribution with unit variances peaked at $\pm \Delta$ with $\Delta= 2.0$ and the model $p_\theta(x)$ is a Gaussian with mean zero and same variance as $p_\theta(x)$. (Top) $p_\mathrm{data}$ and $p_\theta$ for $\Delta =2$. (Bottom)  $D_{KL}(p_\mathrm{data} || p_{\theta})$ (Data-Model) and  $D_{KL}(  p_\theta || p_\mathrm{data})$ (Model-Data) as a function of $\Delta$. Notice that  $D_{KL}(p_\mathrm{data} || p_{\theta})$ is insensitive to placing weight in the model distribution in regions where $p_\mathrm{data} \approx 0$ whereas $D_{KL}(  p_\theta || p_\mathrm{data})$ punishes this harshly.}
\label{fig:KL-fig1}
 \end{figure}

\begin{figure}[t]
\includegraphics[width=0.9\columnwidth]{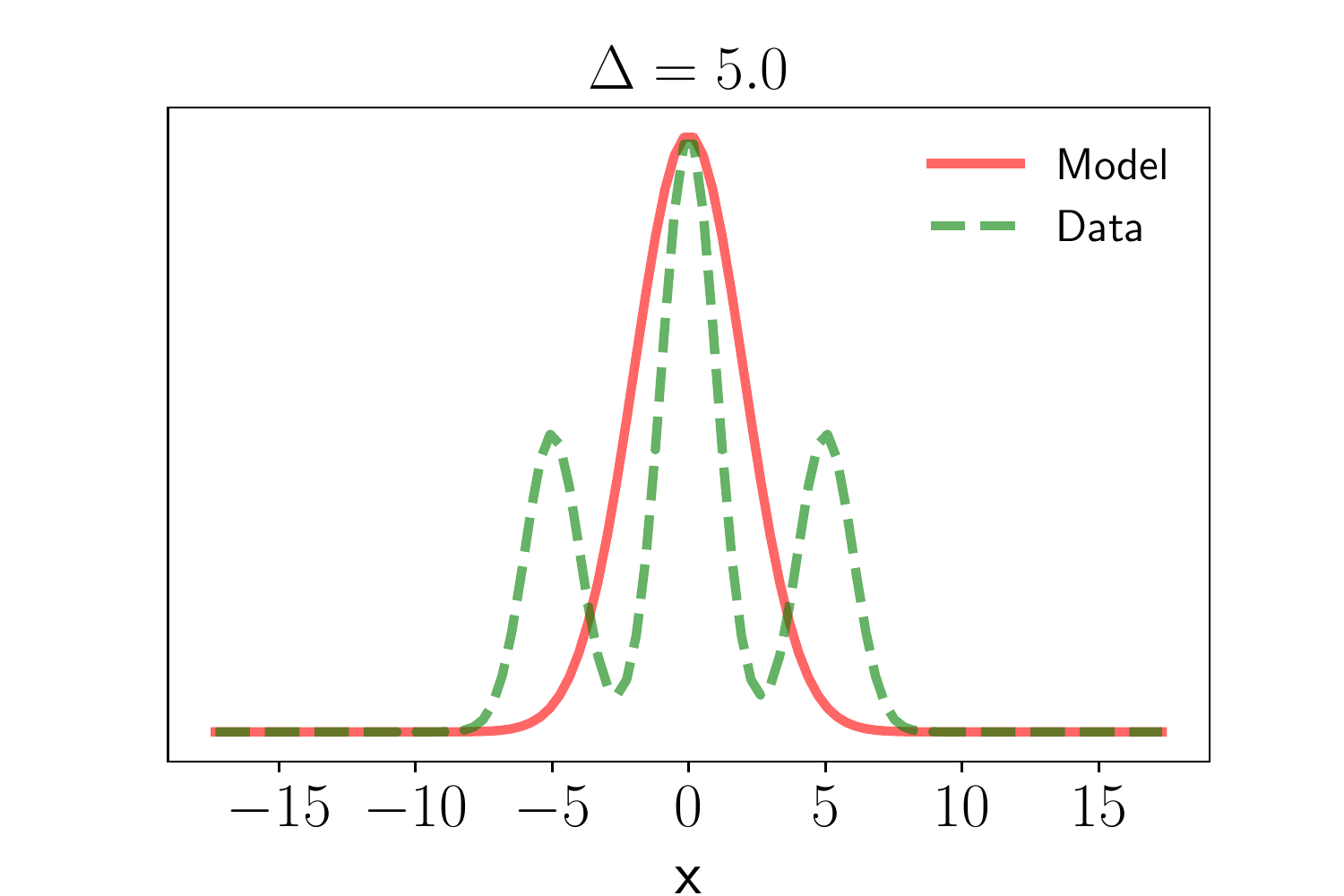}
\includegraphics[width=0.9\columnwidth]{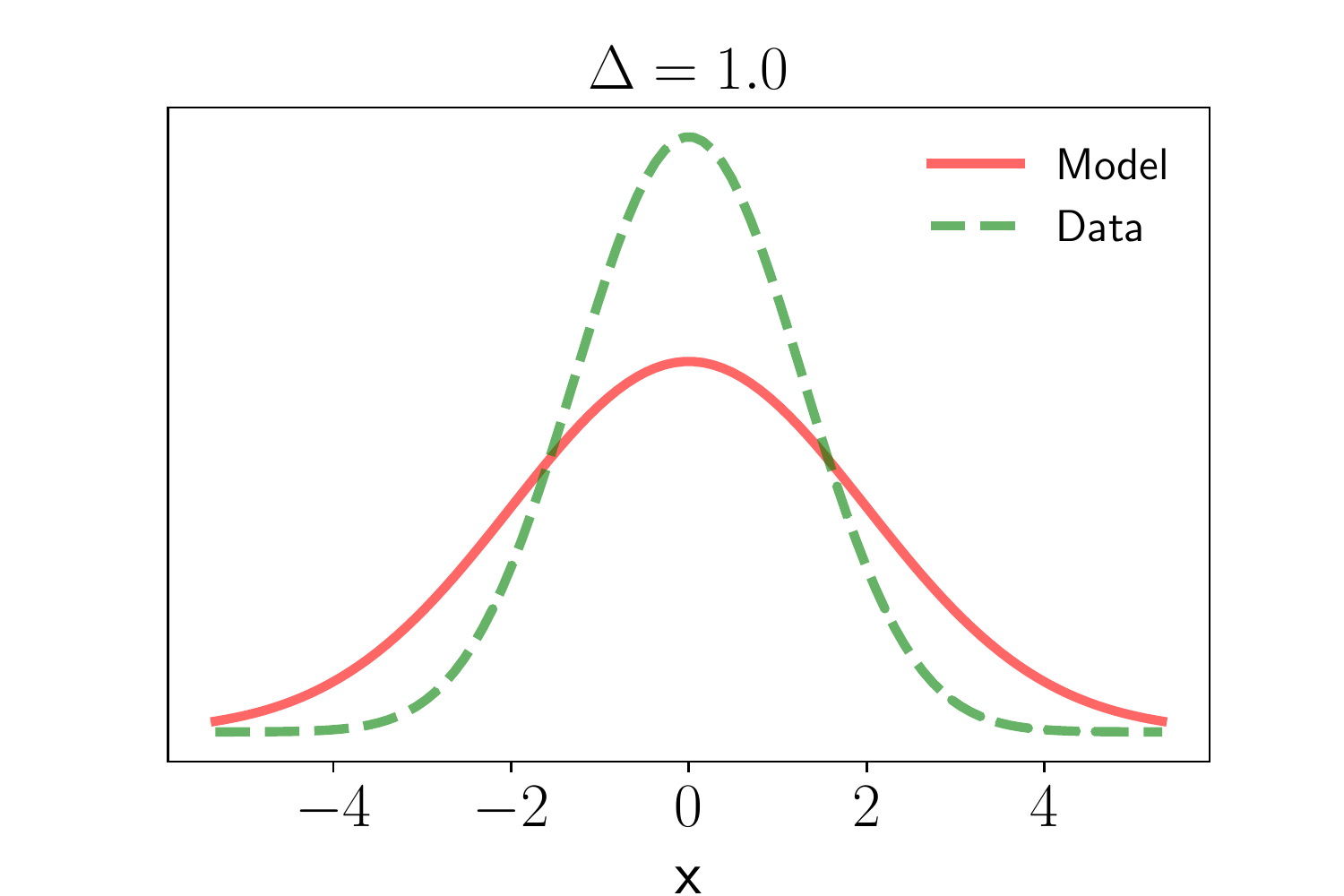}
\includegraphics[width=0.9\columnwidth]{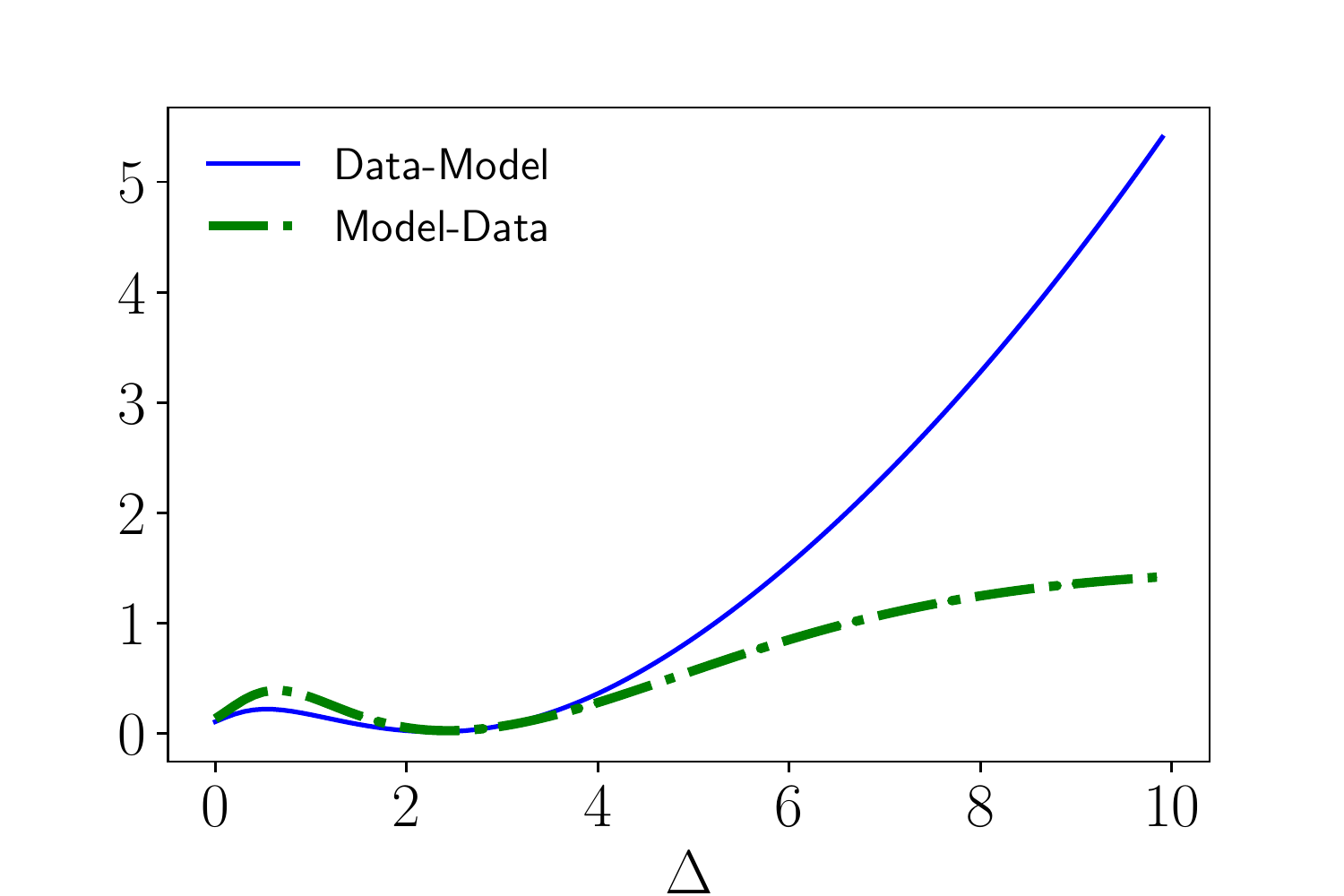}
\caption{ KL-divergences between the data distribution $p_\mathrm{data}$ and the model $p_\theta$. Data is drawn from a Gaussian mixture of the form
$p_\mathrm{data}=0.25 \mathcal{N}(-\Delta)+0.25 *\mathcal{N}(\Delta)+ 0.5 \mathcal{N}(0)$ where $\mathcal{N}(a)$ is a normal distribution with unit variance centered at $x=a$.  $p_\theta(x)$ is a Gaussian with $\sigma^2=2$. (Top) $p_\mathrm{data}$ and $p_\theta$ for $\Delta =5$. (Middle) $p_\mathrm{data}$ and $p_\theta$ for $\Delta =1$. (Bottom)  $D_{KL}(p_\mathrm{data} || p_{\theta})$ [Data-Model] and  $D_{KL}(  p_\theta || p_\mathrm{data})$ [Model-Data] as a function of $\Delta$. Notice that $D_{KL}(  p_\theta || p_\mathrm{data})$ is insensitive to placing weight in the model distribution in regions where $p_\theta \approx 0$ whereas  $D_{KL}(p_\mathrm{data} || p_{\theta})$ punishes this harshly .} 
\label{fig:KL-fig2}
\end{figure}

The Kullback-Leibler (KL)-divergence plays a central role in many generative models. Developing an intuition about KL-divergences is one of the keys to understanding why adversarial learning has proved to be such a powerful method for generative modeling. Here, we revisit the KL-divergence with an eye towards understanding GANs and motivate adversarial learning. The KL-divergence measures the similarity between two probability distributions  $p({\bf x})$  and  $q({\bf x})$. Strictly speaking, the KL divergence is not a metric because it is not symmetric and does not satisfy the triangle inequality. 

Given two distributions, there are two distinct KL-divergences we can construct:
\begin{align}
D_{KL}(p || q) &= \int \mathrm d {\bf x} p({\bf x}) \log \frac{p({\bf x})}{q({\bf x})} \\
D_{KL}(q || p) &= \int \mathrm d {\bf x} q({\bf x}) \log \frac{q({\bf x})}{p({\bf x})}.
\end{align}
 A related quantity called the Jensen-Shannon divergence,
\begin{equation}
D_{JS}(p,q) = \frac{1}{2} \left[ D_{KL}\left(p \bigg \| \frac{p + q}{2}\right) + D_{KL}\left(q \bigg \| \frac{p + q}{2}\right) \right] \nonumber
\end{equation}
does satisfy all of the properties of a squared metric (i.e., the square root of the Jensen-Shannon divergence is a metric).  An important property
of the KL-divergence that we will make use of repeatedly is its positivity: $D_{KL}(p || q)\ge 0$ with equality if and only if $p({\bf x})=q({\bf x})$ almost everywhere.

In generative models in ML, the two distributions we are usually concerned with are the model distribution $p_{\theta}({\bf x})$ and the data distribution $p_\mathrm{data}({\bf x})$. We of course would like these models to be as similar as possible. However, as we discuss below, there are many subtleties about how we measure similarities that can have large consequences for the behavior of training procedures. Maximizing the log-likelihood of the data under the model is the same as minimizing the KL divergence between the data distribution and the model distribution $D_{KL}(p_\mathrm{data} || p_{\theta})$. To see this, we can rewrite the KL divergence as:
\begin{align}
D_{KL}(p_\mathrm{data} || p_{\theta}) 
&= \int \mathrm d {\bf x} p_\mathrm{data}({\bf x}) \log p_\mathrm{data}({\bf x}) \nonumber \\
&- \int \mathrm d {\bf x} p_\mathrm{data}({\bf x}) \log p_{\theta}({\bf x}) \nonumber \\
&= -S[p_\mathrm{data}] - \langle \log p_{\theta}({\bf x}) \rangle_\mathrm{data}
\end{align}
Rearranging this equation, we have
\be
\langle \log p_{\theta}({\bf v}) \rangle_\mathrm{data}=-S[p_\mathrm{data}]-D_{KL}(p_\mathrm{data} || p_{\theta}) 
\ee
The equivalence follows from the positivity of KL-divergence and the fact that the entropy of the data distribution is constant. In contrast, the original formulation of GANs minimizes an upper bound on the Jensen-Shannon divergence between the model distribution $p_\theta({\bf x})$ and the data distribution $p_\mathrm{data}({\bf x})$ \cite{goodfellow2014generative}.

This difference in objectives underlies the difference in behavior between GANs and likelihood based generative models. To see this, we can compare the behavior of the 
two KL-divergences $D_{KL}(p_\mathrm{data} || p_{\theta}) $ and $D_{KL}(p_{\theta} || p_\mathrm{data})$. As is evident from Fig.~\ref{fig:KL-fig1} and Fig.~\ref{fig:KL-fig2}, though 
both of these KL-divergences measure similarities between the two distributions, they are sensitive to very different things. $D_{KL}(  p_\theta || p_\mathrm{data})$ is insensitive to setting $p_\theta \approx 0$ even when $p_\mathrm{data} \neq 0$ whereas $D_{KL}(p_\mathrm{data} || p_{\theta})$ punishes this harshly. In contrast, $D_{KL}(p_\mathrm{data} || p_{\theta})$ is insensitive to placing weight in the model distribution in regions where $p_\mathrm{data} \approx 0$ whereas $D_{KL}(  p_\theta || p_\mathrm{data})$ punishes this harshly. In other words, $D_{KL}(p_\mathrm{data} || p_{\theta})$ prefers models that have a high probability in regions with lots of training data points whereas $D_{KL}(  p_\theta || p_\mathrm{data})$ punishes models for putting high probability where there is no data.

In the context of the above discussion, this suggests that the way likelihood-based methods are most likely to fail, is by improperly ``filling in'' any low-probability density regions  between peaks in the data distribution. In contrast, at least in principle, the Jensen-Shannon distribution which underlies GANs is sensitive both to placing weight where there is data since it has information about  $D_{KL}(p_\mathrm{data} || p_{\theta})$ and to not placing weight where no data has been observed (i.e.~in low-probability density regions) since it has information about $D_{KL}(  p_\theta || p_\mathrm{data})$. 

In practice, $D_{KL}(p_\mathrm{data} || p_{\theta})$ can be calculated easily directly from the data using sampling. On the other hand, $D_{KL}(  p_\theta || p_\mathrm{data})$ is impossible to compute since we  do not know $p_\mathrm{data}({\bf x})$. In particular, this integral cannot be calculated using sampling since we cannot evaluate $p_\mathrm{data}({\bf x})$ at the locations of the fantasy particles. The idea of adversarial learning is to circumnavigate this difficulty by using an adversarial learning procedure. Recall, that $D_{KL}(  p_\theta || p_\mathrm{data})$ is large when the model artificially over-weighs low-density regions near real peaks (see Fig.~\ref{fig:KL-fig1}). Adversarial learning accomplishes this same task by teaching a discriminator network to distinguish between real data points and samples generated from the model. By punishing the model for generating points that can be easily discriminated from the data, adversarial learning decreases the weight of regions in the model space that are far away from data points -- regions that inevitably arise when maximizing likelihood. This core intuition implicitly underlies many adversarial training algorithms (though it has been recently suggested that this may not be the entire story \cite{goodfellow2016nips}).

\subsection{Generative models and adversarial learning }

\begin{figure}[t]
\includegraphics[width=0.9\columnwidth]{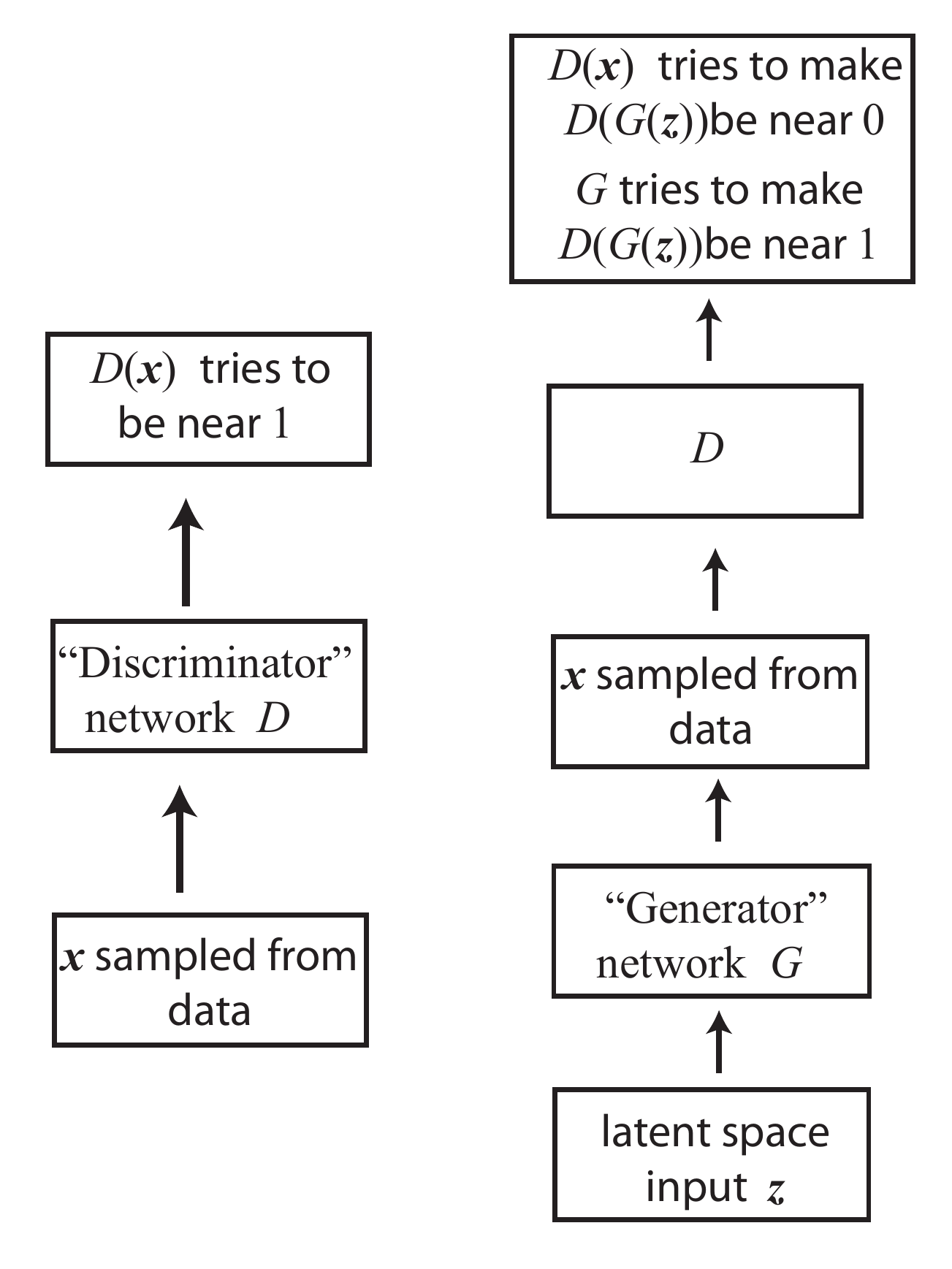}
\caption{A GAN consists of two differentiable functions (usually represented as deep neural networks): a generator function $G({\bf z}; \theta_G)$ that takes as an input a ${\bf z}$ sampled
from some prior on the latent space and outputs a point ${\bf x}$. The generator function (neural network) has  parameters $\theta_G$. The discriminator function $D({\bf x}; \theta_D)$ discriminates between ${\bf x}$ from the data and samples from the model: ${\bf x}= G({\bf z}; \theta_G)$. The two networks are trained by ``playing a game'' where the discriminator is trained to distinguish between synthetic and real examples while the generator is trained to try to fool the discriminator. Importantly, the cost function for the discriminator depends on the generator parameters and vice versa.
} 
\label{fig:GANschematic}
\end{figure}

Here, we give a brief high-level overview of the basic idea behind GANs. The mathematics and theory of GANs draws deeply from concepts in Game Theory such as Nash Equilibrium that are foreign to most physicists. For this reason, a comprehensive discussion of GANs is beyond the scope of the review.  Readers interested in learning more are directed to the comprehensive tutorial by Goodfellow \cite{goodfellow2016nips}. GANs are also notorious for being hard to train. For this reason, readers wishing to play with GANs should also consider the very nice practical discussion entitled ``How to train a GAN'' (affectionately labeled ``ganhacks'') available at \url{https://github.com/soumith/ganhacks}.

The central idea of GANs is to construct two differentiable neural networks (see Fig.~\ref{fig:GANschematic}). The first neural network, usually a (de)convolutional network based on the DCGAN architecture \cite{radford2015unsupervised},  approximates a generator function $G({\bf z}; \theta_G)$ that takes as input a ${\bf z}$ sampled from some prior on the latent space, and outputs a ${\bf x}$ from the model. The second network approximates a discriminator function $D({\bf x}; \theta_D)$ that is designed to distinguish between ${\bf x}$ from the data and samples generated by the model: ${\bf x}= G({\bf z}; \theta_G)$.  The scalar $D(\bf{x})$ represents the probability that ${\bf x}$ came from the data rather than the model  $p_{\theta_G}$. We train $D$ to distinguish actual data points from synthetic examples and the generative network to fool the discriminative network.

To define the cost function for training, it is useful to define the functional
\begin{align}
V(D,G) &=\mathbb{E}_{{\bf x} \sim p_\mathrm{data}} \left(\log D({\bf x}) \right)\nonumber \\
&+  \mathbb{E}_{{\bf z} \sim p_\mathrm{prior}} \left(\log{[1-D(G({\bf z}))]} \right).
\end{align}
In the version of GANs most amenable to theoretical analysis  -- though not the version usually implemented in practice -- we take the cost function for the discriminator and generators to be  $\mathcal{C}^{(G)} = -\mathcal{C}^{(D)}={1 \over 2} V(D,G)$. This choice of cost functions corresponds to what is called a zero-sum game. Since the discriminator
is maximized, we can write a cost function for the generator as
\be
\mathcal{C}(G) = \max_D V(G,D).
\ee
It turns out that this cost function is related to the Jensen-Shannon Divergence in a simple manner  \cite{goodfellow2014generative, goodfellow2016nips}:
 \be
 \mathcal{C}(G) =-\log{4}+2 D_{JS}(p_\mathrm{data},p_{\theta_G}).
 \ee
This brings us back full circle to the discussion in the last section on KL-divergences.

\subsection{ Variational Autoencoders (VAEs) }

\begin{figure}[t]
\includegraphics[width=0.9\columnwidth]{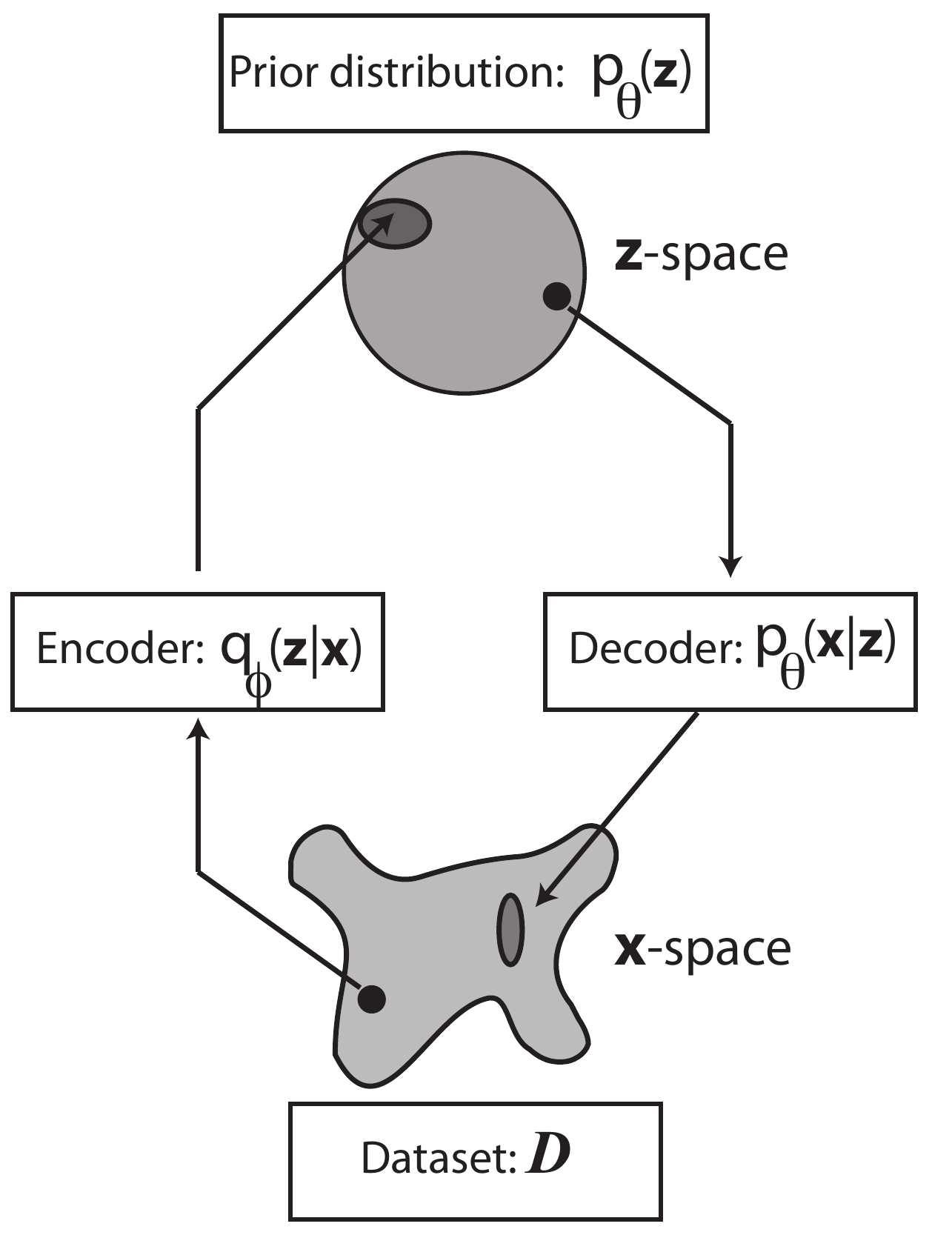}
\caption{VAEs learn a joint distribution $p_\theta({\bf x}, {\bf z})$ between latent variables ${\bf z}$ with prior distribution $p({\bf z})$ and data ${\bf x}$. The conditional distribution $p_\theta({\bf x}|{\bf z})$ can be thought of as a stochastic ``decoder'' that maps latent variables to new examples. The stochastic ``encoder'' $q_\phi({\bf z}|{\bf x})$ approximates the true but intractable $p_\theta({\bf z}|{\bf x})$  -- much like mean-field theories in statistical physics approximate true distributions with analytically tractable approximations.  Figure based on Kingma's Ph.D. dissertation Chapter 2. \cite{kingma2017variational}. }
\label{fig:VAEschematic} 
\end{figure}

We now turn our attention to another class of powerful latent-variable, generative models called Variational Autoencoders (VAEs). VAEs exploit the variational/mean-field theory ideas presented in  Sec.~\ref{sec:varl_MFT} to build complex generative models using  deep neural networks (DNNs). The central idea behind VAEs is to represent the map from latent variables  to  observable variables  using a DNN. The use of latent variables is a common theme in many of the generative models we have encountered in unsupervised learning tasks from Gaussian Mixture Models (see Sec.~\ref{sec:clustering}) to Restricted Boltzmann Machines. However, in VAEs this mapping, $p(\bf{x}|\bf{z}, \theta)$ is much less restrictive and much more complicated since it takes the form of a DNN. This added complexity means we cannot use techniques such as Expectation Maximization to train the model and instead must rely of methods based on backpropagation.

\subsubsection{VAEs as variational models} 

We start by discussing VAEs from a variational perspective. We will make extensive use of the concepts introduced in Sec.~\ref{sec:varl_MFT} and the reader is strongly-encouraged to refresh their memory of this section before proceeding. A VAE is a latent-variable model $p_\theta({\bf x}, {\bf z})$ with a latent variables ${\bf z}$ and observed variables ${\bf x}$. The latent variables are drawn from some pre-specified prior distribution $p({\bf z})$. In practice, $p({\bf z})$ is almost always taken to be a multivariate Gaussian. The conditional distribution $p_\theta({\bf x}|{\bf z})$ maps points in the latent space to new examples (see Fig.~\ref{fig:VAEschematic}). This is often called a ``stochastic decoder'' and defines the generative model for the data. The reverse mapping that gives the posterior over the latent variables $p_{\theta}({\bf z}|{\bf x})$ is often called the ``stochastic encoder''.

A central challenge in latent variable modeling is to infer the posterior distribution of the latent variables given a sample from the data. This can in principle be done via Bayes' rule: $p_\theta({\bf z}|{\bf x}) = \frac{p({\bf z})p_{\theta}({\bf x}|{\bf z})}{p_\theta({\bf x})}$. For some models, we can calculate this analytically. In this case, we can use techniques like Expectation Maximization (EM) (see Sec.~\ref{sec:varl_MFT}). However, in general this is intractable since the denominator requires computing a sum over all configurations of the latent variables, $p_\theta({\bf x}) = \int p_\theta({\bf x},{\bf z}) d{\bf z} = \int p_\theta ({\bf x}|{\bf z})p({\bf z})d{\bf z}$ (i.e.~a partition function in the language of physics), which is often intractable for large models. In VAEs, where the $p_\theta ({\bf x}|{\bf z})$ is modeled using a DNN, this is impossible.
 
A first attempt to address the issue of computing $p({\bf x})$ could be through importance sampling \cite{neal2001annealed}. That is, we choose a proposal distribution $q({\bf z}|{\bf x})$ which is easy to sample from, and rewrite the sum as an expectation with respect to this distribution:
\begin{equation}
p_\theta ({\bf x}) = \int p_\theta({\bf x}|{\bf z}) \frac{p({\bf z})}{q_\phi({\bf z}|{\bf x})}q_\phi({\bf z}|{\bf x}) d{\bf z}.
\end{equation}
Thus, by sampling from $q_\phi({\bf z}|{\bf x})$ we can get a Monte Carlo estimate of $p({\bf x})$. However, this requires generating samples and thus our estimates will be noisy. If our proposal distribution is poor, the variance in the estimate can be very high.

An alternative approach that avoids these sampling issues is to use the variational approach discussed in Sec.~\ref{sec:varl_MFT}. We know from Eq.~\eqref{eq:main_var_eq} that we can write the log-likelihood as 
\begin{equation}
\log p({\bf x}) = D_{KL}(q_\phi(\bd{z}| {\bd x}) \| p_\theta(\bd{z}|\bd{x},\bd{\theta}))-F_{q_\phi}(\bf x),
\end{equation}
where the variational free energy is defined as 
\begin{equation}
-F_{q_\phi}({\bf x}) \equiv \mathbb{E}_{q_\phi({\bf z}|{\bf x})}[\log p_\theta ({\bf x}|{\bf z})] - D_{KL}(q_\phi({\bf z}|{\bf x})|p({\bf z})).
\label{eq:ELBO}
\end{equation}
In writing this term, we have used Bayes rule and Eq.~\eqref{eq:Fptilde}. Since the KL-divergence is strictly positive, the (negative) variational free energy is a lower-bound on the log-likelihood. For this reason, in the VAE literature, it is often called the \emph{Evidence Lower BOund} or ELBO. 

Equation~\eqref{eq:ELBO} has a beautiful interpretation. The first term in this equation can be viewed as a ``reconstruction error'', where we start with data ${\bf x}$, encode it into the latent representation using our approximate posterior $q_\phi({\bf z}|{\bf x})$, and then evaluate the log probability of the original data given the inferred latents. For binary variables, this is just the cross-entropy which we first encountered when studying logistic regression, cf.~Sec.~\ref{sec:log_reg}. The second term acts as a regularizer and encourages the posterior distributions to be close to $p({\bf z})$. By maximizing the ELBO, we minimize the KL-divergence between the approximate and true posterior. By choosing a tractable $q_\phi({\bf z}|{\bf x})$, we make this feasible (see Fig.~\ref{fig:VAEschematic}).

\subsubsection{Training via the reparametrization trick}

\begin{figure*}[t]
\includegraphics[width=0.8\textwidth]{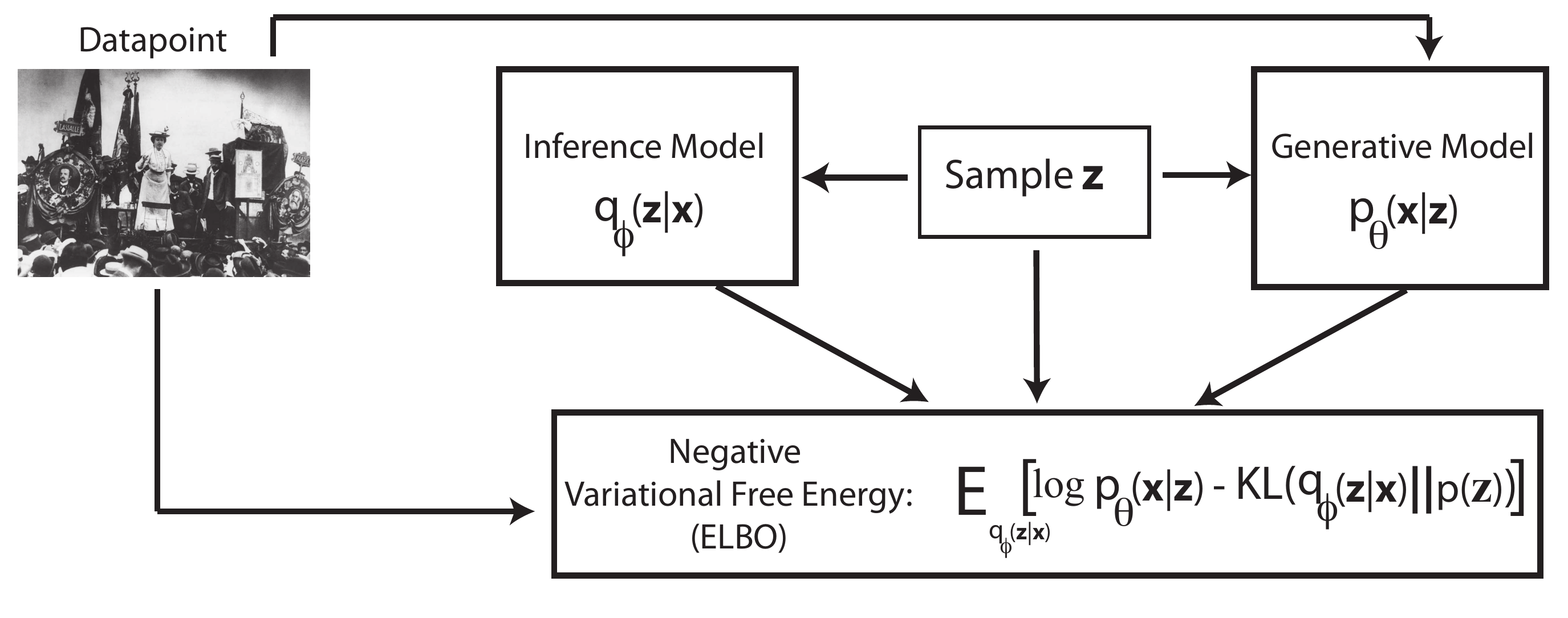}
\label{fig:VAEcomp}
\caption{Schematic explaining the computational flow of VAEs. Figure based on Kingma's Ph.D. dissertation Chapter 2. \cite{kingma2017variational}. }
 \end{figure*}

VAEs train models by minimizing the variational free energy (maximizing the ELBO). Training a VAE is somewhat complicated because we must simultaneously learn two sets of parameters: the parameters $\theta$ that define our generative model $p_\theta({\bf x}, {\bf z})$ as well as the variational parameters $\phi$ in $q_\phi({\bf z}|{\bf x})$. The basic approach will be the same as for all DNN models: we will use gradient descent with the variational free energy as the objective (cost) function. For a dataset $\mathcal{L}$, we can write
our cost function as
\be
\mathcal{C}_{\theta, \phi}(\mathcal{L})=\sum_{{\bf x} \in \mathcal{L}} -F_{q_\phi}({\bf x}).
\ee

Taking the gradient with respect to $\theta$ is easy since only the first term in Eq.~\eqref{eq:ELBO} depends on $\theta$,
\bea
\mathcal{C}_{\theta, \phi}({\bf x}) &=&  \mathbb{E}_{q_\phi({\bf z}|{\bf x})}[\nabla_\theta \log p_\theta ({\bf x}|{\bf z})] \nonumber \\
&\sim& \nabla_\theta \log p_\theta ({\bf x}|{\bf z})
\eea
where in the second line we have replaced the expectation value with a single Monte-Carlo sample ${\bf z}$ drawn from $q_\phi({\bf z}|{\bf x})$ (see Fig.~\ref{fig:VAEcomp}). When $p_\theta({\bf x}|{\bf z})$ is approximated by a neural network, this can be calculated using backpropagation with the reconstruction error as the objective function. 

On the other hand, calculating the gradient with respect to the parameters $\phi$ is more complicated since $\phi$ also appears in the expectation value $\mathbb{E}_{q_\phi({\bf z}|{\bf x})}$. Ideally, we would like to also use backpropagation to calculate this as well. It turns out that this can be done by a simple change of variables that often goes under the name the ``reparameterization trick'' \cite{kingma2013auto, rezende2014stochastic}. The basic idea is to change variables so that $\phi$ no longer appears in the distribution we are taking an expectation value with respect to. To do this, we express the random variable ${\bf z} \sim q_\phi({\bf z}|{\bf x})$ as some differentiable and invertible transformation of another random variable $\boldsymbol{\epsilon}$:
\be
{\bf z}=g(\boldsymbol{\epsilon}, \phi, {\mathbf x}),
\ee
where the distribution of $\boldsymbol{\epsilon}$ is independent of ${\bf x}$ and $\phi$. Then, we can replace expectation values over $q_\phi({\bf z}|{\bf x})$ by expectation values over $p_{\boldsymbol{\epsilon}}$
\be
\mathbb{E}_{q_\phi({\bf z}|{\bf x})}[f({\bf z})]= \mathbb{E}_{p_{\boldsymbol{\epsilon}}}[f({\bf z})].
\ee
Evaluating the derivative then becomes quite straight forward since
\be
\nabla_\phi \mathbb{E}_{q_\phi({\bf z}|{\bf x})}[f({\bf z})]\sim \mathbb{E}_{p_{\boldsymbol{\epsilon}}}[\nabla_\phi f({\bf z)}].
\ee
Of course, when we do this we still need to be able to calculate the Jacobian of this change of variables 
\be
d_\phi({\bf x}, \phi)=\mathrm{Det}\bigg |{\partial {\bf z} \over \partial \boldsymbol{\epsilon}}\bigg |
\ee
since
\be
\log q_\phi({\bf z}|{\bf x})= \log p(\boldsymbol{\epsilon})-\log{d_\phi({\bf x}, \phi)}.
\ee
Since we can calculate gradients, we can now use backpropagation on the full the ELBO objective function (we return to this below when we discuss concrete architectures and implementations of VAE).

One of the problems that commonly occurs when training VAEs by performing a stochastic optimization of the ELBO (variational free energy) is that it often gets stuck in undesirable local minima, especially at the beginning of the training procedure \cite{bowman2015generating, sonderby2016ladder, kingma2017variational}. The underlying reason for this is that the ELBO objective function can be improved in two qualitatively different ways corresponding to each of the two terms in Eq.~\eqref{eq:ELBO}: by minimizing the reconstruction error or by making  the posterior distribution $q_\phi({\bf z}|{\bf x})$ to be close to $p({\bf z})$ (Of course, the goal is to do both!). For complex datasets, at the beginning of training when the reconstruction error is extremely poor, the model often quickly learns to make $q({\bf z}|{\bf x}) \approx p({\bf z})$ and gets stuck in this local minimum. For this reason, in practice it is found that it makes sense to modify the ELBO objective
to use an optimization schedule of the form
\be\label{eq:VAE-obj-practice}
 \mathbb{E}_{q_\phi({\bf z}|{\bf x})}[\log p_\theta ({\bf x}|{\bf z})] -\beta D_{KL}(q_\phi({\bf z}|{\bf x})|p({\bf z}))
 \ee
where $\beta$ is slowly annealed from $0$ to $1$  \cite{bowman2015generating, sonderby2016ladder}. An alternative regularization is the ``method of free bits'': modifying the objective function of ELBO to ensure that on average $q_\phi({\bf z}|{\bf x})$ has at least $\lambda$ natural units of information about $p({\bf z})$ (see Kingma Ph.D thesis \cite{kingma2017variational} for details) . 

These observations hints at the more general connection between VAEs and information theory that we turn to in the next section.

\subsubsection{Connection to the information bottleneck}

There is a fundamental connection between the variational autoencoder objective and the information bottleneck (IB) for lossy compression~\cite{tishby2000information}. The information bottleneck imagines we have input data $x$ that is correlated with another variable of interest, $y$, and we are given access to the joint distribution, $p(x,y)$. Our task is to take $x$ as input and compress it in such a way as to retain as much information as possible about the relevance variable, $y$. To do this, Tishby et al. propose to maximize the objective function

\begin{equation}
L_{IB}=I(y;z) - \beta I(x;z)
\end{equation}
over a stochastic encoding distribution $q(z|x)$, where $z$ is our compression of the input, and $\beta$ is a tradeoff parameter that sets the relative preference of compression and accuracy, and $I(y;z)$ is the mutual information between $y$ and $z$. Note that we choose a slightly different but equivalent form of the objective relative to Tishby et al.. This objective is only known to have a closed-form solution when $x$ and $y$ are jointly Gaussian~\cite{chechik2005information}. Otherwise, the optimization can be performed through a Blahut-Arimoto type iterative update scheme~\cite{arimoto1972algorithm, blahut1972computation}. However, this is only guaranteed to converge to a local optimum. A significant difficulty in implementing IB is that it requires knowledge of the joint distribution $p(x,y)$ and that we must be able to compute the mutual information, a notoriously difficult quantity to estimate from samples. Hence, IB has in recent years been utilized less than it might otherwise.

To address these problems, variational approximations to the IB objective function have been developed~\cite{alemi2016deep,chalk2016relevant}. These approximations, when applied to a particular choice of $p(x,y)$ give the same objective as the variational autoencoder. Here we follow the exposition from Alemi et al.\cite{alemi2016deep}. To see this, consider a dataset of $N$ points, $x_i$. We set $x=i$ and $y=x_i$ in the IB objective, similar to~\cite{slonim2005estimating,strouse2017deterministic}. We choose $p(i)=1/N$ and $p(x|i) = \delta(x-x_i)$. That is, we would like to find a compression of the data that preserves information about data point location while reducing information about data point identity.

Imagine that we are unable to directly work with the decoder $p(x|z)$. The first approximation replaces the exact decoder inside the logarithm with an approximation, $q(x|z)$. Due to the positivity of KL-divergence, namely,
\bea
D_{KL}(p(x|z)||q(x|z))\ge 0 \nonumber\\
\Rightarrow \int dx\, p(x|z)\log p(x|z) \ge \int dx\, p(x|z)\log q(x|z),\nonumber\\
\eea 
we have
\bea\label{eq:IBfirst}
I(x;z) &=& \int dx dz \,p(x) p(z|x) \log\left(\frac{p(x|z)}{p(x)}\right)\nonumber\\
 &\geq& \int dx dz \,p(x) p(z|x) \log q(x|z) + H_p(x)\nonumber \\
&\geq&\int dx dz\, p(x) p(z|x) \log q(x|z),
\eea
where $H_p(x)\ge 0$ is the Shannon entropy of $x$. This quantity can be estimated from data samples $(i,x_i)$ after drawing from $p(z|i)=p(z|x_i)$. Similarly, we can replace the prior distribution of the encoding, $p(z)=\int dx \,p(x) q(z|x)$ which is typically intractable, with a tractable $q(z)$ to get
\begin{equation}\label{eq:IBsecond}
I(i;z) \leq \frac{1}{N}\sum_i \int dz \,p(z|x_i) \log\frac{p(z|x_i)}{q(z)}
\end{equation}
Putting these two bounds Eqs.~\eqref{eq:IBfirst}and~\eqref{eq:IBsecond} together and note that $x=i$ and $y=x_i$, we get an upper bound for the IB objective that takes the same form as the VAE objective Eq.~\eqref{eq:VAE-obj-practice} we saw earlier:
\bea
L_{IB} &=& I(x;z) - \beta I(y;z) \nonumber\\ \label{eq:IBupperbound-first}
&\leq& \int dx \,p(x) \mathbb{E}_{p(z|x)}[\log q(x|z)] \\
&- &\beta \frac{1}{N}\sum_i D_{KL}(p(z|x_i)|q(z)).
\eea
Note that in Eq.~\eqref{eq:IBupperbound-first} we have a conditional distribution of $x$ given $z$ but not their joint distribution inside the expectation, which was the case in Eq.~\eqref{eq:VAE-obj-practice}. This is due to that we dropped the entropy term pertaining to $x$, which is irrelevant in the optimization procedure. In fact, this objective has been explored and is called a $\beta$-VAE~\cite{higgins2016beta}. It's interesting to note that in the case of IB, the variational approximations are with respect to the decoder and prior, whereas in the VAE, the variational approximations are with respect to the encoder.

\subsection{VAE with Gaussian latent variables and Gaussian encoder}

\begin{figure}[t]
\includegraphics[width=0.9\columnwidth]{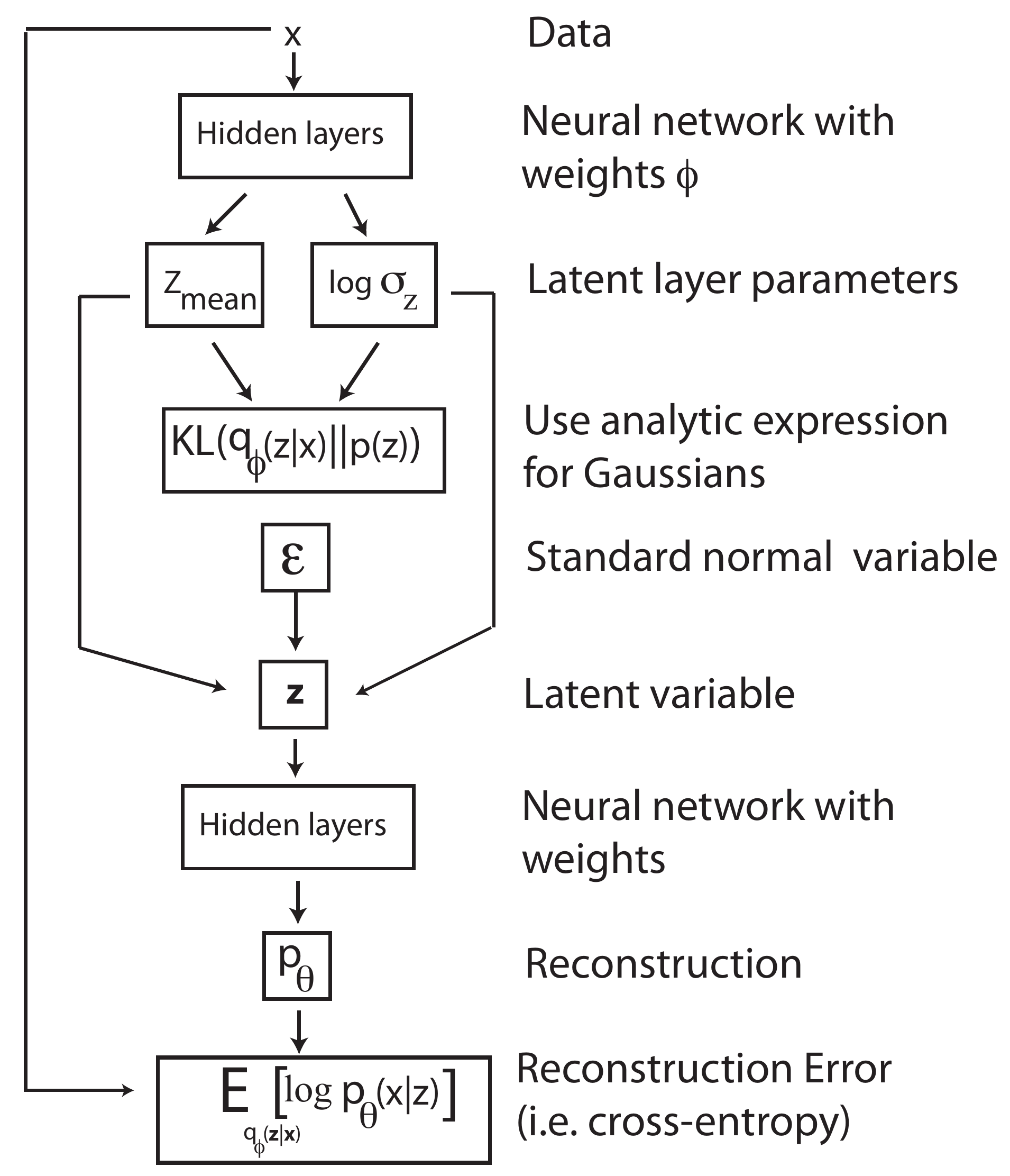}
\caption{Computational graph for a VAE with Gaussian hidden units (i.e. $p({\bf z})$ are standard normal variables $\mathcal{N}(0,1)$ and 
Gaussian variational encoder whose posterior takes the form  $q_\phi({\bf z}|{\bf x}) = \mathcal{N}(\boldsymbol{\mu}({\bf x}),\boldsymbol{\sigma}^2({\bf x}) )$.}
\label{fig:GaussianVAE} 
 \end{figure}

Our discussion of VAEs thus far has been quite abstract. In this section, we discuss one of the most widely employed VAE architectures: a VAE with factorized Gaussian posteriors, $q_\phi({\bf z}|{\bf x})= \mathcal{N}({\bf z}, \boldsymbol{\mu}({\bf x}), \mathrm{diag}(\boldsymbol{\sigma}^2({\bf x})))$ and standard normal latent variables $p({\bf z})=\mathcal{N}(0, {\bf I})$. The training and implementation simplifies greatly here because we can analytically workout the term $D_{KL}(q_\phi({\bf z}|{\bf x})|p({\bf z}))$.

\subsubsection{Implementing the Gaussian VAE}

We now show  how we can combine analytic expressions for the KL-divergence with backpropagation to efficiently implement a Gaussian VAE. We start by first deriving analytic expressions for $D_{KL}(q_\phi({\bf z}|{\bf x})|p({\bf z}))$ in terms of the means $\boldsymbol{\mu}({\bf x})$ and variances $\boldsymbol{\sigma}^2({\bf x})$. This is just a simple exercise in Gaussian integrals. For notational convenience, we drop the ${\bf x}$-dependence of
the means $\boldsymbol{\mu}({\bf x})$, variances $\boldsymbol{\sigma}^2({\bf x})$, and $q_\phi({\bf x})$. A straight-forward calculation gives
\bea
\int d{\bf z} q_{\phi}({\bf z})\log p({\bf z}) &=& \int \mathcal{N}({\bf z}, \boldsymbol{\mu}({\bf x}), \mathrm{diag}(\boldsymbol{\sigma}^2({\bf x}))) 
\log{\mathcal{N}(0, {\bf I})}\nonumber \\
&=& -{J \over 2} \log{2\pi} -{1 \over 2} \sum_{j=1}^J (\mu_j^2+\log{\sigma_j^2}),
\eea
where $J$ is the dimension of the latent space. An almost identical calculation yields
\be
\int d{\bf z} q_{\phi}({\bf z})\log q_\phi({\bf z})=-{J \over 2} \log{2\pi} -{1 \over 2} \sum_{j=1}^J (1+\sigma_j^2).
\ee
Combining these equations gives
\begin{align}
-D_{KL}(q_\phi({\bf z}|{\bf x})|p({\bf z}))={1 \over 2} \sum_{j=1}^J \left (1+\log{\sigma_j^2({\bf x})}-\mu_j^2({\bf x}) -\sigma_j^2({\bf x})\right).
\end{align}
This analytic expression allows us to implement the Gaussian VAE in a straight forward way using neural networks. The computational graph for this implementation is  shown in Fig.~\ref{fig:GaussianVAE}. Notice that since the parameters are all compositions of differentiable functions, we can use standard backpropagation algorithms to train VAEs.

\subsubsection{VAEs for the MNIST dataset}

\begin{figure}[t]
\includegraphics[width=1.0\columnwidth]{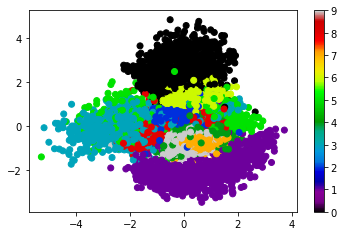}
\label{fig:VAE-MNIST-latent}
\caption{Embedding of MNIST dataset into a two-dimensional latent space using a VAE with two latent dimensions (see \href{https://physics.bu.edu/~pankajm/MLnotebooks.html}{Notebook 19} and main text for details.) Data points are colored by their identity [0-9].} 
\end{figure}

In  \href{https://physics.bu.edu/~pankajm/MLnotebooks.html}{Notebook 19}, we have implemented a VAE using  Keras and trained it using the MNIST dataset. The basic architecture is the one describe above. All figures were generated with a VAE that has a latent space of dimension $2$. The architecture of both the encoder and decoder is a Multi-layer Perceptron (MLPs) -- neural networks with a single hidden layer. For this example, we take the dimension of the hidden layer for both neural networks to be $256$. We trained the VAE  using the RMS-prop optimizer for 50 epochs.

We can visualize the embedding in the latent space by plotting ${\mathbf z}$ of the test set and coloring the points by digit identity [0-9] (see Figure \ref{fig:VAE-MNIST-latent}). Notice that in general, digits that are similar end up being closer to each other in the latent space. However, this is not always the case (see bright green points for example). This is a general feature of these low-dimensional embeddings and we saw a similar phenomenon when we examined t-SNE in Section \ref{sec:dim-red}.

\begin{figure}[t!]
\includegraphics[width=1.0\columnwidth]{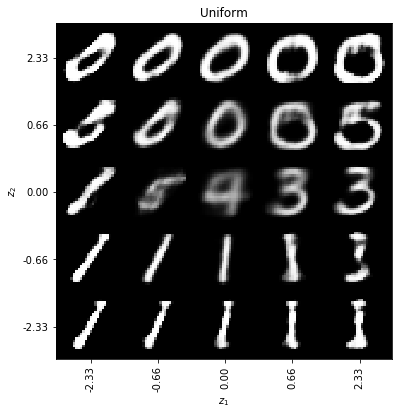}
\includegraphics[width=1.0\columnwidth]{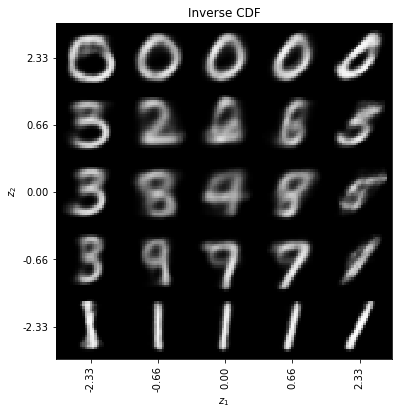}
\label{fig:VAE-MNIST-fantasy}
\caption{(Top) Fantasy particle generated by uniform sampling of the latent space ${\bf z}$. (Bottom) Fantasy particles generated by uniform sampling of
probability $p({\bf z})$ mapped to latent space using the inverse Cumulative Distribution Function (CDF) of the Gaussian. }
\end{figure}

The real advantage that VAEs offer over embeddings such as t-SNE is that they are generative models. Given a set of examples, we can generate new examples -- or fantasy particles as they are commonly called in ML -- by sampling the latent space ${\bf z}$ and then using the decoder to map these latent variables to new examples. The results of this procedure are shown in Figure \ref{fig:VAE-MNIST-fantasy}. In the top figure, we sample the latent space uniformly in a $5 \times 5$ grid. Notice that this results in extremely similar examples through much of the latent space. The underlying reason for this is that uniform sampling does not respect the underlying Gausssian structure of the latent space ${\bf z}$. In the bottom figure, we perform a uniform sampling on the probability $p({\bf z})$ and mapped this back to the latent space using the inverse Cumulative Distribution Function (CDF) of the Gaussian. We see that the diversity of the generated examples is much higher for this sampling procedure. 

This example is indicative of a more general problem: once we have learned a generative model how should we sample latent spaces \cite{white2016sampling}. This is especially important in high-dimensional spaces where direct visualization is not possible. Often certain directions in the latent space can have different meanings. A particularly striking visual illustration is the ``smile vector''  that interpolates between smiling and frowning faces \cite{white2016sampling}. 

\subsubsection{VAEs for the 2D Ising model}

\begin{figure}[t!]
\includegraphics[width=1.0\columnwidth]{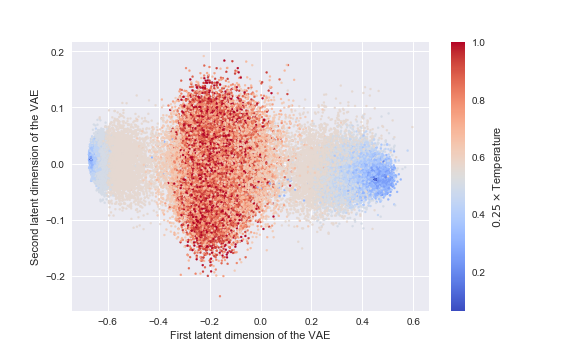}
\includegraphics[width=1.0\columnwidth]{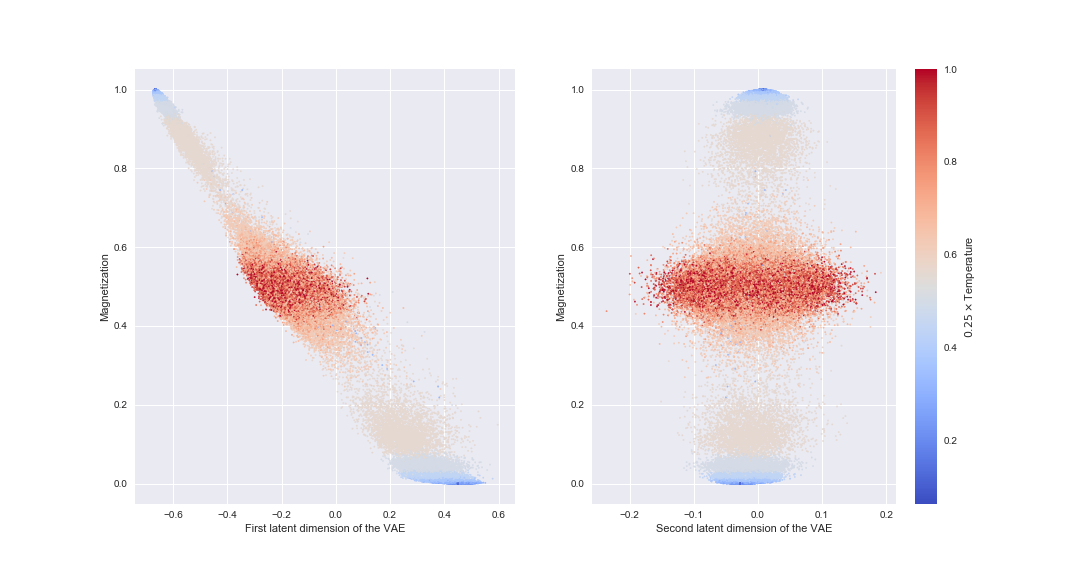}
\label{fig:VAE-Ising-Latent}
\caption{(Top)  Embedding of the Ising dataset into a two-dimensional latent space using a VAE with two latent dimensions (see \href{https://physics.bu.edu/~pankajm/MLnotebooks.html}{Notebook 20} and main text for details.) Data points are colored by the temperature each sample was drawn at. (Bottom) Correlation between the  latent dimensions
and the magnetization for each sample. Notice that the first principle component corresponds to the magnetization.  }
\end{figure}

In \href{https://physics.bu.edu/~pankajm/MLnotebooks.html}{Notebook 20}, we used an almost identical architecture (though coded in a slightly different way)  to train a VAE on the Ising dataset discussed through out the review. The only differences between the two VAEs are that the visible layer of the Ising VAE now has 1600 units (our samples are $40 \times 40$ instead of the $28 \times 28$ MNIST images) and we have changed the standard deviation of the Gaussian of the latent variables $p({\bf z})$ from $\sigma=1$ to $\sigma=0.2$. 

We once again visualize the embedding learned by the VAE by plotting ${\mathbf z}$ and coloring the points by the temperature at which the sample was drawn (see Figure~\ref{fig:VAE-Ising-Latent} top). Notice that the latent space has learned a lot of the physics of the Ising model. For example, the first VAE dimension is just the magnetization (Fig.~\ref{fig:VAE-Ising-Latent} bottom). This is not surprising since we saw in Section \ref{sec:dim-red} that the first principal component of a PCA also corresponded to the magnetization. 

\begin{figure}[t!]
\includegraphics[width=0.9\columnwidth]{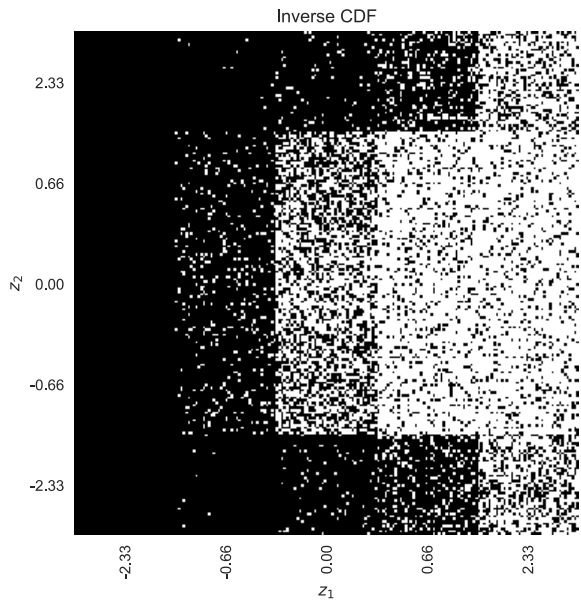}
\caption{Fantasy particles for the Ising model generated by uniform sampling of
probability $p({\bf z})$ mapped to latent space using the inverse Cumulative Distribution Function (CDF) of the Gaussian. }
\label{fig:VAE-Ising-fantasy}
 \end{figure}

 We now ask how well the VAE can generate new examples (see Fig.~\ref{fig:VAE-Ising-fantasy}). We see that the examples look quite different from real Ising configurations -- they lack the large scale patchiness seen in the critical region. They mostly turn out to be unstructured speckles that reflect only the average probability that a pixel is on in a region. This is not surprising since our VAE has no spatial structure, has only two latent dimensions, and the cost function does not know about ``correlations between spins'' : there is very little information about correlations in the binary cross-entropy which we use to measure reconstruction errors. The reader is encouraged to play with the corresponding notebook and generate examples as we change the latent dimension and/or choose modified architectures such as decoders based on CNNs instead of MLPs.
 
This example also shows how much easier it is to discriminate between labeled data than it is to learn how to generate new examples from an unlabeled dataset. This is true in all spheres of machine learning.  This is also one of the reasons that generative models are one of the cutting edge areas of modern Machine Learning research and there are likely to be a barrage of new techniques for generative modeling in the next few years.

%% file: sections/outlook.tex
In this review, we have attempted to give the reader the intellectual and practical tools to engage with  Machine Learning (ML), data science, and 
parts of modern statistics. We have tried to emphasize that ML differs from ordinary statistics in that the goal is to predict rather than to fit. For this
reason, all the techniques discussed here have to navigate important tensions that lie at the heart of ML. The most prominent instantiation of these inherent tradeoffs is the bias-variance tradeoff, which is perhaps the only universal principle in ML. Identifying how these tradeoffs manifest in a particular algorithm is the key to constructing and training powerful ML methods.

The immense progress in computing power and the corresponding availability of large datasets ensure that ML will be an important part of the physics toolkit. In the future, we expect ML to be a core competency of physicists much like linear algebra, group theory, and differential equations. We hope that this review will play some small part toward this aspirational goal. 

We wrote this review to provide a relatively concise introduction to ML using ideas and language familiar to physicists (though the review ended up being almost twice the planned length). In writing the review, we have tried to accomplish two somewhat disparate tasks. First, we have tried to highlight more abstract and theoretical considerations to show the unity of ML and statistical learning. Many ML techniques can be understood by starting with some key concepts from statistical learning (MLE, bias-variance tradeoff, regularization) and combining them with core concepts familiar from statistical physics (Monte-Carlo, gradient descent, variational methods and MFT). Despite the high-level similarities between all the methods presented here, the way that these concepts manifest in any given technique is often quite clever and understanding these ``hacks'' is the key to understanding why some ML techniques turn out to be so powerful and others not so much.  ML, in this sense, is as much an art as a science. Second, we have tried to give the reader the practical know-how to start using the tools and concepts from ML for immediately solving problems. We believe the accompanying \href{https://physics.bu.edu/~pankajm/MLnotebooks.html}{python notebooks} and the emphasis on coding in python have accomplished this task.

\subsection{Research at the intersection of physics and ML}

We hope the review catalyzes more research at the intersection of physics and machine learning. Here we briefly highlight a few promising research directions. We note that this list is far from comprehensive.
\begin{itemize}

\item{ {\bf Applying ML to solve physics problems}. One theme that has reoccurred through out the 
review is that ML is most effective in settings  with well defined objectives and lots of data.  For this reason, we expect ML to become a core competency of data rich fields such as high-energy experiments and astronomy. However, ML may also prove to be useful for helping further our physical understanding through data-driven approach to other branches of physics that may not be immediately obvious, such as quantum physics~\cite{dunjko2017machine}. For example, recent works have used ideas from ML to investigate disparate topics such as non-local correlations~\cite{canabarro2018machine}, disordered materials and glasses~\cite{schoenholz2017combining}, electronic structure calculations~\cite{grisafi2017symmetry} and numerical analysis of ferromagnetic resonances in thin films~\cite{tomczak2018ferromagnetic}, designing and analyzing quantum materials by integrating ML with existing techniques such as Dynamical Mean Field Theory (DMFT) \cite{arsenault2014machine}, in the study of inflation~\cite{rudelius2018learning}, and even for experimental learning of quantum states by using ML to aid in quantum tomography~\cite{rocchetto2017experimental}. For a comprehensive review of ML methods in seismology, see \cite{qingkai2018machine}. }

\item{{\bf Machine Learning on quantum computers}. Another interesting area of research that is likely to grow is asking if and how quantum computers can help improve state-of-the art ML algorithms~\cite{schuld2015introduction,arunachalam2017survey,ciliberto2017quantum,schuld2017implementing,benedetti2016quantum,benedetti2017quantum,daskin2018quantum,rebentrost2017_quantum,mitarai2018quantum,perdomo2017opportunities,bromley2018batched,innocenti2018supervised,schuld2018quantum}. Concrete examples that seek to extend some of the basic ideas and methods we introduced in this review to the quantum computing realm include: algorithms for quantum-assisted gradient descent~\cite{kerenidis2017quantum,rebentrost2016quantum}, classification~\cite{schuld2017quantum_ensembles}, and Ridge regression~\cite{yu2017quantum}. Interest in this field will undoubtedly grow once reliable quantum computers become available (see also this recent review ~\cite{dunjko2017machine} ).}

\item{{\bf Monte-Carlo Methods}. An interesting area that has seen a renewed interest with Bayesian modeling is the development of new Monte-Carlo methods for sampling complex probability distributions. Some
of the workhorses of modern Machine Learning -- Annealed Importance Sampling (AIS) \cite{neal2001annealed} and Hamiltonian or Hybrid Monte-Carlo (HMC) \cite{neal2011mcmc} -- are intimately related to physics. As pointed out by Radford Neal, AIS is just the Jarzynski inequality \cite{jarzynski1997nonequilibrium} as a Monte-Carlo method and HMC was developed by physicists and exploits Hamiltonian dynamics to improve proposal distributions.}

\item{{\bf Statistical physics style theory of Deep Learning}. Many techniques in ML have origins in statistical physics. Yet, a physics-style theory of Deep Learning remains elusive. A key question is to ask when and why these models manage to generalize well. Physicists are only beginning to ask these questions \cite{advani2017high,mehta2014exact, saxe2013exact, shwartz2017opening}. But right now, it is fair to say that the insights remain scattered and a cohesive theoretical understanding is lacking.}

\item{{\bf Biological physics and ML}. Biological physics is generating ever more datasets in fields ranging from neuroscience to evolution and immunology. It is likely that ML will be an important part of the biophysics toolkit in the future. Many of the authors
of this review were inspired to engage with ML for this reason.}

\item{{\bf Using ideas from physics to develop new ML algorithms}. Many of the core ideas of ML from Monte-Carlo techniques to variational methods have their origin in physics. There has been a tremendous amount of recent work developing tools to understand physical systems that may be of potential use to ML. For example, in quantum condensed matter techniques such as DMRG, MERA, etc. have enriched both our practical and conceptual understandings \cite{vidal2007entanglement,white1992density, stoudenmire2012studying}. It will be interesting to figure how and if these numerical methods can be translated from a physics to a ML setting. There are tantalizing hints that this is likely to be a fruitful direction \cite{stoudenmire2016supervised, stoudenmire2018learning, han2017unsupervised}.}
\end{itemize}

\subsection{Topics not covered in review}
Despite the considerable length of the review, we have had to make many omissions for the sake of brevity.  It is our hope and belief that after reading this review the reader will have the conceptual and practical knowledge to quickly learn about these other topics. Among the most prominent topics missing from this review are:
\begin{itemize}
\item{ {\bf Temporal/Sequential Data}. We have not covered techniques for dealing with temporal or sequential data. Here, too there are many connections with statistical physics. A powerful class of models for sequential data called Hidden Markov Models \cite{rabiner1989tutorial} that utilize dynamical programming techniques have natural statistical physics interpretations in terms of transfer matrices  (see \cite{mehta2011statistical} for explicit example of this). Recently, Recurrent Neural Networks (RNNs) have become an important and powerful tool for dealing with sequence data \cite{Goodfellow-et-al-2016}. RNNs generalize many of the ideas discussed in the DNN section to deal with temporal data.}

\item{{\bf Reinforcement Learning}. Many of the most exciting developments in the last five years have come from combining ideas from reinforcement learning with deep neural networks \cite{sutton1998reinforcement,mnih2015human}. RL traces its origins to behaviourist psychology, when it was conceived as a way to explain and study reward-based decision making. RL was put on solid mathematical grounds in the 50's by Richard Bellman and collaborators, and has by now become an inseparable part of robotics and artificial intelligence. RL is a field of Machine Learning, in which an agent learns how to master performing a specific task through an interaction with its environment. Depending on the reward it receives, the agent chooses to take an action affecting the environment, which in turn determines the value of the next received reward, and so on. The long-term goal of the agent is to maximise the cumulative expected return, thus improving its performance in the longer run. Shadowed by more traditional optimal control algorithms, Reinforcement Learning has only recently taken off in physics~\cite{chen_14_ML,reddy2016infomax,lamata2017basic,cardenas2017generalized,dunjko2017super,ramezanpour2017_opt,neukart2017quantum,sriarunothai2017speeding,melnikov2017active,bukov_17RL,august_18,foesel_18,zhang2018automatic,niu2018universal,albarran2018measurement,bukov2018reinforcement,sweke2018reinforcement,chen2019manipulation}. Of particular interest are biophysics inspired works that seek to use RL to understand  navigation and sensing in turbulent environments \cite{masson2009chasing, reddy2016learning, colabrese2017flow,vergassola2007infotaxis}. }

\item{{\bf Support Vector Machines (SVMs) and Kernel Methods}. SVMs and kernel methods are a powerful set of techniques that work well when the amount of training data is limited \cite{burges1998tutorial}. The mathematics and theory of SVM are very different from statistical physics and for this reason we chose not to include them here. However, SVMs and kernel methods have played an extremely important role in ML and are worth understanding.}
\end{itemize}

\subsection{Rebranding Machine Learning as ``Artificial Intelligence''}

The immense scientific progress in ML has also been accompanied by a massive public relations effort centered around Silicon Valley. Starting with the success of ImageNet (the most prominent early use of GPUs for training large models) and the widespread adoption of Deep Learning based techniques by the Silicon Valley companies, there has been a deliberate move to rebrand modern ML as ``artificial intelligence'' or AI (see graphs in \cite{katz2017manufacturing}). Recently, computer scientist Michael I. Jordan (who is famously known for his formalization of variational inference, Bayesian network, and expectation-maximization algorithm in machine learning research) cautioned that ``\emph{This confluence of ideas and technology trends has been rebranded as ``AI'' over the past few years. This rebranding is worthy of some scrutiny}''\cite{jordan2018artificial}.

AI, by design, is an ambiguous term that mixes aspirations with reality. It also conflates the statistical ideas that form the basis of modern ML with the more commonplace notions about what humans and behavioral scientists mean by intelligence (see \cite{lake2017building} for an enlightening and important modern discussion of this distinction from a quantitative cognitive science point of view as well as \cite{dreyfus1965alchemy} for a surprisingly relevant philosophy-based critique from 1965). 
 
Almost all the techniques discussed here rely on optimizing a pre-specified objective function on a given dataset. Yet, we know that for large, complex models changing the data distribution or the goal can lead to an immediate degradation of performance. Deep networks have poor generalizations to even a slightly different context (the infamous Validation-Test set mismatch). This inability to abstract and generalize is a common criticism lobbied against branding modern ML techniques as AI \cite{lake2017building}. For all these reasons, we have chosen to use the term Machine Learning rather than artificial intelligence through out the review.
 
This is far from the first time we have seen the use of the term artificial intelligence and the grandiose promises that it implies. In fact, the early 1950's and 1960's as well as the early 1980's saw similar AI bubbles (see this interesting summary by Luke Muehlhauser for Open Philanthropy \cite{Muehlhauser2016What}).  These AI bubbles have been followed by what have been dubbed ``AI Winters'' \cite{mcdermott1985dark}.

The ``Singularity'' may not be coming but the advances in computing and the availability of large data sets likely ensure that the kind of statistical learning frameworks discussed are here to stay. Rather than a general artificial intelligence, the kind of techniques presented here seem to be best suited for three important tasks: (a) automating prediction from lots of labeled examples in a narrowly-defined setting (b) learning how to parameterize and capture the correlations of complex probability distributions, and (c) finding policies for tasks with well-defined goals and clear rules. We hope that this review has given the reader enough conceptual tools to start forming their own opinions about reality and hype when it comes to modern ML research. As Michael I. Joran puts it, ``\emph{...if the acronym ``AI'' continues to be used as placeholder nomenclature going forward, let’s be aware of the very real limitations of this placeholder. Let’s broaden our scope, tone down the hype and recognize the serious challenges ahead}"\cite{jordan2018artificial}.

\subsection{Social Implications of Machine Learning}

The last decade has also seen a systematic increase in the use and deployment of Machine Learning techniques into new areas of life and society. Some of the readers of this review may currently be (or eventually be) employed in industrial settings that seek to harness ML for practical purposes. However, caution is in order when applying ML. Without foresight and accountability, the scale and scope of modern ML algorithms can lead to large scale unaccountable and undemocratic outcomes that can reinforce or even worsen existing inequality and inequities. Mathematician and data scientist turned social commentator Cathy O'Neil has dubbed the indiscriminate use of these Big Data techniques  ``Weapons of Math Destruction'' \cite{o2017weapons}.

When ML is used in a social context, abstract statistical relationships have real social consequences. False positives can mean the difference between life and death (for example in the context of ``signature drone strikes'') \cite{mehta2015big}. ML algorithms, like all techniques, have important limitations and should be employed with great caution. It is our hope that ML practitioners keep this in mind when working in social settings.

All algorithms involve inherent tradeoffs in fairness, a point formalized by computer scientist Jon Kleinberg and collaborators in a very interesting recent paper \cite{kleinberg2016inherent}. It is far from clear how to make algorithms fair for all people involved. This is even more true with methods like Deep Learning that are hard to interpret. All ML algorithms have implicit assumptions and choices reflected in the datasets we use to the kind of functions we choose to optimize. It is important to remember that there is no `` view from nowhere''  \cite{katz2017manufacturing,adam2006artificial} -- all ML algorithms reflect a point of view and a set of assumptions about the world we live in. For this reason, we hope that ML practitioners and data scientists will take the time to consider the social consequences of their actions. For example, developing a Hippocratic Oath for data scientists is now being considered \cite{simonite2018should}. Doing no harm seems like a good start for making sure that we harness ML for the benefit of all members of society.

%% file: sections/dataset.tex
\subsection{Ising dataset}
The Ising dataset we use throughout the review was generated using the standard Metropolis algorithm to generate a Markov Chain. The full dataset consist of $16\times10000$ samples of $40\times40$ spin configurations (i.e. the design matrix has 160000 samples and 1600 features) drawn at temperatures $0.25, 0.5,\cdots 4.0$. The samples are drawn for the Boltzmann distribution of the two-dimensional ferromagnetic Ising model on a $40\times40$ square lattice with periodic boundary conditions.

\subsection{SUSY dataset}

The SUSY dataset was generated by Baldi et al \cite{baldi2014searching} to explore the efficacy of using Deep Learning for classifying collision events. The dataset is downloadable from the \href{https://archive.ics.uci.edu/ml/datasets/SUSY}{UCI Machine Learning Repository}, a wonderful resource for interesting datasets. Here we quote directly from the paper:

\begin{quote}
The data has been produced using Monte Carlo simulations and contains events with two leptons (electrons or muons). In high energy physics experiments, such as the ATLAS and CMS detectors at the CERN LHC, one major hope is the discovery of new particles. To accomplish this task, physicists attempt to sift through data events and classify them as either a signal of some new physics process or particle, or instead a background event from understood Standard Model processes. Unfortunately we will never know for sure what underlying physical process happened (the only information to which we have access are the final state particles). However, we can attempt to define parts of phase space that will have a high percentage of signal events. Typically this is done by using a series of simple requirements on the kinematic quantities of the final state particles, for example having one or more leptons with large amounts of momentum that is transverse to the beam line ( pT ). Here instead we will use logistic regression in order to attempt to find out the relative probability that an event is from a signal or a background event and rather than using the kinematic quantities of final state particles directly we will use the output of our logistic regression to define a part of phase space that is enriched in signal events. The dataset we are using has the value of 18 kinematic variables ("features") of the event. The first 8 features are direct measurements of final state particles, in this case the  pT , pseudo-rapidity, and azimuthal angle of two leptons in the event and the amount of missing transverse momentum (MET) together with its azimuthal angle. The last ten features are functions of the first 8 features; these are high-level features derived by physicists to help discriminate between the two classes. You can think of them as physicists attempt to use non-linear functions to classify signal and background events and they have been developed with a lot of deep thinking on the part of physicist. There is however, an interest in using deep learning methods to obviate the need for physicists to manually develop such features. Benchmark results using Bayesian Decision Trees from a standard physics package and 5-layer neural networks and the dropout algorithm are presented in the original paper to compare the ability of deep-learning to bypass the need of using such high level features. We will also explore this topic in later notebooks. The dataset consists of 5 million events, the first 4,500,000 of which we will use for training the model and the last 500,000 examples will be used as a test set.
\end{quote}

\subsection{MNIST Dataset}

The MNIST dataset is one of the simplest and most widely used Machine Learning Datasets. The MNIST dataset consists of hand-written images of numerical characters $0-9$ and consists of a training set of 60,000 examples, and a test set of 10,000 examples \cite{lecun1998gradient}. Information about the MNIST database and its historical importance can be found at Yann Lecun's wedsite: \url{http://yann.lecun.com/exdb/mnist/}. A brief description from the website:
\begin{quote}
The original black and white (bilevel) images from NIST were size normalized to fit in a 20x20 pixel box while preserving their aspect ratio. The resulting images contain grey levels as a result of the anti-aliasing technique used by the normalization algorithm. the images were centered in a 28x28 image by computing the center of mass of the pixels, and translating the image so as to position this point at the center of the 28x28 field.
\end{quote}
The MNIST is often included by default in many modern ML packages.